\documentclass[11pt]{article}
\usepackage{graphicx}
\usepackage{color}
\usepackage{titlesec}
\usepackage{amssymb}
\definecolor{hbl}{rgb}{0.153,0.498,0.835} 
\definecolor{dbl}{rgb}{0,.15,0.46}              

\titleformat{\chapter}[display]
{\color{hbl}\bfseries\boldmath\huge}{\chaptertitlename\
\thechapter}{20pt}{\Huge}
\titleformat{\section}
{\color{hbl}\bfseries\boldmath\Large}{\thesection}{1em}{}

\newcommand{\muJy}{$\mu$Jy}
\newcommand{\mujy}{$\mu$Jy}

\newcommand{\sqdeg}{$\Box^{\circ}$}
\newcommand{\ndeg}{$^{\circ}$}
\newcommand{\lum}{$\mathrm{L}_\mathrm{1.4GHz}$}

\newcommand{\wh}{W/Hz}

\newcommand{\hi}{H\mbox{\,\sc i}}

\newcommand{\msun}{\mbox{${\rm M}_\odot$}}

\newcommand{\kms}{\mbox{$\rm km\, s^{-1}$}}

\def\sun{\hbox{$\odot$}}
\def\degree{\nobreak\ifmmode{^\circ}\else{$^\circ$}\fi}

\def\grtsim{\mathrel{\hbox{\rlap{\hbox{\lower2pt\hbox{$\sim$}}}\raise2pt\hbox{$>$}}}}
\def\lesssim{\mathrel{\hbox{\rlap{\hbox{\lower2pt\hbox{$\sim$}}}\raise2pt\hbox{$<$}}}}
\newcommand{\ie}{{\it i.e.}}

\newcommand{\etal}{{et~al.}}
\newcommand{\rd}{{R\&D}}
\newcommand{\skai}{{SKA$_1$}}
\newcommand{\skaii}{{SKA$_2$}}
\newcommand{\noi}{\noindent}

\newcommand{\hsp}{\vspace{1cm}}

\newcommand{\mnras}{MNRAS}
\newcommand{\apj}{ApJ}
\newcommand{\apjs}{ApJS}
\newcommand{\apjl}{ApJL}
\newcommand{\aj}{AJ}
\newcommand{\aap}{A\&A}

\newcommand{\zaa}{Zentrum f\"ur Astronomie and Astrophysik, Technische Universit\"at Berlin}
\newcommand{\airub}{Astronomisches Institut der Ruhr-Universit\"at Bochum}
\newcommand{\aifa}{Argelander-Institut f\"ur Astronomie, Universit\"at
  Bonn}
\newcommand{\mpifr}{Max-Planck-Institut f\"ur Radioastronomie Bonn}

\newcommand{\ubie}{Fakult\"at f\"ur Physik,  Universit\"at Bielefeld}
\newcommand{\jbremen}{Jacobs University Bremen}
\newcommand{\zarm}{ZARM, Universit\"at Bremen}
\newcommand{\fau}{Physikalisches Institut der Friedrich-Alexander Universit\"at Erlangen-N\"urnberg}
\newcommand{\tguf}{Institut f\"ur Theoretische Physik, Goethe-Universit\"at Frankfurt}
\newcommand{\mpa}{Max-Planck-Institut f\"ur Astrophysik Garching}
\newcommand{\mpe}{Max-Planck-Institut f\"ur extraterrestrische Physik
  Garching}
\newcommand{\eso}{European Southern Observatory Garching}
\newcommand{\ugoett}{Institut f\"ur Astrophysik der Georg-August-Universit\"at G\"ottingen}
\newcommand{\hhstw}{Hamburger Sternwarte, Fachbereich Physik,
  Universit\"at Hamburg}
\newcommand{\mpiaei}{Max-Planck-Institut f\"ur Gravitationsphysik
  (Albert-Einstein-Institut) Hannover}
\newcommand{\mpia}{Max-Planck-Institut f\"ur Astronomie Heidelberg}
\newcommand{\ita}{Institut f\"ur Theoretische Astrophysik, Universit\"at Heidelberg}
\newcommand{\mpik}{Max-Planck-Institut f\"ur Kernphysik Heidelberg}
\newcommand{\tpi}{Theoretisch-Physikalisches Institut, Friedrich-Schiller-Universit\"at Jena}
\newcommand{\juelich}{Forschungszentrum J\"ulich}
\newcommand{\ukoeln}{I. Physikalisches Institut der Universit\"at zu K\"oln}
\newcommand{\mps}{Max-Planck-Institut f\"ur Sonnensystemforschung Katlenburg-Lindau}
\newcommand{\lmum}{Universit\"atssternwarte der Ludwig-Maximilians-Universit\"at M\"unchen}
\newcommand{\oldenburg}{Institut f\"ur Mathematik und Naturwissenschaften,
  Carl von Ossietzky Universit\"at Oldenburg}
\newcommand{\aip}{Leibniz-Institut f\"ur Astrophysik Potsdam}

\newcommand{\mpiaeip}{Max-Planck-Institut f\"ur Gravitationsphysik
  (Albert-Einstein-Institut) Potsdam}
\newcommand{\tat}{Theoretische Astrophysik, Eberhard Karls Universit\"at T\"ubingen}

\newcommand{\stwtt}{Th\"uringer Landessternwarte Tautenburg}

\newcommand{\fhgise}{Fraunhofer-Institut f\"ur Solare Energiesysteme ISE Freiburg}

\begin{document}
%
%
%
\sffamily
\vspace{5cm}
\pagecolor{white}
\color{black}
\pagestyle{empty}

\parbox{\textwidth}{
{\bfseries\boldmath \Huge \color{hbl} Pathway to the Square Kilometre Array \vspace{0.5cm}}\\
\noi {\LARGE \color{hbl} The German White Paper}\\
}

\vspace{-1.5cm}
\parbox{\textwidth}{\hspace{-2.1cm}
\parbox{\textwidth}{\includegraphics[scale=0.24,angle=-55]{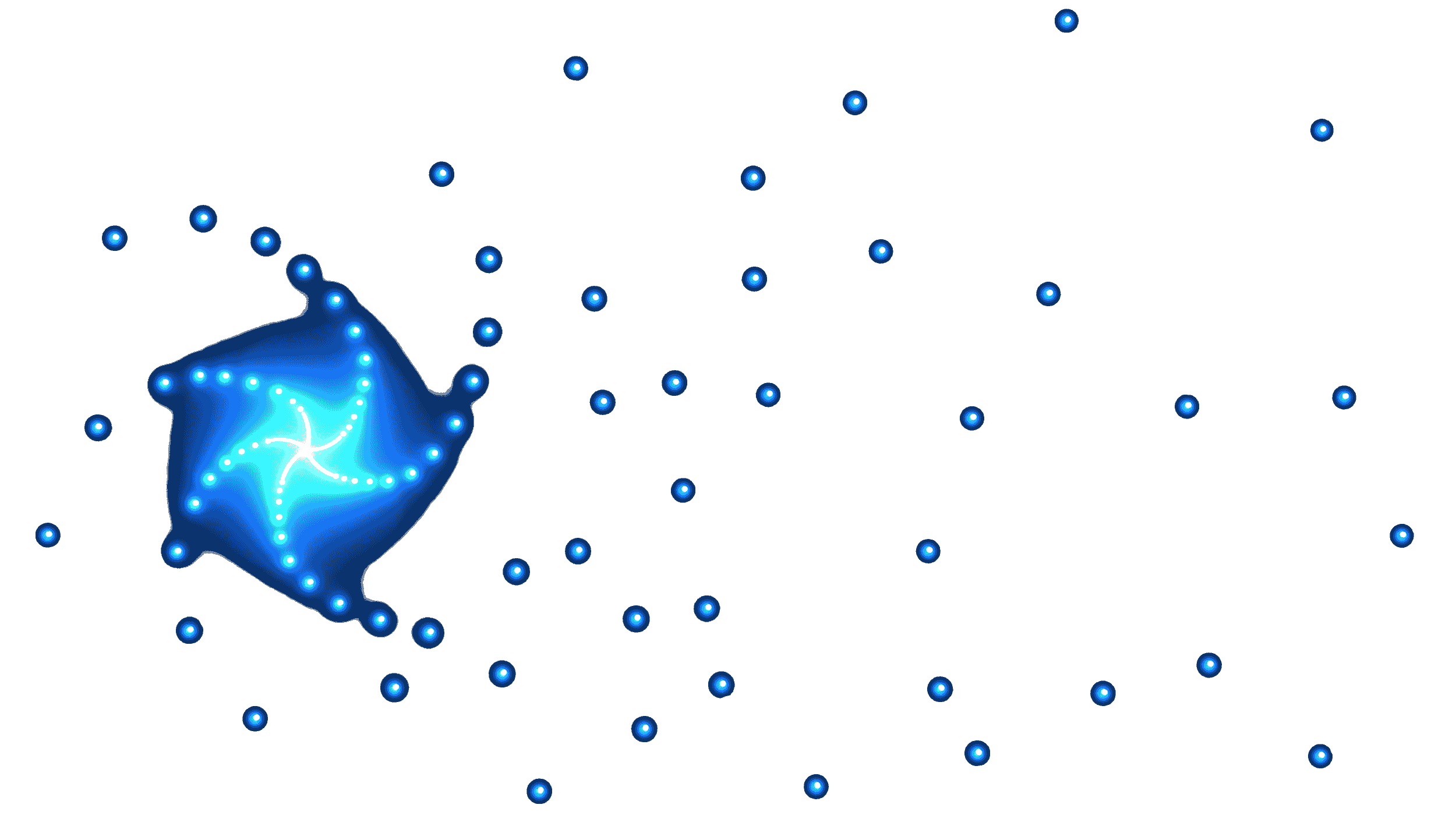}}
\hspace{-4.5cm}
\vspace{-13cm}
}

\parbox{\textwidth}{\hspace{10cm}
\parbox{\textwidth}{\includegraphics[scale=0.22,angle=0]{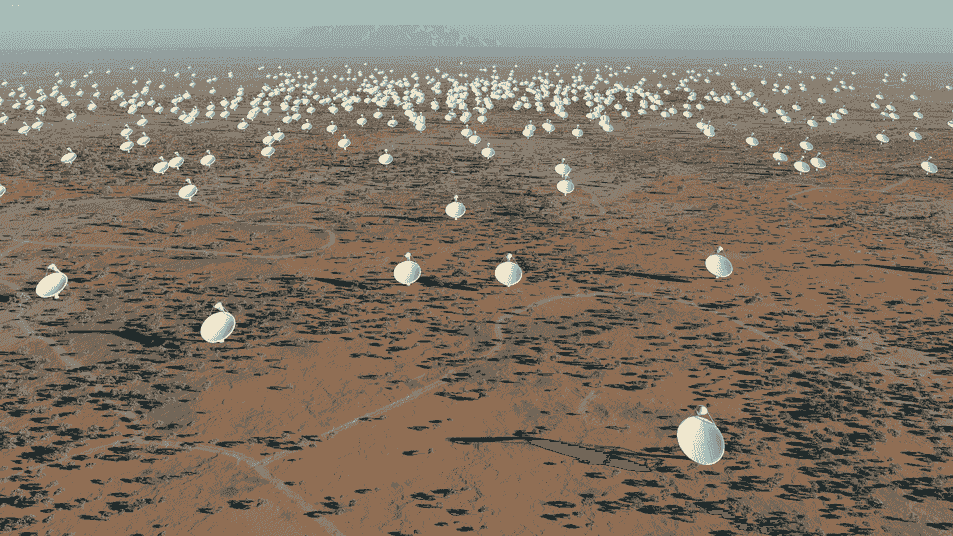}}
\vspace{-1cm}
}

\parbox{\textwidth}{\hspace{8cm}
\parbox{\textwidth}{\includegraphics[scale=0.22,angle=0]{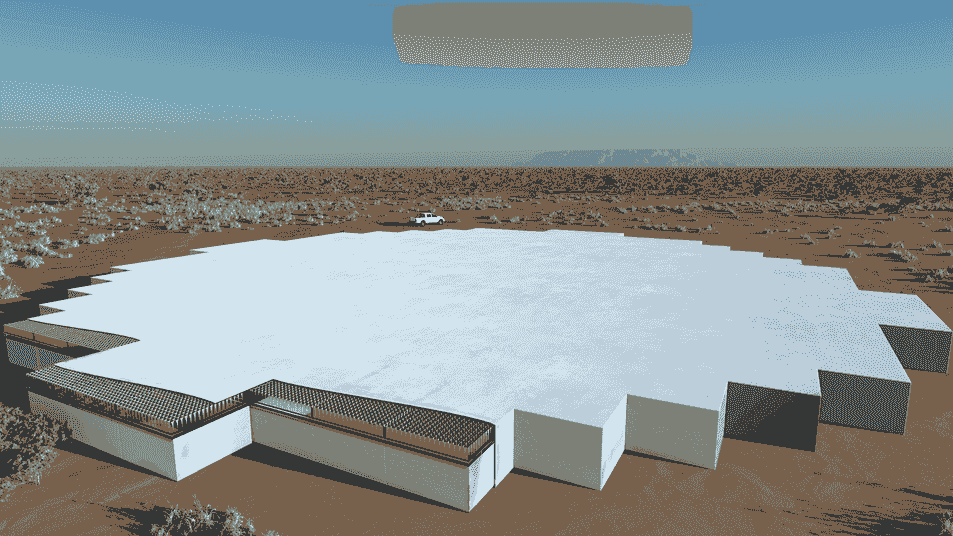}}
\vspace{-1cm}
}

\parbox{\textwidth}{\hspace{6cm}
\parbox{\textwidth}{\includegraphics[scale=0.22,angle=0]{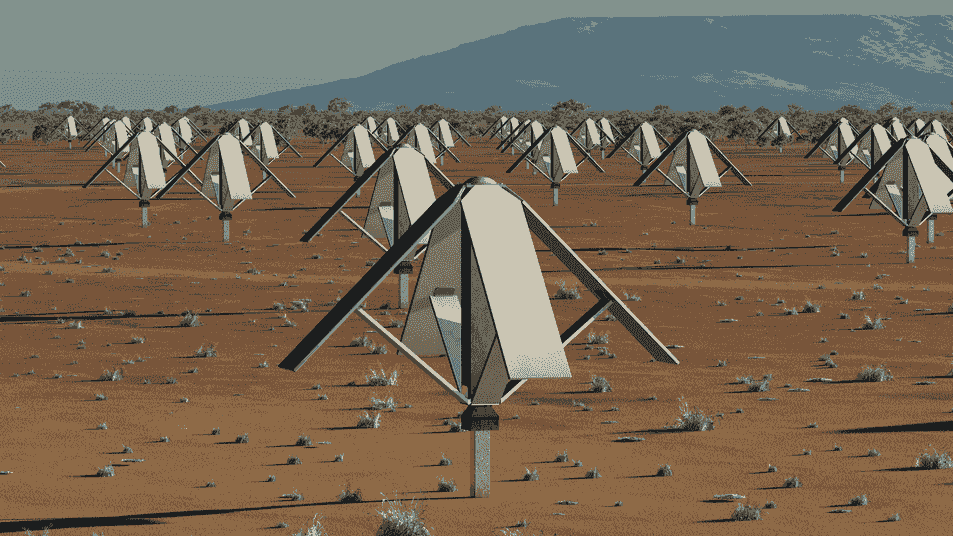}}
\hspace{-7cm}}

\vspace{-3.8cm}
\parbox{\textwidth}{
{\large
\noi \underline{Editors}\\
\noi  Hans-Rainer Kl\"ockner \\
\noi Michael Kramer\\
\noi  Heino Falcke\\
\noi  Dominik Schwarz\\
\noi Andreas Eckart\\
\noi Guinevere Kauffmann\\
\noi  Anton Zensus\\
}}

\newpage
\pagecolor{white}
\parbox{0.8\textwidth}{\vspace{18.34cm}

\noi Published by the Max-Planck-Institut f\"ur Radioastronomie
(MPIfR), Bonn, Germany

\bigskip

\noi Copyright: MPIfR, 2012

\smallskip

\noi Reproduction is permitted, provided the source is
acknowledged. Neither the MPIfR nor any person that contributed to
this ``White Paper'' is responsible for the
use by others of the information contained in this publication or can
be held responsible for any errors that may remain in the text.

\bigskip

\noi Cover:

\noi Cover images show an artist impression of the Square Kilometre
Array (SKA) configuration and of the different antenna types used in
the SKA. 

\noi Images: SPDO/Swinburne Astronomy Productions}

%
%
%
\newpage
\vspace{5cm}
\pagestyle{empty}

\color{black}

\parbox{\textwidth}{
{\bfseries\boldmath \Huge \color{hbl} Pathway to the Square Kilometre Array \vspace{0.5cm}}\\
\noi {\LARGE \color{hbl} The German White Paper}\\
}

\vspace{5cm}

\parbox{0.8\textwidth}{
\noi Editors:\\
\noi  Hans-Rainer Kl\"ockner, Michael Kramer, Heino Falcke, Dominik
Schwarz, Andreas Eckart, Guinevere Kauffmann, Anton Zensus
}


\vspace{0.8cm}

\parbox{0.88\textwidth}{
\normalsize
\noindent {\normalsize Contributions from: }

\noi {\sffamily \tiny
{\sffamily\bf F.~Aharonian} [\mpik],
{\bf T.G.~Arshakian} [\ukoeln],
{\bf B.~Allen} [\mpiaei],
{\bf R.~Banerjee} [\hhstw],
{\bf R.~Beck} [\mpifr],
{\bf W.~Becker} [\mpe],
{\bf D.J.~Bomans} [\airub],
{\bf D.~Breitschwerdt} [\zaa],
{\bf M.~Br\"uggen} [\hhstw],
{\bf A.~Brunthaler} [\mpifr],
{\bf B.~Catinella} [\mpa],
{\bf D.~Champion} [\mpifr],
{\bf B.~Ciardi} [\mpa],
{\bf R.~Crocker} [\mpik],
{\bf R.-J.~Dettmar} [\airub],
{\bf D.~Engels} [\hhstw],
{\bf T.~En{\ss}lin} [\mpa],
{\bf H.~Enke} [\aip],
{\bf T.~Fieseler} [\juelich],
{\bf L.~Gizon} [\mps],
{\bf E.~Hackmann} [\zarm],
{\bf B.~Hartmann} [\jbremen],
{\bf C.~Henkel} [\mpifr],
{\bf M.~Hoeft} [\stwtt],
{\bf L.~Iapichino} [\ita],
{\bf D.~Innes} [\mps],
{\bf C.~James} [\fau],
{\bf J.~Jasche} [\aifa],
{\bf D.~Jones} [\mpik],
{\bf V.~Kagramanova} [\oldenburg],
{\bf G.~Kauffmann} [\mpa],
{\bf E.~Keane} [\mpifr],
{\bf J.~Kerp} [\aifa],
{\bf H.-R.~Kl\"ockner} [\mpifr],
{\bf K.~Kokkotas} [\tat],
{\bf M.~Kramer} [\mpifr],
{\bf M.~Krause} [\mpe],
{\bf M.~Krause} [\mpifr],
{\bf N.~Krupp} [\mps],
{\bf J.~Kunz} [\oldenburg],
{\bf C.~L\"ammerzahl} [\zarm],
{\bf K.J.~Lee} [\mpifr],
{\bf M.~List} [\zarm],
{\bf K.~Liu} [\mpifr],
{\bf A.~Lobanov} [\mpifr],
{\bf G.~Mann} [\aip],
{\bf A.~Merloni} [\mpe],
{\bf E.~Middelberg} [\airub],
{\bf J.~Niemeyer} [\ugoett],
{\bf A.~Noutsos} [\mpifr],
{\bf V.~Perlick} [\zarm],
{\bf W.~Reich} [\mpifr],
{\bf P.~Richter} [\aip],
{\bf A.~Roy} [\mpifr],
{\bf A.~Saintonge} [\mpe],
{\bf G.~Sch\"afer} [\tpi],
{\bf J.~Schaffner-Bielich} [\ita],
{\bf E.~Schinnerer} [\mpia],
{\bf D.~Schleicher} [\ugoett],
{\bf P.~Schneider} [\aifa],
{\bf D.J.~Schwarz} [\ubie],
{\bf A.~Sedrakian} [\tguf],
{\bf A.~Sesana} [\mpiaeip],
{\bf V.~Smol\v{c}i\'{c}} [\aifa],
{\bf S.~Solanki} [\mps],
{\bf R.~Tuffs} [\mpik],
{\bf M.~Vetter} [\fhgise],
{\bf E.~Weber} [\fhgise],
{\bf J.~Weller} [\lmum],
{\bf N.~Wex} [\mpifr],
{\bf O.~Wucknitz} [\aifa],
{\bf M.~Zwaan} [\eso],
}
}

\vspace{4cm}

\hsp
\newpage
\pagestyle{empty}
\noi {\bf }
 
\newpage
\setcounter{page}{1}
\pagestyle{plain}
\renewcommand{\thepage}{\Roman{page}}

\tableofcontents

\newpage
\pagestyle{empty}
\noi{\bf }

\newpage
\pagestyle{plain}

\section*{Vorwort}
\addcontentsline{toc}{section}{Vorwort}{}{}

\bigskip

\noi Liebe Freunde und F\"orderer astrophysikalischer Forschung in Deutschland,

\bigskip
\bigskip

\noi als vor nunmehr 10 Jahren die astronomische Gemeinschaft in
Deutschland die DFG-Denkschrift ``Status und Perspektiven der Astronomie
in Deutschland 2003\,--\,2016'' formulierte, sah sie ihre Forschung am
Beginn einer ``goldenen Epoche'', ein Gedanke, der in der 2007
ver\"offentlichten ``Science Vision'' der europ\"aischen
ASTRONET-Initiative aufgegriffen und weitergef\"uhrt wurde. Und es
wurde nicht zu viel versprochen! Extrasolare Planeten werden heute
routinem\"a{\ss}ig entdeckt, die Parameter, die die k\"unftige
Entwicklung des Kosmos bestimmen, sind mit einer Genauigkeit von besser
als 10\% bestimmt, und Studien im hochrotverschobenen Universum
dringen in die ``dark ages'', den Zeitraum zwischen der Rekombination
und dem Erscheinen der weitest entfernten Galaxien und Quasare
vor. Anekdotenhaft kann dieser Erfolg daran gemessen werden, dass in
keiner Dekade so viele Astronomie-orientierte Nobelpreise in der Physik
vergeben wurden wie in den vergangenen zehn Jahren und zahlreiche
neue, hochrangige und hochdotierte internationale Preise initiiert wurden (mit
bemerkenswertem Erfolg von Forschern aus Deutschland).

\smallskip

\noi Ma{\ss}geblich f\"ur diese Entdeckungen war und ist die
Verf\"ugbarkeit der gro{\ss}en, boden- und weltraumgest\"utzten
internationalen Observatorien wie das ``Very Large Telescope'' der ESO,
die interferometrische Zusammenschaltung von Radioteleskopen zum VLBI,
das Weltraumteleskop Hubble oder wie Satellitenobservatorien in anderen
Wellenl\"angenbereichen wie Chandra, XMM oder Herschel, wobei all diese
Observatorien zunehmend synergetisch genutzt werden (als Beispiel
seien insbesondere die ``deep fields'' erw\"ahnt). Aufseiten der Theorie
werden diese Unternehmungen durch hochaufgel\"oste und hochkomplexe
Supercomputersimulationen erg\"anzt und verbunden.

\smallskip

\noi Trotz dieser gewaltigen Fortschritte sind viele zentrale Fragen erst
angerissen worden: die Entdeckung eines erd\"ahnlichen Exoplaneten in der
habitablen Zone steht weiter aus, die ``dark ages'' liegen nach wie vor 
gr\"o{\ss}tenteils im Dunkeln und der dominante Teil des Universums --
23\% dunkle Materie, 73\% dunkle Energie -- ist nach wie vor von
mysteri\"oser Komposition.

\smallskip

\noi Sowohl die DFG-Denkschrift wie auch die ASTRONET ``Infrastructure
Roadmap'' haben deshalb die n\"achste Generation gro{\ss}er Observatorien
am Boden und im Weltall identifiziert, die in den n\"achsten Jahren
ihren Betrieb aufnehmen und die Epoche der ``Pr\"azisionsastronomie''
initiieren werden. Dies beginnt 2013 mit der Astrometrie-Mission GAIA
und f\"uhrt mit der Inbetriebnahme des E-ELT der ESO sowie des
Euclid-Satelliten der ESA in die n\"achste Dekade. Wie bereits bei den
Vorg\"angermissionen werden diese Projekte zahlreiche technologieorientierte ``spin-offs''
weit abseits der Astronomie generieren.

\smallskip

\noi Als nunmehr zeitlich letztes in dieser Serie der gro{\ss}en
internationalen Observatorien steht nun die Weichenstellung zur
Realisierung des ``Square Kilometre Arrays'' an, um im
Multi-Frequenz-Portfolio die kritische L\"ucke im Bereich langer
Wellenl\"angen zu schlie{\ss}en. Das SKA soll nach derzeitiger Planung
2024 in den Regelbetrieb \"ubergehen.  Dieses Wei{\ss}buch (``White
Paper'') ist eine Momentaufnahme der deutschen
SKA-Forschungslandschaft und erl\"autert die verschiedenen
Forschungsthemen und -gebiete, an denen Wissenschafter in Deutschland
interessiert sind. Das Portfolio beeindruckt hierbei nicht nur durch
die Vielfalt und Tiefe der Fragestellungen von der Sonnenphysik bis
hin zur Kosmologie, sondern auch durch die Diversit\"at der Autoren --
sie reicht von Theoretikern \"uber Astroteilchenphysikern bis hin zu
beobachtenden Astronomen verschiedener Spezialisierung.  Hierbei
sticht ins Auge, dass die traditionell als Radioastronomen
bezeichneten Wissenschaftler eher eine Minderheit
darstellen. Ein wichtiger Aspekt sind auch technologische
Herausforderungen, z.B. die Handhabung von gro{\ss}en Datenmengen
oder von Technologien im Bereich der erneuerbaren Energien.

\smallskip

\noi Das wichtigste Kapitel bleibt jedoch im Wei{\ss}buch naturgem\"a{\ss}
ungeschrieben: Neue, unerwartete Entdeckungen und daraus resultierende
neue Forschungsfelder, die sich erst durch die Nutzung der neuen
Observatorien ergeben.\\

\vspace{0.25cm}
\noi Prof. Dr. Matthias Steinmetz\\
\noi Vorsitzender des Rats deutscher Sternwarten

\newpage
\pagestyle{empty}
\noi{\bf }

\newpage
\pagestyle{plain}
\section*{Zusammenfassung: Das Square Kilometer Array -- ein
  Technologie-Teleskop der Superlative}

\addcontentsline{toc}{section}{Zusammenfassung: Das Square Kilometer Array -- ein
  Technologie-Teleskop der Superlative}{}{}

Das SKA wird als einziges globales Projekt mit bedeutender
internationaler Partnerschaft in der ESFRI-Liste aufgef\"uhrt und
erhielt zusammen
mit dem E-ELT die h\"ochste Priorit\"at als k\"unftige
bodengebundene astronomische Einrichtung.

Das SKA wird mit einer effektiven Empfangsfl\"ache von einer
Million Quadratmeter (m$^2$) das mit Abstand gr\"o{\ss}te
Radioteleskop sein, das in einem Frequenzbereich von 70\,MHz bis
10\,GHz, oder h\"oher, operieren wird.  Das SKA wird in zwei Phasen
gebaut werden.  Es wird erwartet, dass erste wissenschaftliche
Ergebnisse im Jahr 2019 erzielt werden k\"onnen, w\"ahrend die volle
Inbetriebnahme des gesamten Teleskops f\"ur 2024 geplant ist.  Der
niederfrequente Teil des Observatoriums (70\,--\,500\,MHz) wird an dem
hierf\"ur optimalen Standort in West-Australien gebaut, w\"ahrend die
mittel- und hochfrequenten Teile im s\"udlichen Afrika errichtet
werden. Dort werden sie den s\"udafrikanischen SKA-Vorl\"aufer MeerKAT
erg\"anzen und k\"onnen mit den gro{\ss}en Radioteleskopen in Europa,
hier insbesondere das 100-m Effelsberg Radioteleskop in Deutschland,
verbunden werden und das europ\"aische ``Very Long Baseline
Interferometry''-Netzwerk erheblich erweitern.

Das SKA wird das weltweit f\"uhrende ``Imaging und
Survey''-Teleskop sein, das aufgrund einer Kombinaton aus
beispielloser Vielseitigkeit und Empfindlichkeit neue T\"uren zu neuen
Entdeckungen aufsto{\ss}en wird. Es wird sich in einer einzigartigen
Landschaft neuer Einrichtungen wiederfinden, mit denen das Universum
in elektro-magnetischen und anderen Wellenl\"angen erforscht
wird. Dabei wird das SKA nicht nur selbst faszinierende Entdeckungen
erm\"oglichen, sondern auch eine au{\ss}ergew\"ohnliche
Komplementarit\"at bei der Beantwortung unserer fundamentalsten
offenen Fragen herstellen. In der Tat wird das SKA Astronomen in die
Lage versetzen, Erkenntnisse in allen wissenschaftlichen
Schl\"usselfragen zu erlangen, wie z.B.~nach der Bildung und
Entwicklung der ersten Sterne und Galaxien nach dem Urknall, nach der
Natur der Schwerkraft und unserer Vorstellung von Raum und Zeit, nach der kosmischen
Geschichte des neutralen Wasserstoffs, nach der Rolle des kosmischen
Magnetismus und m\"oglicherweise auf die Frage nach
au{\ss}erirdischem Leben. Insbesondere wird uns das SKA erlauben,
folgendes zu untersuchen:

\begin{itemize}

\item[--]{{\bf \small Das dunkle Zeitalter}, durch die Erforschung
der Bildung der ersten Strukturen/K\"orper zu einer Zeit, als
das Universum gerade von einem vorwiegend neutralen Zustand in den heutigen
ionisierten Zustand \"uberging, etwa eine Milliarde Jahre nach dem Urknall;}

\item[--]{{\bf \small Die Galaxienentwicklung, Kosmologie und Dunkle Energie}, 
durch Untersuchungen der Anordnung, der Verteilung und der
Eigenschaften von Galaxien, z.\ B. durch Messungen des neutralen Wasserstoffs;}

\item[--]{{\bf \small Tests der Schwerkraft in starken 
Gravitationsfeldern}, in denen Pulsare verwendet werden, um 
Einsteins allgemeine Relativit\"atstheorie und unsere Vorstellung von Raum und
Zeit auf die Probe zu stellen;}

\item[--]{{\bf \small Die Wiege des Lebens},
durch das Studium der bewohnbaren Bereichen von protoplanetaren
Scheiben und der Suche nach pr\"abiotischen Molek\"ulen, die auf
primordiale erd\"ahnliche Bedingungen hindeuten.}

\end{itemize}

\medskip

\noi Die Radioastronomie hat einige der gr\"o\ss{}ten Entdeckungen des
20. Jahrhunderts produziert, die mit nicht weniger als vier
Nobelpreisen f\"ur Physik belohnt wurden.  Einige dieser Ergebnisse
sind die kosmische Hintergrundstrahlung, Pulsare, Gravitationswellen,
dunkle Materie, Quasare, Schwarze L\"ocher, Molek\"ule und
relativistische Plasma-Jets. Von zentraler Bedeutung f\"ur diese
Entdeckungen waren technologische Innovationen, die die Grenzen des
Machbaren in r\"aumlicher, zeitlicher und spektraler
Aufl\"osung verschoben haben. 

\noi Mit dem Bau des SKA wird mit dieser Tradition der
Innovation fortgefahren, durch die Kombination von grundlegenden
Neuentwicklungen in der Radiofrequenztechnologie,
Informationstechnologie und dem Supercomputing. Dies wird vergleichbare
Anstrengungen in anderen elektromagnetischen und
nicht-elektromagnetischen Wellen-l\"angenbereichen zum Universum komplementieren.  Die
Information aus jedem dieser Bereiche ist einzigartig und wichtig, und
nur durch die Kombination all dieser Informationen k\"onnen
wir hoffen, die physikalischen Prozesse im Universum verstehen
und einige der gro{\ss}en Fragen der Wissenschaft
beantworten zu k\"onnen. Tats\"achlich macht es die gebr\"auchliche (obwohl
nie wirklich g\"ultige) Einteilung der Wissenschaftler in Physiker,
Hoch\-energiephysiker, optische Astronomen, Radioastronomen und
dergleichen unm\"oglich, die modernen erfolgreichen Forschungs-projekte dieser Tage
korrekt zu beschreiben. In den verschiedenen Beitr\"agen dieses ``Wei{\ss}buchs'' zeigt sich,
dass die einzelnen Astronomen, Institute und Forschungskooperationen
heutzutage boden-und weltraumgest\"utzte Instrumente im gesamten
\mbox{elektromagnetischen} Spektrum verwenden, nicht-elektromagnetische
Fenster nutzen, Supercomputing f\"ur numerische Simulationen und
Datenanalyse verwenden und eng mit Teilchenphysikern
zusammenarbeiten. Die Diskussion der Ergebnisse erfolgt hierbei anhand
wissenschaftlicher Fragestellungen und nicht im Rahmen
enggefasster Definitionen spektraler Bereiche.

\bigskip

\noi Das SKA ist ein unersetzlicher Teil einer globalen Anstrengung, das
Universum, seine Grundgesetze, Herkunft und  Entwicklung
zu verstehen. Als eines der gro{\ss}en weltweiten Observatorien wird das
SKA ein unverzichtbares Werkzeug im gesamten Portfolio der deutschen
astronomischen Gemeinschaft sein. Au{\ss}erdem werden f\"ur  das SKA 
L\"osungen erarbeitet, um gemeinsam die \"ubergreifenden
technologischen Herausforderungen zu meistern, wie zum Beispiel das
Problem der Handhabung, Verwaltung und Analyse erstaunlich gro{\ss}er
Datenmengen -- ein Problem, dass alle neuen Einrichtungen der
Spitzenforschung betrifft.

\smallskip

Die technischen Herausforderungen und die anstehenden
Anforderungen an die Datenverarbeitung werden die Art und Weise, wie
astronomische Forschung heutzutage gemacht wird, grundlegend ver\"andern. Um
das SKA zu verwirklichen, brauchen wir einen revolution\"aren Bruch
mit dem konventionellen Design von Radioteleskopen. Das SKA wird
insbesondere im Bereich der Informations- und Kommunikationstechnologie
die technische Entwicklung vorantreiben.  Spin-off-Innovationen in
diesen Bereichen werden auch f\"ur andere Gebiete in Industrie und
Wissenschaft von Nutzen sein, die ebenfalls gro{\ss}e Datenmengen von
geografisch verteilten Quellen verarbeiten m\"ussen.  Die enormen
Energieanforderungen des SKA bieten au{\ss}erdem eine Chance, die
Technologieentwicklung im Bereich der skalierbaren Erzeugung
erneuerbarer Energien, deren Verteilung sowie Speicherung bei
gleichzeitiger Reduzierung des Verbrauchs zu beschleunigen.  Bereits
heute verf\"ugt die deutsche Gemeinschaft \"uber das ben\"otigte Know-how zur
Planung, Konstruktion, Inbetriebnahme und zur wissenschaftlichen
Ausbeutung eines solchen Observatoriums.  Dies wird
demonstriert durch den Bau des ``Internationalen LOFAR Teleskops''
(ILT) und seiner deutschen Stationen, das bis heute den gr\"o{\ss}ten
und modernsten SKA-Pathfinder darstellt. Aus diesem Grunde k\"onnte
die deutsche SKA-Gemeinschaft erheblich zum SKA und dessen Design
beitragen.

\smallskip

Der Bauabschnitt SKA Phase\,1 wird wissenschaftlich wie technisch
eine Teilmenge der SKA Phase\,2 sein. Diese Planung erlaubt es, die
technische Reife der verschiedenen Systemkomponenten zu \"uberpr\"ufen
und m\"oglicher-weise entsprechende \"Anderungen w\"ahrend der Bauphase des
SKA vorzunehmen.  Die Zielkosten f\"ur das komplette SKA sind auf 1,5
Milliarden Euro festgelegt worden, w\"ahrend die Kosten f\"ur Phase\,1 auf 350 Millionen Euro begrenzt werden sollen.  Die
Kostensch\"atzungen des gesamten SKA-Projekts basieren auf einem
Kalkulationsmodell, das speziell f\"ur das SKA entwickelt wurde und
die Einhaltung der Kosten im Detail und im Ganzen erm\"oglicht.  Damit
ist das SKA einzigartig im Vergleich zu anderen internationalen
Wissenschaftseinrichtungen bez\"uglich der Kontrolle der endg\"ultigen
Kosten. Durch den modularen Charakter des SKA als ein ``Synthese
Array'' ist es zum Beispiel m\"oglich, bei einer Kostenreduktion
(wenn wissenschaftlich begr\"undbar) die Gesamtplanung durch eine
\"Anderung der m\"oglichen Antennenanzahl entsprechend zu ver\"andern.

\bigskip

\noi Die weltweiten Anstrengungen zur Ausarbeitung der
wissenschaftlichen Ziele und der technischen Spezifikationen f\"ur die
n\"achste Generation von Radioteleskopen bestehen seit 1993.  Seitdem
wurde das Konzept und die Technologie-Entwicklung f\"ur das SKA von
einem internationalen Konsortium unternommen, das rund 55
Institutionen in 19 L\"andern umfasste. Deutschland war von Beginn an
Teil dieser Konsortia. Im Jahr 2011 haben sieben nationale
Regierungs-und Forschungsorganisationen aus Australien, China,
Italien, den Niederlanden, Neuseeland, S\"udafrika und dem Vereinigten
K\"onigreich eine formale SKA-Organisation gegr\"undet, der vor kurzem
Kanada und Schweden beigetreten sind, die als ein unabh\"angiges,
nicht-gewinnorientiertes Unternehmen nach britischem Recht aufgestellt
wurde. Ziel der Organisation in dieser ``Vor-Konstruktions-Phase'' ist
es, die Beziehungen zwischen den internationalen Partnern zu
formalisieren und die Leitung des SKA-Projekts zu zentralisieren. Die
Finanzierung dieser Organisation ist bis 2016 gesichert. Danach soll
das Projekt in die Bauphase \"ubergeleitet werden, die von einer
neuen unabh\"angigen Organisation durchgef\"uhrt werden k\"onnte und
sollte.

\bigskip

\noi Dieses ``Wei{\ss}buch'' versucht, die wissenschaftlichen
Interessen an dem SKA-Projekt in Deutschland zu erfassen. Die
\"uberwiegende Mehrheit der Beitr\"age befasst sich mit astronomischen
Fragen, aber weitere wichtige Themen wie High Performance Computing,
``Daten-Handling, Management und Mining'' sowie Energieversorgung sind
von gro{\ss}er Bedeutung f\"ur die deutsche SKA-Gemeinschaft. Eine
konservative Sch\"atzung der Gr\"o{\ss}e der deutschen
``SKA-Community'' ergibt eine Zahl von mehr als 400 Einzelpersonen,
die direkt vom SKA profitieren w\"urden.  Um die deutschen
SKA-Aktivit\"aten zu organisieren und koordinieren, hat das ``German
Long Wavelength Consortium'' (GLOW) eine SKA-Arbeitsgruppe gegr\"undet,
die als verbindendes Element zwischen der astronomischen
Gemeinschaft, der SKA-Gemeinschaft insgesamt sowie den Industrie- und
politischen Partnern dienen wird. Dies ist notwendig, da das SKA ein
einzigartiges ``Welt-Teleskop'' sein wird, das eine strategische
Allianz von technischen, kommerziellen und wissenschaftlichen
Partnern darstellen und gleichzeitig eine hochkar\"atige
Forschungs- und Entwicklungeinrichtung sein wird.  Investitionen in das
SKA w\"urde einen Wissenstransfer f\"ur die Industrie und die Forschung mit
Hilfe einer High-Tech-Einrichtung bedeuten, die in einem weltweiten
Netzwerk operiert.

\smallskip

Es ist von gr\"o{\ss}ter Bedeutung, dass Deutschland sich bei den
wichtigen Forschungseinrichtungen wie das SKA, E-ELT und das Cherenkov
Telescope Arrays (CTA) engagiert und seine Bem\"uhungen weiter
verst\"arkt. Als solches muss es eine aktive Rolle bei den
Entscheidungen im SKA-Projekt einnehmen, um die Interessen der
beteiligten deutschen Gemeinschaften zu sichern und die
Wettbewerbsf\"ahigkeit (und oft f\"uhrende Rolle) der
astrophysikalischen Forschung sowie technologischen Entwicklung in
Deutschland zu erhalten.

\newpage
\pagestyle{empty}
\noi{\bf }

\newpage
\setcounter{page}{1}
\pagestyle{empty}
\renewcommand{\thepage}{\arabic{page}}

\section*{Foreword}
\addcontentsline{toc}{section}{Foreword}{}{}

\bigskip

\noi Dear Friends and Supporters of astrophysical research in Germany,

\bigskip
\bigskip

\noi When the astronomical community in Germany composed the DFG
memorandum `` Status and Perspectives of Astronomy in Germany 2003\,--\,2016'', now 10 years ago, it considered its research to be at the
beginning of a ``golden era'' -- an idea that was picked up and
elaborated on in 2007 in the ``Science Vision'' of the European
ASTRONET initiative. And, they did not exaggerate! Extrasolar
planets are today routinely discovered, the parameters that dictate
cosmic evolution are determined with an accuracy of better than
10\,\%, and studies of the high-redshift universe penetrate the
``dark ages'', the period between the recombination and the appearance
of the most distant galaxies and quasars. This success story can be
gauged by the fact that in no other decade have there been more
astronomy-related Nobel Prizes in Physics than in the last ten years,
and many new, high-level and financially substantial awards have been
established (with remarkable success of researchers from Germany).

\smallskip

\noi To a large extent these discoveries have been made possible by
continued access to the large international, ground- and space-based
observatories like the Very Large Telescope of ESO, the
interferometric combination of radio telescopes for VLBI, the Hubble
Space Telescope, and satellite observatories at other wavelength
regimes such as Chandra, XMM, or Herschel. Interestingly, all these
observatories are increasingly used synergistically, with the ``deep
fields'' as particularly good examples. On the theory side, these efforts
are supplemented by, and combined with, high-resolution and highly
complex super-computer simulations.

\smallskip

\noi Despite this tremendous progress, many central questions have
thus far only been touched upon: an Earth-like exoplanet in the
habitable zone remains to be discovered, the ``dark ages'' are still
mostly in the dark, and the composition of the dominant part of the
Universe, i.e.~23\,\% dark matter and 73\,\% dark energy, is still a
mystery.

\smallskip

\noi Both the DFG memorandum as well as the ASTRONET ``Infrastructure
Roadmap'' have therefore identified the next generation of large
observatories on the ground and in space, which will become
operational in the next years and which will initiate an era of
``precision astronomy''. This will start in 2013 with the astrometric
mission GAIA and leads with the commissioning of the E-ELT of ESO and
the Euclid satellite of ESA into the next decade. As with the
predecessor missions, these projects will result in numerous
``spin-offs'' of technological nature far away from astronomy.

\smallskip

\noi Now it is time to pave the way for the last in this sequence of
great international observatories, namely to make the ``Square
Kilometre Array'' a reality in order to close the critical gap in the
multi-frequency portfolio at long wavelengths. The completed SKA is
planned to go into regular operation in 2024.  This ``white paper''
describes and explains the different research interests and
applications of scientists in Germany. The presented portfolio is not
only impressive in the variety and depth of the issues, from solar
physics to cosmology, but also in the diversity of the authors --
ranging from theoreticians via astro-particle physicists to
observational astronomers of different specialisation. Only a minority
can be considered to be radio astronomers in the traditional
sense. Important aspects are also the technological challenges, e.g.~the
handling of large amounts of data and technologies in the field of
renewable energies.

\smallskip

\noi Naturally, the most important chapter in the white paper remains
unwritten: new, unexpected discoveries and resulting new research
fields that will only be created by the use of the new observatories. \\

\vspace{0.25cm}
\noi Prof. Dr. Matthias Steinmetz\\
\noi Chairperson of the Council of German Observatories (RDS)      

\newpage
\pagestyle{empty}
\noi{\bf }

\newpage
\pagestyle{plain}
\section{Preamble}

\noi The Square Kilometre Array (SKA) is the most ambitious radio telescope
ever planned. With a collecting area of about a square kilometre, the
SKA will be far superior in sensitivity and observing speed to all current
radio facilities. The scientific capability promised by the SKA and
its technological challenges provide an ideal base for interdisciplinary
research, technology transfer, and collaboration between universities,
research centres and industry.  The SKA in the radio regime and the
European Extreme Large Telescope (E-ELT) in the optical band are on
the roadmap of the European Strategy Forum for Research
Infrastructures (ESFRI) and have been recognised as the essential
facilities for European research in astronomy.

Within the next few years the SKA project will move into a
construction phase and important decisions are necessary regarding the
telescope design, the infrastructure, and the technical
developments. It is of vital importance to both the German science
community and industry that Germany will be fully engaged in the
SKA project  in order to secure German interests
in research and development (R\&D) and its scientific usefulness.

\medskip

\noi This ``White Paper'' outlines the German science and R\&D interests in
the SKA project and will provide the basis for future funding
applications to secure German involvement in the Square Kilometre
Array.

\bigskip

\section{Executive summary: The Square Kilometre Array -- a technology telescope of superlatives}

\noi The SKA is listed as the only global project in the ESFRI
with significant international partnership and, together with the
E-ELT, it received the highest priority for future ground-based
astronomical facilities. 

The SKA will be a giant radio telescope with
an effective collecting area of one-million square\ metres\ (m$^2$)
that will operate between 70\,MHz to 10\,GHz, or more. The SKA will be
built in two phases, first science is expected in 2019 and it is
planned to be fully operational by 2024. The low-frequency part of the
observatory will be built at the optimal site in Western Australia,
while the mid- and high-frequency part will be built in Southern
Africa where it can be merged with the SKA precursor MeerKAT and
connected to the large European and German radio telescopes,
enormously extending the European very long baseline interferometry
network.

The SKA will be the world's premier imaging and surveying telescope
that, with a combination of unprecedented versatility and sensitivity,
will open up new windows of discovery space. It will find itself
placed in a unique combination of new facilities exploring the
electromagnetic and other windows into the Universe, making not only
fascinating discoveries on its own but also providing exceptional
complementarity in the exploration of the Universe's most fundamental
open questions. With these capabilities the SKA will provide
astronomers insight into the key science questions such as, the
formation and evolution of the first stars and galaxies after the
``big bang'', the nature of gravity, the history of neutral hydrogen,
the role of cosmic magnetism, and possibly life beyond Earth. In
particular the SKA enables us to probe:

\smallskip

\begin{itemize}
\item[--]{{\bf \small the dark ages}, by
studying the formation of the first structures/bodies at a time when
the Universe made a transition from largely neutral to its ionised
state today;}

\item[--]{{\bf \small galaxy evolution, cosmology, and dark energy},
by investigating the assembly, the distribution, and the properties of
galaxies, e.g. through measurements of neutral hydrogen;}

\item[--]{{\bf \small strong field tests of gravity}, by using pulsars
to challenge Einstein's general relativity and the nature of space and
time;}

\item[--]{{\bf \small cradle of life}, by studying the habitable segment
of proto-planetary disks and to search for prebiotic molecules that
indicates primordial Earth-like conditions.}

\end{itemize}

\bigskip

\noi Radio astronomy has produced some of the greatest discoveries of
the 20th century that have been rewarded with no less than four Nobel
Prizes for physics. Some of these findings are the cosmic background
radiation, pulsars, gravitational waves, dark matter, quasars, black holes,
molecules, and
relativistic plasma jets. Central to these discoveries have been
innovations in technology pushing the observational frontiers of
sensitivity as well as spatial-, temporal- and
spectral-resolution. The SKA will carry on this tradition of
innovation by combining fundamental developments in radio frequency
technology, information technology and high-performance computing,
hence complementing comparable efforts in other electromagnetic and
non-electromagnetic windows to the Universe. The information of each
window is unique and important and only by combining all information
from all observational windows can we hope to understand the physical
processes at work in the Universe. Indeed, the commonly used (though
never really valid) division of scientists into physicists, high
energy physicists, optical astronomers, radio astronomers, and the like
fails to represent the successful research endeavours performed these
days. As also indicated by the various contributions of this ``White
Paper'' individual astronomers, institutes and research collaborations
today use ground-based and space-based facilities across the
electromagnetic spectrum, exploit non-electromagnetic windows, use
high-performance computing facilities for numerical simulations and
data analysis, and interact closely with particle physicists, all at the
same time. In particular, they discuss their results in terms of
science questions, rather than using a narrow definition of spectral
windows.\\

\noi The SKA is an irreplacable part of this global endeveour to explore the
Universe and its fundamental laws, its origin and its fate. As one of
the great global observatories, the SKA will provide an indispensable
tool in the portfolio of the German astronomy community as a whole,
sharing also solutions to common technological challenges, such as the
handling, management and analysis of a staggering amount of data that
is common to most many future top facilities.

The technical challenges and the upcoming data products will
dramatically change the way astronomical research is done today
and to make the SKA a reality it demands a revolutionary break from
the traditional framework of radio telescope design. The SKA will
drive technology development particularly in information and
communication technology. Spin-off innovations in these areas will
benefit other systems in industry and science that process large
volumes of data from geographically dispersed sources. The energy
requirements of the SKA also present an opportunity to accelerate
technology development in scalable renewable energy generation,
distribution, storage, and demand reduction. The German community has
already developed expertise, in the design, construction,
commissioning, and scientific exploration, of the ``International
LOFAR Telescope'' the largest and most advanced SKA pathfinder to
date, and therefore could significantly contribute to the SKA and its
design.

The SKA Phase\,1 will be built first and will be a scientific and
technological subset of the SKA Phase\,2. This setup allows evaluation
of the technical maturity of various components of the system and will
guide the decisions made in the course of a full SKA.  The target cost
of the full SKA has been set to be 1.5 Billion Euros (``1.5 Milliarden
Euro''), whereas the cost of Phase\,1 will be capped to 350 Million
Euros. The cost estimates of the entire SKA project are based on a
costing model especially developed for the SKA allowing for detailed
cost modelling in order to minimise the uncertainties of the final
expenses. The SKA is unique compared to the international flagship
facilities in science in terms of controlling the final fiscal cost. Due to
the modular nature of the SKA, as a ``synthesis array'', a cost
reduction, if scientifically justified, is possible by alternating the
number of telescope stations to match a
final targeted cost figure.\\

\noi Since 1993 a worldwide effort has been established to develop the
scientific goals and technical specifications for the next generation
of telescope. Since then the concept and technology development of the
SKA has been undertaken by an international SKA consortium that
includes some 55 institutions in 19 countries. Germany has
participated and was involved in this endeavour since the beginning.
In 2011, seven national governmental and research organisations from
Australia, China, Italy, the Netherlands, New Zealand, South Africa,
and the United Kingdom, provided the base to form a formal SKA
organisation. Recently, Sweden has joined the organisation, which 
is an independent, not-for-profit
company that has been established to formalise relationships with
international partners and to centralize the leadership of the SKA
project. The funding of this organisation is secured until 2016 at
which point the construction phase will start. It is foreseen that the
construction is executed by a new independent organisation that will
also run the telescope.\\

\noi This ``White Paper'' aims to capture the current science interests in
the SKA project in Germany. The vast majority of contributions addresses astronomical
questions, but further important subjects like high performance
computing, data handling, management and mining, as well as energy
provision are subjects of great importance to German scientists. A
conservative estimate of the size of the German SKA community results
in more than 400 individuals that would benefit from the SKA.  In
order to organize and coordinate the German SKA activities, the German
Long Wavelength Consortium (GLOW) has established an SKA Working Group
that will serve as the binding element between the radio astronomical
community, the SKA community at large as well as industrial and
political partners. This is necessary since the SKA will be a
``world-telescope'' that build a strategic alliance enabling
technical, commercial, and scientific partnerships and research and
development in a high-profile project. Furthermore, investing in the
SKA would provide knowledge transfer for industry and accademia in a
high-tech facility that operates in a worldwide network.

It is of utmost importance that Germany preserve and strengthen
its position in the flagship facilities like the SKA, E-ELT, and the
Cherenkov Telescope Array (CTA), and takes an active role in the
decisions to be made in the SKA project to safeguard the German
communities interests and the competitiveness (and often leading role)
of astrophysical research in Germany.


\bigskip

\section{Radio astronomy and its role in understanding the  Universe}

\noi {\bf \small Astronomy} is an observational science that connects
to the human spirit and its curiosity, which drives us forward to
explore and venture into new worlds. Indeed, modern astronomy has much
to offer. It has dramatically changed over the last century, opening
new observational windows to the Universe. Astronomy is essential,
since \\ 

\hspace{-0.5cm}
\parbox{\textwidth}{ \parbox{0.5\textwidth}{\includegraphics[width=0.51\textwidth]{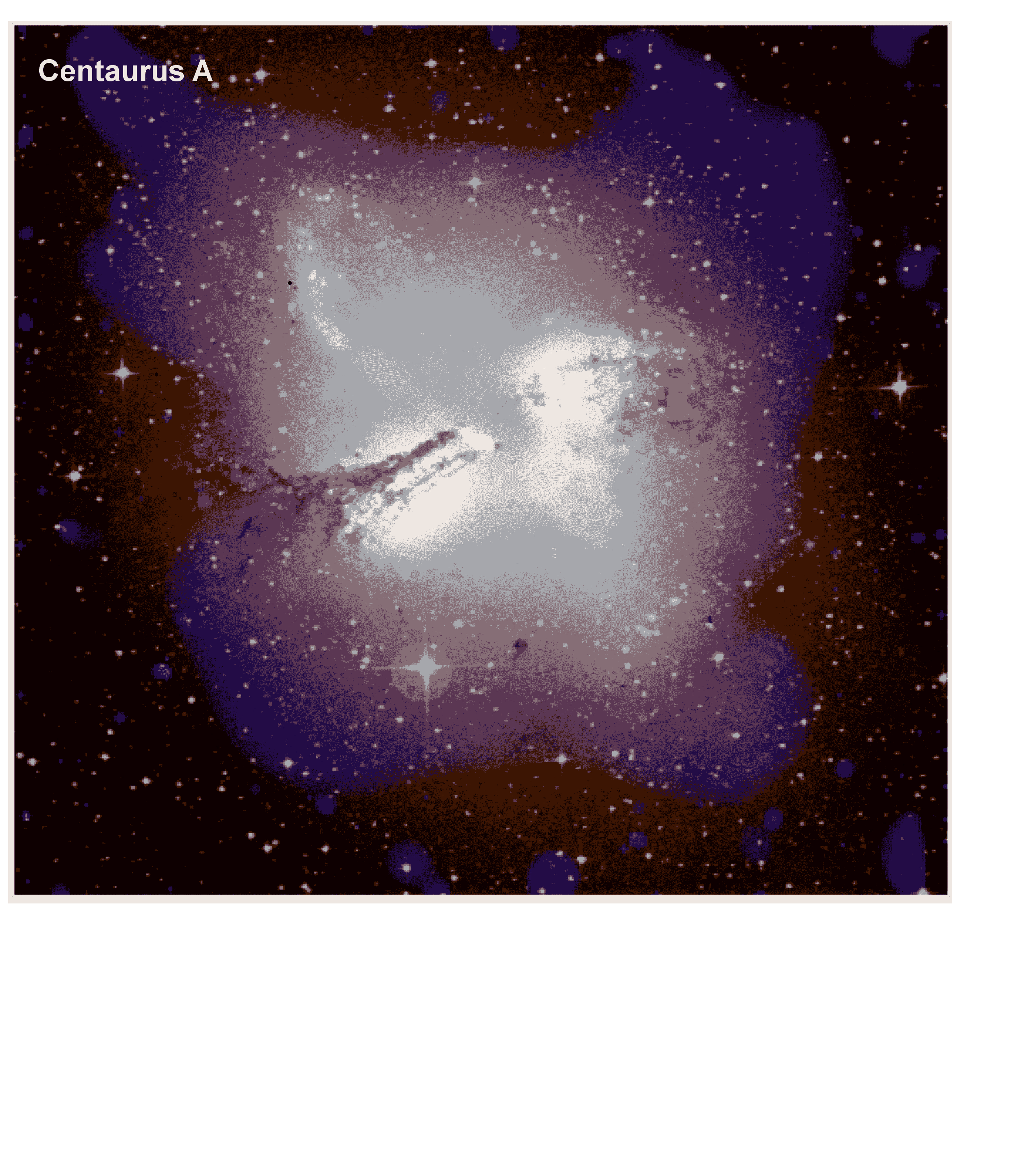}\vspace{-10.8cm}

\hspace{8.5cm}
\parbox{0.5\textwidth}{\includegraphics[width=0.505\textwidth]{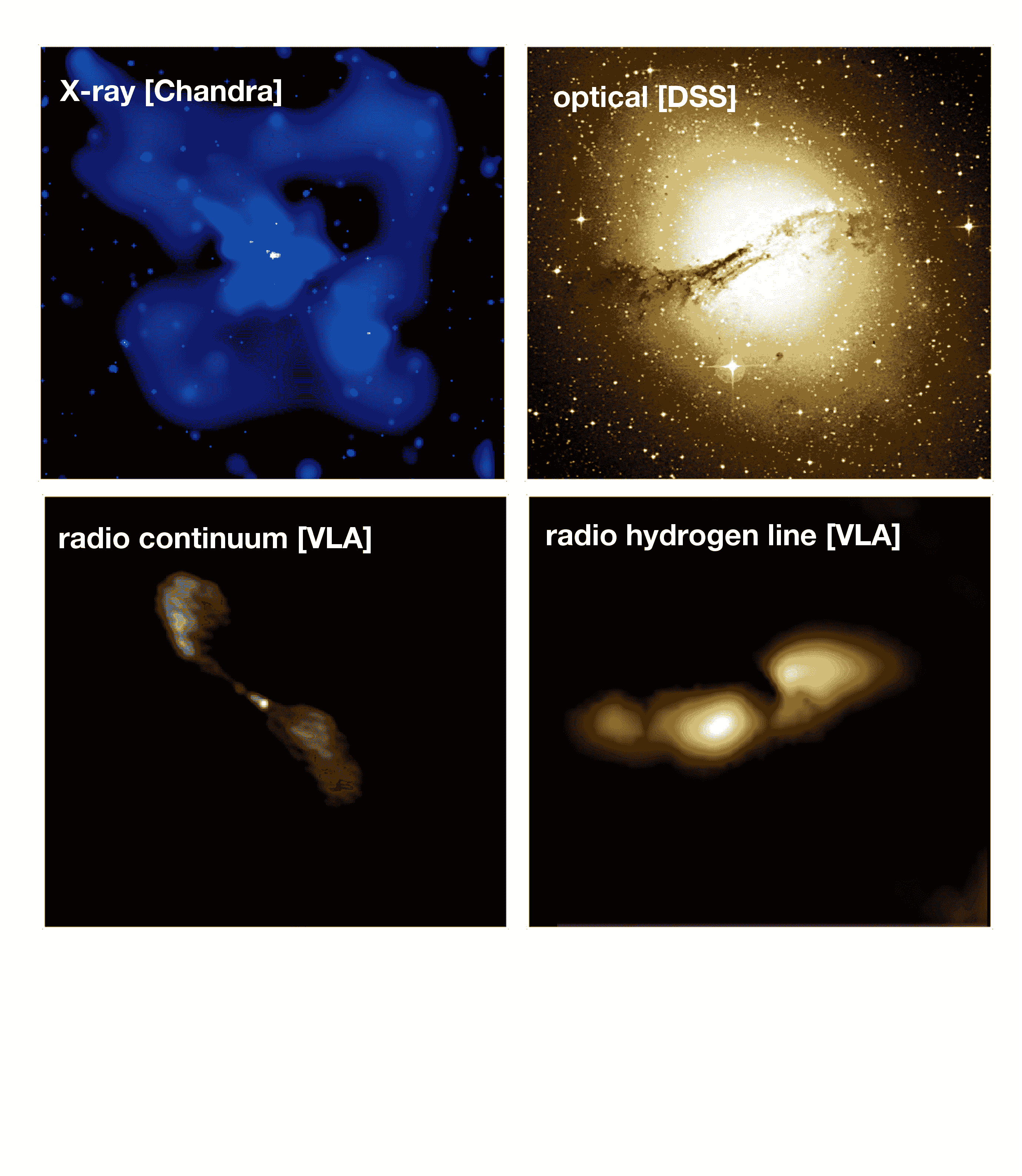}\vspace{-1.8cm}}

\hspace{0.6cm} 
\parbox{0.9\textwidth}{Figure I: Composite image of the
    extragalactic source ``Centaurus A'' observed from short wavelengths (X-ray) to long
  wavelengths (radio). In particular, the radio emission maps out various
  components of the galaxy. The radio continuum emission 
  traces the outflow-jet structure of the galaxy whereas the neutral
  hydrogen emission traces the kinematics of the disk like structure
  seen in the optical image. Furthermore, by using the hydrogen line
  emission the distance to the galaxy itself can be determine via the
  redshift relation. (Image credits: see page~\pageref{imcred})}\vspace{0.3cm}}}

\noi many of the scientific tests, experiments or hypotheses proposed
by theory can only be examined in the cosmic laboratory as many
conditions cannot be replicated on Earth.  In particular radio
astronomy, developed since the 1930s plays an important part and is
now a multi-disciplinary science that connects to quantum and high
energy physics, general relativity, information technology, chemistry,
electrical and mechanical engineering, communication, optics, material
science, mathematics, and much more.

Depending on physical conditions, cosmic signals are created in
several ways, mostly by matter that emits, reflects, or absorbs energy
across the electromagnetic spectrum ranging from very short
wavelengths (e.g. gamma rays, X-rays) to very long wavelengths
(e.g. infrared, radio waves). Each part of this information is unique
and important, and only by combining all information from all
observational windows can we hope to understand the physical processes
at work. Indeed, it can be shown that multi-wavelength papers, in
particular the combination of optical and radio observations, are on
average more influential than other publications (Tremble \& Zaich
2006). This is a lesson that is very much adopted by the astronomical
and physical community as a whole. Hence, dividing scientists into
physicist, high energy physicists, optical astronomers, radio
astronomers and the like, is a very much a misrepresentation of the
research successfully performed these days. As a result, the classical
division between users of certain facilities has disappeared.
Individual astronomers, institutes and research collaborations today
use ground-based and space-based facilities across the electromagnetic
spectrum, exploit non-electromagnetic windows, use high-performance
computing facilities for numerical simulations and data analysis,
interact closely with particle physicists, all at the same time -- in
particular, they discuss their results in terms of science questions,
rather than using an out-dated narrow definition of spectral windows.

For instance, it is clear that the most burning questions about the
nature of dark matter and dark energy can only be solved by combining
studies in the radio-, optical-, and high-energy windows, paired with
results from particle accelerators and vast simulations using the best
super-computers in the world. Similarly, all future science facilities
have one property in common: they will produce a staggering amount of
data that needs to handled, managed and analysed. Solutions to this
common problem are likely to come from working together towards a
science infrastructure that can handle the storage and, also, energy
requirements.

\hspace{-1.8cm}
\parbox{\textwidth}{\parbox{0.9\textwidth}{\includegraphics[width=1.15\textwidth]{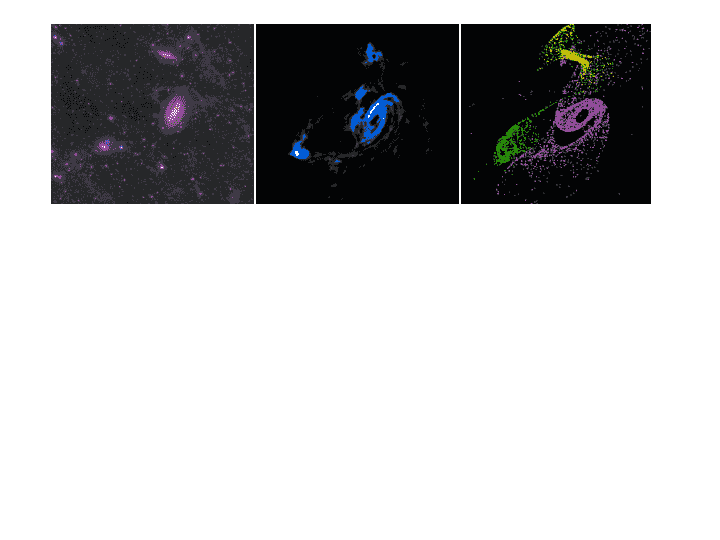}\vspace{-9cm}}

\hspace{1.95cm} 
\parbox{0.9\textwidth}{Figure II: An example of modern astronomy,
  combining multi-wavelength observations with computer
  simulations. Here shown are the interacting galaxies M\,81, M\,82,
  and NGC\,3077 in the optical (left) and in neutral hydrogen (\hi ,
  middle). The interaction is only observable by the atomic \hi\ gas
  that connects all three galaxies by filaments. The numerical
  simulation (right) demonstrates that all \hi\ filaments are the
  results of tidal interaction among these galaxies. (Image credits:
  see page~\pageref{imcred}) }\vspace{0.5cm}}

An example of the required combination of information across the
electromagnetic spectrum is illustrated in Figure\ I which shows a
composite image of a galaxy and the contributing observations at
different wavelengths. Apart from tracing different processes, quite
often, energy from one part of the electromagnetic spectrum will be
blocked by interstellar matter, interplanetary dust, or Earth's
atmospheric constitutes, while energy from other parts of the spectrum
will be transmitted freely to the astronomical observatory. For
instance, cosmic radio emission has been shown to be less limited by
these effects and is even observable during day time. Moreover, even
though radio photons may be not very energetic, they often result from
highly energetic processes, revealing objects that are otherwise not
visible. For example Figure\, II shows galaxy mergers who's energetic
processes are only visible with radio observations. Furthermore, the
figure demonstrates the interplay of observations and simulations in
modern astronomy. Without doubt the use of radio astronomical
techniques will continue to play an important part in this joint
endeavour to understand the cosmos and the fundamental laws that
govern it.  In the accordance with these efforts, despite humble
beginnings, radio telescopes have delivered some of the arguably most
important discoveries of the last 100\,years.\\

\noi When radio astronomy was ``born'' in the 1930s, its inventor Karl
Jansky was employed at Bell Laboratories, studying the disturbance of
radio telephone links. His findings that these were of
extraterrestrial origin triggered further ``radio'' work and, by the
early 1950s, radio astronomy had discovered radio emission from the
Milky Way, the Sun, and from very distant galaxies. These discoveries
challenged the theoretical models of the time, and there were no
accepted theories in place to explain these phenomena.  However, it
was not until the 1960s that radio astronomy began to play its part in
the ``big science area'' on a large, transformational scale (Verschuur
2007). The resulting discoveries were many, including signals from the
beginnings of all time and space (i.e. ``big bang'' background
radiation at 3\,Kelvin), and have been rewarded with no
less than four Nobel Prizes for physics. The discoveries include:
\begin{itemize}
\item[--] the cosmic microwave background (CMB), as the remaining signal from
  the big bang,
\item[--] Pulsars, as the remnants of massive stars whose existence proves
the validity of quantum mechanical laws in space,
\item[--] the existence of gravitational waves using binary pulsars in
  cosmic clock experiments,
\item[--] the existence of gravitational lenses that bend light to form
  multiple images of distant super-massive black holes,
\item[--] the evidence for ``Dark Matter'' in the rotational properties of 
galaxies (neutral hydrogen),
\item[--] the existence of complicated molecules in space that may well
be related to the formation of life (carbon monoxide, ammonia),
\item[--] the first extra-solar planets.
\end{itemize}

\smallskip

\noi As indicated earlier, these discoveries address or have led to
fundamental questions that are not exclusive to radio astronomy but
which initiated multi-disciplinary research about our origin and fate and that of 
the Universe, as well as the fundamental laws of physics.
Given this important complementarity of the different waveband
observations, it is does not come as a surprise that both the Hubble
Space Telescope (HST) and the radio interferometer Very Large Array
(VLA) remain as the two astronomical facilities with the largest number
of publications and citations of a single facility.  But it is also true 
that consistently, publications combining radio and optical
data are extremely well cited and typically more influential than
single-waveband papers (see e.g.\ Trimble
\& Zaich 2006, Trimble \& Ceja 2008). The SKA will follow
this tradition by providing an indispensable tool in the
portfolio of the German astronomy community as a whole.

\bigskip

\section{The SKA in the astronomical landscape}

The strategic plan for European astronomy described in the ``
  ASTRONET Roadmap'' identifies the SKA, together with the European
Extremely Large Telescope (E-ELT), as the highest priority project for
ground-based astronomy due to the potential for fundamental
breakthroughs in a very wide rage of scientific fields.  Besides the
SKA and the E-ELT three other projects were considered scientifically
outstanding but in narrower fields and with lower budgets (see
``ASTRONET Roadmap''). In descending order of priority identified, these are the
European Solar Telescope (EST), the Cherenkov Telescope Array (CTA)
and the neutrino detector KM3NET -- all which have significant German
involvement or leadership.

On the same timescale as the realisation of SKA Phase 1 and, certainly,
Phase 2, further instruments will be operational or are in
preparation, such as ALMA, JWST, ELTs, ATHENA, SPICA, CTA LIGO,
eLISA/NGO, LSST, Euclid/JDEM, CMBPOL, GAIA, KM3NeT and the LHC
(running in CERN) all covering different wavelength or energy
regimes. The SKA provides an important and crucial radio component in
the pursuit of the burning science topics that are, for instance,
captured by the science questions formulated in the ``ASTRONET Science
Vision'', which are:
\begin{itemize}
\item[--]{\em What is the origin and evolution of stars and planets?}
\item[--]{\em Do we understand the extremes of the Universe?}
\item[--]{\em How do galaxies form and evolve?}
\item[--]{\em How do we fit in?}
\end{itemize}
As described in the science contributions to this ``White Paper'', the
German interests cover all these areas, and indeed, the SKA will not
only contribute to them, but SKA observations will often play a or
even {\em the} decisive role in answering them. A good example is the
SKA's role in answering the questions about the nature of dark matter
and dark energy. The is already a large German involvement and
interest in instruments and projects that address these questions, such
as the Sloan Digital Sky Survey (SDSS), WiggleZ, 4MOST, LSST, HETDEX,
Pan-Starrs, or DES, others space-based missions such as Euclid.  All
of them are operating in the optical or near-infrared. In the radio, several SKA
pathfinder telescopes are also preparing continuum radio surveys
(e.g. LOFAR MSSS, LOFAR Tier1, WODAN, EDU; Raccanelli \etal\ 2011),
while the SKA itself will survey a billion galaxies in redshifted
hydrogen, which will automatically provide distances, so that the
Universe can be conveniently sliced in redshift bins. This information
promises to provide unprecedented and pivotal answers in determining
the equation-of-state of dark energy.\\

\noi In order to understand the role of the SKA for science in the
next decades see Figure\,III
or it is perhaps even more enlightening to recast the SKA contribution in
the following themes, namely
\begin{description}
\item[The laws of nature: Fundamental Physics \& Cosmology] including
\begin{description}
\item[Gravity:]    Is general relativity  our last word in
  understanding gravity? \\
   What happens in strong gravitational fields? \\
   What are the properties of gravitational waves? \\
   What is ``Dark Matter''? \\
   What is ``Dark Energy''? 
\item[Magnetism:]
    What is the origin of cosmic magnetism? \\
    How did it evolve? \\
    What is its role in the formation and evolution 
    of stars and galaxies?
\item[Strong \& weak forces:]
   What are the properties of matter? \\
   What is the nuclear equation-of-state?
\end{description}
\item[Origins: Galaxies across cosmic time, Galactic neighbourhood, 
               stellar \& planetary formation] including
\begin{description}
\item[Galaxies and the Universe:] 
    How did the Universe emerge from its Dark Ages? \\
    How did the structure of the cosmic web evolve? \\
    Where are most of the metals throughout cosmic time? \\
    How were galaxies assembled?
 \item[Stars, Planets, and Life:]
    How do planetary systems form and evolve? \\
    What is the life-cycle of the interstellar medium and stars?
    (biomolecules)? \\
    Is there evidence for life on exoplanets (SETI)? 
\end{description}
\end{description}
German involvement in related projects is wide-spread and other
observatories like LOFAR, ALMA, JWST or the E-ELT and dedicated
projects such as LSST and EUCLID will superbly contribute to the
pursuit of these science goals. But, in many cases, the science is
also unique to the radio band, as for instance in the study
of cosmic magnetism. 
\label{ppra}

\vspace{0cm}
\hspace{0.2cm}
\parbox{\textwidth}{\parbox{0.9\textwidth}{\includegraphics[width=16.0cm]{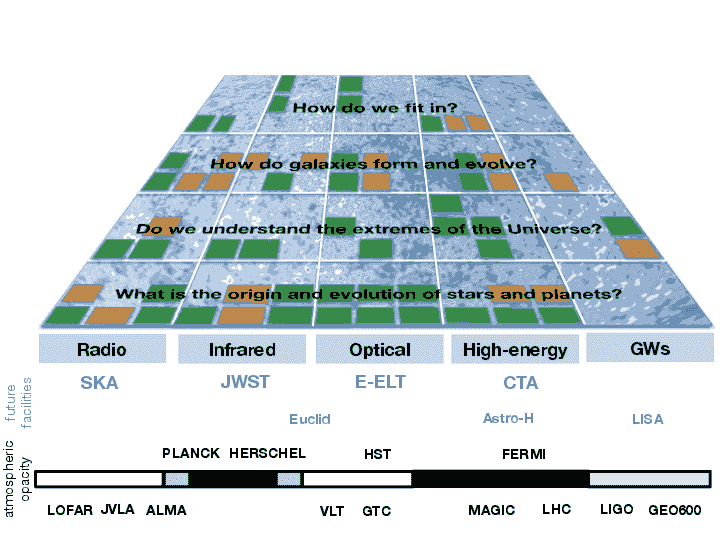}\vspace{0cm}}

\hspace{-0.1cm} 
\parbox{0.9\textwidth}{Figure\,III: Overview of future and recently developed
  or upgraded astronomical facilities. The upcoming flagship
  facilities have been related to the science questions and the recommendations of the ``ASTRONET
  Science Vision''. The science vision indicates 6\,--\,7 science goals
for each main research area, which has been divided into two classes:
Essential (green rectangles), implying that the
scientific goal can-not be reached without this facility, and
complementary (orange rectangles), where the facility will provide very beneficial
information to reach the scientific goal.}\vspace{0.5cm}}\\

The complementarity of the SKA, however, goes far beyond the
exploitation of the Universe using the classical electromagnetic
windows. Naturally, the SKA will also provide excellent matching
insight when studying the gravitational wave sky or when connecting to
the Cosmos with particle physics or Cherenkov telescopes. On one hand,
radio astronomical techniques provide the only evidence so far that
gravitational waves exist, and hence they allow us to study similar
but also additional sources expected to be observed with ground-
and space-based gravitational wave detectors. The detectors {\em
  GEO600, Advanced LIGO} and {\em VIRGO} will observe gravitational
waves from merging compact binary systems like double neutron stars or
binary black hole systems, while {\em LISA/NGO} will observe, for
instance, extreme mass-ratio inspirals. By comparison, the SKA will
detect and study the gravitational waves from super-massive binary
systems (see also contributions in this ``White Paper''). Hence, the
SKA is a logical step towards securing Germany's leading role in
gravitational wave astronomy.

Furthermore, radio photons are those with the lowest energies, but
they often result from the most energetic processes known. Radio
emission is therefore indeed often associated with those sources that
are visible at the highest energies with Gamma-ray satellites like
{\em FERMI} or ground-based Cherenkov-telescopes. Some of the
particles observed with such detectors have energies a hundred million
times greater than that achievable by terrestrial accelerators, and
their observation raises several questions; How can cosmic
accelerators boost particles to these energies? What is the maximum
energy achievable by galactic sources such as supernova remnants,
neutron stars, or micro-quasars? How do they propagate through the
Universe?  Does the cosmic ray energy spectrum extend beyond the
maximum energy a proton can maintain when travelling over large cosmic
distances, as they would eventually collide with the omnipresent
microwave background? (questions obtained from
http://www.aspera-eu.org). These questions can be addressed by new
facilities like the {\em Cherenkov Telescope Array} (CTA) where
Germany is also playing a leading role. Indeed, many of the sources
for the photons and particles observed with the CTA are indeed
uniquely observable with radio telescopes and, in particular, with the
SKA. Pending the siting decision for the CTA, there is a large chance
that both SKA and CTA will be collocated in Southern Africa, providing
unparalleled possibilities in studying the high energy sky. A prime
example of such a successful combination of experiments is given by
the LOPES array (LOFAR PrototypE Station;
http://www.astro.ru.nl/lopes/), that operates radio antennas and
photomultipler technology, from particle physics, to perform
coincidence measurements of air showers that are produced by cosmic rays
when they enter the Earth's atmosphere.

A fundamental question that requires the interplay of radio
astronomy, optical astronomy and particle detectors and accelerators
is, of course, that about the nature of dark matter.  First evidence for dark
matter has been obtained from the kinematics of stars in the Galaxy as
revealed by ground-based optical observations in the first third of
the 20th century (Oort 1932) and the kinematics of galaxies in
clusters (Zwicky 1933). Since then, dark matter has become the keystone
of the standard cosmology model based on much wider evidence than
optical astronomy alone, in particular from hydrogen observations of
galaxies in the radio regime. Based on the latest measurements from
the {\em Wilkinson Microwave Anisotropy Probe} (WMAP; satellite
operating at infrared- and radio frequencies) only 4\,\% of the Universe
is made of ordinary matter. Whereas 73\,\% of the cosmic energy budget
seems to consist of dark energy and 23\,\% of dark matter.

The ultimate answer on the nature of dark matter will likely come from
the observation the exotic particles that constitute dark matter. These
particles may be first observed in subterranean laboratories, by the
planned detectors recording the nuclear recoils due to the impact of
dark matter particles (``direct detection''). Alternatively, signs of
dark matter particles may arise as products of their annihilation in
celestial bodies and may be detected by Gamma-ray telescopes
(e.g. the Cherenkov Telescope Array) at ground level or in space, by 
{\em neutrino telescopes} deep
underwater or in ice, or by cosmic ray spectrometers in space
(``indirect detection''). Discoveries based on particle physics
technology will have immediate consequences on our understanding of
the Universe and, vice versa, discoveries in radio astronomy by the SKA
will have fundamental impact on theories of the infinitely small.
Some of the subjects where the SKA will impact in particle physics is
discussed in Section\ \ref{fuphy} on fundamental physics.\\

\noi In summary, like all new major facilities the SKA will not only
boost scientific output it will also impact on astronomy and natural
science in general.  In addition, the massive increase in sensitivity
and survey speed will not only detect every active radio galaxy in the
Universe but it will also trigger a revolution in how
research is done in radio astronomy that will reverberate in
other areas of astrophysics and fundamental science.
Overall, Germany's contribution
to the SKA will help cement a leading position for Germany's
astronomy community.\\

\newpage

\section{The German SKA community \& the GLOW consortium}
\label{glow}

\noi In order to estimate the size of the German science community
that would benefit from the SKA, three different parts to that
community are considered. Universities and research institutes that
have a radio-astronomical division or sub-division, institutes with
radio-astronomical interests or that use radio astronomical
instruments, and institutes that would conduct science using the
SKA. These different groups are considered and their numbers are discussed
below.

In the first group, the number of people can be estimated
directly, whereas for the second group individual scientists can be 
counted, who either directly work on radio data or make use of
science results obtained by current radio facilities. In particular,
people involved in the ``GLOW-Konsortium'' can be named. The
third group reflects the fact that the SKA will be a general purpose
science facility that will be accessible as a general purpose observatory.
The third group therefore made out of scientists, who do not
consider themselves as astronomers, but who make use of the information
and the data provided by the SKA to study fundamental science questions.

The example of ALMA, however, underlines impressively our statement in
the previous section that it is not appropriate anymore to divide the
astronomical community according to wavebands.  Indeed, the number of
authors of ``ALMA Cycle-0'' proposals employed in Germany exceeds the
number of people who would be considered as radio astronomers or
``classical'' ALMA users. It shows that if a new instrument promises
to provide new answers to scientific questions, the distinction into
the classical observing bands (e.g. optical-, infrared-, or
radio-bands) does not hold anymore as, in fact, nowadays a
multi-disciplinary approach is needed to answer our science
needs. According to our definition above, the current ALMA users could
and should be added to our second group of potential SKA
users. Following
these considerations, we proceed as outlined.

\noi {\bf \small Radio-astronomical institutes or sub-divisions:~~}The
following institutes host a ``Radioastronomical Group'' that either do
research in radio astronomy, technical development, or develop/run a radio
observatory (in alphabetic order).

{\small
\begin{itemize}
\item[--] \airub
\item[--] \mpifr
\item[--] Geod\"atisches Institut der Universit\"at Bonn (GIUB is part
  of FGS)
\item[--] Wettzell Observatorium (Forschungsgruppe Satellitengeod\"asie
  (FGS))
\item[--] 1. Physikalisches Institut der Universit\"at zu K\"oln
\item[--] \aip
 \end{itemize}
}

\noi The total size of the group of $\sim$\,270 people is based on
counting institute member with a tenture track, postdoctoral, or
student position. In addition to this number one could add people
working in the ESO ALMA ARC's, which would
result in a total of $\sim$\,280 people .\\

\noi {\bf \small Institutes with radio-astronomical interests:~~} In this
group are institutes that make use of radio facilities or run an
observatory without technical research and development. In particular
institutes of the German Long-Wavelength Consortium (GLOW) are mentioned
here:

{\small
\begin{itemize}
\item[--] \ubie
\item[--] \airub\ (not counted again, as counted already above)
\item[--] \aifa 
\item[--] \mpifr\ (not counted again, as counted already above)
\item[--] \jbremen
\item[--] \mpa
\item[--] Exzellenz Cluster ``Universum'' Garching
\item[--] \hhstw
\item[--] \juelich
\item[--] 1. Physikalisches Institut der Universit\"at zu K\"oln (not counted again, as counted already above)
\item[--] \aip\ (not counted again, as counted already above)
\item[--] \stwtt
 \end{itemize}
}

\noi From these institutes an additional $\sim$\,30 people have been counted. \\

\noi Surveying the publicly available observing schedules of radio
telescopes like Effelsberg, EVN, Parkes, VLA and Westerbork a further
set of 10\,--\,20 people has been identified who work actively on radio
data (e.g. colleagues from the ``Landessternwarte Heidelberg'' or
the ``Max-Planck-Institut f\"ur Astronomie'' are very active
users). We consider this latter number as a conservative lower estimate.\\

\noi {\bf \small Institutes with interests in SKA research and
  development:~~}The SKA will not only be a general purpose
observatory, but with its technical challenges and large-volume data
products it will also be a mega-science infrastructure that will
dramatically change the way research is achieved. The requirements of
the SKA will push toward new research areas in eScience, data
handling, as well as digital electronics, and energy consumption and
solar energy storage. Despite this wide appeal, at the current time an
estimate of the science (and industry) community related to those
questions is difficult to give. However, as an extremely conservative
lower bound for the size of the third group of the SKA community, we
can at least count those colleagues who are interested in fundamental
science questions related to research and development in physics and
astronomy, who also contributed to this ``white paper''. Obviously, the
real number is likely to be larger by a factor of a few, at least. For
now, we count:

{\small
\begin{itemize}
\item[--] \zaa
\item[--] \zarm
\item[--] \fau
\item[--] Fraunhofer-Institut f\"ur Solare Energiesysteme ISE Freiburg
\item[--] \tguf
\item[--] \ugoett
\item[--] \mpik
\item[--] \tpi
\item[--] \mps
\item[--] \lmum
\item[--] \oldenburg
\item[--] \mpiaeip
\item[--] \tat
 \end{itemize}
}

\noi A conservative measure for the number of people who have not been taken
into account in any of the previous groups would be of the order of
(at least) 50~people.\\

\noi {\bf \small In conclusion~}this community is far larger than
if one only counts the ``classical'' radio-astronomical institutes in
Germany.  Adding the previous estimates and parts of the ALMA
community, which partially overlaps, a conservative evaluation of the
size of the {\bf  \small German SKA community} results in
more than {\bf \small 400 individuals}.\\

\subsection{The German Long Wavelength Consortium and its
SKA Working Group {\scriptsize [M. Hoeft]}}

\noi The German Long-Wavelength Consortium (GLOW) aims to foster radio
astronomy in Germany. Twelve German astronomy institutes form a
consortium for exchanging knowledge, bundling efforts and coordinating
activities in radio astronomy. Initially focussing on low-frequency
radio astronomy and its exploitation with the {\em Low Frequency
  Array} (LOFAR), GLOW has recently formed an SKA working group that
will act in a role as an organising entity, seeking to coordinate the
overall German SKA interests. Given the impact of the SKA on areas
outside radio astronomy or astronomy as a whole, GLOW offers
membership in this working group to all interested parties from
science and industry.

LOFAR with its core in the
Netherlands, is a unique telescope for very long wavelengths.
GLOW member institutes strive to enhance LOFAR by building and
operating LOFAR stations in Germany and by participating in the
LOFAR Key Science projects (KSPs). GLOW coordinates these
LOFAR-related activities. Similarly, GLOW will provide a framework
for coordinating SKA related activities in Germany.

GLOW is an open consortium which can be joined by any institute
which subscribes to the ``Memorandum of Understanding'' (MoU).  As an initial
step into the LOFAR project the German ``LOFAR white paper''
(eds. Br\"uggen, et al. 2005) was compiled in 2005, summarising
the scientific and technical interests in the LOFAR telescope. In 2006
ten institutes, which aimed to extend LOFAR into Germany by
building several stations, formed the German Long Wavelength
Consortium. Among the founding institutes were ``major players'' in
radio astronomy such as the ``Max-Planck-Institut f\"ur
Radioastronomie'' (MPIfR, Bonn) and also ``newcomers'' such as the
Th\"uringer Landessternwarte (TLS, Tautenburg).  At this time only the
MPIfR had started to build a LOFAR station while other institutes were
planning to do so. Currently, five stations have been completed and
are in operation. A sixth station is in preparation. LOFAR crucially
relies on information and communication technologies. The
Forschungszentrum J\"ulich (FZJ) is interested in technological
aspects of LOFAR, hence, it also joined GLOW. The GLOW members
assemble annually for an intense exchange with regard to scientific
and technological topics. An Executive Committee, a Science Working
Group, and a Technical Working Group conduct the GLOW activities
between the annual meetings.

When the German institutes joined LOFAR they also established two
new KSPs, namely one on ``Cosmic Magnetism'' and one on ``Solar
Physics and Space Weather''. The two KSPs significantly enhance
the scope of LOFAR, e.g. the Cosmic Magnetism KSP uses Rotation
Measure Synthesis for studying magnetic fields in the cosmos.
Hence, suitable methods need to be implemented in the LOFAR
data analysis pipelines. A substantial amount of man-power is
needed for developing the software, conducting commissioning
observations, and analysing the huge amounts of scientific
data. GLOW has coordinated efforts for writing proposals. Most of
them have been granted, in particular, a DFG research group (FOR
1254, ``Cosmic Magnetism'') and projects in the framework of the
German Verbundforschung (D-LOFAR I+II). The
``Verbundforschungs''-grants also allowed two of the German
LOFAR stations to be built.

The International LOFAR Telescope (ILT) is based on National
Consortia. Therefore, a crucial task of GLOW is to provide the
link between the ILT and station owners and KSP members in
Germany. For instance, GLOW nominates a representative for the
Board of ILT and balances contributions to LOFAR and scientific returns for
station owners. Another crucial German contribution to LOFAR --
besides building and operating stations -- is an archiving facility
offered by FZJ to the LOFAR Long Term Archive.

GLOW also strives to educate young scientists in radio
astronomy. To this end two schools have been organised (2010 in
Hamburg and 2012 in Bielefeld) at which students were
introduced to radio interferometry by lectures and hands-on
tutorials. Moreover, GLOW members organize scientific meetings focussing
on low frequency radio astronomy, e.g. the International LOFAR
Workshop in Hamburg (2008) and Splinter meetings at the annual
meetings of the German Astronomical Society.

Currently, most activities of GLOW are related to LOFAR. However,
GLOW has a more general mission, namely to foster radio
astronomy. Consequently, GLOW has established a coordinating SKA
Working Group that will serve as the binding element between the
radio astronomical community, the SKA community at large as well
as industrial and political  partners.

\newpage
\pagestyle{empty}
\noi{\bf }

\newpage
\pagestyle{plain}
\section{The SKA project}

\subsection{History, Governance, Timeline \& Top level management
 structure}

\noi Since the early 1980s the need for a bigger telescope has been
discussed in the radio astronomical community triggering e.g. the
construction of the Giant Metrewave Radio Telescope (GMRT) in India. 
Around 1990 these plans were developed by considering even larger
arrays (Wilkinson 1991; Noordam et al. 1991). Since then the Square
Kilometre Array has evolved over the years from a purely ``hydrogen
array'' observing at frequencies of 1.4\,GHz and below, to a
multi-facetted science facility covering a frequency range from about 50\,MHz
to 10\,GHz or possibly 25\,GHz, capable of answering many of
the major questions in modern
astrophysics and cosmology.\\

\noi In September 1993 the International Union of Radio Science (URSI)
established the Large Telescope Working Group to begin a worldwide
effort to develop the scientific goals and technical specifications
for a next generation radio observatory.  Subsequent meetings of the
working group provided a forum for discussing the technical research
required and for mobilising a broad scientific community to cooperate
in achieving this common goal.  In 1997, eight institutions from six
countries (Australia, Canada, China, India, the Netherlands, and the
USA) signed a ``Memorandum of Agreement''
(MoA) to cooperate in a technology
study programme leading to a future very large radio telescope.  In 2000
a ``Memorandum of Understanding'' (MoU) to establish the ``International Square
Kilometre Array Steering Committee'' (ISSC) was signed by
representatives of eleven countries (Australia, Canada, China,
Germany, India, Italy, the Netherlands, Poland, Sweden, the United
Kingdom [UK], and the USA). This was superseded by a MoA
to collaborate in the development of the Square Kilometre
Array which came into force in 2005 and which has been
extended until 2007.  It made provision to expand
the ISSC to 21 members (7 members from each: Europe, USA, and the Rest of the
World) and to establish the International SKA Project Office
(ISPO). 

A further international collaboration agreement for the SKA programme was
drawn up in 2007, which became effective on 1 January 2008. It was
signed by the European, US, and Canadian SKA Consortia, the Australian
SKA Coordination Committee, the National Research Foundation in South
Africa, the National Astronomical Observatories in China, and the
National Centre for Radio Astrophysics in India. This agreement
established the SKA Science and Engineering Committee (SSEC) as a
replacement to the ISSC. The SSEC acts as the primary forum for
interactions and decisions on scientific and technical matters for the
SKA among the signatories to the International Collaboration
Agreement. A further agreement was drawn up in 2007, a MoA
to establish the SKA Program Development Office (SPDO).\\

\noi In 2011 a call was launched to host the SPDO
headquarter. Germany, the Netherlands and the UK applied. Due to
the very strong support by their government the United Kingdom was
chosen to host the SKA headquarter at Jodrell Bank near
Manchester. In April 2011, the Founding Board was formed.
The Board members, Germany, the UK, the Netherlands, Australia,
New Zeeland, South Africa, China, France and Italy carried out
the preparatory work for the SKA Organisation which was formed in November 2011.
 At that time, seven national governmental and research
organisations announced the formation of the SKA Organisation, as an
independent not-for-profit company (Ltd.)  established to formalise
relationships with international partners and centralise the
leadership of the Square Kilometre Array (SKA) telescope project. The
founding signatories are Australia (Department of Innovation,
Industry, Science and Research), China (National Astronomical
Observatories, Chinese Academy of Sciences), Italy (National Institute
for Astrophysics [INAF]), the Netherlands (Netherlands Organisation
for Science Research [NWO]), New Zealand (Ministry of Economic
Development), South Africa (National Research Foundation [NRF]) and
the United Kingdom (Science and Technology Facilities Council [STFC])
to fund the project in the period leading up to the construction phase
which will start in 2016. In March 2012 Canada via the National
Research Council (NRC) joined the SKA organisation as a full member
and will make financial contributions to the project. In
April 2012 India joined the SKA organisation as the first associate
member. Sweden joined the SKA organisation as a full
member in June 2012. Overall the SKA organisation is the legal entity that will
design, built and operate the SKA.


\noi In early 2012 the SKA organisation received the site evaluation report
and started the process on a resolution to
select a site. The two sites that proposed to host the SKA were:
\begin{itemize}
\item[--]{South Africa (SA) partnering with Namibia, Botswana, Kenya,
    Madagascar, Mauritius, Mozambique and Zambia.}
\item[--]{Australia, together with New Zealand.}
\end{itemize}

\smallskip

In the site decision process the members of the SKA organisation
which are bidding to host the SKA did not take part. Therefore, of
the SKA Organisation the following members were eligible to vote:
Canada, China, Italy, the Netherlands, and the United Kingdom.
The site decision of the SKA was announced in May 2012:
the majority of the members of the SKA Organisation is in favour of a
dual-site implementation of the SKA. The evaluation report concludes
that a scientifically and technically viable approach is to split the
SKA array in frequency space. The long term solution for the ``full''
SKA will therefore be that the low-frequency array will be hosted
in Australia and New Zealand whereas the mid- and high-frequency array
will be placed in South Africa. However during the first phase of the
SKA (\skai ) both sides will host different technical realisations of
the dish array of the SKA. A full account of the various
telescope types and the realisation of the SKA at both sites is
discussed in Section~\ref{tele}.

While the split-site decision certainly keeps the whole global SKA
science and technology community fully engaged, it offers also a
number of advantages. Firstly, the large investments in the SKA
pre-cursors, MeerKAT and ASKAP, now become integral part of SKA Phase
1. This promises to keep the costs for the first phase under
control, as MeerKAT itself already represents a significant fraction
of Phase\,1's collecting area. Similarly, the low-frequency part
of the SKA will now be co-located at the MWA site, which is arguably
also the better site for very low-frequency radio astronomy. It was
also decided that the final approval for Phase\,2 construction depends
on the continuing pristine radio interference-free conditions of the sites. Having two sites
prepared for Phase\,2 constructions promises a reliable path into the
future, as one site can always act as a back-up site for the other.
Finally, it should be noted that the plan for South Africa to host the high- and
mid-frequency array of the SKA will open up an additional science
possibility for Germany (and Europe in general) as it will be possible
to use very long baseline interferometry, 
connecting the Effelsberg telescope (the biggest single dish telescope
in Europe) with the SKA. It is also worh noting that the European
Parliament has called for greater collaboration with Africa in the
field of radio astronomy, following its adoption of ``Written
Declaration\ 45'' on science capacity building in Africa.

\bigskip
\bigskip

\parbox{\textwidth}{ \parbox{0.9\textwidth}{ \includegraphics[width=0.95\textwidth,
    angle=0]{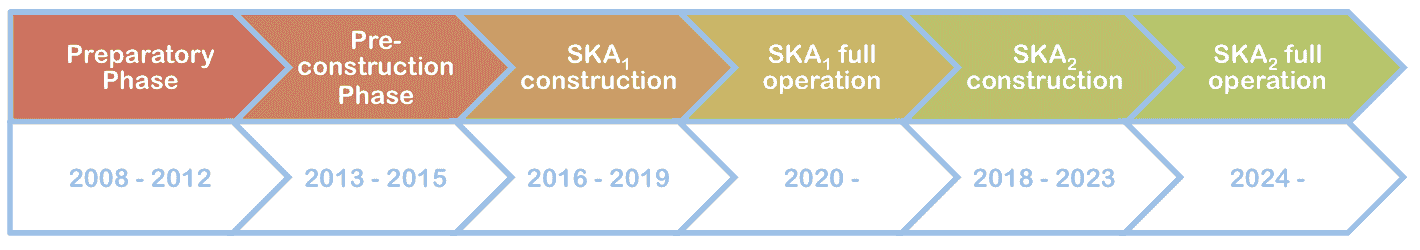}

\hspace{0.3cm}
\parbox{0.9\textwidth}{Figure IV: SKA overall project timeline}
\vspace{0.0cm}}}\\

\noi The overall SKA project timeline of the next decade can be
divided into six major periodes (see Figure\ IV). The upcoming
pre-construction phase allows the employment of a formal project
management structure, enabling a top-level description of work and a
breakdown structure of work packages (WP). Furthermore, it will
establish the science specifications and system requirements for SKA
Phase\,1 (\skai ). In addition, the work in the first two periods
(``Preparatory Phase and Pre-Construction Phase'') will be carried out on
the baseline design and the ``Advanced Instrumentation Program'' (AIP)
with the aim of preparing the international project for the start of
construction. The construction of the SKA is planned to start in 2016
allowing for the first science results within this decade. The
schedule to extend the frequency coverage to 25\,GHz at the high end
of frequency range would define the third phase in the construction of
the SKA, but up to this point in time there is no defined schedule for
SKA$_3$. A full overview of the individual programmes within the SKA
project timeline can be found in
the appendix (page~\pageref{appen}).\\

\noi The work of the SKA organisation is overseen by its ``Board of
Directors'' which has the authority to appoint senior staff, to decide
on budgets, to admit new project partners to the organisation and
to direct the work of the global work package consortia. Every SKA
member of the organisation appoints two representatives to the board
of directors. Furthermore, the top level management structure of the
SKA organisation includes the ``director general'', ``the office of
the SKA organisation'' (``the office'' in the following) and various
advisory committees. The SKA project will have a strong central office
with management and system design authority, that will not only
contract the work of major telescope subsystems but also  coordinate a
small number of work package contractors. The work package contractors
themselves will be consortia of participating organisations (PO) and
industrial partners, but could also be individual companies or POs. On
overview on the management structure is illustrated in Figure\ V. The
subjects of the individual work packages cover the following 11 areas:
management of the pre-construction phase and management support, such
as quality assurance, configuration management, and procurement;
science; overall system; maintenance and support; dish sub-systems;
aperture array sub-systems; signal transport and network; signal
processing; software and computing; power; site and
infrastructure. \\

\hspace{0.2cm}
\parbox{\textwidth}{ \parbox{0.9\textwidth}{ \includegraphics[width=0.9\textwidth,
    angle=0]{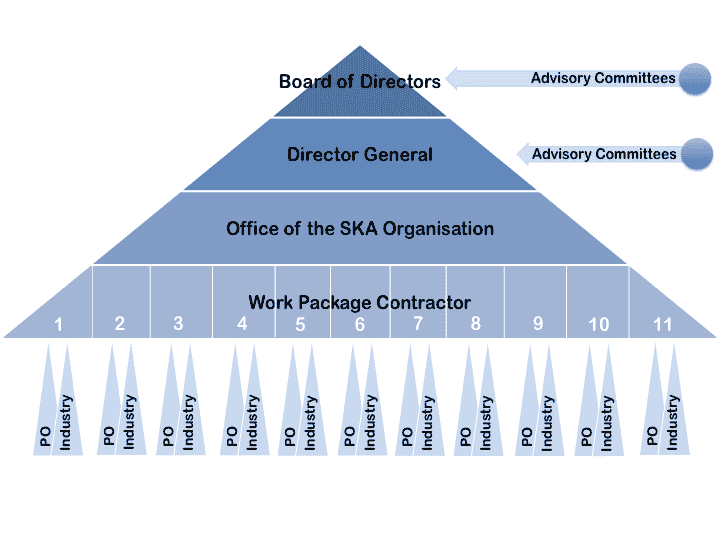}\vspace{-1.3cm}

\hspace{-0.1cm}
\parbox{0.9\textwidth}{Figure V: Schematic overview of the management
  structure of the SKA organisation. Contributions to the SKA project from third
  parties are organised in work packages. The full description of the
  work packages and their different subjects and are discussed in the text. }
}
  \hspace{5.5cm} \parbox{6cm}{}}\\

\bigskip

\noi In general, the timeline and the overall project management indicates that
important decisions will be made in the next years that strongly
influence the technical and scientific capability of the SKA and an
early involvement of Germany in these decisions is needed to secure
their interests in research and development. \\

\subsection{The telescope}
\label{tele}

\noi The SKA will be the World's premier imaging and surveying
telescope with a combination of unprecedented versatility and
sensitivity.  The SKA will continuously cover most of the frequency
range accessible from the ground, ranging from 70\,MHz to 10\,GHz in
the first and second phase of construction and offers the possibility,
at a later stage, to extend the frequency range up to 25\,GHz to
close the frequency gap to ALMA and to be able to observe water masers,
ammonia, etc.  Having
a collecting area of a million square meters (1\,000\,000\,m$^2$), it
will be about 10\,--\,100 times more sensitive than the largest single dish
telescope in Arecibo (305\,m diameter, Puerto Rico), and fifty times
more sensitive than the currently most powerful interferometer, the
Karl G. Jansky Very Large Array (JVLA, in Socorro/USA). The enormously
wide field of view (FoV) is another major advancement, ranging from
200 square degrees at 70\,MHz to at least 1 square degree at 1.4\,GHz
(for reference the full moon covers 0.25 square degrees).  Such a wide
field of view will enhance the speed of observations for large parts
of the sky, particularly at the lower frequencies, allowing for survey
speeds which will be ten thousand to a million times
faster than what is possible today.\\

\bigskip
\bigskip

\parbox{\textwidth}{
\hspace{0cm}
\parbox{\textwidth}{ \parbox{5cm}{\includegraphics[scale=0.277]{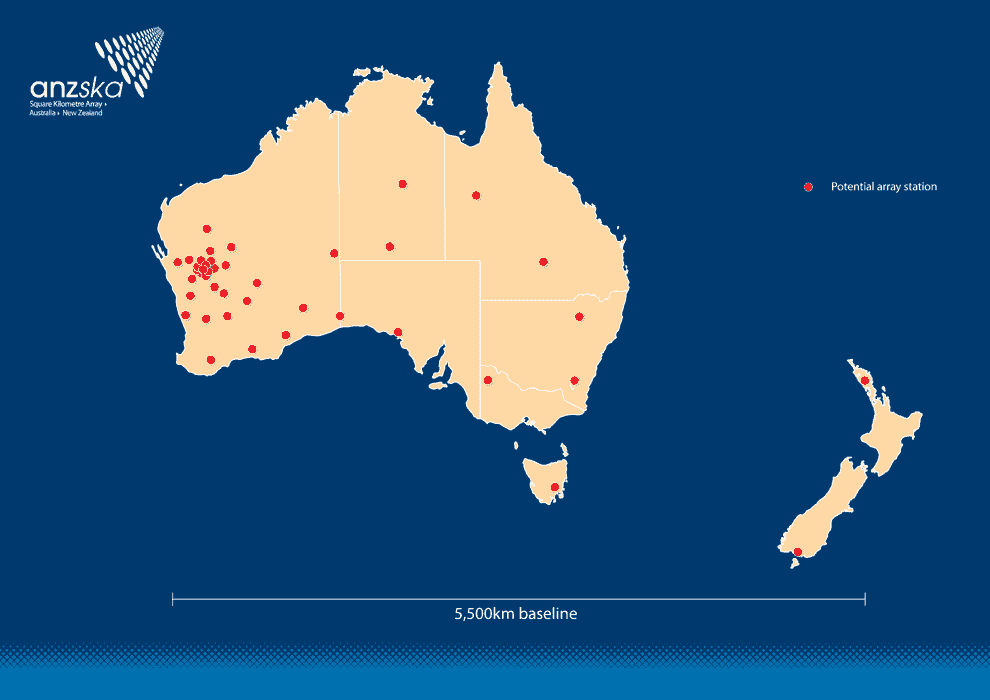}\includegraphics[scale=0.65625]{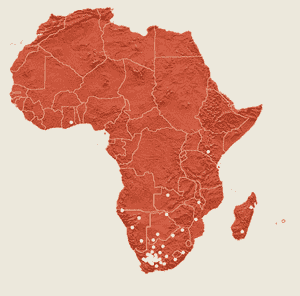}\vspace{0.5cm}}

\hspace{0.1cm} 
\parbox{0.9\textwidth}{Figure VI: Overview of the overall SKA
    configuration. The SKA will have 3 central stations that one built of 
    a core of 1\,km and an inner region of 5\,km in extent. The
    individual antenna stations are distributed along 5 spiral arms
    within a region of 180\,km. Further stations will be distributed
    most likely along 3 spiral arms up to a distance of 3000\,km from the
    core. The different configurations are needed to answer the
    differently motivated science questions, which need to detect
    weak and diffuse radio emission, but also allow for observations at
    high angular resolution. (Image credits:
  see page~\pageref{imcred})}
\vspace{0cm}}
}

\bigskip
\bigskip

\noi In order to achieve both high sensitivity and high-resolution
images of the radio sky, the antennas of the
SKA will be densely distributed in the central region of the array
(``core'' and ``inner region'') and then logarithmically positioned in
groups along five spiral arms, such that each group becomes more
widely spaced further away from the centre. This configuration will
make up the SKA up to distances of several hundreds of kilometres in
diametre ($\sim$\,200\,km)\footnote{Note that in \skai\ this
  configuration may only range from 50 to 100\,km in
  diametre.}. Beyond this configuration a possible layout of 3~arms, also logarithmically
spaced, is anticipated and will extend the SKA baselines up to several
thousands of kilometres. Depending on the type of antennas the
individual groups of the SKA will be made of either single dipoles or
small arrays of dipoles, or single small telescopes. This setup
allows a continuous frequency coverage from 70\,MHz to 10\,GHz. The SKA
high frequency dishes will be either equipped with single pixel
feeds or, at later stage in the project, with phased array feeds
(PAF) to archive a larger FoV for this antenna type. For an overview of
the configuration and the antennas types see the figures.\\

\noi Combining the signals from a single antenna type creates a telescope
with a collecting area of about one square kilometre. For example, in case
of the high frequency array the SKA would require of the order of
$\sim$\,6000 high frequency dishes each of 15\,m in diametre to assemble a
collecting area of 1 million m$^2$.\\

\bigskip
\bigskip

\parbox{\textwidth}{

\hspace{0cm}
\parbox{5cm}{ 
\includegraphics[scale=0.065]{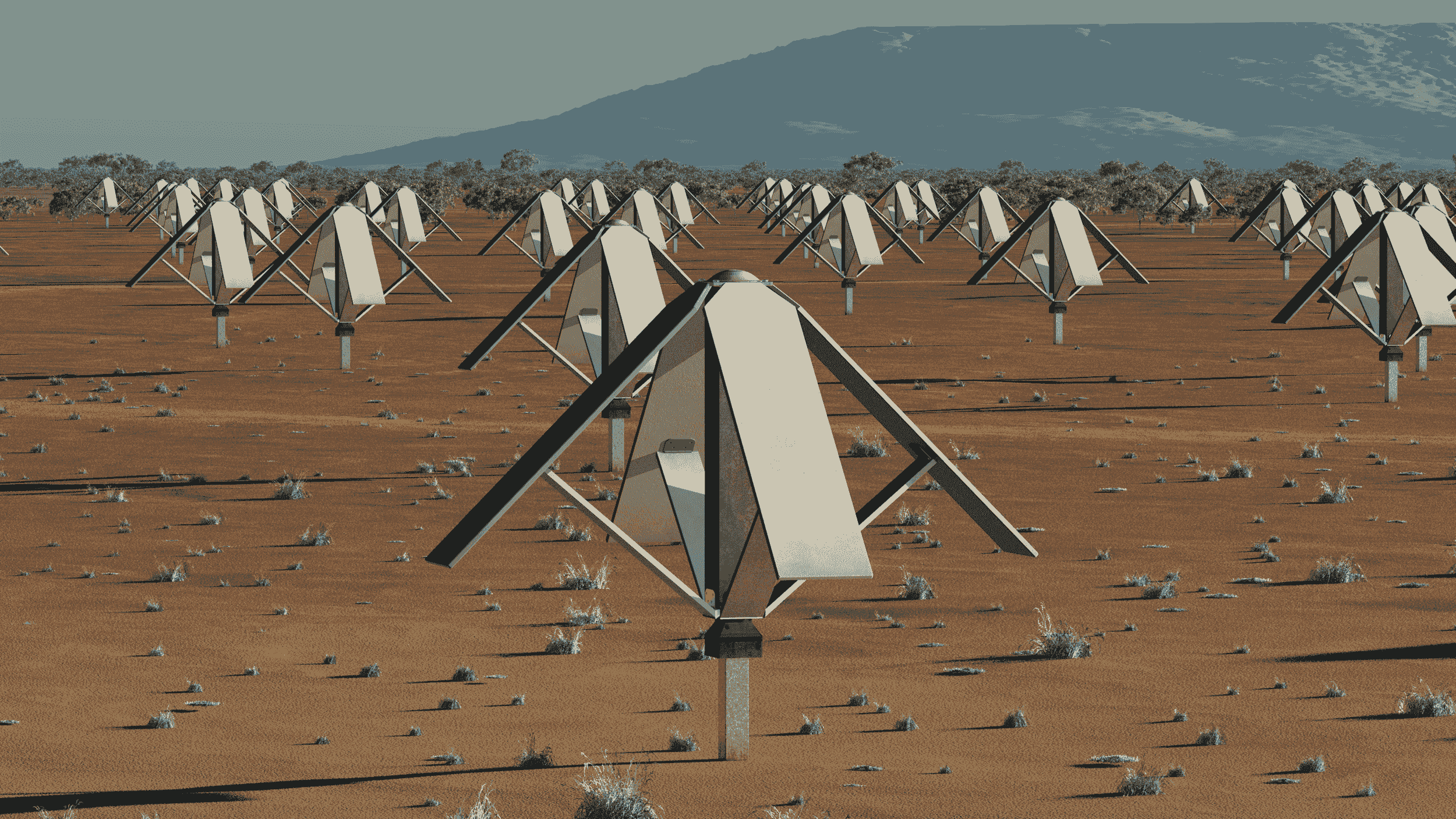}\includegraphics[scale=0.065]{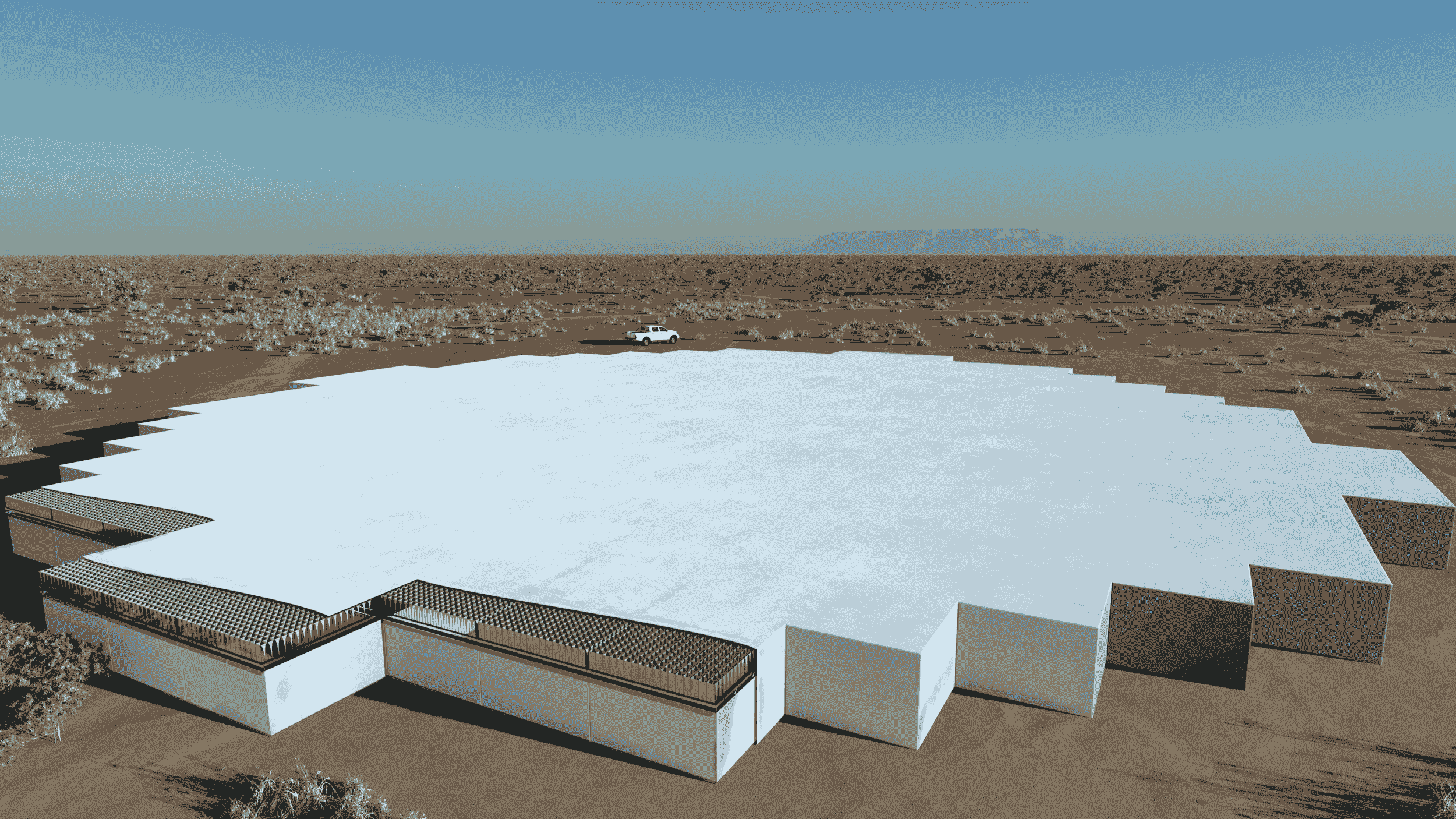}\includegraphics[scale=0.065]{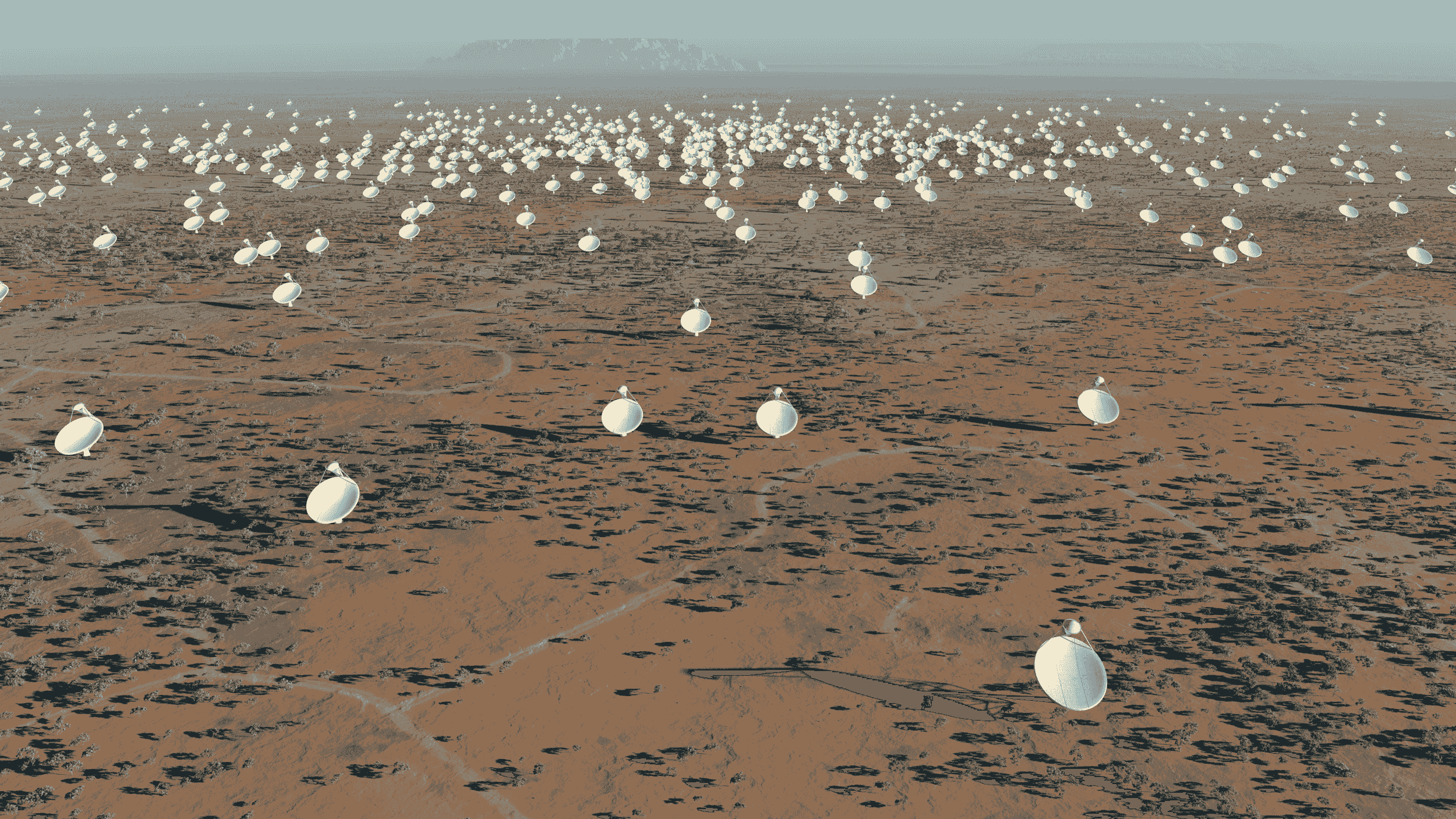}\vspace{0.5cm}}

\hspace{0.2cm}
\parbox{0.9\textwidth}{Figure VII: The three different antenna 
  types (``radio wave receptors'') of the SKA. From left to right:  sparse aperture array (low:
  70\,--\,500\,MHz) used in SKA Phase\,1 \& 2 , dense aperture array (mid:
  500\,--\,1000\,MHz) used in SKA Phase\,2 only, and high frequency
  dishes (high: 500\,MHz\,--\,{\small [3\, GHz \skai ]} 10\,GHz) used in SKA Phase\,1 \& 2. Note that
  the frequency ranges and in particular the edge frequencies of each
  band will vary in the design process. (Image credits:
  see page~\pageref{imcred})}
\vspace{0cm}}

\bigskip
\bigskip
\bigskip

\parbox{\textwidth}{

\hspace{0.1cm}
\parbox{\textwidth}{
\parbox{5cm}{  
\includegraphics[scale=0.11]{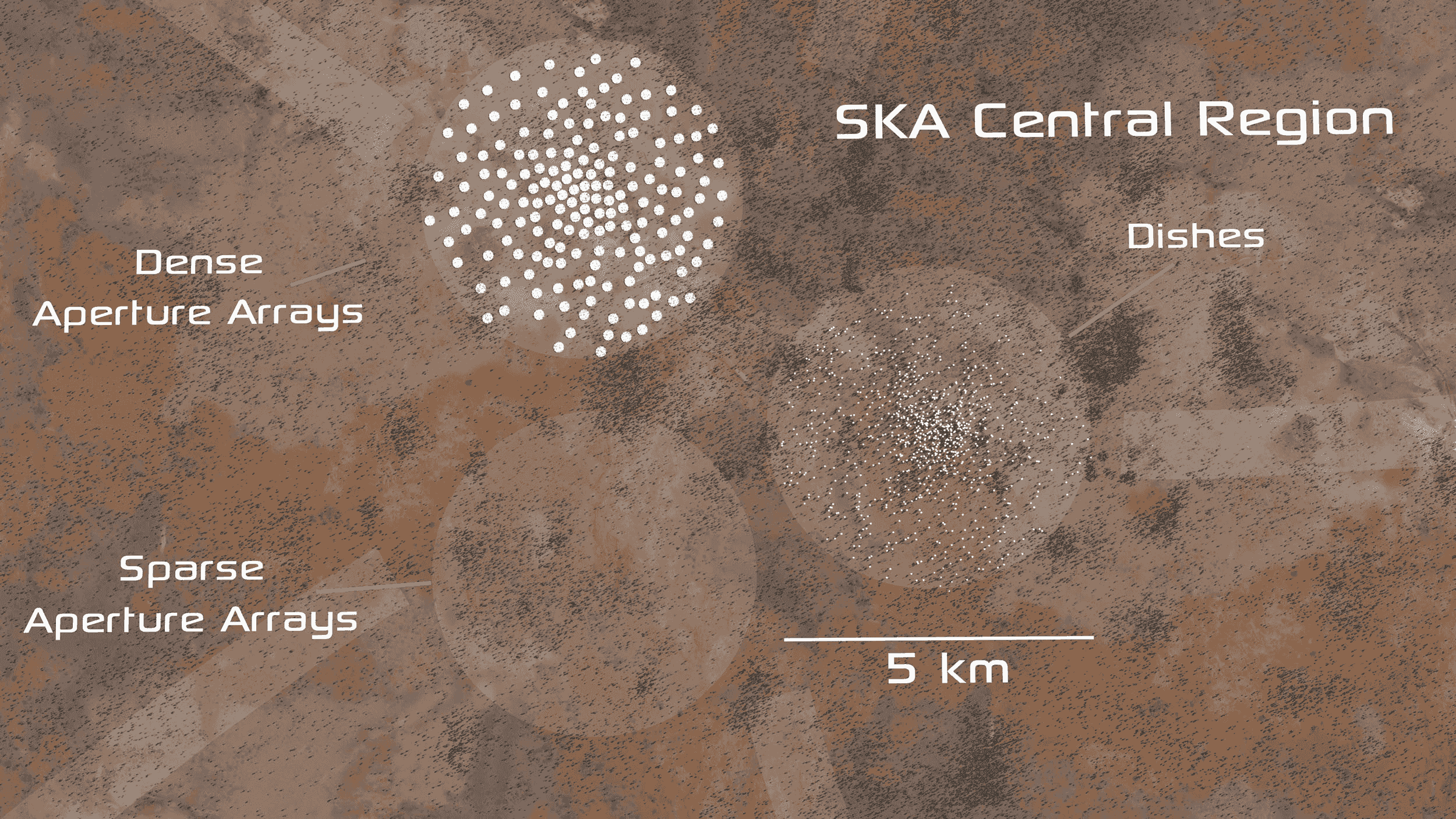}\includegraphics[scale=0.189]{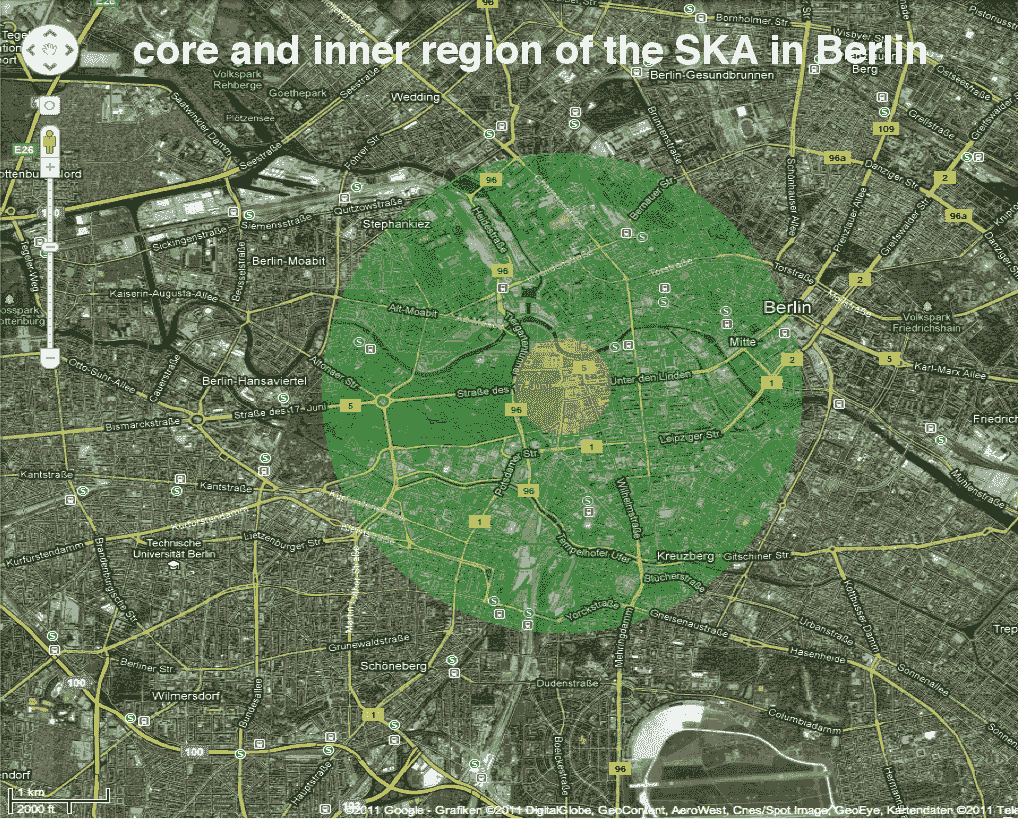}\vspace{0.5cm}}

\hspace{0cm}
\parbox{0.9\textwidth}{Figure VIII: Left:~An example layout of the central region of the SKA. Each of the three
central stations are made up of a  ``core'' and an ``inner region''. The ``core'' and ``inner region'' distinguish themselves by different
distributions of the antennas, a very dense region within 1\,km making up the very
  sensitive core and a less dense distribution within 5\,km.
Right: An example of the size and extent of one SKA central station
  superimposed on Berlin. The station has been centred on the ``Deutsche
  Bundestag'' and would cover parts of Berlin Mitte (including the BMBF), Sch\"oneberg,
  Kreuzberg, and Moabit. The orange region indicates the dense core
  with a diametre of 1\,km, whereas the green area displays a region
  of 5\,km. (Image credits:
  see page~\pageref{imcred})}
\vspace{0cm}}
}

\bigskip
\bigskip
\bigskip

\noi {\bf \small Realising the SKA telescope in two phases \skai\ and
  \skaii :~~} The construction of a radio telescope with a collecting
area approaching one million square metres across a wide frequency
range is a major undertaking and is planned to be implemented in
phases in order to ensure technical readiness and to spread and
control the cost impact.  Therefore, the SKA will be built in two
distinct phases of which the SKA Phase\,1 (\skai ) is a scientific and
technological subset of SKA Phase\,2 (\skaii ), the full SKA.  Of the
three antenna types only the sparse aperture array and the dish array
will be built in \skai . Each of these sub-arrays will make up 10\,\%
of the full collecting area of the SKA.

In general, radio interferometers (aperture synthesis telescopes) have
the advantage over other astronomical facilities (e.g. telescopes operating in the optical band solid mirror are needed) that
they can be built out of blocks (e.g. single dishes or sub-arrays) and
these building blocks can be integrated into a running system at any
possible stage and physical position in the array. Such a setup will
allow for an iterative process in which feedback loops ultimately
trade performance and cost against science return. Furthermore, the
science projects and the technical maturity of various components will
guide the design process from \skai\ to \skaii\ and allows significant
progress toward the major science goals of the full SKA.  Some key
aspect to be defined for \skaii\ is the relative proportion of: dishes
(with or without PAF), sparse aperture arrays,
and the dense aperture arrays in the final system.\\

\noi In this light, developments in the Advanced Instrumentation
Program (AIP) will be continued into the pre-construction phase in
order to develop and elaborate on the technical readiness of PAFs on
dishes, ultra-wideband single pixel feeds on dishes, and the
mid-frequency aperture arrays. This process may also lead to the
construction of a stand-alone demonstrator in the
pre-construction phase e.g. for the mid-aperture array, or opening up
the opportunity for individual research groups to fund new systems on
existing instrumentation relevant to the SKA. The conceptional and
technical outcome of this verification programme will provide a good
understanding of the system performance, in large volume manufacturing, in
deployment and maintenance costs, and risk assessment in order to
have mature and cost-effective components incorporated into the
final SKA design.  Therefore, the design of \skai\ should also allow
for the possibly including PAFs or ultra-wideband feeds, as modular
sub-systems, on the \skai -dishes to enhance the science impact of
\skai .\\

\noi A prototype of a full system design is expected, once the overall SKA
system design and costing exercise is completed at the end of 2012,
prior to the detailed engineering design in the pre-construction phase
(Garrett et al. 2010).

\medskip
\noi The \skai\ timeline is planned as follows:

\begin{minipage}[t]{0.85\textwidth}
-- telescope system design, prototyping and costing (2010\,--\,2012)

\smallskip

-- detailed engineering design \& pre-construction phase (2013\,--\,2015)

\smallskip

-- \skai\ construction, commissioning \& early science observations (2016\,--\,2019)

\smallskip

-- \skai\ Advance Instrumentation Programme (AIP) decision (2016)

\smallskip

-- \skai\ full operation (2020)

\end{minipage}

\bigskip

\noi It should be noted that the \skai\ array already represents the essential radio component of a
suite of ``Origins'' and ``Fundamental Physics \& Discovery'' instruments
now being planned or under construction, including facilities like
ALMA, JWST, ELTs, SPICA, CTA LIGO, eLISA/NGO, LSST, Euclid/JDEM, CMBPOL, GAIA, KM3NeT
and the LHC (running in CERN).\\

\bigskip

\noi {\bf \small The dual-site implementation of the SKA
  telescope:~~}In May 2012 the majority of the members of the SKA
organisations were in favour of a dual-site implementation model
which has been proposed by the Site Options Working Group (SOWG). The
SOWG work shows that a scientifically justified and technically viable
approach is possible, and concludes that in \skai , viable dual-site
implementations exists that not only maintain, but add to, the
scientific appeal of the first stage of the SKA. Taking this into
account the SKA organisations
agreed on the following dual-site implementation:\\

\hspace{0.3cm}
{\footnotesize
\begin{tabular}{l l}
\begin{tabular}{l}
\\
\\
\\
Phase\,1\\
\end{tabular} & 

\begin{tabular}{|p{6cm} |p{6cm}|}
\hline
& \\ 
{\bf Australia} & {\bf South Africa}\\
& \\\hline
& \\
 \skai\  low frequency sparse aperture array & \skai\ mid frequency dish array with \\
& single pixel feed \\
\skai\  mid frequency dish array with & \\
phased array feeds (PAF)  & \\
& \\ \hline
\end{tabular} \\

\begin{tabular}{l}
Phase\,2\\
\end{tabular} & 
\begin{tabular}{|p{6cm} | p{6cm} |} 
& \\
\skaii\  low frequency sparse aperture array & \skaii\  mid/high
 frequency dish array with\\
& single pixel feed \\
& \\
& \skaii\  mid frequency dense aperture array\\
& \\ \hline 
\end{tabular} \\
\end{tabular}}

\bigskip

\noi For an overview on the various frequency ranges and telescope
types see the Figure\ VII on the previous pages.

\subsection{Cost estimation}

\noi This section briefly reviews the cost modelling effort in the SKA
project. For a project such as the SKA, true costing will be
influenced very strongly by political, economic, and commercial
considerations. While every cost model attempts to include some aspects
of commercial drivers (such as economies of scale), some of the
uncertainties can be very large. Therefore, the main focus of
the SPO's work during the pre-construction phase is to firmly establish 
the overall cost envelope, using the detailed input from specialists
and potential suppliers. Finally, the SKA's advantage of being
an interferometer allows one to essentially constrain the cost by
building out the array to only such distances and collecting area,
that are consistent with the initial cost estimates.

\noi The first cost modelling methodologies are described in SKA memos\footnote{The SKA Memo series is available via the SKA homepage and its
  www address can be found on page~\pageref{refs}.} 92 and
93 and permit the analysis of the relative cost/performance of
the realisations of the SKA system as design parameters are varied. The
modelling engine has been developed in a high-level programming
language (python) and aspects of system design and details of the cost
model are implemented as modules of the costing engine. The
modelling also includes a formal uncertainty analysis of these
results. However, the results of course depend on the reliability of
the input models and do not allow for wholesale changes in the
underlying system design. Therefore the initial cost modelling provides
a better guide to how costs scale using different technology routes
rather than providing a full capital cost figure.

The logical structure of the costing of a specific system design is
determined by summing the costs of components by following the signal
path through the system. Appropriate cost models are adopted for
components such as: --~a cost performance model (e.g. cost as a
function of dish diameter and surface accuracy), --~a simple financial
model (e.g. cost as a function of purchase date), --~a model for
expected performance as function of time (e.g. the cost per operation
within computing hardware dropping with Moore's law), and --~economies
of scale.  However a number of aspects were partially or not
included at all in the early cost modelling. These include: non recurrent expenditure
(e.g. development costs), infrastructure, project delivery, management
costs, and operations. Importantly, these constraints mean that it is
not possible to consider, for example, total cost of ownership when
considering different technology options.

However the initial approach provides a very flexible, robust costing engine
with excellent error/consistency checking, Monte Carlo modelling,
facilitates cost reduction exercise, and ``what if'' analysis. The disadvantages are
that the underlying cost modules are not easily accessible to the
user and input from domain experts is more difficult to obtain and to
incorporate. Based on the initial approach {\it a new cost modelling engine
is currently in development}, that will provide cleaner access to different parts
of the model by domain experts and designers. To further facilitate
this, new interfaces are being developed that clearly separates the tools
and the models. Furthermore, the new system will also be able to
support many advanced realisations of the system design.\\

\noi The underlying philosophy of the {\it new costing strategy} is that
cost estimation will be an ongoing and iterative process throughout
the design, the development and the construction stages of the SKA. The confidence
levels attributed to cost estimates are predicated on both the
maturity of the SKA design and the substantiating evidence in support
of the cost estimate. Therefore, in the pre-construction phase confidence levels
in cost estimations for \skai\ will be high as the design matures to the
point that formal invitation to tender documentation is 
prepared. Further work will be required to increase confidence levels
in cost estimations for \skaii .  The work involved in estimating the
project costs will include studies and comparison of costs against
analogous systems as well as verification against precursors and
pathfinders (e.g. MeerKAT, ASKAP, LOFAR). It will furthermore include,
extraction of cost estimations from POs and industry and the
analysis, assessment and administration of that data.  As the project
moves forward, cost estimates will be gathered at opportunities such
as design- and project-reviews. These data will be integrated into the full
SKA estimate with the aim to continuously refine the
uncertainty in the cost estimate.

\medskip

\noi At the moment, the costing for SKA Phase\,1 is being updated, since the engineering
work is expected to generate input to the costing models within the
preparatory- and pre-construction phase. In any case,  the design of
interferometers allows to break down the arrays into building
blocks of individual receptors (e.g. dishes) and therefore reducing the
cost by reducing the number of building blocks could be an option to
fix a targeting cost. However this option will influence the 
sensitivity of the \skai\ and, if required, needs to be evaluated with the
scientific scope of \skai .

The capital cost of the SKA Phase\,1 is fixed at 350~MEuro, including
a significant element of contingency. The targeted cost of the full
SKA Phase\,2 system is estimate to 1.5 Billion Euro (``1.5 Milliarden
Euro''), once there are good estimates of costs for all aspects of the
SKA project a trade-off analyses will be done at all system levels ensuring a
detailed cost calculation, including uncertainty estimates. \\

\noi {\bf \small Risk Strategy and Risk Management:~~} As an advanced
technology project, the SKA faces many risks and uncertainties but
also opportunities, the identification and the management of
uncertainties and opportunities will have to be a focus within the
project management scheme in order to fulfil the project goals. 

Risk management is already an integral part of the management of the
project in the ``Preparatory Phase'' and will continue to occupy a central
position throughout the project lifecycle.  Currently,
the risk management programme is being established at all levels and
within all domains of the project. It will be maintained and managed
by the project managers and system engineers. As the SKA project
progresses, the nature of the risks and the context and environment in
which they will have to be managed will change, and therefore this
risk management process will be reviewed and adapted on a regular
basis. An overall and integrated risk register/database will be
established and maintained and will be the central source of
information on the status of risks and will be used for management
as well as for internal and external communication purposes. The level and
detail of these communications will vary and it will be the
responsibility of ``the office'' management to ensure that the correct level
of communication is achieved with relevant stakeholders. Regular
risk focus events will be conducted such as: brainstorming
sessions, risk assessment workshops, work
sessions during technical design reviews, external review(s) of risk
register/database, structured or semi-structured interviews with
experts, using knowledge, expertise and experience of team,
failure mode and effect analysis (FMEA), and fault tree analysis (FTA).\\

\noi In combination with other aspects such as the management and
engineering processes, this setup of risk management enhances
visibility into the project activities, strengthens decision making
and facilitates the project goals.

\bigskip

\subsection{The SKA key science case}
\label{sec:ss}

The expected science return of an observatory like the SKA will be
overwhelming. To maximize it, the expected science needs to be
captured and translated into science requirements/specifications. This
has been achieved at different
times in the life of the project (e.g. SKA Memo 3, 45, 83; the memo
series can be found via the link provided in Section\ \ref{refs}). The
international community has developed a full ensemble of SKA
experiments and a detailed and compelling science case for the SKA, as
described in detail in New Astronomy Reviews, volume 48 (Carilli \&
Rawlings 2004).

\smallskip

\noi Following an extensive period of consultation with the
international community, the SKA science case has been built around five key
science areas that are either unique to the SKA or where the SKA
will provide the essential knowledge to answer them. Based on these
topics five Key Science Projects (KSPs) have been identified, each of
which contain unanswered questions in
fundamental physics,  astrophysics, or cosmology. \\

\noi The following describes the major objectives of each of the
KSPs. Additional targets may be found in e.g. the science case, the
SKA
Memo 100 and references therein.\\

\noi{\bf \small Probing the dark ages} will focus on the formation of the first
structures/bodies at a time when the Universe made the transition from
a largely neutral
to its largely ionised state today.\\

\hspace{0.4cm}
\begin{minipage}[t]{0.85\textwidth}
-- Mapping out redshifted neutral hydrogen (\hi ) from the epoch of
reionisation (EoR) --

\smallskip

The SKA will use the emission of neutral hydrogen to trace the most
distant objects in the Universe. The energy output from the first
energetic stars and the jets launched near young black holes (e.g. in quasars)
started to heat the neutral gas, forming bubbles of ionised gas as
overall structure emerged. This is called the epoch of reionisation and it
should be possible to map out the signatures of this exciting transition
phase. The frequency range of the SKA
will allow us to detect hydrogen up to redshifts of 20, and therfore
enables us to trace the evolution of the EoR starting with the
footprint of the transition up to a neutral to an ionized Universe, and hence
provide a critical test of our present-day cosmological model.
\end{minipage}\\

\bigskip
\noi{\bf \small Galaxy Evolution, Cosmology and Dark Energy} will aim to
probe the assembly, the distribution and the properties of
the Universe's fundamental constituent, galaxies.  \\

\hspace{0.4cm}
\begin{minipage}[t]{0.85\textwidth}
-- Dark energy via baryonic oscillations traced by the 3-d galaxy
distribution in the Universe --

-- Galaxy evolution as a function of cosmic time (\hi\ observations in
emission and absorption) --

\smallskip

The expansion of the Universe is currently accelerating. This
phenomenon is not understood but described as the result of a
mysterious ``Dark Energy'' that vastly dominates the energy content
of the Universe. One important method of distinguishing between the
various explanations for dark energy is to compare the distribution of
galaxies at different epochs in the evolution of the Universe to the
distribution of matter at the time when the cosmic microwave
background (CMB) was formed. Small distortions in the distribution of
matter, called baryon acoustic oscillations (BAO), should persist from
the era of CMB formation until today. Tracking if and how these
distortions change in size and spacing over cosmic time can then tell
us if one of the existing models for dark energy is correct or if new
theories are needed. For this purpose a deep all-sky survey is needed
to detect hydrogen emission from Milky Way-like galaxies out to
redshifts of about 1 and hydrogen emission and absorption from other
galaxy types up to redshifts of 3 and beyond (redshift is equivalent
to lock back time)\footnote{The todays age of the Universe is
  $\sim$\,13.8\,Giga-years (Gyrs) and observing galaxies at a redshift
  of unity allows to look back in time and investigate the Universe
  at an age of $\sim$\,5.9 Gyrs. At a redshift of 3 the Universe has an age
  of 2.1\,Gyrs.}. Using a ``1-billion galaxy survey'' the SKA will
``slice'' the Universe into different redshift (time) intervals and
hence will reveal a comprehensive picture of the Universe's history.

The same data set will provide unique information about the evolution
of galaxies, how the hydrogen gas was concentrated to form galaxies,
how quickly it was transformed into stars, and how much gas  galaxies
 from intergalactic space acquire during their lifetime.
\end{minipage}

\bigskip
\newpage

\noi{\bf \small Strong Field Tests of Gravity} aims to probe fundamental physics,
challenge general relativity and the nature of space and time.\\

\hspace{0.4cm}
\begin{minipage}[t]{0.85\textwidth}
-- Tests of theories of gravity via binary pulsars with neutron star and
black hole companions --

-- Detection of nano-Hertz gravitational radiation using pulsar timing
arrays --

\smallskip

Pulsars are ideal probes for experiments in the strong gravitational
fields around black holes. We expect that almost all pulsars in the
Milky Way (those pointed towards Earth) will be detected with the
SKA, while several 100s of bright pulsars will be detected in nearby
galaxies. The SKA will search for and find radio pulsars orbiting black holes
(stellar black holes as well as Sgr\,A$^*$) that we can use to trace the
extremely curved space with high precision, and hence enable us to
probe the limits of General Relativity that would not be possible
otherwise.

Regular high-precision observations with the SKA of a network of
pulsars with periods of milliseconds opens the way to the detection of
gravitational waves with wavelengths of many parsecs, as expected,
for example, from two massive black holes orbiting each other with a
period of a few years (resulting from galaxy mergers in the early
Universe). When such a gravitational wave passes by the Earth, the
nearby space-time changes slightly at a frequency of a few nHz
(about 1 oscillation per 5\,--\,20\,years). This wave can be detected as
apparent systematic delays and advances of the clock-like pulsar in
particular directions relative to the wave propagation on the sky.
\end{minipage}

\bigskip

\noi{\bf \small The Origin and Evolution of Cosmic Magnetism} aims to
understand how magnetic fields form, evolve over cosmic times, stabilise
galaxies, influence the formation of stars and planets, and regulate
stellar activity. \\

\hspace{0.4cm}
\begin{minipage}[t]{0.85\textwidth}
-- The rotation measurement  grid --

\smallskip

Synchrotron radiation and Faraday rotation have revealed magnetic
fields in our Milky Way, nearby spiral galaxies, and in galaxy
clusters, but little is known about magnetic fields in the
intergalactic medium. Furthermore, the origin and evolution of
magnetic fields is still unknown. The SKA will measure the Faraday
rotation towards several tens of million polarised background sources
(mostly quasars), allowing us to derive the magnetic field structures
and strengths of the intervening objects, such as, the Milky Way,
distant spiral galaxies, clusters of galaxies, and in intergalactic
space. This will provide essential information to interprete the
apparent source locations of high energy sources observed with
particle detectors.
\end{minipage}

\bigskip

\noi{\bf \small The Cradle of Life} will investigate all aspects of astrobiology.\\

\hspace{0.4cm}
\begin{minipage}[t]{0.85\textwidth}
-- Planet formation in proto-planetary disks --

\smallskip

The SKA will be able to probe the habitable segment of proto-planetary
disks and to detect the thermal radio emission from centimetre-sized
``pebbles'', which are thought to be the first step in assembling
Earth-like planets.  The prebiotic chemistry - the formation of the
molecular building blocks necessary for the creation of life - occurs
in interstellar clouds long before that cloud collapses to form a new
planetary system. Therefore observing prebiotic molecules as a probe for
primordial Earth-like conditions and generating movies of planet
growth and hence informing on our origins is in reach with the SKA. 
\end{minipage}

\bigskip

\noi In addition to the KSPs the {\bf \small Exploration of the
  Unknown} is very much a theme that will be pursued by constructing
the SKA as a multi-purpose observatory.  If history is any example,
the most exciting science done with the SKA may not come from
answering the known science questions listed above, but by addressing
new questions raised by unexpected discoveries enabled by a paradigm
change of observations made possible with the SKA.\\

\noi The large German interests in the SKA science scopes were
established and recognised very early in the project and resulted in
the leadership of SKA KSP programmes and key science topics. In
addition to these research programmes,  Chapter\ \ref{gss} provides a
full record of the scientific interests of the German SKA community.

\subsubsection{The major science goals for SKA Phase\,1}

The SKA will undergo a significant construction phase. While many
aspects of the SKA Key Science projects require the capability of the
full SKA,  Phase\,1 of the SKA will already provide a world-class facility
that is unrivalled in its capabilities. For this reason and thanks to the
modular nature of radio interferometers, important science topics that
do not require the sensitivity, angular resolution nor frequency
coverage of the full array have been identified as ``headline science
topics''  for \skai . This ensures early breakthrough science in the
project, but the truly transformational science return across
all the Key Science projects will need the full capabilities of
the \skaii .\\

\noi The following {\bf \small headline science topics}  which drive the technical
specifications for the SKA Phase\,1 have been identified:

\begin{itemize}	
\item[--] Understanding the history and role of neutral hydrogen in the
  Universe from the dark ages to the present-day.

\item[--] Detecting and timing binary pulsars and spin-stable millisecond
  pulsars in order to test theories of gravity (including General
  Relativity and quantum gravity), to detect gravitational waves
  from cosmological sources, and to determine the equation of state of
super-dense matter.
\end{itemize}

\noi In case of the {\bf \small neutral hydrogen} (\hi ) the expected transition from  \skai\
to \skaii\ science is such that the \skai\ will make the first
statistical detection 
of the epoch of reionisation  (EoR) and delineate the \hi\ content in galaxies in the post-EoR
Universe. The statistical measurements of the power
spectra of galaxies between redshift z\,=\,7 and z\,=\,0 will rival, and potentially
exceed that of other techniques, and it will provide the only
measurements of large-scale structure in the dark ages (z\,=\,13 to 20).

In the EoR and dark ages studies, the chief
\skaii\ science driver will be to move from the limited resolutions and
sky areas observable with \skai\ to higher resolution (that needed
to map ionised structures directly associated with quasars and
star-forming galaxies) over a large fraction of the sky, requiring the
significant ($\sim$\,10\,--\,100) planned gains in mapping speed.  This will
provide a definitive, in parts cosmic-variance-limited, map of the EoR
and ``dark ages'', just as WMAP has provided and Planck will soon provide
for the CMB allowing for a battery of new cosmological tests.

In the post-EoR Universe, the main science driver will be
to use the results of the advanced instrumentation programme  (AIP) to,
in going from \skai\ to \skaii , enhance
the mapping speed of the z\,$\sim$\,2 Universe by a factor $\sim$\,100\,--\,10\,000
(depending on the adopted AIP technology). This will allow the ``all-sky''
and thresholded ($\sigma >$\,5) ``billion galaxy'' surveys needed to address key questions such as neutrino mass, that is
measurable to the lowest limit allowed by particle physics experiments
at \skaii\ sensitivity, and sub-per-cent accuracy on the dark energy
$w$ parameter; both are provided by \skaii\ galaxy power spectra (in
several independent redshift bins) achieving high
signal-to-noise-ratio on features due to Baryon Acoustic
Oscillations (BAO), and allowing marginalization over galaxy bias through
accurate measurement of velocity-space distortions. \\

\noi In the {\bf \small pulsars} case, the main science driver for
going from \skai\ to \skaii\ is to achieve the full planned increase
in sensitivity. For pulsar surveys, \skaii\ will then deliver the full
census of $\sim$\,30\,000 normal and $\sim$\,3000 millisecond pulsars in
our galaxy, with the concomitant increase in chances of finding the
rare ``holy grail'' systems, and almost certainly the first known
pulsar-black hole system. Such systems can be used to make definitive
tests of the ``Cosmic Censorship Conjecture'' and the ``No-Hair
Theorem''. In this context, the highest-frequency coverage of the
\skaii\ is crucial to detect pulsars in the vicinity of the Galactic
Centre and allows for precision timing of pulsars orbiting Sgr\,A$^{*}$. Pulsar timing experiments will also probe the equation of
state of nuclear matter at extreme densities. The increase in timing
precision of \skaii\ over \skai\ is the main science driver here
because the sensitivity increase needs to allow timing (every
$\sim$\,10\,--\,20 days) of all the millisecond pulsars to the
$\sim$\,100\,ns time-of-arrival precision needed for their use in a
Pulsar Timing Array (PTA). The science that can be done with such as ``\skaii -PTA'' can go far beyond detection of a cosmic background of
gravitational waves (that should be already within reach of \skai ):
e.g. experiments to measure or constrain the spin and mass of the graviton; and,
crucially, the ability to pinpoint individual gravitational wave
sources, with this capability requiring astrometry to provide
distances, and hence long (few-1000\,km) \skaii\ baselines.\\

\noi Appart from the major science goals, the \skai\ offers the possibility
for the {\bf \small magnetism} KSP to directly image close-by galaxies and to
piggy-back on the EoR surveys in order to map rotation measures
from individual galaxies or our Milky Way at low spatial
resolution. Whereas the additional sensitivity and resolution of
\skaii\ (cf. \skai ) will enable far denser grids of rotation measures
to be observed, allowing higher spatial sampling of the magnetic field
structure in both external galaxies and clusters of galaxies, as well
as the Milky Way. In addition it will provide a sensitive probe of the
small scale structure in the outflows of nearby giant radio
galaxies, AGN and protostellar objects, as well as increasing
sensitivity to magnetic structure in the high redshift Universe.
These considerations are crucial not only for magnetism science but
also for successful completion of other Key Science, such as
constraining the EoR where accurate characterisation and correction of
polarised foregrounds is key. In addition, the magnetic field KSP
places high demands on the calibration and the polarisation
performance of the full SKA such that it will impact on the other
KSPs.\\

\noi In the case of the {\bf \small cradle of life} KSP the system
specifications of the SKA Phase\,1 offers only very limited science
return. The main science driver of this KSP is to probe the
terrestrial planet zones in the nearest circumstellar disks in order
to map out the distribution of complex organic and potentially
biological molecules. Therefore this KSP requires the largest
frequency coverage (potentially up to 25\,GHz) and the highest
sensitivity capabilities of the SKA. This enables us to detect the
molecules and together with the best possible angular resolution, on
(sub-)milliarcsec scale (which requires dishes separated by a few
thousand of kilometres), enables us to probe the terrestrial planet
zones in the nearest circumstellar disks in order to generate movies of
planet growth and hence provide information about our origins.\\

\newpage

\section{German involvement in technical and scientific pathfinder
  programmes of the SKA}

\noi In the different phases of the SKA project various 
organisations have been or are currently involved in precursor and
organisation studies such as SKADS, PrepSKA, GO-SKA and in pathfinder
telescope projects like LOFAR, ASKAP, or MeerKAT in order to investigate
many of the technical design issues underlying the SKA system design,
and to develop some of the science themes and techniques necessary for
the SKA. In addition, the SKA-relevant design greatly benefits from
the knowledge that is generated by the newly upgraded
facilities like the JVLA, eMERLIN, and eEVN.

\medskip

\noi Here a basic overview of these SKA studies and the
pathfinder telescopes with German involvement is given.

\bigskip

\subsection{SKA pathfinder programmes}

\subsubsection{The SKA design study (SKADS)}

\noi SKADS was an EC funded design study for 
the SKA running from 2005 nominally for 4 years, which was completed
at the
end of 2009. SKADS received EC-FP6 10.4\,MEuro funding with national
matching such that was brought to a total of $\sim$\,38\,MEuro.
Participation included many European countries: The
Netherlands, United Kingdom, Germany, France and Italy, plus
contributions from Australia, South Africa and Canada. 

SKADS covered multiple aspects of SKA research and design including
science simulations, configuration, communications, and costing plus
technical development and technology road mapping to implement
mid-frequency phased aperture arrays. The scientific outcomes of
SKADS are 43 scientific memos and an internationally well-recognised
simulation of the radio Universe (SKA Simulated Skies S$^3$). This is 
a set of computer simulations of the radio and (sub)millimeter
Universe primarily dedicated to the preparation of the SKA and its
pathfinders. The physical outcome of SKADS was the construction
and testing of three demonstrators: 1-\ EMBRACE -- a 144\,m$^2$ aperture
array with RF beam-forming; 2-\ 2-PAD -- a 9\,m$^2$ entirely digital aperture
array tile; and 3-\ BEST -- a focal line installation on the Northern Cross
radio telescope in Italy (Medicina).  The results from SKADS have been highly influential
for the SKA and are discussed in parts in the mid-term SKADS science
conference proceedings (Kl\"ockner et al. 2006), in detail in the
final conference proceedings (Torchinsky et al. 2009) and in the ``SKADS
white paper'' (Faulkner et al. 2010).

\medskip

\noi In summary, SKADS was very successful at taking advanced
engineering concepts and reviewing potential implementations of the
SKA to maximise the scientific output. SKADS also formed a major
collaboration between many countries and institutions, which is
continuing in the PrepSKA phase of the project. The principal
conclusion of SKADS is that an SKA implementation based on aperture
arrays operating from 70\,MHz to 1.4\,GHz observing frequency, with dish
based receivers above 1\,GHz is achievable, affordable and a highly
desirable solution in the SKA timeframe.

\bigskip

\subsubsection{Preparatory phase proposal for the SKA (PrepSKA)} 

PrepSKA is a consortium that consists of 24 partners from around the
world including Germany and acts as a coordinating body for much
of the technical and policy work currently underway for the SKA. It is
a 35\,MEuro programme running from 2008 to possibly the end of 2012, that
has been funded by the EC with 5.5\,MEuro.

There are several issues that need to be addressed before construction
of the SKA can start, these are: design, location, legal framework and
governance, procurement and funding. This preparatory study for the
SKA is designed to address all these points. Furthermore, PrepSKA is
coordinating and integrating worldwide R\&D work to develop the costed
design for Phase\,1 of the SKA (\skai ). The main deliverable will be an
implementation plan forming the basis of a funding proposal to
governments to start construction. The principal objectives of PrepSKA
are:

\noi 1) to produce a  deployment plan for the full SKA, and a detailed
costed system design for Phase\,1 of the SKA. The technical development 
work is being carried out in Work Packages that are informed by the SKA Precursor and Pathfinder projects and
the Design Studies;

\noi 2) to further characterise the two candidate
SKA sites in Southern Africa and Australia and to analyse the
various risks associated with locating the SKA at each of
the sites;

\noi 3) to develop options for viable models of governance and the
legal framework for the SKA during its construction and operational
phases; 

\noi 4) to develop options for
how the SKA should approach procurement and how it should involve
industry in such a global project;

\noi 5) to investigate all
aspects of the financial model required to ensure the construction,
operation and, ultimately, the decommissioning of the SKA;

\noi 6) to demonstrate the impact of the SKA on society, the economy
and knowledge.

\noi 7) to integrate all of the activities, reports and
outputs of the various working groups to form an SKA implementation plan.

\bigskip

\subsubsection{Global organisation for the SKA  (GO-SKA)}

\noi The global project organisation for the SKA is a coordination
action funded by the European Commission and designed as a follow-up
of PrepSKA whose aim was to
formulate an implementation plan that forms the basis of a funding
proposal to governments to start the construction of the SKA. The
decisions and the ensuing developments will have a significant impact on
the organisation of the SKA project and raise new topics to be
investigated, in order to narrow down and implement the governance,
funding and procurement options delivered by the PrepSKA policy work
packages. The purpose of GO-SKA is to investigate and provide guidance
at policy-level to the SKA Organisation, so that it will be optimally
prepared for the construction and operation of the SKA in 2016.  The
project period is from 2011 to the end of 2014 and has a total cost of
1.2\,MEuro, the EU will contribute 0.9\,MEuro. Whereas PrepSKA
has assembled the best options for the SKA, GO-SKA will focus on the
further development and implementation during the next stage of the
SKA Project. The principal objectives of GO-SKA will be:

\noi to broaden and strengthen the involvement of funding agencies and governments around the globe;

\noi to establish world-wide partnerships between industry and the SKA;

\noi to prepare the establishment of global governance for the SKA organisation;

\noi to develop strategies to further define the conditions by which
non-scientific benefits from large-scale research infrastructures can
best be integrated into investment decision-making.

\medskip

\noi Germany is partner in this project via the MPIfR as the leader of
work package 5: ``Developing SKA as a tool to address global
challenges''.  Their main responsibility is to identify global
challenges and main innovation drivers of the SKA that have a direct
impact on society and socio-economic structures. In order to reach
this goal a strategic forum is going to be build and established.

\bigskip

\subsection{SKA pathfinder telescopes}

\subsubsection{e-MERLIN (UK)}
e-MERLIN is a cm-wavelength telescope array, spanning 217\,km
baselines connected by a new dark fibre network. It was the first
full-time array to be connected at 10\,Gb/sec using a combination of
specially installed fibre cable (90\,km in total) and trunk dark fibre
(600\,km in total) leased from a number of providers. SKA pathfinding
activity includes: 1)~Low-cost techniques of cable installation and
the experience of procuring, managing and maintaining this dark fibre
network are valuable for the SKA.  The data transmission equipment
follows the JVLA/ALMA design (in which Manchester/e-MERLIN staff
participated), but e-MERLIN has extended this approach to more than
ten times the maximum link used in these arrays, by using multiple
amplification/regeneration sites.  e-MERLIN will provide useful
experience in operating this type of link over these distances. 2)~As
part of the SKADS project e-MERLIN has been used to demonstrate
optical phase transfer links over $>$\,100 km and with multiple hops,
allowing this approach to be extended to hundreds of km, if
required. The system is based on an optical implementation of the
pulsed $\sim$\,1\,GHz band link system designed for MERLIN, over the
installed e-MERLIN optical
fibre network. 3)~e-MERLIN science tested for the SKA in the
high-resolution study of the SKA populations  using gravitational
lensing and in the high-resolution follow-up of radio transients.

\medskip

\noi German scientists are involved in the commissioning and the
legacy proposal programme of e-MERLIN.

\bigskip

\subsubsection {e-European VLBI network  (eEVN)}
The EVN is a consortium of institutes
operating an interferometric network of radio telescopes on a
global scale from Europe to China to Puerto Rico and South Africa. The
eEVN is a development programme to transfer data in real-time
from the remote EVN telescopes to the central processing facility
via optical fibre cables – to replace the ``traditional''
implementations of VLBI in which the data are first recorded
at the telescopes on magnetic media (tapes or discs) and
then physically delivered to the processor. Over the last
four years, fibre links have been established to most of the
EVN telescopes. Data rates achieved currently are 1 Gbits/sec per VLBI
station, but the expectation is that this will increase to 16
Gbits/sec/station.  Its SKA knowledge transfer will be: 1)~The eEVN
will serve as a test-bed for the SKA for long distance signal
transmission across national boundaries. 2)~eEVN science
pathfinder for the SKA in separating AGN and starburst emission in
distant sources and obtaining ultra-high-resolution
follow-up of radio transients.

\medskip

\noi The Effelsberg telescope in
Germany is the biggest telescope in Europe and German scientists and
engineers have played a major role in the development of the eEVN array.

\bigskip

\subsubsection{Karl G. Jansky Very Large Array (JVLA)}

\noi The JVLA is a 27-element array of 25-m diameter dishes located in
Socorro, New Mexico. Technical areas of interest to the SKA are:
High-rate data transmission (120\,Gbits/s from each of 27 antennas)
over an internally installed and maintained fibre network. SKA
pathfinding activity includes: 1) 2:1 bandwidth receivers with low
system noise temperatures and high polarisation purity; 2) RFI-tight
designs and detailed radio frequency interference (RFI)-testing
protocols to prevent self-interference that can limit the dynamic
range of a large array; 3) Data archiving and default image production
in real-time or near-real-time; 4) Wide field-of-view imaging by means
of new imaging algorithms that involves parallel processing; 5) Remote
operations, including dynamic scheduling of the array in short
scheduling blocks (an hour and shorter) to take optimal advantage of
atmospheric conditions; 6) JVLA science pathfinding for SKA includes the
first surveys for galaxies in the EoR via low CO transitions and
studies of
cosmic magnetism.

\medskip
 
\noi German scientist make frequent use of the JVLA which was
discussed in the section ``German SKA community''   and are principle
investigators of large surveys such as THINGS
(http://www.mpia-hd.mpg.de/THINGS/) or COSMOS
(http://www.mpia-hd.mpg.de/COSMOS/).

\bigskip

\subsubsection{Aperture tile in focus (APERTIF)}

\noi The ``APERture Tile In Focus'' system (APERTIF) is a Phased Array
Feed (PAF) that is being developed for the Westerbork Synthesis Radio
Telescope (WSRT) in the Netherlands to increase its survey speed by
a factor $\sim$\,20.  APERTIF will operate in the frequency range from
1000 to 1750\,MHz, with an instantaneous bandwidth of 300\,MHz, a
system temperature of 55\,K and an aperture efficiency of 75\,\%. The
goal is to generate 37 beams on the sky for an effective field of view
of 8 square degrees.  The current horn feeds have a 30\,K system
temperature, 55\,\% aperture efficiency and 160\,MHz bandwidth.  The
PAF will reduce the sensitivity of a single beam observation compared
to the current horn feeds, but in terms of survey speed this is more
than compensated by the 37 times larger field of view.  Each PAF
consists of a dual polarised antenna arrays of 121 tapered slot
elements with Low Noise Amplifiers and a Uniboard-based digital
beamformer. The APERTIF correlator may be based on Uniboard
technology.  APERTIF is an important SKA pathfinder in the following
ways: 1)~Demonstrating the capabilities for PAF technology on an
existing, well-characterized telescope array; 2)~Measurements with the
first APERTIF prototypes (called DIGESTIF) already demonstrate the
unique capabilities of PAFs in practice: wide field of view (scan
range), low system temperature, excellent illumination efficiency,
synthesis imaging and a significant reduction of the reflector-feed
interaction; 3)~APERTIF provides a clear performance and cost
benchmark for PAF technology. 4)~APERTIF science will include
medium-deep \hi\ surveys aimed at measuring evolution in the number
density and cosmological
bias of the \hi\ population.

\medskip

\noi German scientists are involved in the Expression of Interest
(EoI) programme of the APERTIF-WSRT project. Furthermore, MPA and ASTRON scientists
have now teamed up to carry out the ``WSRT Bluedisk project'' which
can be regarded as a pilot study for upcoming APERTIF surveys.

\bigskip

\subsubsection {Australian  SKA  pathfinder  (ASKAP)}
ASKAP the Australian SKA Pathfinder is a CSIRO project being built at
the Murchison Radio-astronomy Observatory (MRO), Australia's 
SKA location.  It is a facility to trial wide-field-of-view
high-dynamic-range technologies for the SKA, deliver cutting edge
science, and develop the MRO as a world-class observatory. ASKAP will
be operational by mid-2013, while early hardware is already available to
test technology and science since 2011.  Many of the technical
developmental aspects to be addressed in the SKA design are included
in the ASKAP design: 1)~The inexpensive sky-mount telescopes represent
one possible configuration and allow one to directly measure the
effects of parallactic rotation of the image.  Aspects of cooling and
other environmental effects are being investigated; 2)~The phased
array feed (PAF) represents one possible implementation for the SKA.
ASKAP provides a platform to trial PAFs in an array configuration,
and the group is working with other teams around the world to jointly
develop the best PAF for science; 3)~Given the vast amounts of data to
be transported, ASKAP signal transport and network developments are
directly relevant to the SKA.  ASKAP is looking at a variety of
signals over optical fibre as well as networking protocols. 4)~The
ASKAP digital system represents architectures and developments
directly related to the SKA.  Digital hardware studies to scale the
beamformer for SKA cost and energy goals will be pursued, as well as
correlator architectures for the SKA; 5)~ASKAP software, calibration,
imaging and temporal solutions will have direct relevance to the SKA.
Scaling studies to the SKA are also being pursued; 6)~Power and
communication solutions appropriate for the SKA are being investigated
and implemented; 7)~System engineering, life-cycle studies, operations
planning, logistics engineering management plan and cost models are
all being developed and evaluated; 8)~ASKAP will deliver cutting edge
science to inform the SKA science case and development.  Examples:
large sky area, low-redshift \hi\ surveys to measure the galaxy power
spectra and hence cosmological biases of the galaxy populations; large
sky area continuum surveys to trace the evolution of black holes to
high redshift, and star-forming galaxies to moderate redshift. The
``Science Survey Team'' process instituted by CSIRO has been an effective
way to involve the broader
international community effectively at an earlier stage.

\medskip

\noi German scientists are involved in science simulations and the
planned legacy programmes
of the ASKAP project.

\bigskip

\subsubsection {MeerKAT}

\noi MeerKAT will be built in South Africa and is an array of
 60 13.5-metre offset-feed dishes with single pixel 
wideband feeds, and is therefore closely related to a
 major component of the SKA baseline design. 

\smallskip

\noi Technologies being developed and used for MeerKAT that
have direct relevance to the SKA include: 1)~One-piece
moulded reflectors fabricated using composite materials. 
Costing and performance information resulting from the design,
construction and operation of  these composite dishes will be
made available to the international project; 2)~Electromagnetic
modeling of feed and dish optics.  Simulations of the
MeerKAT feed and dish optics aimed at optimising Ae/Tsys and
sidelobe  characteristics will be made available to the
international project; 3)~High fidelity single pixel octave
bandwidth digital receivers.  

\smallskip 

The specific components and
technologies relevant to the SKA include: novel feed horns,
OMTs and LNA coupling, low cost, low maintenance and high
reliability cryogenic systems based on  Stirling cycle
refrigerators, integrated RF chain systems, wide bandwidth
ADCs, temperature  stabilization, and RFI shielding. Packet-switched architectures for radio astronomy signal
processing applications. 

\smallskip

MeerKAT will test the
scalability of the CASPER packet-switched architectures for
digital  signal processing applications for radio
astronomy. 4)~Calibration, imaging and time-domain
processing. 5)~System  Engineering:  life-cycle  studies, 
operations  planning,  logistics  engineering  management plan
and cost models. 6)~MeerKAT  will  deliver  science  on  the  pathway
to  the  SKA.  Examples are:  deep  \hi\  surveys  to  establish the
cosmological evolution in \hi\ to redshift z\,$\sim$\,1.4 and, potentially
piggybacked, a  deep  continuum  survey  to  reach  (with  stacking)
the  populations  likely  to  dominate  SKA  surveys;  pulsar  timing
for  fundamental  physics  as  precursor  to  SKA  pulsar  KSP. The
call  for  large  proposals  issued  by  SKA  South  Africa  has motivated
early  involvement  by  a  broad  international community. 

\medskip

\noi German scientists are involved in early science proposals and the
legacy proposal programmes of the MeerKat project, including a
PI-ships for selected key science projects. In addition, German
scientists are leading a
pilot project on the MeerKAT precursor telescope KAT--7.

\bigskip

\subsection{Low frequency array [LOFAR] 
  {\scriptsize [M. Hoeft]}}
\label{subslofar}

\noi LOFAR is the newest European radio telescope being
constructed by ASTRON in the Netherlands. It is a low-frequency
aperture array operating at two largely unexplored frequency bands:
15\,--\,80\,MHz and 110\,--\,240\,MHz.  LOFAR leads the way for a new generation
of radio telescopes, like the SKA.  It will consist of about
fifty antenna fields each comprising a multitude of small and cheap
antennas without moving parts. The signal of all antennas is digitally
processed and combined at each field. Resulting data are send via fast
links to a supercomputer in Groningen (NL). Since the data processing
is fully implemented in software, LOFAR is an extremely flexible
instrument. For instance LOFAR can be operated as interferometer with
a large field of view but it is also used to detect cosmic rays.

\noi LOFAR will consist of at least of 40 stations in the Netherlands, 6
stations in Germany, and one station in each of United Kingdom, France
and Sweden. More international LOFAR stations are planned in the UK,
Poland, Italy and Ukraine. The first German station was completed in
2009 next to the 100-m Effelsberg radio telescope, the second in
Tautenburg (Th\"uringen) was completed later that year, the third
German stations near Garching (Unterweilenbach) in 2010, the fourth
and fifth stations in Bornim near Potsdam and in J\"ulich in 2011. The
sixth station is planned near Hamburg. 

\noi The large collecting area (about 0.1 square kilometre) and the variety
of operating modes is reflected by the diversity of the six ``Key Science
Projects'', which drive the design of the hardware and the software of
LOFAR:
\begin{itemize}
\item[--]The epoch of reionisation
\item[--]Extragalactic surveys
\item[--]Transient radio phenomena and pulsars
\item[--]High energy cosmic rays
\item[--]Cosmic magnetism (German leadership; http://www.mpifr-bonn.mpg.de/staff/rbeck/MKSP/mksp.html)
\item[--]Solar physics and space weather (German leadership; http://www.aip.de/groups/osra/sksp/)
\end{itemize}

\noi Thus far almost forty stations have been built and validated in the
Netherlands, five in Germany, and one each in the UK and France. LOFAR
operates in a largely unexplored frequency regime, it has a very large
field of view, the operation needs to extremely flexible to account
for transient phenomena, and it produces tremendous amounts of ``raw''
data after correlation. Hence, LOFAR needs newly software tools new
strategies for calibration. Currently, the hardware and software are
being commissioned. First scientific results verify the high quality
of the data obtained with LOFAR.

\medskip

\noi The German LOFAR activities (www.lofar.de) are organised
and coordinated by the GLOW consortium, which is described in
Chapter\ \ref{glow}.

\bigskip

\subsubsection{LOFAR\  --\  long baselines {\scriptsize [O. Wucknitz]}}

\noi Long baselines of low-frequency arrays pose particular calibration
challenges. Beyond baseline length of approximately 30\,km, the stations probe
independent patches of the ionosphere, which introduces considerable
phase fluctuations that have to be accounted for. At the same time the
signal is reduced on long baselines, because most sources (targets and
calibrators) start to be resolved.

\medskip

\noi LOFAR will have 11 ``international stations'' outside of the Netherlands,
with baselines up 1000\,km and more. The flux density detected on
international baselines is typically at least an order of magnitude
lower compared to short baselines. Together with the rapid phase
fluctuations (in time and frequency), this implies that more
sophisticated calibration schemes have to be applied. Even standard
fringe-fitting techniques developed for VLBI are not sufficient for
LOFAR because of the large dispersion in the ionosphere and the strong
differential Faraday rotation between stations.

\medskip

\noi The development of the long baselines for LOFAR is organised by the Long
Baseline Group at the AIfA, University of Bonn. Preliminary
simplified methods are currently used to analyse commissioning data
and produce the first science results on long baselines. As of
autumn 2011, long-baseline results have exclusively originated from
these efforts.  The implementation of long-baseline calibration
methods as part of the general pipeline will be pursued by the same
team together with a postdoc/developer appointed at AIfA.

\bigskip
\bigskip
\bigskip

\parbox{\textwidth}{

\hspace{-0.4cm}
\parbox{5cm}{  \includegraphics[width=0.6\textwidth, angle=0]{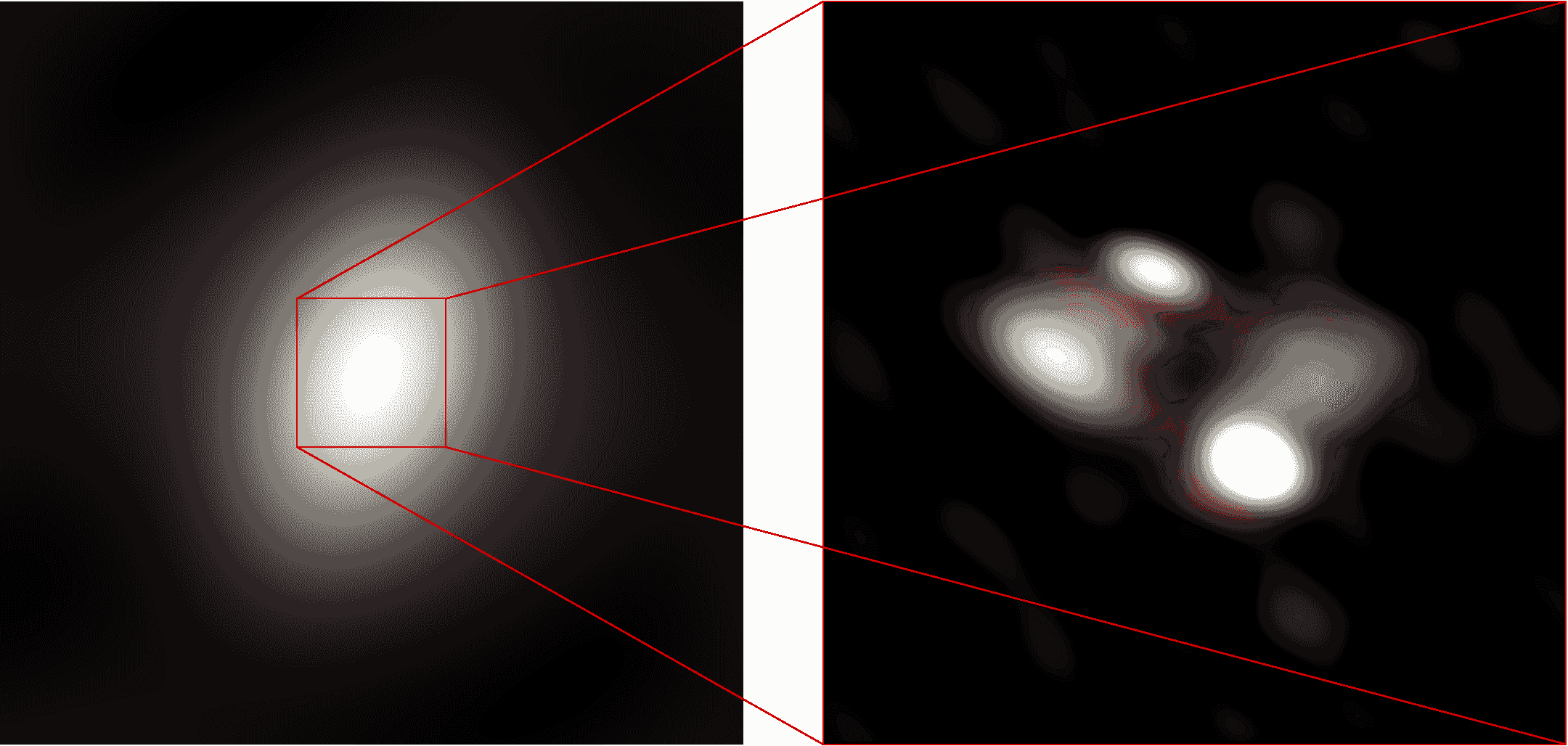}}
\hspace{6cm}
\parbox{6cm}{Figure 1$_{\rm Wu}$: LOFAR image of the extragalactic source 3C\,196
 showing the radio emission at 30\,--\,80\,MHz. Left: Radio emission of
 3C\,196 using only the stations in the Netherlands (35\,x\,22\,arcsec
 angular resolution).  Right: Radio emission of 3C\,196 including all
 Dutch and international stations (1.5\,x\,0.9\,arcsec angular
 resolution).}
}\\

\bigskip
\bigskip
\bigskip

\parbox{\textwidth}{

\hspace{-0.4cm}
\parbox{5cm}{  \includegraphics[width=0.6\textwidth, angle=0]{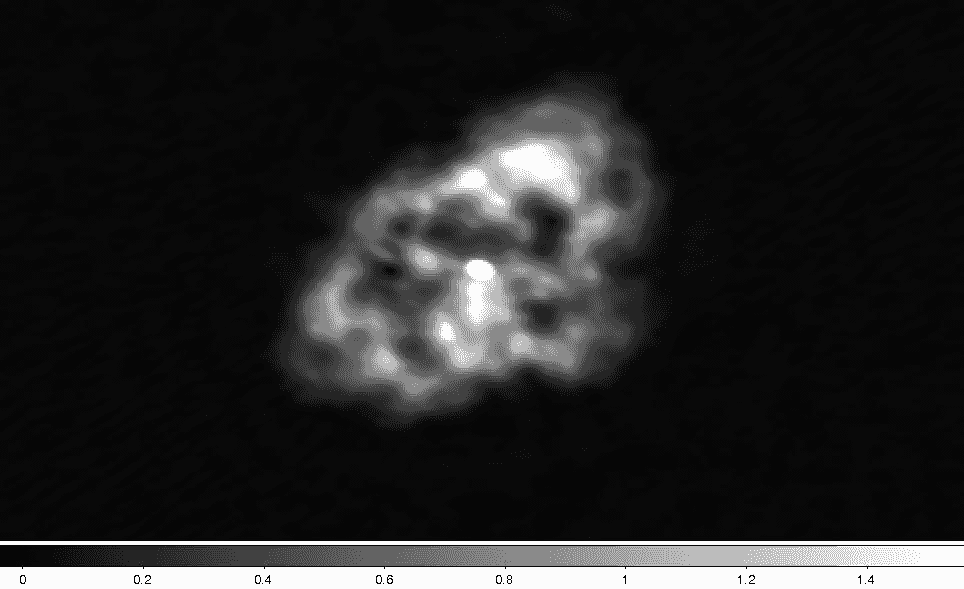}}
\hspace{6cm}
\parbox{6cm}{Figure 2$_{\rm Wu}$: The first LOFAR image of the Crab Nebula, the calibration only
  made possible by the international baselines.}
}

\bigskip
\bigskip
\bigskip

\parbox{\textwidth}{

\hspace{-0.4cm}
\parbox{5cm}{\includegraphics[width=0.6\textwidth, angle=0]{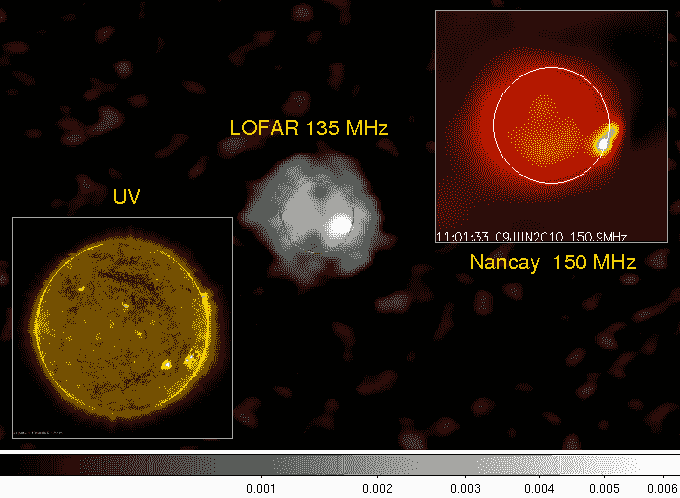}}
\hspace{6cm}
\parbox{6cm}{Figure 3$_{\rm Wu}$: The first LOFAR images at 135\,MHz of the Sun, compared to
  radio observations at 150\,MHz (Nan\c cay, France) and an ultra-violet
  image of the Sun.}
}\\

\bigskip

\newpage
\pagestyle{empty}
\noi{\bf }

\newpage
\pagestyle{plain}

\section{The German astrophysical science interests}
\label{gss}

\noi The SKA will be one of the fundamental pillars in the exploration
of the Universe.  With its enormous collecting area, that will be
around 10\,--\,100 times more sensitive than todays largest single dish
telescope, and with its survey speed, allowing to survey the sky a
million times faster than what is possible today, the SKA will
significantly impact on current science programmes and offer
opportunities in exploring new science scopes. To gather the
German community and to guarantee that their scientific interests are
well represented on a national and international level, a call for
contributions was initiated in autumn 2011 and a splinter session
was organised at the AG Tagung in Heidelberg 2011 (``A fresh view
of the radio sky: science with LOFAR, SKA, and its pathfinders'').

In order to allow for an unbiased approach to the science potential of
the SKA no particular guidelines have been given and therefore the
nature of these contributions are widely spread, ranging from general
overviews of specific science areas of interest, to direct observing
experiments, up to descriptions of new concepts of future scientific
and technology studies.

\subsection{Cosmology}

\subsubsection{Cosmology with the SKA {\scriptsize [D.J. Schwarz]}}

\noi {\bfseries\boldmath\small Introduction:~~}The SKA will allow us to study the Universe’s inventory of neutral
hydrogen (\hi ) and trace its evolution from the dark ages (long before
the very first stars and galaxies light the sky) to the present
ages. In doing so, the SKA will put cosmology on more fundamental
grounds. Since the turn of the millennium we speak about precision
cosmology, but so far that term is only justified for the observations
of the spectrum and the intensity anisotropies of the cosmic microwave
background (CMB) radiation. The SKA will bring the study of large-scale
structures in the Universe to the realm of precision cosmology.

\noi Hydrogen is the most common atom in the Universe. The SKA will map
the 21\,cm emission from neutral hydrogen throughout the visible and
dark Universe and is therefore able to trace atomic matter in the most
remote places of the Universe.  This is due to

\begin{itemize}
\item[--] a large sky coverage (approximately 2\,$\pi$)
\item[--] a deep sky coverage (\hi\ clouds shall be observed out to redshifts
  of 20 to 30)
\item[--] a broad frequency coverage (70\,MHz to 25\,GHz)
\end{itemize}

\noi The SKA will allow us to probe cosmic \hi\ in two ways. It will map the
\hi\ density field (non-thresholded SKA), similar to how we map the CMB
temperature anisotropies. But in contrast to the CMB, the SKA will resolve
the \hi\ density field in three dimensions (two angles and a
redshift). This is most suited for studying the dark ages of the
Universe when most of the ordinary matter in the Universe is neutral
hydrogen.

\noi Alternatively we might use SKA observations to generate an \hi\ 
galaxy redshift survey (thresholded SKA). This second possibility is
well suited for the visible Universe at redshifts below 6, as counting
point sources is less sensitive to foreground and RFI issues. The SKA is
likely to be the first telescope that will be able to map \hi\ 
intensities at redshifts $>$ 6 and will be the first radio telescope to
produce a huge galaxy redshift survey.

\noi Besides observing the \hi\  inventory, the SKA will also be well suited
for an exquisite continuum radio survey.

\noi Before highlighting some of the cosmological issues attacked by
the SKA, let us shortly review the status of modern cosmology.

\bigskip

\noi {\bfseries\boldmath\small Status of modern cosmology and the role
  of the SKA:~~}Modern cosmology rests upon
the cosmological principle, which states that the Universe is
isotropic and homogeneous at the largest scales. A further key
assumption is the validity of Einstein’s theory of general
relativity. Both assumptions will be tested by the SKA.

\noi The observed isotropy of the cosmic microwave background (CMB)
radiation at a temperature of 2.7\,K, the relic radiation from the epoch
of the formation of the first atoms in the Universe, holds in 1 part
in 1000. If we account the observed dipolar temperature anisotropy (3
mK) to the proper motion of the Local Group and remove some foreground
emission of the Milky Way, isotropy holds to in 1 part in 100\,000.
The remaining tiny temperature anisotropies (of order 10 $\mu$K) are
believed to be due to quantum fluctuations generated during an early
epoch of cosmological inflation and have been studied by several
ground, balloon borne and space CMB missions (among them COBE, Toco,
Boomerang, Maxima, WMAP, ACT, SPT and Planck).

\noi The analysis of the CMB allows us to measure the curvature of
space and to model the matter and energy content of the Universe
(Komatsu \etal\ 2011). It
turns out that the Universe is spatially flat, in agreement with the
prediction from cosmological inflation, but surprisingly it turns out
that we are missing most of the Universe’s matter and energy density
in our labs. Only 4\,\% of the Universe seems to be made out of atomic
matter, up to one per cent might be contributed by neutrinos – the
most elusive particles known by particle physics so far. The vast
amount in the universal energy budget seems to stem from a
cosmological constant or a component called dark energy (70\,\%) and
about 25\,\% are attributed to dark matter. The presence of dark energy
originally has been inferred from the observation of supernovae of
type Ia at redshifts up to order unity. These supernovae are
standardizeable candles and can be used to measure the acceleration
and rate of the expansion of the Universe.

\noi The search for dark matter and dark energy is one of the most
urgent problems in modern physics and cosmology. The SKA will contribute
to this search by further constraining the properties of dark energy
and dark matter. Sections\ 8.1.2, 8.2.5, and 8.2.6 describe further aspects of the
contribution of the SKA to the solution of the dark energy puzzle and
Section\ 8.5.1 elaborates on the identification of dark matter. 

\noi The tiny temperature anisotropies in the CMB reflect density
fluctuations, which seed the formation of the large-scale structure of
the Universe. Due to the action of gravity these tiny fluctuations
grow and eventually collapse to from gravitationally bound structures
like sheets, filaments and clusters of galaxies, in between them void
regions of enormous extend (of the order of 100 Mpc). The study of
large-scale structure has so far be driven by optical galaxy redshift
surveys, pioneered by the CfA slices, more recently by the 2dF (two
degree field) galaxy redshift survey, the Sloan Digital Sky Survey
(SDSS), WiggleZ, and is currently continued by BOSS. A large number of
surveys is in preparation, many of them ground based, such as 4MOST,
LSST, HETDEX, Pan-Starrs, or DES, others space based such as Euclid.
All of them in the optical or near-infrared. In the radio, several SKA
pathfinder telescopes are preparing continuum radio surveys (e.g. LOFAR
MSSS, LOFAR Tier1, WODAN, EDU; Raccanelli \etal\ 2011).

\noi Galaxy redshift surveys so far allowed us to measure the power
spectrum of the galaxy distribution and to identify the scale of the
baryon acoustic oscillations (BAO) – a relic of the plasma
oscillations in the primordially hot Universe. Tracing the power
spectrum as a function of the redshift (or even better as a function
of time) would allow us to directly measure the growth of structure
and use that to constrain the properties of dark matter and neutrinos,
as well as dark energy and would allow us to test the action of
gravity.  Identification of the BAO as a function of redshift (or
time) allows us to measure the geometry and expansion history of the
Universe. The \hi\ galaxy surveys that Apertif, Meerkat, and ASKAP are going
to provide will give us a first glimp (Camera \etal\ 2012) on the
possibilities the SKA will offer. Sections 8.1.3 and 8.1.4 illuminate
in several aspects of the study of large-scale structure. 

\noi A prediction of cosmological inflation is the Gaussian
distribution of primordial density fluctuations.  Tiny departures from
Gaussianity are expected from the non-linear growth of structures and
from models of cosmological inflation that involve several dynamical
degrees of freedom and/or depart significantly form the slow-roll
regime. A survey covering cosmologically large volumes is able to test
both aspects (see Section 8.5.2 for more details).
 
\noi Modern cosmology is based on exquisite observations of the CMB, which
provide information from redshifts of order 1000, large-scale
structure surveys, SN and cluster studies, providing information on
redshifts up to order unity so far. There are pieces of information
form deep surveys and quasar surveys out to redshifts of about 6, but
we miss information from the Universe at redshifts $>$\,6. This is due to
the fact that, according to our current model, there was almost no
visible light at those redshifts.

\noi We know from the so-called Gunn-Peterson test that the present
Universe is ionized out to redshifts of about 6. However the small
optical depth of the CMB tells us that this cannot hold true up to a
redshift of 1000 and we believe that the Universe was in fact filled
with neutral hydrogen between a redshift of 1000 and somewhere between
15 to 30. This epoch is the so-called dark age of the Universe. An
instrument that can map \hi\  at high redshifts is what is needed to
reveal to origin of stars, galaxies etc. We expect that the very
first stars and galaxies produce UV light that eventually reionizes
the Universe by a redshift around 10. However, there might also be
other mechanisms of reionization like the injection of energy by
annihilation or dark matter particles or by the decay of metastable
particles (Natarajan \& Schwarz 2009). These and more aspects concerning the dark ages and the
epoch of reionisation (EoR) are elaborated in Section 8.1.6. 

\noi Returning to the cosmological principle, the second part of it,
homogeneity, is not tested very well. One of the key challenges of
observational cosmology is to devise means to test it. The problem is,
that the most naive method to directly test homogeneity, would be to observe
the Universe from another place, which is not possible at all. Thus we have to rely
on information from other places in the Universe that sample a past
light cone different from ours. The observation of local temperatures
at other places offers such a possibility. The spin temperature
observable via the 21\,cm brightness allows us to think about such a
test. Alternatively also molecular lines can be used, to measure the
CMB temperature at other places in the Universe.

\noi High-fidelity observations of the CMB have revealed several anomalies
at the largest angular scales (Bennett \etal\ 2011, Copi \etal\ 2011).
 Among them a lack of angular correlation on scales larger
60\,degrees, compared to the prediction of inflationary cosmology. Such a
lack of power might point towards a non-trivial topology of the
Universe or to short epoch of inflation, such that we are able to
actually observe the pre-inflationary Universe at the largest
scales. The SKA will be able to further investigate the physics of these
anomalies and confirm or rule out the primordial origin of these
anomalies.

\noi A new class of fundamental test will be come possible by means of the superb 
frequency resolution of the SKA, which will allow us to directly see
the expansion of the Universe based in the redshift drift of
individual objects. Thus the SKA will be able to do real time cosmology,
as described in more detail in Section 8.1.7.

\bigskip

\noi {\bfseries\boldmath\small Continuum radio surveys at SKA:~~}We still know surprisingly
little about the vicinity of the Local Group. A badly understood issue
is the motion of the Local Group with respect to the CMB. A dipole
component in the CMB is commonly interpreted as the Doppler effect due
to the motion of the Local Group through the Universe at a speed of
600 \kms . This hypothesis cannot be tested by CMB observations, the
only way to test it, is to confirm it via another background which is
supposedly at rest w.r.t. the CMB. The background of radio galaxies
out to a redshift of order 1 would be such a reference frame (Blake \&
Wall 2002, Crawford 2009).

\noi For this analysis a sample of about a billion point sources from
a continuum SKA survey over all of the observable sky could bin down
the dipole component at an accuracy matching the CMB dipole. As
redshift information is not necessary for this test, this measurement
could already be completed by the \skai . 

\noi Number counts and measurements of the angular auto-correlation
will allow us to constrain cosmological parameters, but will be less
powerful than the thersholded \hi\ survey (see below). A useful and
powerful probe will be the cross-correlation of a continuum survey
with Planck's CMB maps enabling us to probe dark energy and modified
gravity via the integrated Sachs-Wolfe effect (Raccanelli \etal\
2011). We expect that errors on the dark energy equation of state and
modified gravity parameters can be constrained at the few per cent
level with an exquisite control
of systematic errors, due to the broad frequency coverage of the SKA.

\bigskip

\noi {\bfseries\boldmath\small Flux limited \hi\ radio surveys at SKA (thresholded SKA):~~}The \hi\  point source
power spectrum at several redshift shells could directly reveal the
onset of cosmic acceleration from the study of the growth rate of
large-scale structures and constrain our models of dark energy. The
measurement of the BAO scale with an unseen accuracy will allow us to
measure the geometry and the expansion rate of the Universe. A
combination of SKA results and CMB results from Planck is expected to
constrain all cosmological parameters of the concordance model and a
few new ones (such as the dark energy equation of state) at well below
the per cent level. This will enable us to confirm or rule out the
cosmological constant with high confidence. However, the full power of
the SKA will only be obtained with \skaii .

\noi Beyond the two-point correlation function, the SKA is ideal to
search for huge structures like voids of order 100\,Mpc and more, that
could be the reason for several anomalous cold spots observed by WMAP
(Bennett \etal\ 2011) and look for counterparts of the SDSS Great Wall, a 400 Mpc long
structure that fills about a third of the volume of the SDSS. As
structures that large have been a surprise and have not been predicted
by the concordance model of cosmology, it is
important to find out how frequent they are in the Universe.

\bigskip

\noi {\bfseries\boldmath\small \hi\ intensity mapping with the SKA
  (non-thresholded SKA):~~}In order to study the intergalactic medium
in the late Universe, the epoch of reionisation and the dark ages of
the Universe, a three dimensional intensity mapping of the 21\,cm line
will allow us to trace all structures that contain neutral hydrogen
(see e.g. Paciga \etal 2011). We hope to learn about the very first
stars, the very first galaxies, understand how the central galactic
black holes form and further constrain the properties of dark matter.
Dark matter annihilations and decays would affect the spin temperature
during the dark ages, to name just two effects that could not be
observed otherwise.  The intensity mapping of the 21\,cm line promises
an enormous information gain, but is challenged by astrophysical (and
terrestrial) foregrounds (similar to the CMB).  First steps will be
possible with \skai , but to fully explore the potential of the SKA,
\skaii\ will be required to complete this task.

\noi There are many more cosmological issues to address, let me just
mention the genesis of cosmic magnetic fields (see Sections\ 8.1.3 and
8.1.6).

\noi The technological challenge to handle and analyse the huge
cosmological data volume will require a world wide and concentrated
effort, which will be worth its return, as our understanding of the
Universe will certainly be very different from today once the SKA has been
realised.\\

\parbox{0.9\textwidth}{
\noi{References:}\\
\noi{\scriptsize Abdalla F.A., Rawlings S., 2005, MNRAS, 360, 27;
  Abdalla F.A., Blake C., Rawlings S., 2010, MNRAS 381, 1313; 
Bennett C.L., \etal , 2011, ApJS, 192, 17;
Blake C., Wall J., 2002, Nature, 416, 150;
Camera \etal\ 2012arXiv:1205.1048;
Copi C.J., \etal , 2010, Adv. Astron. 2010, 847541 ;
Crawford F., 2009, ApJ, 692, 887;
Komatsu \etal , 2011, ApJS, 192, 18;
Natarajan A., Schwarz D.J., 2009, Phys. Rev. D 80, 043529;
Paciga G., \etal , 2011, MNRAS, 413, 1174;
Raccanelli A., et al., 2011, preprint  arXiv:1108.0930}}\\

\subsubsection{Dark energy with the SKA {\scriptsize [J. Weller]}}

\noi One of the largest puzzles in modern cosmology is the explanation of
the observed accelerated expansion of the Universe (Riess \etal\ 1998;
Perlmutter \etal\ 1997). The simplest way to incorporate accelerated
expansion into Einstein’s equations of general relativity is the
introduction of a cosmological constant, whose energy density can be
associated with a constant vacuum energy, which comprises about 75\,\%
of the energy budget in the Universe. However this requires an extreme
fine-tuning of initial conditions to 120 orders of magnitude. Hence,
dynamical models, which introduce a new scalar field, are considered
(Wetterich 1988; Ratra 1988; Zlatev 1998). In general these models
allow at late times a slow variation of the energy density in the
Universe, which is be encoded in the equation of state of the
associated fluid. One of the holy grails of observational cosmology
today is the measurement of this equation of state, expressed as the
ratio of pressure and density in the fluid, $w= p/r$.  Further possibly
explanations of the observed accelerated expansion, are that
Einstein’s theory of relativity requires extensions at very large
distances. This can be achieved by higher order gravity models
(Starobinsky 1980) or for example with the introduction of extra
dimensions (Dvali 2000).  These models typically leave an imprint how
large-scale structures form over time.

\noi In recent years it has emerged that one of the most promising
ways to measure cosmological parameters are so called ``Baryon
Acoustic Oscillations'' observed in the matter distribution in the
Universe (Eisenstein 2005). These features are imprints in for example
the galaxy distribution from the time before z\,=\,1100, when baryonic
matter and photons, were tightly coupled together and allowed the
propagation of pressure waves. At the time of decoupling some regions
in space were overdense in baryons and some underdense. This
influenced the distribution of dark matter, which seeds the
large-scale matter and galaxy distribution, in a sense that there is a
bump at a scale of 110\,Mpc (Eisenstein 2005).

\noi The SKA can measure these baryon acoustic oscillations by observing the
galaxy distribution with the 21\,cm line. So far the observation of the
21\,cm line of distant galaxies has been nearly impossible because of
the lack of sensitivity. The SKA will change this situation with its large
collecting area. Since the wavelength of the 21\,cm is redshifted due to
the expansion of the Universe, the SKA will allow drawing a 3 dimensional
picture of the cosmic web. The SKA has the potential to find a billion
galaxies out to redshift z\,=\,1.5. For example, if we parameterize the
equation of state with two parameters: $w_0$ for the value today and $w_a$
the linear evolution, these parameters can be measured together with
the information provided by the Planck cosmic microwave background
measurement, to 5\,\% resp. 15\,\% accuracy (Abdalla 2005). In addition
the SKA is the best suited instrument to map out the history of the
equation of state (Tang \etal\ 2011), This in turn will allow to severely
constrain the space of possible dark energy models. As mentioned above
another possible explanation of accelerated expansion is the extension
of gravity on large scales. Via the measurement of redshift space
distortions (Guzzo 2008), the SKA will also be able to address this
question. In summary the SKA will most likely be the most powerful and
precise instrument to reveal the nature of dark energy.\\

\parbox{0.9\textwidth}{
\noi{References:}\\
\noi{\scriptsize Riess A., \etal\ 1998, AJ, 116, 1009; 
Perlmutter S.,  et al. 1997, ApJ, 483, 565;
Ratra B., Peebles P.J.E., 1988, PR, D37, 2406;
Wetterich C., 1998, Nucl. Phys., B302, 668;
Zlatev. I., Wang L., Steinhardt P.J., 1999, PRL, 82, 896;
Starobinsky A.A., 1980, Phys. Lett, B, 91, 99;
Dvali G., Gabadadze G., Porrati M., 2000, Phys. Lett. B, 485, 208;
Eisenstein D.J., et al, 2005, ApJ, 633, 560;
Abdalla F.B., Rawlings S., 2005, MNRAS, 360, 27;
Tang J., Abdalla F.B., Weller J., 2011, MNRAS, in press, preprint astro-ph/0807.3140;
Guzzo L. et al., 2008, Nature, 451, 541
}}\\

\subsubsection{Large-scale structure {\scriptsize [M. Br\"uggen]}}
\vspace{0cm}

\noi One of the fundamental problems in modern cosmology is the fact
that stars, neutral atomic and molecular gas, and the diﬀuse hot gas
within clusters of galaxies account for only a third of the baryon
density in the local Universe as predicted from Big Bang
nucleosynthesis (Fukugita \etal\ 1988) and fluctuations in
the microwave background (Spergel et al. 2003). Some fraction of the
missing baryons lies in the Lyman-alpha forest at low redshift
(e.g. Penton \etal\ 2000), but much is believed to reside in
a warm (T\,=\,10$^{5-7}$\,K), low-density intergalactic medium (Cen \& Ostriker
1999, Dave \etal\ 2001, Cen \etal\ 2001, Tripp \etal\ 2000).  

\noi The temperature and low density of the WHIM leave it nearly
undetectable in emission. There are three approaches by which the SKA
can detect the WHIM and the cosmic web: 

\begin{itemize}
\item[--]{map the distribution of low
column density neutral hydrogen. Although the gas in these
inter-galactic filaments is moderately to highly ionized, QSO
absorption lines have shown that the surface area increases
dramatically in going down to lower \hi\  column densities. With moderate
integration times, the necessary resolution and sensitivity can be
achieved out to distances beyond the Virgo cluster. When combined with
targeted optical and UV absorption line observations, the total
baryonic masses and enrichment histories of the cosmic web will be
determined over the complete range of environmental
over-densities (Braun 1994). }

\item[--]{use
giant pulses from ``Crab-like'' pulsars in nearby galaxies to measure the
dispersion across the local filaments and voids of the cosmic web. The
SKA can detect giant pulses from pulsars in more than 30 bright
galaxies within 7\,Mpc of the Milky Way. As described by Lazio et
al. (2004) the sensitivity of the SKA
will allow for the detection of giant pulses originating in galaxies
as distant as the Virgo cluster. It will also be possible to detect
giant pulses from extragalactic pulsars on the other site of the Local
Void, allowing for measurements for the first time of the baryon
density of voids in the cosmic web. While this technique is limited to
a relatively small number of line of sights determined solely by the
number of giant pulses within a given galaxy, it offers the advantage
of directly measuring the electron density of the baryons in the
cosmic web, something ultraviolet and X-ray absorption line studies
can only infer.}

\item[--]{utilize the tremendous sensitivity and large field of view of the
    SKA to map the cosmic web via imaging of diffuse synchrotron
    emission arising from the infall of baryons onto the large-scale
    structure of the Universe. The formation of the large-scale
    structure of the Universe is thought to be marked by large-scale
    shocks as baryons accrete onto collapsing structures. Typical
    shock velocities of 500\,--\,1000 km/s can be expected for reasonable
    cosmological properties. Such infall velocities are sufficiently
    high that the infalling particles can be accelerated to total
    energies of 10$^{18}$\,--\,10$^{19}$ eV. In the presence of even a weak
    magnetic field in which the energy density of the magnetic field
    accounts for only 1\,\% of the total post-shock energy density, the
    growth of structure should be accompanied by the emission of
    diffuse synchrotron radiation coincident with accretion shocks
    (e.g. Vazza et al. 2010). Thus, the detection of diffuse
    synchrotron emission associated with these external shocks offers
    us the opportunity to accurately map the cosmic web, while at the
    same time measuring the electron density and energy distribution
    and inferring the strength of primordial magnetic fields
    associated with large-scale structure (Br\"uggen \etal\ 2005,
    Dolag \etal\ 2005).}

\end{itemize}

\noi One of the key techniques used to obtain information about
the strength and structure of cluster magnetic fields is the analysis
of Faraday rotation from polarised radio sources located behind and
within clusters. The sources' intrinsic polarisation need not be
known, as the effect can be observed as a characteristic
wavelength-dependent rotation measure (RM) signature. Observations of
a few nearby clusters have established the presence of magnetic fields
with typical strengths of few micro Gauss ($\mu$G) in non-cool core clusters and in
excess of 10\,$\mu$G in the centres of cool core clusters (Carilli \&
Taylor 2002, Feretti \& Giovannini 2008, Bonafede et
al. 2010). Detailed high-resolution RM images of radio galaxies in
merging and cooling-core clusters indicate that the RM distribution is
characterised by patchy structures of a few kpc in size (e.g.
Enßlin \& Vogt 2003, Murgia et al. 2004, Guidetti et al. 2008, Laing et
al. 2008).  

\noi As shown in Krause et al. (2009), in a shallow 100~hours survey, the
SKA array will detect over a million clusters with at least one
background source each, and allow detailed field structure
determination ($>$1000 clusters with more than 100 background sources
each) with a deep survey. If the cosmological evolution of the
rotation measures is proportional to (1 + z)$^n$, the SKA would be
able to measure n to an accuracy of 0.3 (Krause et al. 2009). Compared
to the few RMs known for a few nearby clusters today, this will
revolutionize our knowledge of magnetic fields in large-scale
structure. It will allow us to study the physics of cosmic plasmas, by
for example probing the field geometry inside clusters. This can test
models for anisotropic thermal conduction in the intra-cluster medium,
as well as the distribution of cosmic rays.

\noi Provided the electron densities can be measured via the
Sunyaev-Zel'dovich effect at high redshift, we can expect to follow
the build-up of cosmic magnetism with the SKA.

\noi Finally, the SKA will be able to solve the mystery of
radio halos and relics. Radio halos and relics are Mpc-scale, diffuse
synchrotron source in galaxy clusters whose origin is unclear. Their
existence indicates the presence of magnetic fields and relativistic
electrons. Radio halos are always found in clusters with merging
signatures, and their power at 1.4\,GHz correlates with the X-ray
luminosity of the host cluster (Liang et al. 2000). The origin of
radio halos is still debated (see e.g. Pfrommer et al. 2008, Cassano
et al 2009). Models proposed so far can be divided into two classes: --~re-acceleration models: in which electrons are re-accelerated by
turbulence in-situ through second-order Fermi mechanism (Petrosian et
al. 2001, Brunetti et al. 2001); --~secondary models: in which
electrons originate from hadronic collisions between the long-living
relativistic protons in the ICM and thermal ions (Dennison et
al. 1980).\\

\parbox{0.9\textwidth}{
\noi{References:}\\
\noi{\scriptsize
Bonafede A., \etal , 2010, A\&A, 513, 30;
Braun R., 1994, New Astronomy Reviews, 48, 1271;
Br\"uggen M., \etal , 2005, ApJ, 631, 21;
Brunetti G., \etal , 2001, NewA, 6, 1;
Carilli C.L., Taylor G.B., 2002, ARAA, 2002, 40, 319;
Cassano R., \etal , 2009, arXiv:0902.2971;
Cen R., Ostriker J.P., 1999, ApJ, 514, 1;
Cen R., \etal , 2001, ApJ, 559, 5;
Dave R., \etal , 2001, ApJ, 552, 473;
Dennison B., 1980, ApJ, 239, 93;
Dolag K., \etal , 2005, MNRAS, 364, 753;
En{\ss}lin T., Vogt , 2003, A\&A, 401, 835;
Feretti L., Giovannini G., 2008, arXiv:astro-ph/0703494;
Guidetti D., \etal , 2008, A\&A, 483, 699;
Krause M. \etal , 2009, MNRAS, 400, 646; 
Lazio J.,\etal , 1994, New Astronomy Reviews, 48, 1439;
Laing R., \etal , 2008, MNRAS, 391, 521;
Liang H., \etal , 2000, arXiv:astro-ph/0012166;
Murgia M., \etal , 2004, A\&A, 424, 429;
Penton S.V., Shull J.M., Stocke  J.T., 2000, ApJ, 544, 150;
Petrosian V. , 2001, ApJ, 557, 560;
Pfrommer T.,\etal , 2008, 385, 1211;
Spergel D.N., \etal , 2003, ApJS, 148, 175;
Tripp T.M., Savage B.D., Jenkins E.B., 2000, ApJ, 534, 1;
Vazza F., Brunetti G., Gheller C., Brunino R., 2010, New Astronomy,
15, 695}}\\

\subsubsection{Baysion cosmography {\scriptsize [J. Jasche]}}
\vspace{0cm}

\noi According to the current cosmological paradigm, all observable
structures in the Universe stem from microscopic primordial quantum
fluctuations generated at the very beginning of the Universe. In the
following \(\sim\)13.8 billion years of cosmic history, these seed
perturbations grew via gravitational amplification and formed the
presently observed matter distribution consisting of clusters,
filaments and voids. The processes of cosmic structure formation have
been governed by a variety of exciting physics ranging from quantum
field theory, general relativity to the dynamics of collision less
dark matter and dark energy as well as the behavior of baryons in the
formation of galaxies and stars. All these processes left their
imprints on the three dimensional matter distribution of the Universe.
For this reason, mapping and analyzing the cosmic large-scale
structure from observations has the potential to answer outstanding
questions of modern cosmology, such as the nature of a possible dark
matter and dark energy component and also provide us with a powerful
laboratory to test fundamental physics.  However, contact between
theory and observation cannot be established directly since
observational data is subject to a variety of systematic effects and
strongly coupled uncertainties, such as noise, survey geometry, galaxy
biases, redshift space distortions, selection and light cone effects.
The Bayesian large-scale structure inference framework HADES
(HAmiltonian Density Estimation and Sampling) addresses these problems
and permits unprecedentedly accurate inference of cosmological
information and corresponding uncertainties from probes of the large
scale structure (Jasche et al. 2010a, Jasche et al. 2010b, Jasche \&
Wandelt 2011). Specifically,
the HADES cosmography framework provides high precision maps of the
three dimensional matter distribution of the Universe in the linear
and non-linear regime together with corresponding uncertainties and
statistical properties of the large-scale structure. This is achieved
by exploring the highly non Gaussian and non linear large-scale
structure posterior distribution via an efficient implementation of a
hybrid Markov Chain Monte Carlo method (Jasche et al. 2010a). The
application of the HADES algorithm to the Sloan Digital Sky galaxy
survey demonstrates the ability of Bayesian cosmography to build
precise and detailed maps of the cosmological large-scale structure
(see Figure 1$_{\rm Ja}$ for a slice through the three
dimensional large-scale structure inferred from the Sloan Digital Sky
Survey, SDSS).

\bigskip
\bigskip

\parbox{\textwidth}{
\hspace{1cm}
\parbox{5cm}{  \includegraphics[width=0.48\textwidth, angle=0]{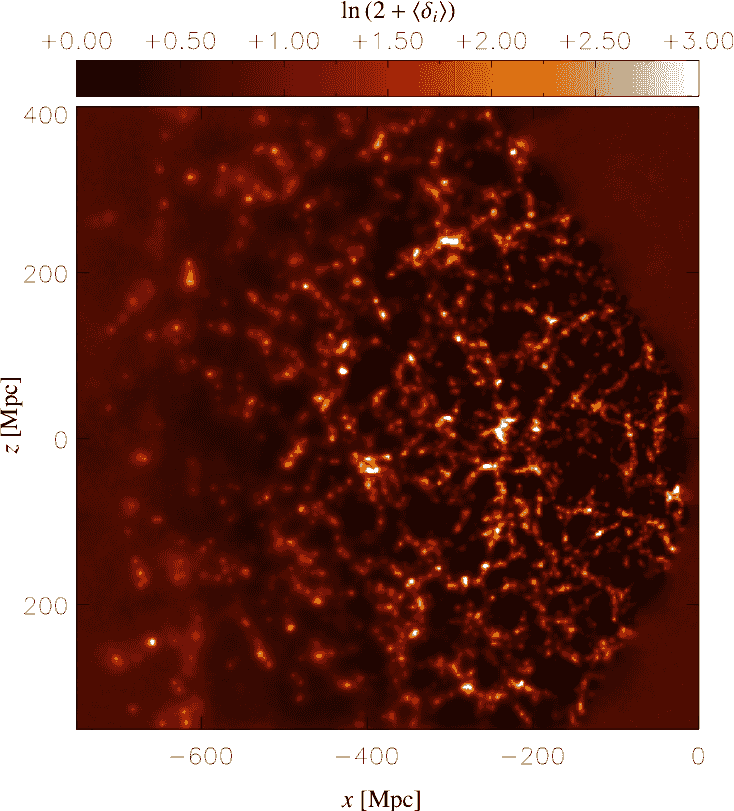}\vspace{-8.4cm}}

\hspace{10cm}
\parbox{6cm}{Figure 1$_{\rm Ja}$: Slice through the three dimensional ensemble mean density
  field, estimated from 40\,000 density samples. The plot demonstrates
  the results of the application of the HADES algorithm to the SDSS Data Release 7 (for details see Jasche et al. 2010b). As can be seen, the large-scale structure consisting of filaments, clusters and voids, has been recovered to great detail.}
\vspace{3.5cm}}\\

\noi These three dimensional maps constitute the basis for a variety
of subsequent scientific analyses. In particular, these maps enable us
to study the processes of structure and galaxy formation in the linear
and non-linear regimes and will help to identify the relation between
the properties of galaxies and their large-scale structure
environments. Furthermore, the maps can be used to predict
observations in complementary cosmological probes such as the Cosmic
Microwave Background (CMB) or weak lensing (for an example see see
Figure 2$_{\rm Ja}$). In particular, the possibility of straightforward
non-Gaussian and non-linear error propagation will significantly aid
the detection of weak signals such as the integrated Sachs-Wolfe
effect in the CMB.

\parbox{\textwidth}{
\hspace{1cm}
\parbox{5cm}{  \includegraphics[width=0.48\textwidth, angle=0]{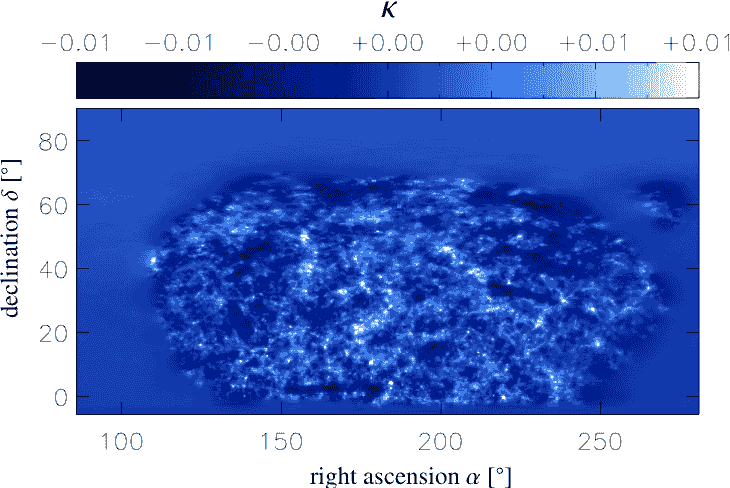}\vspace{-4.5cm}}

\hspace{10cm}
\parbox{6cm}{Figure 2$_{\rm Ja}$: Weak lensing convergence \(\kappa\) projection of the density field, inferred from the SDSS, onto the sky. This plot serves as an example for the possibility to predict physical observations in complementary data sets from the inferred density fields. Furthermore, this plot also demonstrates the degree of detail to which the Bayesian cosmography framework HADES recovers the three dimensional large-scale structure.}
}

\bigskip
\bigskip

\noi In general, scientific results obtained with this cosmography
framework, permit to constrain the nature of the initial conditions
from which cosmic structures formed; extract information about the
nature of dark energy and dark matter; illuminate the relationship
between the distribution of dark matter and the different types of
galaxies detected in surveys; and provide the community with
reconstructions of the cosmic density and velocity fields including a
detailed treatment of the uncertainties inherent in such
reconstructions.\\

\parbox{0.9\textwidth}{
\noi{References:}\\
\noi{\scriptsize Jasche J., Kitaura F.S., 2010, MNRAS, 407, 29;
Jasche J., Kitaura F.S., Li C., En{\ss}lin T., 2010, MNRAS, 409, 355;
Jasche J., Wandelt  B.D., 2011, preprint astro-ph 1106.2757
}}\\

\subsubsection{Turbulence and magnetic dynamo in the cosmological large-scale structure {\scriptsize [L. Iapichino]}}
\vspace{0cm}

\noi The formation and evolution of the large-scale structure, which
proceeds through the process of hierarchical clustering, plays a key
role for the conversion of potential gravitational energy into kinetic
and internal energy in galaxy clusters and groups. In this process,
cluster mergers induce bulk motions in the intra-cluster medium
(henceforth ICM), with velocities up to $1000\ {\rm km\ s^{-1}}$
(e.g. Paul \etal\ 2011). The shearing instabilities associated with
the merger events and the ubiquitous shock waves in the ICM are thus
expected to stir the gas and make the flow turbulent.

\noi Turbulence is an important link between the thermal history of
the cosmic baryons and the non-thermal phenomena, such as cosmic ray
acceleration and amplification of magnetic fields. In particular, the
small-scale, turbulent dynamo is often invoked in the framework of the
evolution of galaxy clusters (Br\"uggen \etal\ 2005, Subramanian
\etal\ 2006). The resulting
magnetic fields have magnitudes of the order $1 - 10\ \mu$G in the
ICM, and $0.1\ \mu$G or less in the cluster outskirts and filaments
(Ryu \etal\ 2008).

\noi Observationally, magnetic fields in the cosmological large-scale
structure can be probed by Faraday rotation measures (RM) of
background sources. This approach is already extensively used for the
study of cluster cores (Murgia \etal\ 2004, Bonafede et al. 2010) and
has been proposed for the study of the magnetisation of the cosmic web
using upcoming instruments, like the SKA and its precursors (for
example, Akahori \& Ryu 2010).

\noi We argue that the SKA will be able to probe the magnetic fields
in the outer regions of galaxy clusters in much more detail than
currently possible, because it will trace the outer regions by using
background galaxies and their measured Faraday rotation (Krause \etal\
2009). These regions are extremely interesting, because they start
being explored only recently by deep X-ray observations using {\it
  Suzaku} (Urban \etal\ 2011, Simionescu \etal\ 2011). Hydrodynamical
simulations (Burns \etal\ 2010, Vazza \etal\ 2011) suggest that the
cluster outskirts are not efficiently settled in hydrostatic
equilibrium and have a substantial turbulent pressure
support. However, the volume-filling factor of turbulence in these
regions is still debated (Iapichino \etal\ 2011).

Recent analytical estimates (Iapichino \& Br\"uggen 2012) show that, if the outer ICM
is turbulent, a moderate level (10 to 30 per cent) of non-thermal
pressure support can amplify the magnetic field to values
around $2.5$\,$\mu$G at a central distance of $0.5\ R_{\rm vir}$. This
estimate, combined with a simple model for the RM dispersion in
idealised clusters (Felten 1996), predicts values for $\sigma_{\rm RM}$
of the order of $\sigma_{\rm RM} = 10$\,${\rm rad\ m^{-2}}$. This
prediction is within reach of future SKA observations, which
might probe the amplification of magnetic fields that are related to
propagating merger shocks and diffuse radio emission in the cluster
outskirts (radio relics). The SKA will be able to shed
light on non-thermal processes in regions of the ICM whose study is,
both theoretically and observationally, otherwise very challenging or
even impossible.\\

\parbox{0.9\textwidth}{
\noi{References:}\\
\noi{\scriptsize Akahori  T., Ryu  D., 2010, ApJ, 723, 476;
Bonafede A., et~al., 2010, A\&A, 513, A30;
Br{\"u}ggen M., et~al. 2005, ApJL, 631, L21;
Burns J.O., Skillman S.W., {O'Shea} B.W., 2010, ApJ, 721, 1105;
Felten J.E., 1996, ASPC Conference Proceedings, 88, 271;
Iapichino L., Br\"uggen M., 2012, MNRAS, in press: 1204.2455;
Iapichino L., et~al., 2011, MNRAS, 414, 2297;
Krause M., \etal , 2009, MNRAS, 400, 646;
Murgia M., et~al., 2004, A\&A, 424, 429;
Paul S., et~al., 2011, ApJ, 726, 17;
Ryu D., et~al., 2008, Science, 320, 909;
Simionescu A., et~al., 2011, Science, 331, 1576;
Subramanian K., Shukurov A., Haugen N.E.L., 2006, MNRAS, 366, 1437;
Urban O., et~al., 2011, MNRAS, 414, 2101;
Vazza F., et~al., 2011, A\&A, 529, A17
}}\\

\subsubsection{Neutral hydrogen and the epoch of reionisation {\scriptsize
    [B. Ciardi]}}
\vspace{0cm}

The epoch of reionisation (EoR) sets a fundamental benchmark in cosmic structure formation, 
corresponding to the birth of the first luminous objects that act to ionize the neutral 
intergalactic medium (IGM). Observations at near-IR wavelengths of absorption by the IGM in 
the spectra of the highest redshift quasars (e.g. Becker et al. 2001, Fan et al. 2006, Mortlock et al. 2011) 
and the radio wavelengths of the electron scattering optical depth for the cosmic microwave 
background (Komatsu et al. 2011) imply that we are finally probing into this key epoch 
of galaxy formation at z\,$>$\,6.

\smallskip

\noi Theoretical modeling based on the above existing observational constraints 
give the impression conveyed in the following figure of
an IGM that becomes filled with growing ionized bubbles on a range of size
scales. With the passage of time, the bubbles finally punch through the walls
separating each other, to leave behind an almost fully ionized IGM, which remains
such to the present. 

\smallskip

\noi The SKA and the pathfinder telescopes which will precede it will 
provide critical insight into the EoR in a number of ways.
The ability of these telescopes to perform a study of the neutral IGM in 
neutral hydrogen (21\,cm) emission/absorption will be a unique probe of the process of reionization, and is recognized as the 
next necessary and fundamental step in our study of the evolution of large-scale structure and 
cosmic reionization. 
The applications of radio technologies to probing the EoR form the basis for several chapters 
in the book ``{\it Science with a Square Kilometre Array}'' (Carilli
\& Rawlings 2004).

\smallskip

\noi The attention dedicated to 21\,cm studies in cosmology has mainly focused on the feasibility
of tomography, which would ideally provide a 3-d mapping of the evolution of neutral hydrogen
(e.g. Madau \etal\ 1997 Tozzi et al. 2000, Ciardi \& Madau 2003, Mellema et al. 2006, Santos et al.
2008, Lidz et al. 2008, Morales \& Wyithe 2010).                    
This is an extremely exciting prospect which suffers though from several severe difficulties,
the most relevant being foreground and ionospheric contamination and terrestrial interference
(e.g. Shaver et al. 1999, Di Matteo \etal\ 2004, Jeli\'c et al. 2008, Bowman et al. 2009).
As tomography of the reionization history will
require approximately one square kilometre of collecting area, this will
most probably not be feasible by the pathfinders, which will instead concentrate on statistical
studies of the 21\,cm signal.
But more than a statistical analysis, IGM tomography
offers an invaluable tool to discriminate between different ionization
sources and to follow the spatial and temporal evolution of the
reionization process.
Thus, despite the challenge of the observations, also the \skai\ should
include them among its goals.

\smallskip

\noi Similarly, while the construction of radio maps of individual ionized bubbles with the
pathfinder arrays will be feasible at best (if at all) after several years
of observations, the \skai\ will have the sensitivity to deliver their direct 
detection. The ability of the telescope for effective survey of large areas of 
sky will lead to the identification of targets for subsequent follow up at other wavelengths.
For example, the \skai\ will be well suited to the direct observation
of ionized bubbles (giant Str\"omgren spheres) around luminous QSOs in a still
significantly neutral IGM (e.g. Tozzi et al. 2000, Wyithe \etal\ 2005). 
If QSO Str\"omgren spheres can be discovered, and their
structure studied, this will be an invaluable source of information
on the supermassive black hole population at high redshifts, the nature of the
ionizing spectra responsible for the reionization of the Universe
and the abundance of neutral hydrogen in the IGM
at those redshifts.

\bigskip
\bigskip

\hspace{-1.5cm}
\parbox{\textwidth}{
\parbox{\textwidth}{
\parbox{5cm}{  \includegraphics[scale=0.31, angle=-90]{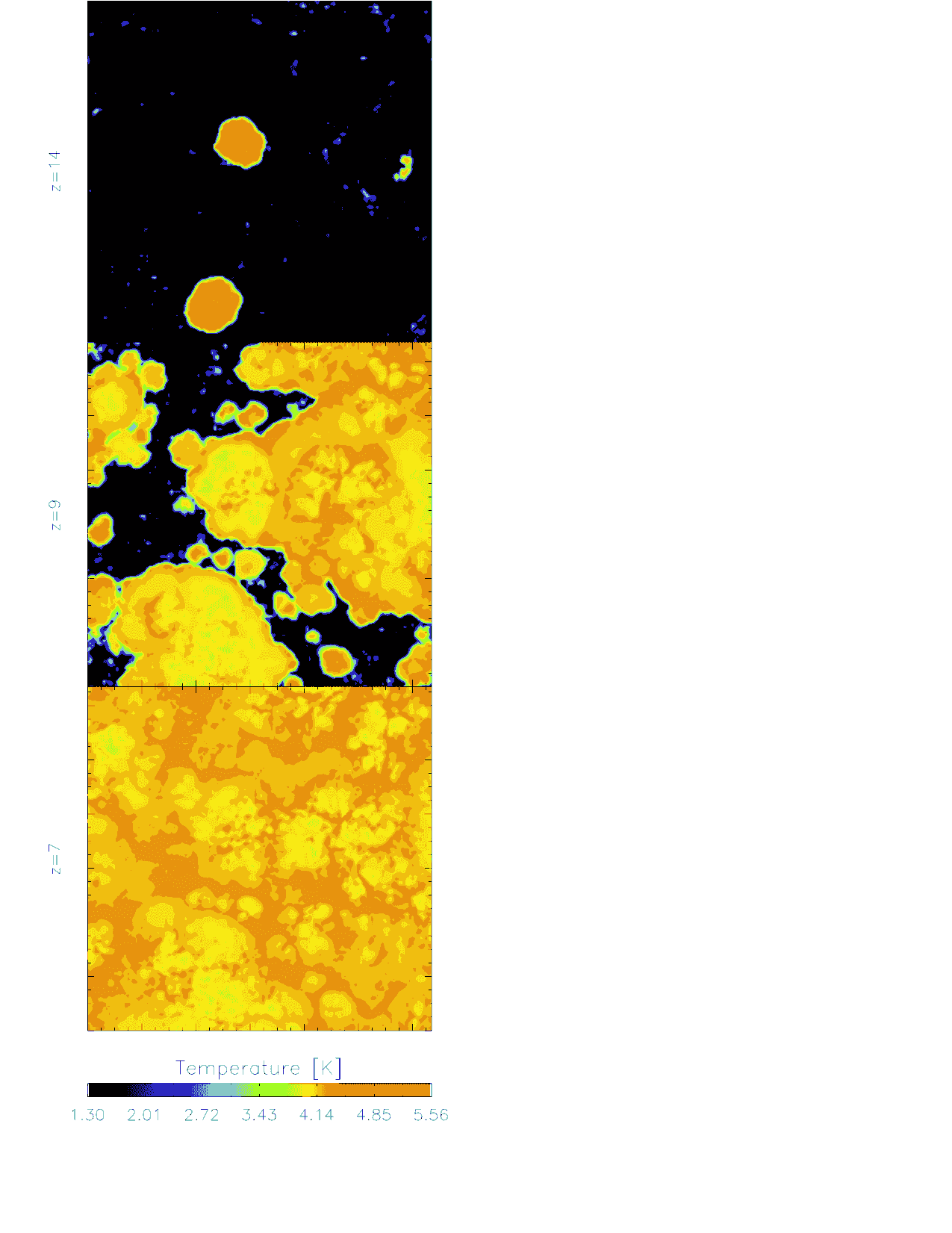}\vspace{-7.6cm}}

\hspace{-0.15cm}
\parbox{5cm}{  \includegraphics[scale=0.31, angle=-90]{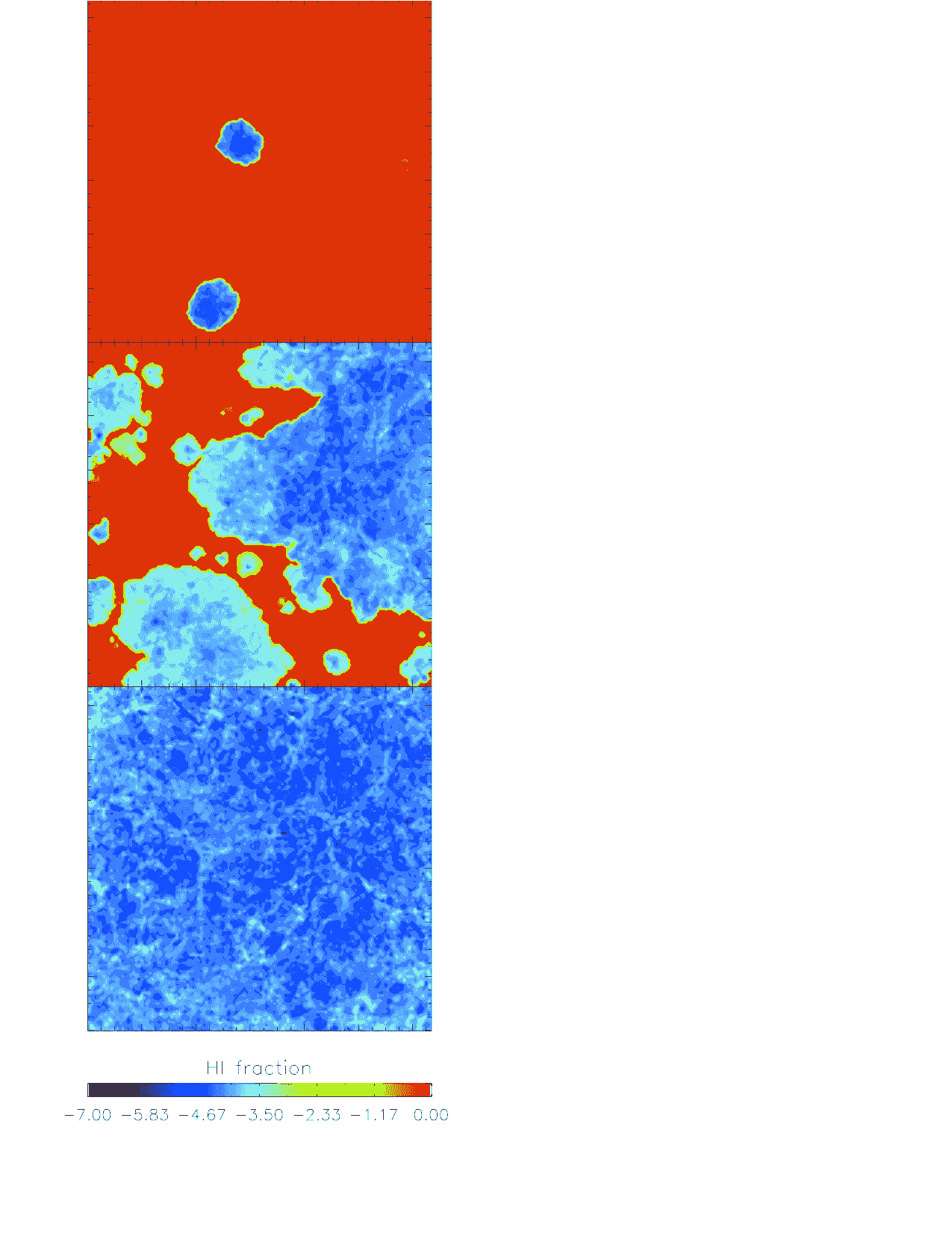}}}

\hspace{1.7cm}
\parbox{0.9\textwidth}{\vspace{-10cm}Figure 1$_{\rm Ci}$: upper
  panel:maps of gas temperature at redshift z\,=\,7 (left panel), 9 (middle
panel) and 14 (right panel) as obtained in simulation ${\mathcal E}$1.2
-- $\alpha$1.8 by 
Ciardi et al. (2012). Each map represents the central slice of the
simulation box.

lower panel: as upper panel for \hi\ fraction.}
\vspace{-3.0cm}}

\noi A valid alternative to tomography is looking at the 21\,cm lines generated in absorption
against high-z radio loud sources by the neutral IGM and intervening collapsed
structures, i.e. to search for the 21\,cm forest (e.g. Carilli \etal\ 2002,
Furlanetto 2006, Xu et al. 2009, Mack \& Wyithe 2011,
Xu \etal\ 2011). Analogous to the extensively studied case of the Lyman-alpha
forest, the 21\,cm forest signal can be detected in the spectra of high-z radio sources
and results from the absorption produced by the \hi\ intervening along the line of sight.
Despite the great challenge posed by the detection of such extremely rare target sources, 
the 21\,cm forest is particularly appealing as it naturally bypasses some the main limitations
expected for 21\,cm tomographyc measurements.\\

\parbox{0.9\textwidth}{
\noi{References:}\\
\noi{\scriptsize Becker R.H., et al., 2001, AJ, 122, 2850;
Bowman J.D., Morales M.F., Hewitt J.N., 2009, ApJ, 695, 183;
Carilli C.L., Gnedin, N.Y., Owen, F. 2002, ApJ, 577, 22;
Carilli C.L., Rawlings S., 2004, New Astronomy Reviews, 48, 979;
Ciardi B., \etal , 2012, MNRAS, 423, 558;
Ciardi B., Madau P., 2003, ApJ, 596, 1;
Di Matteo T., Ciardi B., Miniati F., 2004, MNRAS, 355, 1053;
Fan X., et al., 2006, AJ, 132, 117;
Furlanetto S. R., 2006, MNRAS, 370, 1867;
Jeli\'c V., \etal , 2008, MNRAS, 389, 1319;
Komatsu E., et al., 2011, ApJS, 192, 18;
Lidz A., Zahn O., McQuinn M., Zaldarriaga M., Hernquist L., 2008, ApJ, 680, 962;
Mack, K. J., Wyithe, J. S. B., 2011, ArXiv e-prints;
Madau P., Meiksin A., Rees M. J., 1997, ApJ, 475, 429;
Mellema G., Iliev I.T., Pen U., Shapiro P.R., 2006, MNRAS, 372, 679;
Morales M. F., Wyithe J.S.B., 2010, ARA\&A, 48, 127;
Mortlock D.~J., et al., 2011, Nature, 474, 616;
Santos M.~G., \etal , 2008, ApJ, 689, 1;
Shaver P.A., Windhorst R.A., Madau P., de Bruyn A.G., 1999, A\&A, 345, 380;
Tozzi P., Madau P., Meiksin A., Rees M. J., 2000, ApJ, 528, 597;
Wyithe J.S.B., Loeb A., Barnes D.G., 2005, ApJ, 634, 715;
Xu Y., Chen X., Fan Z., Trac H., Cen R., 2009, ApJ, 704, 1396;
Xu Y., Ferrara A., Chen X. 2011, MNRAS, 410, 2025
}}\\

\subsubsection{A direct measure of the expansion rate of the Universe {\scriptsize
    [H.-R. Kl\"ockner]}}

\noi {\small \bf Abstract:~}In the last recent years cosmology has
undergone a revolution, with precise measurements of the comic
microwave background radiation (CMB), large galaxy redshift surveys,
and the discovery of a recent accelerated expansion of the Universe. All these
details has boosted our understanding of the Cosmos, its evolution and
the models describing it are entering a new phase of precision.

In this light the SKA will provide an opportunity to directly measure
the expansion rate of the Universe via a rather simple experiment by
observing the neutral hydrogen (\hi ) content within and also
toward galaxies at two epochs. Due to the accelerated expansion of the
Universe these signals encounter a shift in redshift space and hence
provide a real time measure of the cosmological expansion rate. The
accuracy of these measurements, together with other
probes of cosmological parameters, allows us to constrain our
current understanding of the Universe and may help to distinguish
between alternative cosmological models.\\

\noi {\small \bf Introduction:~} The Big Bang concept of our Universe
seems to be well established as ``the standard model of cosmology'',
but currently the observational data cannot tighten constrains on the
physics of the very early phase of the Universe. In this current
picture 13.8\,billion years ago the Universe was hot and dense, and it
has expanded and cooled ever since. The content of the universe has
evolved and observations specify a census of the type of matter and
energy in its development. In the very early phase, it began to be dominated
by an energy field with a negative pressure, which drove an early
period of acceleration expansion, ``the inflation'' phase. It was then
dominated by radiation, and later by matter.  Based on this
understanding the expansion of the Universe should decelerate if it is
dominated by baryonic and cold dark matter \footnote{Baryons are
  composite subatomic particles made up of three quarks and participate
  in the strong interaction force. Cold dark matter is a hypothetical
  from of matter that interacts weakly with electromagnetic radiation
  and most of whose particles move slowly compared to light.  Possible
  candidates of ``cold dark matter''-objects are MACHOs, RAMBOs, and weakly
  interacting massive particle (WIMP) and axions. Such constitutes lead to a ``bottom-up''
  scenario of structure formation in the Universe. Dark matter
  is undetectable by its radiation, but its presence can be inferred
  from gravitational effects (e.g. rotation curves in
  galaxies).}. However by using type Ia supernova (SNIa), as standard
candles, a surprising discovery has been made, that the expansion of
the Universe encounters a second epoch of acceleration (Riess et
al. 1998, Perlmutter et al. 1999, this research was awarded a the
Nobel prize in physics 2011).  The reason for the recent accelerated
expansion is still a mystery and points again to a new negative
pressure contribution of the mass-energy field and a possible
modification of Einstein's general relativity. Therefore,
understanding the nature of the current acceleration is the essential
step in our understanding of the inflation phase and the mechanism in
place in this phase. 

\smallskip

\noi Measuring the expansion history of the Universe include
distances, the linear growth of density perturbations, and a
combination of both observed at different epochs. The observations at
different epochs allows for a direct measure of the expansion history,
whereas SNIa surveys, weak lensing (Heavens 2003) and Baryon Acoustic
Oscillations in the galaxy power spectrum (BAO; Wang 2006) are
generally considered to be indirect probes of the acceleration. The
results of the indirect probes are uncertain because the extracted
information requires a priory knowledge of the cosmological model and
even simple parameterisations of dark energy properties can result in
misleading conclusions (e.g. Bassett et al. 2004, Shapiro \& Turner 2006). 

\smallskip

\noi Thus model-independent measurements are needed and using
observations at different epochs, as direct probes, will provide strong evidence on
the expansion history of the Universe.\\

\noi {\small \bf The experiment:~}will make use of the direct change 
of observable properties of galaxies over an extended periods of observing
time. In general, the changes in source brightness, apparent angular size,
and redshift constrains the parameter needed in describing the
expansion history of the Universe (see e.g. Gudmundsson \& Bj\"ornsson
2002). The change of the first two
properties will not be discussed here and only the latter is
considered to be feasible with the SKA.

\smallskip

\noi Already in the 1960s the model-independent approach that measures
the expansion rate directly was first explored by Sandage.  At these
time limitations in technology made these measurements out of reach,
e.g. for a ten years period assuming a $\Lambda$CDM cosmology
($\Omega_{\Lambda}$\,=\,0.7, $\Omega_{m}$\,=\,0.3) the maximum change
in redshift is of the order of $\Delta$z\,=\,$2\,10^{-10}$. This idea
was revisited by Loeb in 1998 discussing the use of Lyman-alpha
absorption lines toward quasars to measure the expansion rate. The
author concluded that the signal might be marginally detectable by
observing approximately 100 quasars over 10 years with a 10-m class
telescope (Loeb 1998). Since then new plans have emerged to build a
giant optical telescope the E-ELT, a 40-m class telescope, with a
specially equipped spectrograph to perform the ``Cosmic Dynamics
Experiment'' (CODEX). The CODEX experiment is based on the Loeb test
and will make use of a specially designed spectrograph to detect the
expansion rate of the Universe via Lyman-alpha absorption lines
(Pasquini et al. 2005; Liske et al. 2008a). Feasibility studies
suggest observations of at least 20 targets for a total observing time
of 4000\,hours to reach a 3.1$\sigma$ detection of the velocity shift
of 2.34\,cm s$^{-1}$ over 20\,years (Liske et al. 2008b). This
experiment is limited by the fact that the Lyman-alpha forest is
only accesible from the ground for z\,$\geq$\,1.7 and it is most
likely that a direct measure of the recent acceleration might not be
possible with the CODEX experiment.

\smallskip

\noi Like the CODEX experiment a similar type of observation can be
achieved with the SKA and instead of observing the Lyman-alpha
forest one could use a \hi\ absorption spectra toward a radio loud active
galaxy at redshift prior to reionisation epoch (Carilli et
al. 2002). In their simulation, the
authors showed that at redshifts above 8 a Cygnus A-type galaxy would
show numerous \hi\ absorption line features in their spectra that are
suitable to measure the velocity shift.  However like the
Lyman-alpha absorption systems these sources are difficult to find
and in case of the radio counterpart have not found yet, but it is not unreasonable to expect the SKA
to be able to find these.

\bigskip

\parbox{\textwidth}{ 
\hspace{0cm}
\parbox{5cm}{ \includegraphics[width=0.6\textwidth,
    angle=0]{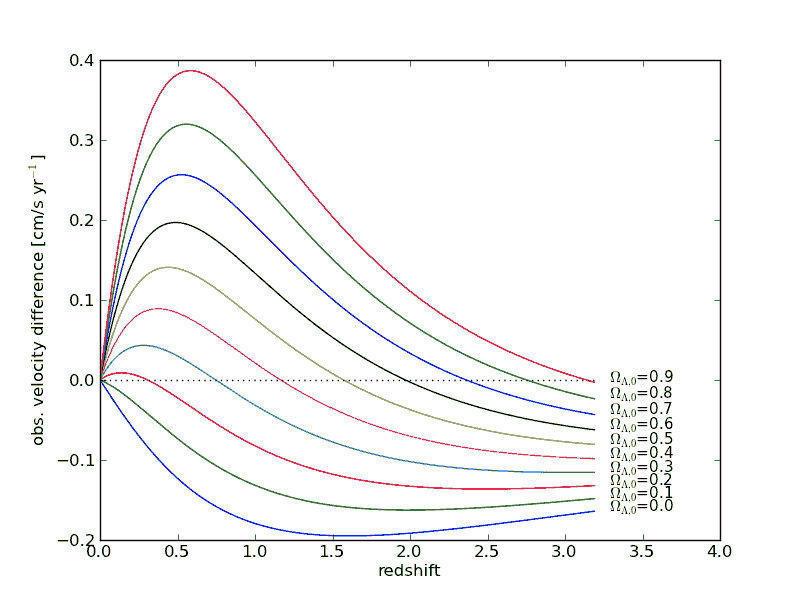}\vspace{-7.35cm}}

  \hspace{10.2cm} 
\parbox{6cm}{Figure 1$_{\rm Kl}$:  The expected velocity shift for various
    $\Lambda$CDM cosmologies ($\Omega_{m}$\,=\,0.27, $\Omega_{\Lambda}$\,=\,0.0 ... 0.9) assuming two epochs of observations within
    1\,year. Note that the velocity shift depends linearly on the
    observing period, therefore for $\Omega_{\Lambda}$\,=\,0.7 a maximum
    velocity shift of $\sim$\,2.5\,cm would be expected for an observing
    period of about ten years.}\vspace{0cm}}

\newpage

\noi Here a new experiment is proposed to directly measure the recent
acceleration of the Universe. Using the spectra, with high spectral
resolution (approx. 0.1\,cm$^{-1}$ per channel), of millions of
galaxies enables us to detect an isotropic and average velocity shift
up to a redshift of 1. Depending on the available number of galaxies per
redshift bin even tracing the functional dependence might be
possible. The figure shows the expected velocity shift within one year
of observing for various cosmological models with $\Omega_{m}$\,=\,0.27
and varying $\Omega_{\Lambda }$.  Depending on the contribution of
$\Omega_{\Lambda }$ a positive velocity shift would be expected up to
redshifts of 3, after that the velocity of the Cosmos will slow
down. Only in the case of the existence of dark energy does the velocity shift show
a pronounced functional dependency up to a redshift of 1\,--\,2 and a
maximum between z\,=\,0 and 0.6. This redshift range is ideal for
the SKA to measure the velocity shift, because the anticipated
sensitivity limits of the SKA allows observations of the \hi\ signal (in
emission) from Galaxy-like systems up to a redshift of 1. Based on the
SKADS simulations, for a flux limit of 3\,$\mu$Jy the expected number
of \hi\ emitting sources at redshift 1 are of the order of 7000 sources
per square degree (Kl\"ockner et al. 2009). Assuming a
``halve-of-the-sky `` \hi\ survey and no evolution of the \hi
-luminosity function the expected number of sources is of the order of
150 Million. This high number of sources will compensate the channel
sensitivity drop of the SKA telescope and therefore enables us to
carry out a statistical detection of the velocity shift. For lower
redshift bins the number of sources will increase and allows for a
further sample of the redshift dependency of the velocity shift. The
statistical detection of the velocity shift would come from, instead
of averaging the individual line features of the Lyman-alpha or
21\,cm forest, by averaging the subtracted 2-epoch spectra of each
individual galaxy.

\noi In general, the \hi\ emission line surveys performed with the SKA offers the
ability to directly measure the recent acceleration of the
Universe. In addition, observations of high redshifted radio sources and
the measurements of the CODEX experiment on the E-ELT could provide
further constrains on the functional dependency of the velocity shift
at larger redshifts,
and enables us to investigate the parameter space of modified
cosmological models of our Universe.\\

\noi {\small \bf Summary:~}A new experiment is proposed using the SKA
to measure the real-time expansion of the Universe using \hi\ emission
line surveys at two different epochs. In addition, a twin experiment
to the CODEX programme in the radio regime was discussed. Such
measurements are still out of reach for current telescopes, but it is
within reach for future facilities such as the SKA or the E-ELT. However
the key issue in this research will be the accuracy to which one can
determine the velocity shifts and which system parameters will cause
systematic effects  (e.g. gravitation shifts due to our planetary system; Kl\"ockner in prep.).\\

\parbox{0.9\textwidth}{
\noi{References:}\\
\noi{\scriptsize 
Bassett B.A., Corasaniti P.S., Kunz M., 2004, ApJL, 617, 1;
Carilli C.L., Gnedin N.Y., Owen F., 2002, ApJ, 577,22;
Gudmundsson E.H., Bj\"ornsson G., 2002, ApJ, 565, 1;
Heavens A., 2003, MNRAS, 343, 1327; 
Kl\"ockner H.-R., \etal , 2009, PoS(SKADS 2009), 009;
Loeb A., 1998, ApJL, 499, 111;
Liske J., \etal , 2008a, The Messenger, 133, 10;
Liske J., \etal , 2008b, MNRAS, 386, 1192;
Pasquini L., \etal , 2005, The Messenger, 122,10;
Perlmutter S., \etal , 1999, ApJ, 517, 565;
Riess A.G., \etal , 1998, AJ, 116, 1009;
Sandage A., 1962, ApJ, 136, 319;
Shapiro C., Turner M.S., 2006, ApJ, 649, 563;
Wang Y., 2006, ApJ, 647, 1
}}

\newpage
\addtocontents{toc}{\protect\newpage}
\subsection{Extragalactic astronomy}

\subsubsection{Neutral hydrogen {\scriptsize [M. Zwaan]}}

\noi What do we know about the evolution of neutral hydrogen in galaxies?
Frankly, compared to the enormous advances made over the last few
decades in the study of the stellar content of galaxies, our knowledge
of the neutral gas evolution is still very limited. Obviously, the
main reason for this is that the \hi\ 21\,cm emission line is extremely
weak and prohibitively long integration times are required on existing
radio facilities to measure the \hi\ content of galaxies at redshifts
beyond z\,$\sim$\,0.2. A few studies have now published \hi\ emission line
detections of a few dozen galaxies at redshifts z\,=\,0.1 to 0.2 (e.g.
Catinella \etal\ 2008, Freudling \etal\ 2011). But for a real
understanding of the role of gas in galaxy evolution, the SKA and its
pathfinders are required.

\noi Understanding the formation and evolution of galaxies can only be
complete by including their most fundamental ingredient:
gas. Primordial hydrogen gas dissipates energy when it falls into dark
matter halos, loses energy to collapse to a neutral hydrogen reservoir
in which denser regions further cool and form molecular clouds, which
in turn produce stars. Much of the properties of galaxies are
determined by the amount of gas they contain, and specifically, how
efficient they are in converting their innate gas content into
stars. The neutral gas is the basic fuel for the build up of stellar
mass.

\noi But the innate gas is only part of the story. Throughout
galaxies' lifetimes, they must continue to accrete gas somehow, as
without tanking for new fuel, they will exhaust their supplies within
a few Giga years (e.g. Hopkins 2008, Bigiel 2011). At the same time,
galaxies also lose part of their neutral gas, either through
interactions with other galaxies and the intra-cluster medium, or
through ionization of the lower column densities and the formation of
molecular gas at the higher column densities. Modeling this delicate
balance of gaining and losing neutral gas is extremely complicated
(e.g. Lagos 2011). In order to understand these processes, it is
essential that we measure the gas content of a large number of
galaxies over a large redshift range and determine how the neutral gas
is distributed over galaxies with different masses as a function of,
e.g. redshift, environment, luminosity etc. Only the SKA will be able
to do just that.

\noi For example, Abdalla et al. (2010) calculate that a one year long
integration with the full SKA, using a field of view of 10\,deg$^2$,
would result in the detection of ~2x10$^4$ galaxies per deg$^2$ per unit
redshift at z\,=\,1. Of course, these numbers depend strongly on the
specifications of the SKA at frequencies between $\sim$\,400 to $\sim$\,1400\,MHz, in
particular the available instantaneous bandwidth and
field-of-view. Also, they are strongly dependent on the assumed model
for the evolution of the \hi\ mass function as function of redshift,
which is exactly what we wish to measure.

\noi At present, the \hi\ mass function is only measured with high
precision in the z\,=\,0 Universe. The largest surveys to date, HIPASS and
ALFALFA, produced consistent \hi\ mass functions over the mass range
10$^{6.5}$ to 10$^{11}$ M$_{\sun}$ (Zwaan et al. 2005, Martin et
al. 2010). However, the survey volumes are still small resulting in
large variations of the measured space density across the sky. Also,
as an example of the current limitations, controversy still surrounds
the magnitude and sign of the density-dependence of the \hi\ mass
function, possibly due to depth and cosmic variance issues with
existing shallow surveys. Compared to HIPASS an ALFALFA, the SKA can
detect low mass galaxies over volumes hundreds time larger and is
therefore much less sensitive to cosmic variance.

\noi Of course, once a large sample of \hi\ selected galaxies is
available, we can go much beyond studying the gas properties of
galaxies. For example, the clustering properties of gas-rich galaxies
can be studied as was done for HIPASS by Meyer et al. (2007). Then,
using the halo occupation distribution formalism, the distribution of
galaxies within dark-matter halos can be studied (Wyithe et al. 2009),
giving insights into how galaxies grow over cosmic time. The large
number of galaxies detected by SKA surveys allow accurate measurements
of the baryon acoustic oscillations signal, which can be compared with
the results of other large-scale facilities (Abdalla et
al. 2010). Also, since 21\,cm surveys deliver \hi\ velocity widths, the
Tully-Fisher relation can be studied over a large range in cosmic
time, using the same method at z\,=\,0 and at higher redshifts. The
velocity widths can also be used to construct the rotational velocity
function, which provides a direct comparison with CDM predictions
(Zwaan et al. 2010, Papastergis et al. 2011).

\noi Going beyond the detection of individual galaxies, stacking
experiments will allow the measurement of statistical gas properties
of galaxies with redshift measurements acquired in the optical (see
Khandai 2011). For example, the stacked 21\,cm signal of Lyman-alpha
detected galaxies at redshifts z\,=\,2 to 3 should be easily
detectable. In ``HI intensity mapping'' the cumulative 21\,cm intensity
fluctuations from large numbers of galaxies are measured, without the
need to catalogue these galaxies individually (see e.g. Chang et al.
2010). The \hi\ power spectrum and the cosmic mass density of neutral
hydrogen can be measured at redshifts beyond where individual galaxies
can be detected.\\

\parbox{0.9\textwidth}{
\noi{References:}\\
\noi{\scriptsize Abdalla F.B., Blake C., Rawlings S., 2010, MNRAS, 401, 743;
Bigiel, F., \etal , 2011, ApJ, 730L, 13;
Catinella B., \etal , 2008, ApJ, 685L, 13;
Chang T., Pen U., Bandura K., Peterson J.B., 2010, Nature, 466, 463;
Freudling W. \etal , 2011, ApJ, 727, 40;
Hopkins A. M., McClure-Griffiths N. M., Gaensler B. M., 2008, ApJ, 682L, 13;
Khandai N., \etal , 2011, MNRAS, 415, 2580;
Lagos C. \etal , 2011, arXiv1105.2294;
Martin A.M., \etal , 2010, ApJ, 723, 1359;
Meyer M. J., \etal , 2007, ApJ, 654, 702;
Papastergis E., Martin A. M., Giovanelli R., Haynes M. P., 2011, arXiv1106.0710;
Wyithe S., Brown M. J. I., Zwaan M. A., Meyer M. J., 2009, arXiv0908.2854;
Zwaan M. A., Meyer M. J., Staveley-Smith L., Webster, R. L., 2005, MNRAS, 359L, 30;
Zwaan  M. A.; Meyer M. J., Staveley-Smith L., 2010, MNRAS, 403, 1969
}}\\

\subsubsection{Observational studies of gas in galaxies {\scriptsize
    [G. Kauffmann, B. Catinella, A. Saintonge]}}

An international team of astronomers [Key members:  Barbara Catinella
(MPA), Silvia Fabello (MPA), Reinhard Genzel (MPE), 
Tim Heckman (JHU),  Guinevere Kauffmann (MPA),
Carsten Kramer (IRAM), Sean Moran (JHU), Amelie Saintonge (MPE), David
Schiminovich (Columbia University), Linda Tacconi (MPE), Jing Wang (MPA)]
have been carrying out two ambitious surveys (GASS and COLD GASS) 
 to measure the atomic and
molecular gas content of around 1000 galaxies with stellar masses greater
than $10^{10} M_{\odot}$ using two of the largest radio telescopes in the
world. The results of this programme will yield precious insight into how
the interplay between gas and star formation shapes the evolution of
galaxies in the local Universe.

Galaxies are well known to divide into two large families: red, old
ellipticals and blue, star-forming spirals.  While this
distinction has been known for a long time, recent work based on the Sloan
Digital Sky Survey (SDSS) has shown that in the local Universe, the
division into these two large families occurs abruptly at a particular
mass and  density. Theoreticians have postulated a
diverse set of  mechanisms to explain the characteristic scales
evident in the galaxy population.  Most of
these mechanisms involve processes that either eject substantial amounts of
gas from the galaxy (often referred to as ``quenching''), or that regulate
the rate  at which gas is able to accrete onto the galaxy from its external
environment.

The aim of the surveys being carried out by the MPA/MPE groups is to gain
new insight into  the physical processes that regulate the present-day growth
of  galaxies with stellar masses greater than
$10^{10} M_{\odot}$   by surveying their gas content.
Neutral hydrogen (\hi ) is the source of material that will {\em
eventually} form stars;  it thus may represent  a  key ingredient in understanding
the rate at which galaxies  are gaining mass by accretion. The molecular
gas, as traced by carbon monoxide (CO) emission, probes  ``birth
clouds'' in which stars are currently forming. By studying the interplay
between atomic gas, molecular gas and young stars, one hopes to gain
insight into internally-driven  processes that regulate the conversion of gas to
stars in galaxies (see figure below).\\

\noi In order to understand how such processes operate across the galaxy population as
a whole, one requires large, {\em unbiased} samples of galaxies.  The
galaxies in the GASS and COLD GASS surveys have been  selected from the
SDSS. The survey was one of the most ambitious
optical surveys of the sky ever undertaken. Over 8 years of operation, it
obtained multi-colour images covering more than a quarter of the sky and
created 3-dimensional maps containing around a million galaxies.  The data
obtained by the SDSS provide a wealth of information about stellar content of 
nearby galaxies.  Multi-colour images
yield information about stellar ages and masses,  
while the emission and
absorption lines in the spectra  allow astronomers to
derive estimates of metallicities and star formation rates, and to assess
whether or not material was accreting onto central supermassive black
holes. Although this data provided a wealth of new information about
stellar populations in nearby galaxies, lack of information about the
associated gas has prevented real progress in disentangling accretion and
quenching processes.

\hspace{3cm}
\parbox{\textwidth}{
\parbox{5cm}{ \includegraphics[width=0.6\textwidth, angle=0]{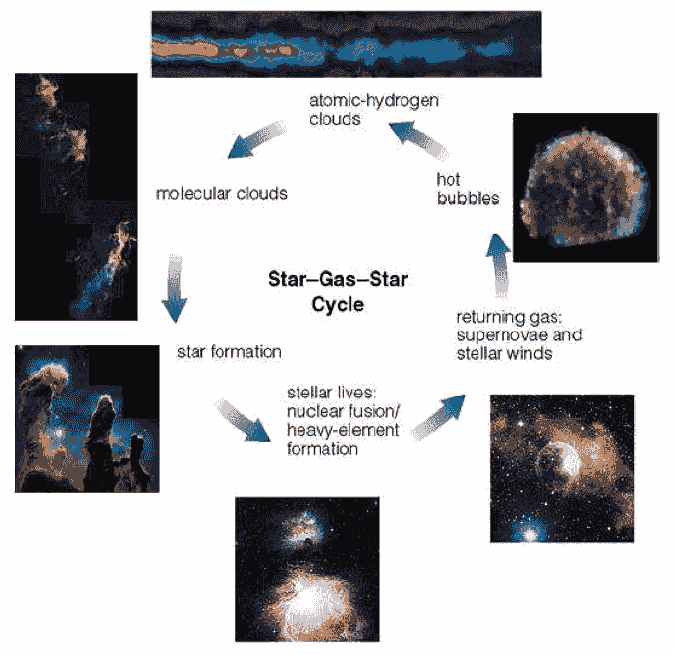}}
\vspace{1cm}
\hspace{-5cm}
\parbox{14cm}{\vspace{11cm}{Figure 1$_{\rm Ka}$: Overview of the star formation cycle in galaxies.}}
}\\

\noi GASS (http://www.mpa-garching.mpg.de/GASS) probes the
relationship between stars and gas by linking  SDSS observations
(which probe the visible light from galaxies), with those from the  space-based Galaxy
Evolution Explorer satellite (which probes light from the youngest stars)
and  from Arecibo, the largest radio telescope in the world.
The Arecibo observations started in 2008 and are on-going. A thousand
targets will be observed  until they are detected or an \hi\ gas mass
fraction limit of few  percent is reached.  A subset of
these targets are being followed up by the IRAM 30\,m telescope in Granada,
Spain, in order to measure molecular gas mass fractions down to the same
limit (see http://www.mpa-garching.mpg.de/COLD\textunderscore GASS/).

\noi The two surveys have already yielded a number of interesting new results,
which have been written up in a series of 9 papers. One of the first
striking results was that  \hi\ gas fraction  can be predicted 
accurately from two optically-derived parameters. The \hi\ fraction increases
in proportion to colour/star formation rate, but it also decreases as a function of
stellar density (Catinella et al. 2010). Disk galaxies are believed to form
from gas that cools within massive dark matter halos. As gas cools, it  loses
pressure support and falls to the centre of the halo until it becomes
rotationally supported.  The smaller the initial angular momentum of the
gas, the more it contracts. Because the star formation rate in a galaxy
increases in proportion to  the density of the gas in its
interstellar medium, denser galaxies use up their available fuel quickly
and become gas-poor. Red, high density galaxies are thus expected to have low \hi\ fractions,
as observed.

\noi Although the majority of galaxies in the sample lie on a tight ``plane''
defined by colour and density, around 10\,\% of the sample deviate significantly 
in having  higher \hi\ content than  inferred from their
optical properties.  These unusually gas-rich galaxies are of considerable
interest, because they may have recently accreted \hi\
from their surroundings.  GASS team members have carried out extensive
investigations of the properties of these objects. One interesting finding
is that unusually \hi -rich galaxies have unusually blue {\em outer} disks
(Wang et al. 2011).  Follow-up long slit spectroscopy on the Multi-Mirror
Telescope in Arizona reveals that the  blue outer disks harbour young
($<$\,1\,Gyr), metal-poor stellar populations (Moran et al. 2010). This lends
credence to the idea that present-day disk galaxies are forming from the
``inside out''. According to theory,   high angular momentum  gas accretes later than low
angular momentum gas. Although this paradigm  has commonly served as the basis
of semi-analytic models of the formation of disks in the context of ``Cold
Dark Matter'' cosmologies, this is the first time that direct supporting 
evidence has been found.

\noi Unusually \hi\ rich galaxies have regular rotation curves and light profiles
that are  symmetric. Their star formation rates are on average  no
higher than similar galaxies without excess gas (Schiminovich et al. 2010).
This suggests that most of the gas is accreted smoothly,  and not in the
form of condensed satellites, which would strongly perturb the disk,
drive gas into the central regions of the galaxy  and
result in ``bursts'' of star formation. 
In normal spirals,  \hi\ gas in the outer disks is probably transported inwards over
timescales of many Gyrs. As gas flows inwards, its  density increases until the
ultraviolet radiation produced by young stars 
is no longer able to penetrate into the densest regions.
These are the conditions under which $H_2$ molecules begin to assemble,
leading to the formation of giant molecular cloud complexes, which are the nurseries in which 
stars are born.

\noi Are the stellar nurseries of all galaxies alike?  What is the fraction of
the available molecular gas that is turned into stars before the birth
cloud is destroyed by energetic output from hot young stars in the form of
radiation and outflows? Does this fraction differ from galaxy to galaxy in
the local Universe, and was it different in galaxies in the early Universe?
These are some of the most topical questions facing astrophysicists who
attempt to understand the physical processes that regulate the rate at
which galaxies form their stars.

\noi Recent results from the COLD GASS survey indicate that the 
molecular gas depletion time (defined as the  time  taken for a galaxy
to exhaust its  supply of molecular gas at its current rate of
star formation) may not be constant, but may vary systematically from one
galaxy to another (Saintonge et al. 2011).  More actively star forming
galaxies  harbour molecular clouds in which star formation is more
efficient. For many years, this was understood to apply
only to  the most extreme star-bursting
galaxies known, the so-called Luminous Infrared Galaxy population.
Most of these galaxies exhibit signs of recent or continuing interactions. 
It was thus hypothesized that their  
interstellar medium properties could be  quite different 
from those of quiescent spirals like our own Milky Way. The
COLD GASS results now link these two populations by showing that
molecular gas depletion times vary smoothly as gas surface density and 
star formation increase.

\noi The main fascination of  galaxies since the time of Edwin
Hubble has been  the intricately interwoven system of correlations or ``scaling
laws'' that relate properties such as mass, size, age and metallicity to
each other.  The GASS and COLD GASS surveys are currently extending this
knowledge to the {\em global} interplay between gas and stars in nearby galaxies. 
The challenge for the future will be to link  phenomena
that operate on vastly different physical scales: from the dense cores of
molecular clouds to the diffuse ionized gas between galaxies. 
Meeting this challenge will undoubtedly require new technologies 
and new surveys.  Our experience in the construction of GASS and
COLD GASS has prepared us well for  many new discoveries to come with
the SKA and its pathfinder telescopes.\\

\noi As a first step the MPA and ASTRON scientists have teamed up
early 2012 to carry out the ``WSRT Bluedisk project''
(http://www.mpa-garching.mpg.de/GASS/Bluedisk/). This project
is studying gas accretion in nearby galaxies by obtaining \hi\ maps of
galaxies where there is evidence of recent growth of the
outerdisk. The observing configuration is similar to upcoming large
surveys that will be carried out using the APERTIF focal plane
array. The Bluedisk survey can be regarded as a ``pilot study''. The
MPA/ASTRON team is using it as a means to develop the necessary
automated data reduction and \hi\ parametrisation software than can
later be applied to the very large samples that will be available from
2015 onwards. In addition, the \hi\ data will be linked to a variety of
multi-wavelength data sets, including the Sloan Digital Sky Survey, UV
imaging from GALEX and IR imaging from WISE.\\

\parbox{0.9\textwidth}{
\noi{References:}\\
\noi{\scriptsize 
Catinella B., \etal , 2010, MNRAS, 403, 683;
Saintonge A., \etal , 2011, MNRAS, 415,61;
Schiminovich D., \etal , 2010, MNRAS. 408, 919;
Moran S.M., \etal , 2010, ApJ, 720,1126;
Wang J., \etal , 2011, MNRAS, 412, 1081
}}\\

\subsubsection{Constraining the neutral gas accretion rates of
low-redshift galaxies with SKA {\scriptsize [P. Richter]}}

\noi One crucial aspect of galaxy formation and evolution concerns
the continuous infall of intergalactic gas onto galaxies.
While it is clear that galaxies do accrete substantial amounts
of gas from intergalactic space to power star formation, the
exact way of how galaxies get their gas is still
a matter of debate. In the conventional sketch of galaxy
formation and evolution gas is falling into a dark matter (DM)
halo and then is shock-heated to approximately the
halo virial temperature (a few 10$^6$\,K, typically),
residing in quasi-hydrostatic equilibrium with the
DM potential well (Rees \& Ostriker 1977). 
The gas then cools slowly through radiation,
condenses and settles into the centre of the potential
where it forms stars as part of a galaxy (``hot mode'' of
gas accretion). It has been argued, however,
that for smaller DM potential wells the infalling gas may radiate
its acquired potential energy at much lower temperatures
(T\,$<$\,10$^{5.5}$\,K, typically), so that one speaks of the ``cold mode''
of gas accretion (White \& Rees 1978). For the
cold mode of gas accretion the star-formation rate of
the central galaxy is directly coupled to its gas-accretion
rate (White \& Frenk 1991).
Numerical simulations indicate that for individual
galaxies the dominating gas-accretion mode depends
on the mass and the redshift (e.g. Keres et al. 2005).
The general trend for z\,=\,0 is that the hot mode of gas accretion 
dominates for massive galaxies with DM-halo masses $>$\,10$^{12}$\,M$_{\rm sun}$,
while the cold accretion mode dominates for galaxies
with smaller DM-halo masses (e.g. van de Voort et al. 2011).

\noi Independently of the theoretically expected gas-accretion
mode of galaxies it is known since a long time
that galaxies at low and high z are surrounded by large
amounts of neutral and ionized gas that partly originates in the
IGM. This material is complemented by neutral and ionized
gas that is expelled from the galaxies as part of galactic
fountains, galactic winds, and from merger processes (see e.g.
Richter 2006). Because the interplay between these
circumgalactic gas components is manifold and the gas
physics of such a turbulent multi-phase medium is
complex, the circulation of neutral and ionized
gas in the inner and outer halos currently
cannot be modeled in full detail in hydrodynamical simulations.
To improve current models of galaxy evolution models it is
of imminent importance to quantify the amount of cool, neutral
gas in and around galaxies from observations and to search
for obsvervational strategies to separate metal-deficient
infalling intergalactic gas from metal-enriched gaseous material
that is circulating in the circumgalactic environment of
galaxies as a result of fountain processes and galaxy mergers.\\

\noi We therefore intent to use the SKA to investigate in detail the 
distribution of neutral hydrogen in the entended halos of 
low-redshift galaxies to constrain the distribution and amount
of cool gas in galaxy halos and to estimate the infall rate of 
neutral gas onto galaxies in the local Universe. 
With its superb sensitivity the SKA represents a particularly 
powerful instrument for this task, as it allows us to map the 
gaseous environments of a very large number of galaxies down to very low
\hi\ column densities of a few 10$^{16}$\,cm$^{-2}$ at sufficient linear 
resolution for the local galaxy population.
Having such a large sample of galaxies is of great importance to 
pinpoint possible differences in the radial distribution of
neutral halo gas in galaxies of different morphological types and in 
different large-scale environments (clusters, groups, isolated 
galaxies). The data will further allow us to study the infall and 
outflow characeristics of neutral and partly ionized gas 
in the inner and outer galaxy halos and to estimate the gas-accretion
rate of individual galaxies, which then can readily compared with
the galaxies' star-formation rates. Finally, deep SKA observations
of the faint gaseous outskirts of galaxies will enable us to 
investigate, how individual galaxies and galaxy groups are 
connected to the intergalactic medium.\\

\noi In summary, the above outlined SKA observations will help to better 
understand the ongoing formation and evolution of galaxies in the 
local Universe and their connection to the cosmic web.\\

\parbox{0.9\textwidth}{
\noi{References:}\\
\noi{\scriptsize 
Kere{\v s} D., \etal , 2005, MNRAS, 363, 2;
Rees M.J., Ostriker J.P., 1977, MNRAS, 179, 541; 
Richter P.,2006, RviMA 19, 31;
van de Voort F., et al., 2011, MNRAS 414, 2458;
White S.D., Rees M.J., 1978, MNRAS, 183, 341;
White S.D., Frenk C.S., 1991, ApJ, 379, 52
}}\\

\subsubsection{Extragalactic water-vapour maser {\scriptsize
    [C. Henkel]}}

\noi {\bfseries\boldmath\small Background:~~}The physical conditions
in active galactic nuclei (AGN) are unique in the cosmos.  Stellar and
gas densities are very large, and enormous amounts of angular momentum
and energy are released as material accretes onto massive black
holes. Stellar and interstellar gas constitute reservoirs of accreting
material.  Studies of the structure, kinematics, and excitation of
this material is the sole means available to investigate massive
compact objects, which are otherwise not directly visible. While stars
close to the massive black holes are difficult to detect because of
obscuration and crowding, emission and absorption by the neutral
atomic, ionized, and molecular components of the interstellar medium
(ISM) may be readily studied at radio wavelengths, applying
interferometry.

\smallskip

\noi The role of the ISM in the vicinity of active nuclei is an important one 
because it feeds the central engines, thus determining their masses and 
angular momenta. In addition, the ISM directly affects the overall appearance 
of AGN. This (1) determines the degree of shielding from various viewing angles 
(recognition of which motivated the formulation of the AGN unification paradigm) 
and (2) results in intense emission of electromagnetic radiation from radio to 
X-ray bands, which has an impact on the ISM structure and energetics in parts 
of the parent galaxies and provides a handle for the study of matter under 
truly extreme conditions (e.g. Morganti et al. 2005).

\smallskip

\noi More than 100 type-2 active galactic nuclei contain known sources of H$_2$O
maser emssion ($\nu_{\rm rest}$\,$\sim$\,22\,GHz). Radio interferometric studies
of these ``nuclear'' masers are the only means by which structures $<$\,1\,pc
surrounding supermassive black holes can be mapped directly. Investigations
of the first few such sources have demonstrated that H$_2$O maser emission 
traces hotspots on Keplerian orbits in warped accretion disks at galactocentric 
radii of 0.1\,--\,2\,pc, surrounding nuclear engines with characteristic masses of 
10$^6$\,--\,10$^8$\,M$_{\odot}$. \\

\noi {\bfseries\boldmath\small Key Science:~~}Interferometric observations of H$_2$O masers allow us to obtain information
of the sub-pc morphology of the ISM surrounding highly obscured AGN
(Miyoshi et al. 1995, Reid et al. 2009, Braatz et al. 2010). Because 
of the high gas density required to support H$_2$O maser action,
disks that cross the lines-of-sight to central engines are ready substitutes
for the obscuring tori featured in the AGN unification paradigm. From the 2-d
images and the kinematical information contained in the spectroscopic 
data, 3-dimensional shapes of nuclear accretion disks can be derived.
Such accretion disks may be warped (e.g. Miyoshi et al. 1995). However,
the warping mechanism (e.g. radiative torques) is not certain. 
Tests of warp models require a sample of sources with different luminosities 
and accretion rates. Another important item is the thickness of the disks, 
which is critical to calculations of accretion rate and identification of 
accretion modes (advective, convective or viscous)

\smallskip
\noi H$_2$O jet-masers may arise not from the circumnuclear tori of AGN but from 
a shocked region at the interface between an energetic nuclear jet and an
encroaching molecular cloud (e.g. Peck et al. 2003). Reverberation measurements 
of the nuclear continuum and H$_2$O line emission can provide information
on the distance between their flaring components on spatial scales even 
well below those of the interferometric measurements. 

\smallskip
\noi Observations of H$_2$O disk-masers also allow us to derive precise dynamical 
masses of the supermassive back holes (BHs), through an analysis of Keplerian 
disks. These masses, compared with those of the galactic bulges, will yield 
essential information on BH - bulge mass correlations, improving our understanding 
of galaxy formation and providing urgently needed data for targets with less 
massive central engines than the otherwise almost exclusively studied giant 
early type galaxies (Greene et al. 2010, Kuo et al. 2011).

\smallskip
As a complement to observations of the cosmic microwave background, an
accurate measurement of the Hubble constant (H$_0$) will provide the best single
constraint on models of dark energy, because H$_0$ is a ``local'' parameter,
referring to a time where dark energy has truly become dominant. This will also
improve our knowledge on the geometry of the Universe, i.e. its observationally 
so far poorly constrained flatness. Making use of the morphology and kinematics 
of the maser disks, direct geometric distances and radial velocities can be 
determined. To give an example: For 100 such galaxies with distances measured 
to an accuracy of $\sim$\,10\,\%, H$_0$ could be determined with a precision of 
$\sim$\,1\,\%. \\

\noi {\bfseries\boldmath\small The impact of the SKA:~~}Because of its enormous sensitivity, the SKA will contribute substantially 
to the detection and mapping of H$_2$O maser sources in AGN. While so far 
only the few strongest sources could be analyzed in detail, with the SKA a 
much larger number of targets can be analyzed. In particular, the SKA will 
be sensitive enough to observe a significant number of masers at high redshift. 
Following Impellizzeri et al. (2008), the mean volume densities and luminosities 
of H$_2$O masers is much higher in the early Universe than locally. With the 
SKA, thus it will be for the first time possible to investigate the evolution 
of nuclear maser properties as a function of redshift.\\

\noi Although the SKA will substantially increase the number of known H$_2$O 
maser sources, most Key Science (apart from statistical studies of detection 
rates for different samples) will require high resolution imaging. The 
simplest solution would be operation of outrigger stations in combination
with the core of the SKA, which itself would best be operated as a phased
instrument. \\

\parbox{0.9\textwidth}{
\noi{References:}\\
\noi{\scriptsize Braatz J.A. \etal , 2010, ApJ 718, 657;
Greene J. \etal , 2010, ApJ 721, 26;
Kuo  C. \etal , 2011, ApJ 727, 20;
Impellizzeri C.M.V. \etal , 2008, Nature 456, 927;
Morganti  R. \etal , 2005, New Astron. 48, 1195;
Miyoshi M. \etal , 1995, Nature 373, 127;
Peck A.B. et al. 2003, ApJ 590, 149;
Reid M. et al. 2009, ApJ 695, 287
}}\\

\subsubsection{Measuring the evolution of the galaxy merger rate with
  extragalactic hydroxyl masers   {\scriptsize
    [A.~Roy, H.-R. Kl\"ockner]}}

\noi {\bfseries\boldmath\small Introduction: } The galaxy population evolves over
cosmological history due either to the evolution of galaxy
luminosities, galaxy number density or both. Density evolution is a
key prediction of models of galaxy formation and one can test it if
one can break the observational degeneracy between luminosity and
density evolution, for example by measuring the rate of galaxy mergers
versus redshift. The environment of mergers provides the conditions
needed for hydroxyl megamaser (OHMM) emission and hence OHMM are perfect
tracers for galaxy mergers. The SKA and its pathfinder telescopes
would be ideal to undertake an OHMM survey to study the evolution of
the merger rate at cosmological distances and this study could be
conducted by mining the data products produced by the SKA \hi\
surveys.\\

\noi {\bfseries\boldmath\small Background: }Extragalactic hydroxyl mega-maser (OHMM)
emission has been detected in the highest luminosity far infrared
(FIR) galaxies (e.g. luminous (LIRG) and ultra-luminous infrared
galaxies (ULIRG)), which are found to have massive star formation
often resulting from mergers or interactions. Their OH main-line (showing
a strong 1667\,MHz and a weaker 1665\,MHz) luminosities are 2 to 4 orders of
magnitude greater than those of Galactic OH masers reaching above
10$^4$\,L$_{\sun}$, which makes them detectable over cosmological
distances. Typical line profiles vary from 100\,\kms\ up to 1000\,\kms\
with broader profiles exhibiting multiple components.

The major surveys for OH megamasers to date are Baan \& Haschick
(1983), Baan (1989 and references therein), Staveley-Smith et
al. (1992), and the Arecibo OH megamaser survey (Darling \& Giovanelli
2000, 2001, 2002), and Kl\"ockner (2004). Fifty-five sources were
detected in earlier targeted surveys of FIR luminous sources. The
targeted survey of Darling \& Giovanelli at Arecibo is the deepest
survey to date and contained 311 targets selected based on IRAS
60\,$\mu$m fluxes. It spanned z\,=\,0.1 to 0.23 and excluded redshifts
near 0.174 due to contamination by Galactic \hi\ emission. Fifty-two OH
MM were found and 1 OH absorber, which gives a detection rate of 16\,\%
at the high luminosity end but this ratio decreases towards lower FIR
luminosities. Typical OH megamasers found had flux densities near the
survey flux-density limit, as expected for any flux-limited survey.

Evolutionary scenarios of galaxy merging rates parameterized by (1 +
z)$^{\rm m}$, with m between 0 (no evolution) and 8 (extreme evolution),
predict a factor of 100 difference in the volume density of masers at
redshift 1 and so are rather easy to distinguish if the survey extends
that far (Briggs 1998). The Arecibo survey went only to z\,=\,0.23 where
the difference in volume density is a factor of five, which could not
be distinguished statistically because of too few detections. These
models can be well constrained given enough OH megamaser detections to
determine the luminosity function with redshift with good
precision. Already, surveys of sub-mm galaxies suggest m = 3.5 at low
redshifts and clearly rule out the no evolution case (Aretxaga et
al. 2007).

The dependence of maser properties on host galaxy properties shows a
strong correlation with IR luminosity, going as ${\rm L}_{\rm OH} \sim
{\rm L}^{2.29\pm 0.1}_{\rm OH}$ (Kl\"ockner 2004) and no correlation
with any other galaxy properties was found. This confirms our
understanding of the basic physical processes and has been recognised
early on in the study of OHMM (Baan 1989). The luminosity function of
Darling \& Giovanelli (2002) was found to have the rather flat
power-law dependence on luminosity of 10$^{-5}$\,(L$_{\rm OH}$ /
L$_{\sun}$ ) - 0.64\,Mpc$^{-3}$\,dex$^{-1}$, and a high upper luminosity
cutoff of 10$^{4.4}$\,L$_{\sun}$.  In contrast, the luminosity function
found by Kl\"ockner (2004) was based on twice the number of sources
and had the steeper power-law index of -1.1 and an exponential cutoff
above 10$^{3.58}$\,L$_{\sun}$. This turnover at the upper end of the LF
is still ill-determined because of the small number of sources, but
also predicts the small number of detections in the local volume ($\sim$\,110 for moderate
evolution).  Surveys at low redshift among luminous FIR galaxies have
been rather complete and have produced about 100 sources with only two
gigamaser sources at redshift z\,$>$\,0.2. This number is consistent with
our predictions for the nearby Universe. This result is also
consistent with the results of the HIPASS survey, where no new masers
were found within redshift window 0.17 and 0.22 (Kanekar \&
Staveley-Smith, personal communication).

Most of the OHMM emission is confined within the nuclear region of
less than a kilo-parsec in size showing starburst and AGN-type nuclear
  phenomena. The clear association of the OH
emission with the infrared emission indicates that OH is a good tracer
of the dusty circum-nuclear environment. This dusty environment plays
an central role in the unification scheme of active galaxies where a
dusty circum-nuclear torus or thick disk is used to explain the
various emission appearance of their nuclear regions (Antonucci \&
Miller 1985).  The existence of such nuclear structures follows from
detailed studies of the OH Megamasers Mrk\,231, Mrk\,273 and III\,Zw\,35,
IRAS\,17208--0014 (Kl\"ockner et al. 2003, Kl\"ockner \& Baan 2004,
Pihlstr\"om et al. 2001, Momjian et al. 2006) showing systematic velocity
gradient of a rotating disk or torus at scale sizes of about a few
hundred of parsecs. The modelling also allows to estimate the enclosed
dynamical masses indicating, in some cases, the presence of a super
massive black hole (SMBH). 

The powerful OHMM in the southern hemisphere, IRAS 20100-4156 (at z\,=\,1.29 and L$_{\rm OH}$ = 10$^{3.96}$\,L$_{\sun}$; Stavely-Smith et
al. 1989) clearly displays both the 1667\,MHz and 1665\,MHz
components. On the other hand, the two highest luminosity sources,
IRAS 14070+0727 (Baan et al. 1992) and IRAS 12032+1707 (Darling \&
Giovanelli 2001) (respectively at z\,=\,0.265 and 0.217 and with L$_{\rm
  OH}$ = 10$^{4.15}$\,L$_{\sun}$ and 10$^{4.13}$\,L$_{\sun}$), show broad and
narrow components that cover a velocity width of more than 2000 \kms
. Therefore, we may find a broader emission profile at higher
redshifts. Such luminous OH megamasers would be detectable out to z\,=\,3.5 assuming a sensitivity achievable already in the pathfinder
telescopes (e.g. 0.2\,mJy
3$\sigma$ in a 100\,kHz channel)  and so megamasers
have the potential to be used for cosmological studies.\\

\noi {\bfseries\boldmath\small Conclusions:} The SKA \hi\ surveys can be used for data
mining to search for extragalactic hydroxyl emission (OH) at
cosmological distances. Based on the current knowledge data mining to
search for the OH mainlines at 1667\,MHz and 1665\,MHz (restframe) seems to be most
promising. The main science driver for OH studies is to measure the rate of galaxy
mergers versus redshift. By doing so one can break the observational degeneracy
between density and luminosity evolution, since galaxy merger events
reduce the volume density of galaxies and so the merger rate is the
first derivative of the galaxy volume density with respect to
time. Models of density evolution take the present comoving volume
density of galaxies and scale as (1 + z)$^{m}$ into the past, with m
between 0 (no evolution) and 8 (extreme evolution), with existing
observations yielding values over the wide range of 3 $<$ m $<$ 8. The
result can place constraints on models of galaxy formation, which
bears on cosmological parameters.

\noi In addition to the main science driver the following studies of OHMM sources
at cosmological distances enables us to:

\begin{itemize}
\item[--] {\bf \small study the nuclear condition} -- Observations to
  image the structure and extent of OH emission relative to the
  continuum emission on angular scales offered by \skai , \skaii\ and
  possibly SKA+VLBI enables us to study the nuclear composition, the
  nuclear kinematics and the extreme physical conditions of the
  nuclear environment.

\item[--]{\bf \small estimate the redshifts of sub-mm galaxies (SMGs)} -- Most of the distant, luminous
systems seen by LABOCA and SCUBA have no optical counterparts and lack
detectable optical or infrared spectroscopic signatures, and so their
redshift distribution and hence their possibly dominant contribution
to star formation history is poorly constrained. Search for CO
emission has been successful in a few and showed them to be massive
and gas-rich systems. However, redshift-blind CO search involves
searching bandwidths of $\sim$\,100\,GHz, which is beyond current
capabilities. Alternatively, Townsend et al. (2001) suggest that
redshifts might be obtainable from OHMM searches in these systems,
assuming that SMGs are also ultra-luminous infrared galaxies with
similar high probability (50\,\%) of producing detectable OH megamaser
emission as in the local ULIRG population with L$_{\rm FIR} > 10^{11-12}$\,L$_{\sun}$. 

\item[--]{\bf \small investigate the evolution in the maser luminosity function} -- The luminosity function
of OHMM might evolve with redshift, for example if cosmic downsizing
also affects the OHMM population. The OHMM population may thus show
density and luminosity evolution independently from the evolution of
the parent ULIRG/SMG population. Given sufficient detections, these
effects may be studied by constructing the luminosity function in
redshift bins and by using the methods used to measure separately the
density and luminosity evolution of QSOs (see Boyle et al. 1988;
Schmidt et al. 1995). Even in the case of no detections at high
redshift, this would rule out most of the evolutionary models used
today for OH Megamaser modelling.

\item[--]{\bf \small determine a possible upper luminosity cutoff} -- It is unknown whether an upper-luminosity
cutoff affects OHMM. The large volume of space surveyed in a
wide and deep SKA survey will vastly improve the chances of detecting rare
high-luminosity megamasers and so explore the presently unexplored
high-luminosity end of the luminosity function. The cutoff luminosity
can be studied by constructing luminosity functions in redshift bins
and looking for changes with redshift. The number of detections
expected in an OH megamaser survey is extremely sensitive to the
(uncertain) location of the cutoff and one could potentially find very
many sources.
\end{itemize}

\parbox{0.9\textwidth}{
\noi{References:}\\
\noi{\scriptsize 
Aretxaga I., et al., 2007, MNRAS, 379,1571;
Baan  W. A., 1989, ApJ, 338, 804;
Baan  W. A., et al., 1992, ApJ, 396, L99;
Baan W. A., Haschick A. D., 1983, AJ, 88, 1088;
Boyle B. J., Shanks T., Peterson B. A., 1988, MNRAS, 235, 935;
Briggs F. H., 1998, A\&A, 336, 815;
Darling J., Giovanelli R., 2000, AJ, 119, 3003;
Darling J., Giovanelli R., 2001, AJ, 121, 1278;
Darling J., Giovanelli R., 2002, ApJ, 572, 810;
Kl\"ockner H.-R., 2004, PhD thesis; University of Groningen;
Kl\"ockner H.-R., Baan W.A., 2004, A\& A, 419, 887;
Kl\"ockner H.-R., Baan W.A., Garrett G.A., 2003, Nature, 421, 821;
Momjian E., Romney J.D., Carilli C.L., Troland T.H., 2006, ApJ, 653, 1172; 
Pihlstr\"om Y.M., \etal , 2001, A\& A, 377, 413;
Robishaw T., Quataert E., Heiles C., 2008, ApJ, 680, 981;
Schmidt M., Schneider D. P., Gunn J. E., 1995, AJ, 110, 68;
Staveley-Smith L., \etal , 1992, MNRAS, 258, 725;
Stavely-Smith, L., \etal , 1989, Nature, 537, 625;
Townsend, R. H. D., \etal , 2001, MNRAS, 328, L17
}}\\

\subsubsection{Gravitational lenses {\scriptsize [O. Wucknitz]}}

\noi Gravitational lenses are unique probes of the Universe from high redshift down
to our cosmic neighbourhood. As predicted by general relativity, light is
deflected by gravitational fields. Depending on the geometry, this effect
leads to weak distortions of image shapes (weak lensing) or even the formation
of multiple images (strong lensing), often with strong magnifications.
Gravitational lensing by the Sun was the first test of general relativity and
still has the potential to study any possible deviations from these
fundamental laws of physics. At the same time the effect is regularly used as
a very versatile tool for a wide area of astrophysical research.\\

\noi {\bfseries\boldmath\small Natural telescopes:~~}Lensing by galaxies or clusters of
galaxies often leads to magnifications by one or even several orders
of magnitude so that they can be seen as natural extensions of our
instruments to increase the resolution significantly. At the same
time, because surface brightness is conserved, they also amplify the
flux densities to boost the sensitivity limits. At the highest
magnifications, a gravitational lens can effectively turn the SKA into
a TSKA (Thousand Square Kilometre Array). Both effects are used
extensively to search for high-redshift sources and to study their
properties as detailed as otherwise
only possible at much smaller distances.

\medskip

\noi{\bfseries\boldmath\small Cosmology:~~}The best-known application of lensing for
cosmology is the measurement of the Hubble constant using time delays
between lensed images. The age of the Universe approximately equals
the inverse Hubble constant and scales directly with the measured time
delays. The only significant uncertainty in this relation is the mass
distribution of the lenses themselves (see below).  Lensing provides a
one-step path to cosmological scales and completely avoids the many
sources of systematical errors that are inherent to classical
distance-ladder methods.  Together with current CMB measurements,
these results can constrain all relevant cosmological parameters.

Weak lensing (see also Section\ 8.2.7), the statistical study of small image distortions, can provide
additional and independent information about cosmological parameters,
particularly about the equation of state of dark energy. So far, weak lensing
has only been studied at optical wavelengths, because radio surveys have not
reached the required combination of source density and resolution. This will
change with the SKA, and pilot studies will already be possible with
LOFAR.

\medskip

\noi{\bfseries\boldmath\small Mass distribution of lenses:~~}The image configuration of
strong lenses provides invaluable information on the mass distribution
(luminous and dark) of the lensing galaxies (or clusters). This is the
only way to determine mass distributions of high-redshift objects and
study the formation of galaxies and clusters in detail. This knowledge
is essential for our understanding of the properties of dark matter.

Subjects of particular interest are the large-scale mass profiles, the
small-scale clumping predicted by CDM structure formation simulations, and the
central density peaks of galaxies and clusters. Lensing is the only way to
study these aspects at high redshifts.

\medskip

\noi {\bfseries\boldmath\small Propagation effects:~~}Lensing can provide several images of
one and the same source, with the same intrinsic spectrum, structure
and polarisation properties. Any observed differences between the
images (e.g. different spectrum or polarisation) can be used to study
differential propagation effects happening on the way, typically in
the lensing galaxy itself.  This makes such studies independent of
arbitrary assumption on the intrinsic properties of the source.
Typical features are dust extinction and reddening in the optical and
scatter broadening, free-free absorption and Faraday rotation in the
radio regime, all of which provide information about the physical
conditions in distant
galaxies.

\medskip

\noi {\bfseries\boldmath\small Radio observations of lenses:~~}Observations at radio
wavelengths offer a number of advantages. They are independent of dust
absorption in the lenses, one of the major source of systematic errors
and selection effects in optical studies. In addition, radio flux
measurements are not affected by microlensing, which is a major
perturbing factor for optical measurements. Because measured flux
ratios are very sensitive to small-scale mass clumping, important
information is lost in the optical.  Instrumentally, radio
interferometry is the only technique providing information on all
scales, from degrees down to milli-arcsec and even below. Luckily,
radio sources show structures on all these scales, so that e.g. lens
mass distributions can be studied on scales from megaparsecs down to
parsecs. This is not possible with any other technique.

So far, no efficient instruments are available to survey large fractions of
the radio sky with the required resolution in the sub-arcsec range to find
lensed radio sources. Only about 40 of such systems are known at the moment,
compared to hundreds in the optical. LOFAR as a pathfinder for the SKA is just
about to reach the required resolution (with international baselines) and
imaging capabilities for large-scale radio lens searches. It has the potential
to increase the number of radio lenses by an order of magnitude. The
experience gained in the preparation and execution of LOFAR lens searches will
be essential to prepare for much larger SKA projects.

Koopmans \etal\ estimated the number of lenses that the SKA can
see in a half-sky general-purpose survey as $10^6$, of which $10^5$ can be
identified easily. Even though the exact numbers will depend on the final
design, it is clear that the SKA will increase the number of known radio
lenses by several orders of magnitude, which will allow us to do new kinds of
science. The properties of the SKA (high resolution and sensitivity, wide
spectral coverage) will make the identification of lenses much easier than
with current instruments. The resolution is sufficient to identify typical
lensing geometries directly, without follow-up observations. The spectral
information will be invaluable to identify which parts of a multiple-image
configurations correspond to the same part of the source, which makes the
modelling much more robust. Algorithms to model the mass distribution and
(unlensed) source structure simultaneously, while accounting for the 
interferometric measurement process properly, are already available (Wucknitz
2004). They are currently being extended and improved for LOFAR and e-MERLIN
projects.

Given the resolution of the SKA and the expected number of lenses, we will not
only obtain a large sample for statistical analyses, but will also find a
number of very exotic lens systems and configurations, e.g. lensing by dwarf
galaxies and many structures distorted by mass-clumps. In the latter case, 
the local mass distribution can be ``mapped'' directly from the observations.\\

\parbox{0.9\textwidth}{
\noi{References:}\\
\noi{\scriptsize 
Koopmans L.V.E., Browne I.W.A., Jackson N.J., 2004, New Astronomy Reviews, 48, 1085;
Wucknitz O., 2004. MNRAS, 349, 1
}}\\

\subsubsection{Weak gravitational lensing with the SKA   {\scriptsize [P. Schneider]}}

\noi The statistical weak gravitational lensing effect caused by the
inhomogeneously distributed matter in the Universe leaves an imprint
on the observable shapes of images of distant galaxies. Among the most
important applications of the weak lensing effect are: 
\begin{itemize}
\item[--] reconstruction of the two-dimensional total (dark + luminous)
  mass profiles of individual galaxy clusters,
\item[--] the determination of the mean mass profile around classes of
  objects, such as galaxies and galaxy groups, to study the
  mass--luminosity relation and to probe for a `universal' density
  profile, 
\item[--] a direct determination of the correlation between luminous
  objects, such as galaxies, and the underlying total mass
  distribution, i.e., the bias of these objects as a function of,
  e.g. luminosity, redshift and object type,
\item[--] the study of the geometry of the Universe and the large-scale
  structure of its matter content, which is used in particular to
  determine cosmological parameters. In fact, the ``Dark Energy Task
  Force'' in the US and the ESA-ESO Working Group on ``Fundamental
  Cosmology'' concluded that weak lensing is potentially the most
  powerful tool to study the equation-of-state of ``Dark Energy''.
\end{itemize}

\noi The principle of weak lensing is to measure correlations between the
ellipticities of distant sources with either the position of
foreground objects (galaxies, groups or clusters), or correlations
between image ellipticities. Based on the fact that no direction is
singled out in the Universe, so that the intrinsic orientation of
sources is statistically isotropic,
these correlations measure the strength
of the accumulated tidal gravitational field along the line-of-sight
to these sources. Since the projected brightness distribution of extended
astronomical sources is not circular, there is a lower bound on the
noise in such measurements, given by the ellipticity dispersion of the
sources. Therefore, to achieve a large signal-to-noise for such
measurements, one needs a high number density of these
sources on the sky. For this reason, weak lensing has been applied almost
exclusively to faint galaxies found in optical wide-field images, as these provide
the highest number density. Furthermore, in order to beat sample
variance, large areas of the sky need to be imaged.  The
state-of-the-art is defined by the CFHTLenS survey, a 5-band imaging
survey over $\sim$\,150$\,{\rm deg}^2$ with a density of galaxies that
can be used for shape measurements of about n\,$\sim$\,20\,arcmin$^{-2}$.
The most ambitious project currently planned for weak lensing is the
ESA Euclid mission, a ``Dark Energy'' satellite which will map half the
sky with a galaxy number density of n\,$\sim$\,35\,arcmin$^{-2}$.

\noi There are several technical difficulties of weak lensing measurements,
in particular the necessity to correct the measured ellipticities for
the influence of the point-spread function. For ground-based
observations, the PSF is typically of the same size as the sources, so
that the PSF causes a strong smearing of the brightness profile of the
images. Corrections to account for this smearing necessarily amplify
the noise in measured ellipticities.  Furthermore, the PSF is in
general not circular, and hence imposes an additional ellipticity to
the images.  To correct for these effects, the PSF, which varies with
the angular position on the sky and is different for exposure to
exposure, needs to be known with very high accuracy. It can be
measured at the location of stars -- i.e., point sources -- but must
then be interpolated to the positions of the individual galaxy images.

\noi A further difficulty lies in the necessity to measure redshifts of the
objects, not only to increase the measurement sensitivity of weak
lensing, but mainly to correct for systematic effects which are due to
intrinsic alignments of galaxy shapes -- for example, the same tidal
field that causes a lensing distortion of a light bundle can lead to
an alignment of galaxies with regards to the orientation of the tidal
field. These systematics can be detected and removed due to their
characteristic redshift dependence, but for this purpose, redshifts
need to be known. For optical weak lensing surveys, this needs to be
done using photometric redshift techniques.

\noi The SKA will open totally new opportunities for weak lensing, mainly
due to five different routes: 
\begin{enumerate}
\item The sensitivity of the SKA will populate
the radio sky with a substantially larger number of (resolved) sources
than is possible with existing or planned optical telescopes, either
from the ground or from space (with the possible exception of the JWST
which, however, has a very small field-of-view and is therefore unable
to conduct efficient weak lensing surveys).  
\item The large
instantaneous field of view will allow for surveys covering
significant fractions of the sky on short time scales. 
\item The PSF
of the SKA is small, so that PSF effects play a substantially smaller
role for the determination of image shapes. This implies that much
smaller corrections to the observed image ellipticities need to be
applied. 
\item The PSF of the SKA is
supposed to be perfectly reconstructible from the locations of the
individual antennas. This will remove the largest current uncertainty in weak
lensing measurements; in particular, one will be able to reliably apply 
more
sophisticated methods for PSF corrections which
appear to be less robust on optical images, most likely due to
remaining uncertainties in the knowledge of the PSF.
\item H{\sc i} measurements will allow us to determine
secure redshifts of (a significant fraction of) the source population,
which will provide a tremendous enhancement over the uncertainties
associated with photometric redshifts. 
\end{enumerate}

\noi Together, these points will render the SKA an enormously powerful weak
lensing machine. Whereas the Euclid mission will be close to the
``perfect'' weak lensing survey possible in the optical wavelength
regime, the SKA will allow for better surveys in terms of image
density, ability to control the PSF, and the ratio of PSF-to-image size
which will lead to far smaller corrections that need to be
applied. Furthermore, the exquisite image quality expected from the
SKA will allow us to measure shape parameters of images beyond the
ellipticity, for example the flexion. Given that flexion turns out to
be notoriously difficult to measure on optical images, mainly due to
uncertainties in the exact profile of the PSF and the relative sizes
of PSF and image, this will provide an
important additional channel for weak lensing measurements.
The as yet unmatched promise of the Euclid mission -- in
particular to measure the equation-of-state parameter $w$ of Dark
Energy to within $\sim$\,1\,\% accuracy -- can probably be topped by
future SKA weak lensing studies.

\smallskip

\noi Finally, in addition to individual objects, the patchy hydrogen
ionization which can be observed with the SKA
provides a large number of statistically isotropic source
screens which in principle can be used for measuring weak lensing
effects to very high redshifts.\\

\subsubsection{Radio continuum as a measure for dust-unbiased star
  formation across the Universe {\scriptsize \mbox{[E.~Schinnerer,} V.~Smol\v{c}i\'{c}]}}

\noi Radio continuum emission is widely used as a dust-unbiased tracer for
measuring the recent massive star formation activity in galaxies -
both in the local and distant Universe. Based on the locally observed
tight relation between infra-red and radio emission, the conversion of
radio flux density into star formation has been derived and widely
applied. However, the underlying physical reasons for this tight
relation still remain elusive. Progress has been slow due to the lack
of adequate data at both mid- to far-infrared wavelengths as well as
the radio regime. In order to reveal the underlying mechanisms causing
the emission and how they are related sensitive images spatially
resolving nearby galaxies are essential to disentangle contributions
from different sources such as young nascent star forming region, HII
regions, and older stellar populations forming the interstellar
radiation field (ISRF).

\noi New observations from the infrared Spitzer and particularly the
European Herschel space observatories have the necessary angular
resolution to allow not only for a spatial separation of the different
environments present in nearby galaxies but also to model the full
spectral energy distribution of the dust emission using realistic
models of the dust composition such as Draine \& Li (2007).  With the
dramatically increased sensitivity and almost contiguous frequency
coverage the Karl G. Jansky Very Large Array (JVLA) can provide the
equivalent data at radio frequencies.  The combined radio and infrared
datasets are prone to shed more light onto the physical bases of the
radio-infrared relation, and thus ultimately, the usefulness of the
radio continuum as a star formation tracer. However, mapping a
moderately sized sample over the full JVLA frequency range is
excessively demanding in observing time.

\bigskip
\bigskip

\parbox{\textwidth}{

\parbox{5cm}{ \includegraphics[ scale=0.34]{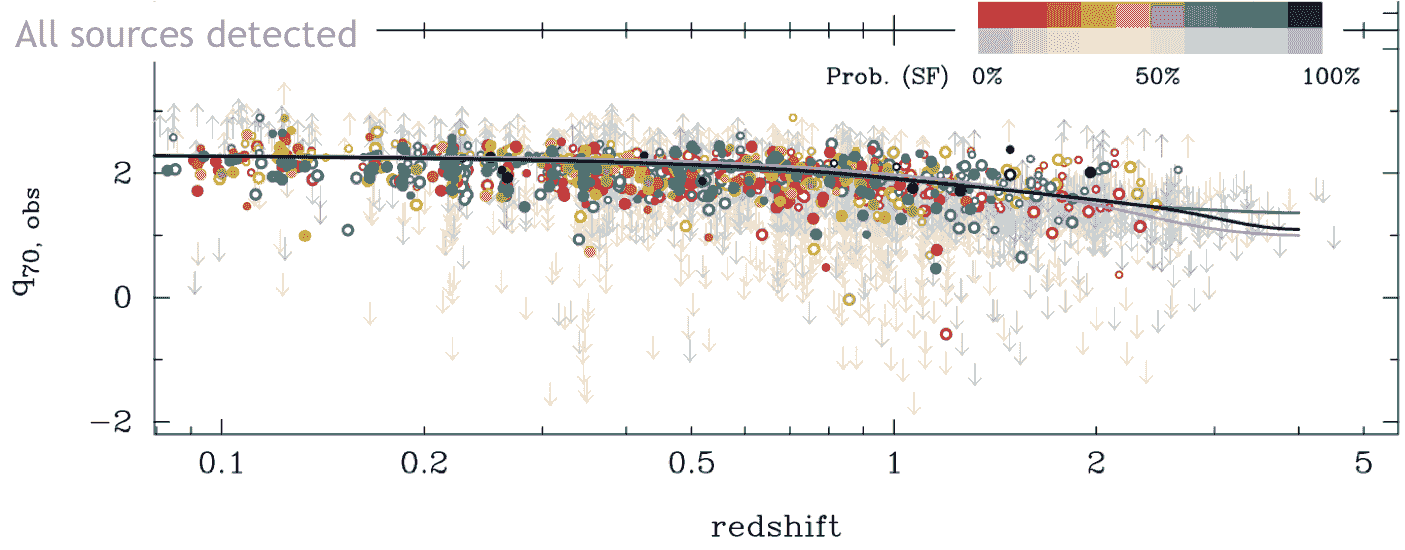}\vspace{0cm}}

\hspace{0.2cm}
\parbox{0.9\textwidth}{Figure 1$_{\rm Sm}$: Monochromatic radio-IR relation as a
  function of redshift (from Sargent et al. 2010a, b). The points show
  the observed, i.e. not K-corrected, 70\,$\mu$m to 1.4\,GHz ratio. The gray and black curves show the expected evolution of the ratio q based on Chary \& Elbaz (2001) SEDs. The filled points are ratios detected at both bands while the colour coding defines the probability that the radio emission of a given object is due to star formation. No evolution of the radio-IR relation out to z\,$\sim$\,3 is obvious.}
\vspace{0cm}
}

\bigskip
\bigskip

\noi New science highlights that have already emerged from such combined
studies is the finding that the radio-infrared relation can become
sub-linear in certain galactic environments such as the inter-arm
regions in a spiral galaxy (e.g. M51, Dumas et al. 2011; M33,
Tabatabaei et al. 2011). The diffusion length of cosmic ray electrons
is not exactly the same in all galaxies but shows a variance based on
the star formation activity suggesting that refreshment of the cosmic
ray electron reservoir is important (Murphy et al. 2006, 2008).
Several authors have shown that the radio-infrared relation typically
breaks below a few 100~pc scales (e.g. Tabatabaei et al. 2011, Dumas et al. 2011) except for the nearby dwarf
galaxy LMC (Hughes et al. 2005). There is tentative evidence now that
the nature of the magnetic field, i.e. its ordered large-scale
structure versus the small-scale turbulence) could play an important
role (Tabatabaei et al. 2012). Clearly, observations of the
3-dimensional structure of the magnetic field would provide
information of an important component of the interstellar
medium. Another aspect is the potentially important role that could be
played by the interstellar gas in enhancing the radio emission as
proposed by Thompson et al. (2006).

While JVLA will clearly make significant contributions to advance our
understanding of the physics underpinning the radio-infrared relation,
only the SKA can obtain high quality data for a representative sample of
nearby galaxies that probe all environment present in the local
Universe. Such a dataset with a spatial resolution of few 10 to 100\,pc
which is close to the constituents of the interstellar medium,
i.e. (giant) molecular clouds, and star forming sites, i.e. HII
region, will provide the information required to test the different
relations that have been put forward recently (e.g. Lacki et
al. 2010).

\parbox{\textwidth}{


\parbox{\textwidth}{\hspace{-1.2cm}
\parbox{5cm}{\includegraphics[scale=0.33,viewport=0 0 1004 620]{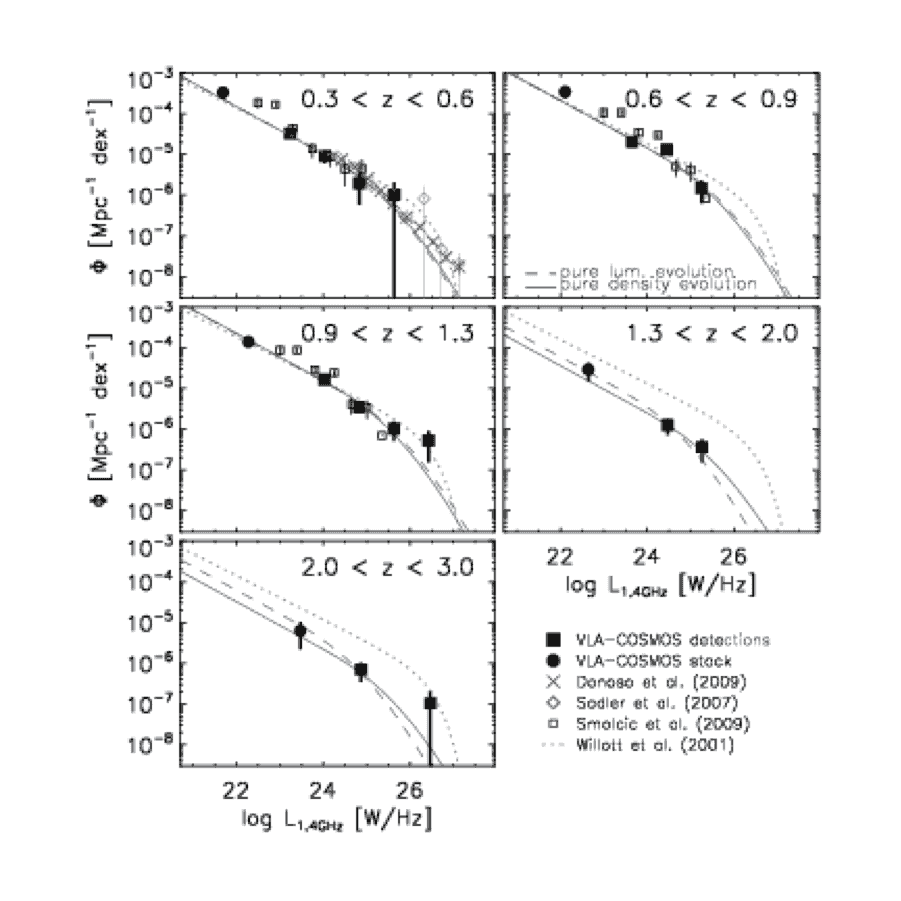}\vspace{-13.8cm}}

\hspace{9.2cm}
\parbox{5cm}{\includegraphics[scale=0.33,viewport=0 0 1004 620]{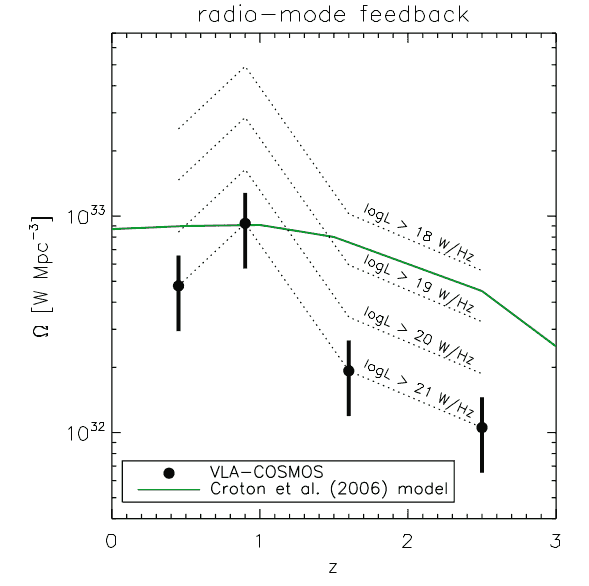}\vspace{3cm}}

\hspace{0.2cm}
\parbox{0.9\textwidth}{Figure 2$_{\rm Sm}$: Radio luminosity functions (LFs) for low-radio power AGN in COSMOS out to
  z$\,=\,3$ (left panel;  Smol\v{c}i\'{c} et al.\ in prep; see also
  Smol\v{c}i\'{c} et al. 2009b). Curves correspond to the analytical
  LFs best fit to the COSMOS data (assuming pure density
  evolution). Right: The comoving volume averaged radio-AGN heating
  done by low-radio power AGN in COSMOS (points and dotted and dashed lines),
  compared to predictions from cosmological models (full line). The
  dotted lines illustrate the strong dependence of the observationally derived curve on the faint end of the radio
    luminosity function. If the LF continues rising with decreasing
    radio luminosity the observationally derived radio-mode heating
    will systematically rise. The way to constrain the faint end of the LF is via deep observations, as
      possible with the SKA.}
\vspace{0cm}}}

\bigskip
\bigskip

A good physical understanding of the relation between radio continuum
emission and star formation is the foundation to widely exploit the
unique angular resolution that will be offered by the SKA compared to
infrared telescopes. In nearby galaxies, ALMA will provide
unprecedented insights into the distribution and nature of the
molecular clouds thus lying the foundation to unveil the onset and
process of star formation within galaxies. In order to uncover the
sites of star formation radio continuum emission can provide the
missing information as the dust continuum will be almost impossible to
be observed across galactic disks even for ALMA at cloud scale
resolution. Hence radio continuum might be the only means to identify
and quantify star formation in nearby galaxies at a resolution of only
a few parsec. These observations are critical to develop a consistent
understanding on how molecular gas is transformed into stars in a
galactic environment and what the main drivers for this process are.\\

\noi The high angular resolution afforded by the SKA in combination with
its order of magnitude increase in sensitivity will allow for the
study of the cosmic star formation history in unprecedented detail. Recent
work revealed a strong relation between the star formation rate
and the stellar mass of galaxies in the local Universe
(e.g. Brinchmann et al. 2004). This relation holds out to high
redshift (z\,$\sim$3), but is evolving in the sense that the average
star formation rate for a given mass is increasing with look-back time
(e.g. Karim et al. 2011). To-date, only the assembly properties of
normal star forming galaxies living at the peak of star formation at
z\,$\sim$2\,--\,3 can be probed via stacking at either radio or infrared
wavelengths. This is preventing deeper insights into the drivers of
star formation. Is it only stellar mass or does the galactic
environment play a role as well? Also the tightness of the relation is
unknown. Therefore probing down to normal star forming galaxies with
star formation rates of a few solar masses per year at high redshifts
will be essential to better constrain the star formation process in
the early Universe. Recently, Karim et al. (2011) showed that the
characteristic mass of the star forming galaxies is basically constant
out to z\,$\sim$3 while they saw tentative evidence that the galaxies
below the characteristic mass might approach an upper limit for the
star formation rate per stellar mass. A possible explanation for this
exciting finding might be the gas accretion rate onto the galaxies
that is renewing the reservoir for star formation. The only means to
push forward on this interesting front of research are ultra-deep
radio observations and the SKA will be the only instrument that has the
required sensitivity at sub-arsecond resolution. In addition, only the
SKA can image the radio continuum emission in distant galaxies well below
sub-kpc resolution providing a direct view onto the sites of star
formation. Knowing the distribution of the star forming sites is key
to understand if galaxy interactions, such as major mergers are
important or if secular evolution of disk-like systems is dominating
the cosmic star formation density. Finally, in conjunction with ALMA
observations of the molecular gas content tests of any locally derived
star formation laws or prescriptions could be tested.\\

\noi Other applications include a full census of the radio-infrared
relation in the local Universe to test the universality of the
detailed studies done on nearby galaxies and a probing of the
radio-infrared relation out to the early Universe. Current studies are
consistent with no redshift evolution of the radio-infrared relation
(e.g. Sargent et al. 2010, see Figure\ 1$_{\rm Sm}$). However, they are biased towards highly
luminous and thus highly star forming systems. Such systems are
expected to harbor strong magnetic fields and could thus minimize the
impact of the CMB onto the emitted radio emission (e.g. Murphy 2010).

\bigskip

\parbox{\textwidth}{

\hspace{-1cm}
\parbox{5cm}{ \includegraphics[ scale=0.35]{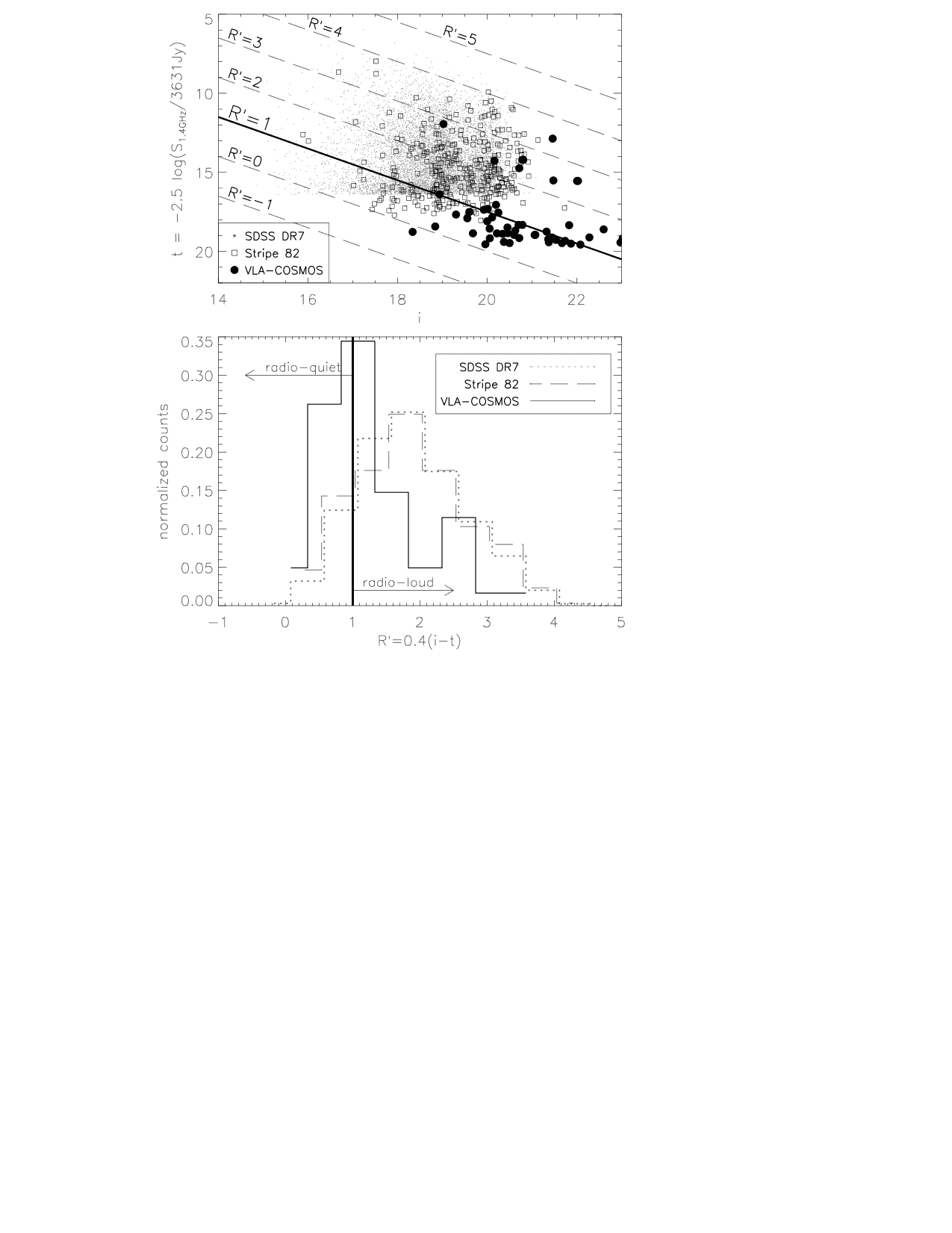}\vspace{-20.2cm}}

\hspace{10.1cm}
\parbox{6cm}{Figure\ 3$_{\rm Sm}$: Top: $i$-band vs.\ 1.4\,GHz
        radio magnitude ($t$) distribution for optically selected
        quasars (i.e.\ broad line AGN) drawn from current
        state-of-the-art surveys: SDSS DR7 - FIRST ($\sim$\,9000\sqdeg
        , $\mathrm{S_{1.4GHz}}\grtsim1$\,mJy; Schneider et al.\ 2010),
        Stripe 82 (92\sqdeg ,$\mathrm{S_{1.4GHz}}\grtsim260~\mu$Jy;
        Hodge et al.\ 2011), and COSMOS (2\sqdeg ,
        $\mathrm{S_{1.4GHz}}\grtsim12~\mu$Jy; Schinnerer et al.\ 2007,
        2010; Lilly et al.\ 2007, 2009). The bottom panel shows the
        distribution of radio loudness, $R'=0.4(i-t)$, for quasars in
        these three surveys.
}\vspace{3cm}}

\bigskip
\bigskip
\bigskip

\noi {\bfseries\boldmath\small Low-power radio-AGN: Relevance for feedback in massive
  galaxy formation}

\noi Within our standard model of galaxy evolution radio-mode AGN
feedback has by now become a standard ingredient in cosmological
models that allows reproducing the observed galaxy properties well
(e.g.\ the galaxies' stellar mass function; Croton et al. 2006, Bower
et al.\ 2006). Radio AGN outflows are thought to be the main source
that heats the halo gas surrounding a massive galaxy, and thereby
quenches its star formation and limits growth to create overly
high-mass galaxies. However, from an observational
perspective this process is far less clear.\\

\noi The first observational support for AGN feedback has been found
by Best et al.\ (2006), who quantitatively showed that in the local
Universe radio outflows may indeed balance the radiative cooling of
the hot gas surrounding elliptical galaxies.  Furthermore, it has been
both theoretically postulated and observationally supported that this
``radio mode'' heating occurs during a {\em quiescent phase of
  black-hole accretion} (presumably via advection dominated accretion
flows), that is reflected in {\em low-power radio AGN activity}
(\lum~$<10^{25}$~\wh ; Evans et al.\ 2006, Hardcastle et al.\ 2006,
2007, Kauffmann et al.\ 2008, Smol\v{c}i\'{c} et al.\ 2009b,
Smol\v{c}i\'{c} 2009, Smol\v{c}i\'{c} \& Riechers 2011). Thus,
studying low-power radio AGN and their evolution is paramount for
understanding galaxy formation!  However, only with the recent advent
of simultaneously deep and relatively wide radio surveys could this
population be comprehensively studied for the first time. The 20\,cm
luminosity function of low-power radio AGN, and its evolution out to
z\,=\,3 (based on VLA-COSMOS data; Schinnerer et al.\ 2004, 2007, 2010;
Smol\v{c}i\'{c} et al.\ 2008, 2009, in prep) provided the first
quantitative insight into the plausibility of ``AGN feedback'' beyond
the local Universe ($0.1\leq z\lesssim3$; see Figure\ 2$_{\rm
  Sm}$). However, these results, crucial for constraining cosmological
models, strongly depend on the faint end of the low-radio power
luminosity function, which is (beyond the local Universe)
unconstrained with current data (see Figure\ 2$_{\rm Sm}$). Furthermore,
to construct the radio luminosity function of such faint sources with
high precision out to the highest redshifts it is paramount to observe
large sky areas to unprecedented depth as
possible only with the SKA.

\bigskip
\bigskip

\parbox{\textwidth}{

\vspace{0cm}
\parbox{\textwidth}{

\hspace{1.8cm}
\parbox{5cm}{ \includegraphics[ scale=0.65]{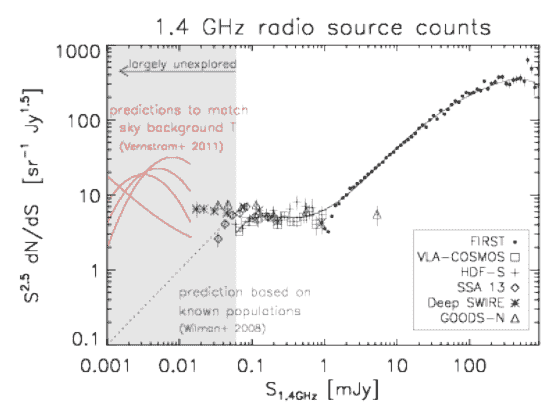}\vspace{0cm}}

\hspace{0.2cm}
\parbox{0.9\textwidth}{Figure 4$_{\rm Sm}$: 1.4\,GHz source count data
        (symbols), and predictions for the lowest flux densities (lines). The red full lines show the
        predictions that produce the background temperature
        necessary to match the ARCADE\ 2 result (Vernstrom et al.\
        2011). In contrast, the blue dotted line shows the generally
        assumed slope of faint radio source counts. The shaded area 
        designates the flux regime largely unexplored by data.
        For example, the faintest point to
        date (Owen \& Morrison 2008), is
        affected by completeness corrections greater than a factor of
        4 as it is drawn from a very small field (Deep SWIRE, 0.44\sqdeg ) with
        non-uniform rms (2.7\,$\mu$Jy/beam only in the centre). Deep SKA observations will settle this issue.
}\vspace{0.5cm}}}\\

\noi {\bfseries\boldmath\small Radio-quiet quasars: Testing the Unified Model for
  AGN:~~} Contrary to low-radio power AGN occurring in a quiescent phase of SMBH
accretion and thought to be responsible for suppressing further stellar
mass growth in massive galaxies (see previous Section), Type~1 (broad
line) AGN (quasars hereafter) reflect the most intense SMBH growth in
galaxies and are thought to possibly quench their star
formation by expelling a fraction of the gas from the galaxy via
quasar winds (so called ``quasar mode AGN feedback''; e.g.\ Hopkins et
al.\ 2006).  The existence of two physically distinct -- radio-loud
and radio-quiet -- quasar populations is a long debated issue that has
far-reaching implications onto astrophysical problems, from unified
schemes of AGN to the evolution of galaxies in the Universe. Although
the quasar radio-loudness ($R'$) distribution has been carefully
studied in many different quasar samples over the past decades
(e.g.\ Strittmatter et al.\ 1980, Ivezi\'{c} et al.\
2002, 2004, White et al.\ 2000, 2007, Cirasuolo et al. 2003, 
Balokovi\'{c} \etal\ 2011), to date there is still no clear consensus on the existence of a bimodality which would imply two physically distinct types of quasars in the Universe (pointing to e.g.\ different SMBH accretion/spin mechanisms and/or geometries; e.g.\ Fanidakis et al.\ 2010). To get a full census of AGN and to understand the role of quasars in galaxy formation and evolution it is paramount to understand the radio-loud and radio-quiet quasar populations.

To date this cosmologically important issue is still open mainly due
to the overwhelmingly high fraction of radio-quiet quasars that are
regularly undetected in radio surveys. As show in Figure\ 3$_{\rm Sm}$
all current radio surveys (even the deepest ones) only ``scratch''
(the loudest end of the) radio-quiet part of the distribution. Hence,
as only $\sim$\,10\,\% of optically selected quasars are radio loud
(e.g.\ Ivezi\'{c} \etal\ 2000), this means that roughly 90\,\% of quasars still remain
  undetected and unexplored at radio wavelengths. Observations of large areas on the sky to depths reachable with the SKA will directly reveal the radio
properties of roughly 90\,\% of optically detected quasars.

\noi{\bfseries\boldmath\small The quest for a missing population of $\mu$Jy radio
  sources:~~} The radio source counts -- the most straight-forward
information drawn from a radio survey and commonly used to predict
source counts in future deeper surveys -- flatten below 1\,mJy and are
generally expected to decrease again at fainter fluxes (see dotted
line in Figure\ 4$_{\rm Sm}$; e.g.\ Hopkins et al.\ 2000, Wilman et al.\
2008). Recent evidence, however, based on a comparison of the sky
brightness temperature measured by the ARCADE~2 experiment and that
derived from the integral of the observed radio source counts
(Vernstrom et al.\ 2011), points instead to a rise of the counts at
these levels (see full lines in Figure\ 4$_{\rm Sm}$ ). This could
possibly be explained by a new IR-faint (thus AGN) radio population so
far missed by existing radio surveys due to its faintness (Vernstrom
et al.\ 2011) and/or a mix of various AGN and star forming populations
(e.g.\ Smol\v{c}i\'{c} et al.\ 2008, Rigby et al.\ 2011). The SKA will
directly test the unconstrained, $<$\,$\mu$Jy, levels of the source
counts.\\

\parbox{0.9\textwidth}{
\noi{References:}\\
\noi{\scriptsize
Best, P.~N., Kaiser, C.~R., Heckman, T.~M., \& Kauffmann, G.\ 2006, \mnras, 368, L67;
Bower, R.~G., Benson, A.~J., Malbon, R., et al.\ 2006, \mnras, 370, 645;
Brinchmann, J., Charlot, S., White, S.~D.~M., et al.\ 2004, \mnras, 351, 1151;
Chary, R., \& Elbaz, D.\ 2001, \apj, 556, 562;
Cirasuolo, M., Celotti, A., Magliocchetti, M., \& Danese, L.\ 2003, \mnras, 346, 447;
Croton, D.~J., Springel, V., White, S.~D.~M., et al.\ 2006, \mnras, 365, 11;
Draine, B.~T., \& Li, A.\ 2007, \apj, 657, 810; 
Dumas, G., Schinnerer, E., Tabatabaei, F.~S., et al.\ 2011, \aj, 141, 41;
Evans, D.~A., Worrall, D.~M., Hardcastle, M.~J., Kraft, R.~P., \& Birkinshaw, M.\ 2006, \apj, 642, 96;
Hardcastle, M.~J., Evans, D.~A., \& Croston, J.~H.\ 2006, \mnras, 370, 1893;
Hodge, J.~A., Becker, R.~H., White, R.~L., Richards, G.~T., \& Zeimann, G.~R.\ 2011, \aj, 142, 3;
Hopkins, A., Windhorst, R., Cram, L., \& Ekers, R.\ 2000, Experimental Astronomy, 10, 419;
Hopkins, P.~F., Somerville, R.~S., Hernquist, L., et al.\ 2006, \apj, 652, 864;
Hughes, A., Wong, T., Ekers, R., et al.\ 2006, \mnras, 370, 363;
Ivezi{\'c}, {\v Z}., et al.\ 2002, \aj, 124, 2364;
Ivezi{\'c}, {\v Z}., et al.\ 2004 Proceedings of IAU Symposium No. 222, p. 525 (also astro-ph/0404487);
Karim, A., Schinnerer, E., Mart{\'{\i}}nez-Sansigre, A., et al.\ 2011, \apj, 730, 61;
Kauffmann, G., Heckman, T.~M., \& Best, P.~N.\ 2008, \mnras, 384, 953;
Lacki, B.~C., \& Thompson, T.~A.\ 2010, \apj, 717, 196;
Lilly, S.~J., Le F{\`e}vre, O., Renzini, A., et al.\ 2007, \apjs, 172, 70;
Lilly, S.~J., Le Brun, V., Maier, C., et al.\ 2009, \apjs, 184, 218;
Murphy, E.~J., Helou, G., Braun, R., et al.\ 2006, \apjl, 651, L111;
Murphy, E.~J., Helou, G., Kenney, J.~D.~P., Armus, L., \& Braun, R.\ 2008, \apj, 678, 828;
Murphy, E.~J.\ 2009, \apj, 706, 482; 
Rigby, E.~E., Best, P.~N., Brookes, M.~H., et al.\ 2011, \mnras, 416, 1900;
Sargent, M.~T., Schinnerer, E., Murphy, E., et al.\ 2010, \apjl, 714, L190;
Schinnerer, E., Carilli, C.~L., Scoville, N.~Z., et al.\ 2004, \aj, 128, 1974;
Schinnerer, E., Smol{\v c}i{\'c}, V., Carilli, C.~L., et al.\ 2007, \apjs, 172, 46;
Schinnerer, E., Sargent, M.~T., Bondi, M., et al.\ 2010, \apjs, 188, 384;
Schneider, D.~P., Richards, G.~T., Hall, P.~B., et al.\ 2010, \aj,
139, 2360; Smol{\v c}i{\'c}, V., Schinnerer, E., Scodeggio, M., et al.\ 2008, \apjs, 177, 14;
Smol{\v c}i{\'c}, V., Schinnerer, E., Zamorani, G., et al.\ 2009a, \apj, 690, 610;
Smol{\v c}i{\'c}, V., Zamorani, G., Schinnerer, E., et al.\ 2009b, \apj, 696, 24;
Smol{\v c}i{\'c}, V.\ 2009, \apjl, 699, L43; 
Smol{\v c}i{\'c}, V., \& Riechers, D.~A.\ 2011, \apj, 730, 64;
Strittmatter, P.~A., Hill, P., Pauliny-Toth, I.~I.~K., Steppe, H., \& Witzel, A.\ 1980, \aap, 88, L12;
Tabatabaei, F.~S., Schinnerer, E., Murphy, E., et al.\ 2011, arXiv:1111.6252;
}}

\parbox{0.9\textwidth}{
\noi{\scriptsize
Thompson, T.~A., Quataert, E., Waxman, E., Murray, N., \& Martin, C.~L.\ 2006, \apj, 645, 186 
Vernstrom, T., Scott, D., \& Wall, J.~V.\ 2011, \mnras, 415, 3641;
White, R.~L., et~al.\ 2000, \apjs, 126, 133;
White, R.~L., Helfand, D.~J., Becker, R.~H., Glikman, E., \& de Vries, W.\ 2007, \apj, 654, 99;
Wilman, R.~J., Miller, L., Jarvis, M.~J., et al.\ 2008, \mnras, 388, 1335
}}\\

\subsubsection{The radio continuum view of galactic scale feedback
{\scriptsize [D.J. Bomans, R.-J. Dettmar]}}

\noi  A key parameter for the formation and evolution of galaxies is the 
stellar and AGN feedback of the (proto-)galaxy onto the gas in the 
galaxy and onto the infalling gas from the circumgalactic or intergalactic 
medium (e.g. Oppenheimer \& Dave 2008).  While most of the large  simulations 
treat this as pure hydrodynamical process, it is clear from observations 
that magnetic fields play a significant role in this  process. 
Observations of the halos of local spiral galaxies with strong  star
formation show ordered magnetic fields in their halos, which are
clearly affected by  outflows/galactic winds (e.g. Tuellmann et
al. 2000, Heesen et al. 2009, Soida et al. 2011). Here an
interaction of a large  scale dynamo in massive galaxies and the
flow pattern of  the gas is at work. Actually, a working dynamo seems to
require the presence of a magnetised halo (Hanasz et al. 2006, 
Gressel et al. 2008) or even a magnetised wind 
(Brandenburg \& Subramanian 2005). 
It is interesting to note, that galaxies at z\,$\sim$\,2
exhibit very high starfomation rates but are generally not merging
galaxies (like the most strongly starforming galaxies  in the low
redshift universe), but rather very gas-rich disk galaxies (e.g. Genzel et
al. 2012). The effect of this spatially very extended starformation
on the magnetic field and outflow structure, its kinematics, and its
interaction with the infalling gas in cold filaments (e.g.  Keres et al 2006, 
Dekel \& Birnboim 2006) is unexplored yet.\\

\noi Surprisingly, dwarf galaxies with significant star formation rate also
show large-scale ordered  magnetic fields in their disks and out into 
their halos (e.g. Klein et al. 1996, Kepley et al. 2010, 2011, Chyzy 
et al. 2000, 2003, 2011). 
In these cases a  different kind of dynamo is needed (Siejkowski et
al. 2010) which sensitively reacts to the local star formation rate.
Simulations clearly show that magnetic field amplification is possible
in strongly starforming dwarf irregular galaxies and that in this case a
significant part of the magnetic flux can escape the potential well
(Siejkowski et al. 2010, 2012).\\

\noi Observations of spiral galaxies with nuclear starbursts and even
galaxies with moderately high starformation rate distributed over the
disk show magnetized outflows and galactic winds. These galaxies
therefore enhance and distribute magnetized plasma into their
surrounding intergalactic medium.  Due to their abundance dwarf-like
galaxies, or more precisely low mass proto-galaxies at high redshift,
for which the local dwarf irregular galaxies can serve as proxies
(e.g. Dekel \& Silk 1986), the magnetized winds of these galaxies
were proposed to fill the intergalactic medium with magnetic fields 
(Kronberg et al. 1999, Bertone et al. 2006, Dubois \& Teyssier 2010).   
The contribution of more massive galaxies has not been explored and on 
the observational side the whole issue is open due to lack of data 
sensitive enough to discuss such effects for local proxies or high 
redshift galaxies.\\

\noi Not only the local wind zones of individual galaxies can be seeded
with magnetic fields, also the collision of magnetized winds can (due
to field line compression) lead to further enhanced  magnetic field.
First hints of such a process are visible in some galaxy groups, 
but conclusive observations are limited by the sensitivity of 
current instruments.\\

\noi The huge plus in sensitivity provided by the SKA will allow to explore
the conditions at lower  starformation rates in search of the threshold
(either locally or globally) for the creation of  magnetized halos,
similar to the threshold for the presence of ionized gas halos
discussed  in Rossa \& Dettmar (2003). The same holds true for 
dwarf galaxies for which the higher sensitivity  will allow for a much 
larger sample size to explore the parameter space (e.g. mass, morphology,
environment, etc.).  The sensitivity and broad wavelength coverage
will enable us to map a much larger extent of polarized (and total power)
radio-continuum emission into the halo and therefore the study cosmic ray 
transport and the interaction of the wind with the surrounding intergalactic 
medium.

In galaxy  groups containing several starforming galaxies, interaction of
the outflows/winds and the effect  of the magnetization of the
intra-group medium and maybe even the intergalactic medium in
filaments of the large-scale structure can be studied, most probably
via RM grid  techniques.   The plus in sensitivity of the SKA implies also
enough signal to noise (at SKA angular resolution) to investigate
the details of the magnetic field at the base of galactic outflows and 
winds, where high ISM density and pressure create special conditions for  
thermal gas kinematics and cosmic ray propagation. 

Finally, the measurement of total power radio continuum at several frequencies
will give estimates of starformation rate and turbulent magnetic field
strength of intermediate to high redshift galaxies (see Murphy 2009).
In deep fields there is even the possibility of reaching the phase of 
galaxy formation.  Again the link to the magnetization of the intergalactic 
medium appears and may be tested with RM grid together with global properties 
and distribution of the starbursting and/or highly starforming galaxy 
population at different redshift intervals.\\

\parbox{0.9\textwidth}{
\noi{References:}\\
\noi{\scriptsize
Bertone S., Vogt C., Ensslin T., 2006, MNRAS, 370, 319;
Brandenburg A., Subramanian K., 2005, Phys. Rep., 417, 1;
Chy{\.z}y  K.T., Beck R., Kohle S., Klein U., Urbanik M., 2000, A\&A, 355, 128;
Chy{\.z}y K.T., Knapik J., Bomans D.J., et al., 2003, A\&A, 405, 513;
Chy{\.z}y K.T., We{\.z}gowiec M., Beck R., Bomans D.J., 2011, A\&A, 529, A94;
Dekel A., Birnboim Y., 2006, MNRAS, 368, 2;
Dekel A., Silk J., 1986, ApJ, 303, 39;
Dubois Y., Teyssier R., 2010, A\&A, 523, A72 
Gaensler B.M., Haverkorn M., Staveley-Smith L., et al., 2005, Science,
307, 1610;
Genzel R., Newman S., Jones T., et al., 2011, ApJ, 733, 101;
Gressel O., Elstner D., Ziegler U.,  R{\"u}diger G., 2008, A\&A, 486, L35; 
Hanasz M., Otmianowska-Mazur K., Kowal G., Lesch H., 2006, AN 327, 469;
Heesen V., Beck R., Krause M., Dettmar R.-J., 2009, A\&A, 494, 563;
Kere{\v s} D., Katz N., Weinberg D.H., Dav{\'e} R., 2005, MNRAS, 363, 2;
Kepley A.A., M{\"u}hle S., Everett J., et al., 2010, A\&A, 712, 536;
Kepley A.A., Zweibel E.G., Wilcots E.M., Johnson K.E., Robishaw T., 2011, ApJ, 736, 139;
Klein U., Haynes R.F., Wielebinski R., Meinert D., 1993, A\&A, 271, 402; 
Klein U., Hummel E., Bomans D.J., Hopp U., 1996, A\&A, 313, 396; 
Kronberg P.P., Lesch H., Hopp U., 1999, ApJ, 511, 56; 
Murphy E., 2009, in: Panoramic Radio Astronomy: Wide-field 1-2 GHz Research 
on Galaxy Evolution, eds.: G. Heald \& P. Serra, 29;
Oppenheimer B.D., Dav{\'e} R., 2008, MNRAS, 387, 577;
Rossa J.,  Dettmar R.-J., 2003, A\&A, 406, 493;
Siejkowski H., Soida M., Otmianowska-Mazur K., Hanasz M., Bomans D.J., 
2010, A\&A, 510, A97;
Siejkowski H., Otmianowska-Mazur K., Soida M., Bomans D.J., Hanasz M., 
2012, in prep.;
Soida M., Krause M., Dettmar R.-J., Urbanik M., 2011, A\&A, 531, A127; 
T{\"u}llmann R., Dettmar R.-J., Soida M., Urbanik M., Rossa J., 2000, 
A\&A, 364, L36 
}}

\subsubsection{Active galaxies -- a science perspectives   {\scriptsize [A. Lobanov]}}

\noi{\bfseries\boldmath\small Outflows an feedback in AGN:~~}The SKA will yield detailed imaging of extended outflows and lobes in
radio galaxies and quasars, providing an excellent tool for probing
physical conditions in low-energy tail of outflowing plasma which is
believed to carry the bulk of kinetic power of the outflow. This will
enable detailed quantitative studies of evolution and re-acceleration
of non-thermal plasma in cosmic objects and provide essential clues
for understanding the power and efficiency of the kinetic feedback
from AGN and its effect on activity cycles in galaxies and
cosmological growth of supermassive black holes. Such studies are
critically needed for making a detailed assessment of the role played
by AGN in the formation and evolution of the large-scale structure in
the Universe.\\

\noi{\bfseries\boldmath\small Galactic mergers and supermassive black
  holes:~~}High-resolution and high-sensitivity radio observations
with the SKA will provide arguably the best AGN and SMBH census up to very
high redshifts. This will enable cosmological studies of SMBH growth,
galaxy evolution, and the role played by galactic mergers in nuclear
activity and SMBH evolution.  Most powerful AGN are produced by
galactic/SMBH mergers. Direct detections of secondary SMBH in
post-merger galaxies are the best way to the evolution of black holes
and galaxies together. Some of the secondary black holes may be
``disguised'' as ultra-luminous X-ray objects accreting at a very small
rate, and hence remaining undetected even in deep radio images at
present.  The SKA would be a superb tool for detecting and classifying
such objects, thus providing an essential observational information
about them SMBH evolution in post-merger galaxies and its influence on
the galactic activity, formation of collimated outflows and feedback
from AGN.\\

\noi {\bfseries\boldmath\small Radio relics and AGN cycles:~~}Nuclear activity in galaxies is believed to be episodic or
intermittent, but relics of previous cycles of nuclear activity are
difficult to detect at centimetre wavelengths because of significant
losses due to expansion and synchrotron emission.  At centimetre
wavelengths, such relics decay below the sensitivity limits of the
present-day facilities within 10\,--\,100 thousand years after the fuelling
of extended lobes stops. The SKA, working below 1\,GHz, would be able to
detect such relics for at least 10 million years after the fuelling
stops, and this would make it possible to assess the activity cycles
in a large number of objects, searching for signs of re-started
activity in radio-loud objects and investigating ``paleo'' activity in
presently radio-quiet objects. This information will be essential for
constructing much more detailed models of evolution and nuclear
activity of galaxies.\\

\subsubsection{Active galactic nuclei  {\scriptsize [A. Merloni]}}

\noi {\bfseries\boldmath\small Introduction:~~}Soon after their discovery in 1963, it was realised that Quasars
(QSOs) were a strongly evolving class of cosmological sources, which
could effectively be used as tracers of the structural properties of
the Universe and the cosmological parameters (Longair 1966). The
physical understanding of QSOs and of their lower-luminosity
counterparts (both generally called Active Galactic Nuclei, AGN) as
accreting supermassive black holes (SMBH) immediately led to
speculations about the presence of their dormant relics in the nuclei
of nearby galaxies, with Soltan (1982) first proposing a method to
estimate the mass budget of SMBH based on the demographics and
evolutionary paths of observed QSO/AGN.  In the last decade, it has
emerged that tight scaling relations between the central black holes
mass and various properties of their host spheroids (velocity
dispersion, $\sigma^*$, stellar mass, M, luminosity, core mass deficit)
characterize the structure of nearby inactive galaxies (Tremaine et
al. 2002). These correlations have revolutionized the way we conceive
the physical link between galaxy and AGN evolution. Together with the
fact that supermassive black holes (SMBH) growth is now known to be
due mainly to radiatively efficient accretion over cosmological times,
taking place during active phases (Marconi et al. 2004; Merloni \&
Heinz 2008), all this led to the suggestion that most, if not all,
galaxies went through a phase of nuclear activity in the past, during
which a strong physical coupling (generally termed ``feedback'') must
have established a long-lasting link between host’s and black hole’s
properties.  From the cosmological point of view, the crucial aspect
is that the growth of supermassive black holes through accretion (and
mergers) is accompanied by the release of enormous amounts of energy
which can be either radiated away, as in Quasars and bright Seyfert
galaxies, or disposed of in kinetic form through powerful, collimated
outflows or jets, as observed in radio galaxies.  Direct evidence of
AGN feedback in action has been found in the X-ray observations of
galaxy clusters showing how black holes deposit large amounts of
energy on kpc scales in response to radiative losses of the cluster
gas (Fabian et al. 2006; Allen et al. 2006). From these (and other)
studies of the cavities, bubbles and weak shocks generated by the
radio emitting jets in the intra-cluster medium (ICM) it appears that
AGN are energetically able to balance radiative losses from the ICM in
the majority of cases (Best et al. 2006). On the other hand, numerical
simulations of AGN-induced feedback have recently shown that
mechanical feedback from black holes may be responsible for halting
star formation in massive elliptical galaxies, explaining the
bimodality in the colour distribution of local galaxies, as well as the
size of the most massive ellipticals (Springel et al. 2005). At a
global level, these models hinge on the unknown efficiency with which
growing black holes convert accreted rest mass into kinetic and/or
radiative power. Constraints on these efficiency factors are therefore
vital for the models and for our understanding of galaxy formation. As
well as a more detailed understanding of the triggering mechanisms of
AGN activity.  Here we discuss a few key contributions that the SKA will
be able to make in the field. The list is far from being exhaustive
and focuses on specific areas of AGN research best served by multi-
wavelength synergetic surveys.\\

\noi{What the SKA will achieve}:
\begin{itemize}
\item[--] A global view of the evolving radio galaxies: AGN population
studies can be used to reach the above mentioned goal by combining the
evolution of the ‘bolometric’ luminosity function AGN, obtained via
optical, IR and X-ray surveys, with that of radio selected AGN. Thanks
to its unprecedented combination of wide area and depth, SKA continuum
surveys will be able to identify ``typical'' (L$^*$) radio AGN well
beyond the epoch of major black hole growth (z\,$\sim$\,2). While the
bolometric AGN luminosity function evolution provides a mean of
computing the evolution of the black hole mass function, the census of
radio AGN will be fundamental in order to assess the overall amount of
energy released by growing black holes in kinetic form (Merloni and
Heinz 2008).

\item[--] The duty cycle of radio activity: By identifying the radio
counterparts of large, statistically significant, and highly
significant populations of active galaxies selected via their
accretion power indicators (optical emission lines, X-ray emission,
etc.). The SKA will be able to unambiguously address the long-standing
question of the radio-loud vs. radio-quiet AGN dichotomy. A
particularly powerful synergy is envisaged with the eROSITA all-sky
X-ray survey (see e.g. Predehl et al. 2010), as an hypothetical SKA
``all-sky'' survey with 1\,\mujy\ sensitivity will be deep enough to
identify essentially all the $\sim$\,10$^6$ X-ray detected AGN in the eROSITA
sky, both obscured and un-obscured, above a flux limit of $\sim$\,10$^{-14}$
erg/s/cm$^2$ (see Padovani 2011).

\item[--] AGN hosts: Where do they get their gas?: The SKA will also allow a
complete census of galaxies by detecting their \hi\ emission out to
z\,$\sim$\,1.5. Ancillary AGN surveys (mainly in IR and X-rays) will then
reveal in which of them nuclear activity is the strongest. For the
first time we will then be able to assess in a systematic way, and
with high statistical accuracy, the physical connection between cold
gas content and nuclear activity in galaxies spanning a wide range of
masses, redshift and star formation activity.\\
\end{itemize}

\parbox{0.9\textwidth}{
\noi{References:}\\
\noi{\scriptsize
Allen et al. 2006, MNRAS, 372, 21;
Best et al. 2006, MNRAS, 368, L67; 
Fabian et al. 2006, MNRAS, 366, 417; 
Longair, 1966, Nature, 211, 949;
Marconi et al., 2004, MNRAS, 351, 159; 
Merloni and Heinz 2008, MNRAS, 388, 1011; 
Padovani et al. 2011, MNRAS, 411, 1547;
Predehl et al. 2010, Proceedings of the SPIE, Volume 7732, pp. 77320U;
Soltan, 1982, MNRAS, 200, 115;
Springer et al. 2005, ApJ, 620, L79; 
Tremaine et al. 2002, ApJ, 574, 740
}}\\

\subsubsection{Disentangling AGN from starbursts: VLBI wide field imaging   {\scriptsize [E. Middelberg]}}

\noi Since the invention of the VLBI technique in the 1960s, observations
using it have almost exclusively been limited to the study of
carefully selected, small samples of objects. There are two
fundamental reasons for this: first, the need to record the raw data
on tape or disk for later correlation, and second, the limitation of
the field of view because of limited processing power.  Whilst both
effects also affect connected-element interferometers, the long
baselines and consequently high spatial resolution of VLBI
observations limit the field of view to around one
arcsecond. Therefore traditional VLBI observations are helplessly
unsuitable for surveys of large fractions of the sky.

\noi However, with the advent of more and more powerful computers at their
disposal, astronomers have begun to image sources approximately 10
years ago. Whilst imaging large numbers of objects is still far away
from being mainstream, wide-field VLBI observations have become
easier, and calibration has been streamlined and improved. The SKA is
designed to be a survey telescope, and it is expected that its
enormous computing power will easily facilitate production of
wide-field VLBI data. The scientific impact of such observations can
be summarised as follows.

\smallskip

{\bfseries\boldmath\small Identifying AGN:~~}AGN make up a significant fraction of the faint
radio sky. Whilst they dominate the radio source population at high
flux densities, their incidence, activity, and origin at fainter flux
density levels are not understood. How many radio-emitting AGN are
there at sub-mJy levels, how did they evolve into their observed form,
and what is their activity -- these questions can not yet be answered
satisfactorily. Because of the relatively high brightness temperatures
required for a detection, VLBI observations are excellent AGN
filters. Furthermore, they provide the high resolution to image
sub-kpc-scale structures at any redshift, yielding further clues about
the interplay between the AGN and its host galaxy. The enormous
sensitivity of the SKA, combined with baselines of up to 3000\,km and
GHz frequencies will enable detection of much fainter AGN, at larger
distances, to study the role of AGN throughout the Universe.

\smallskip

{\bfseries\boldmath\small Star formation and AGN activity:~~}Particularly at high redshift and
faint flux density levels, both AGN and star formation processes are
likely to be important in a large fraction of galaxies. Such ``hybrid''
objects are amongst the most interesting wide-field VLBI observations
can reveal, since they defy the common wisdom of a star-formation/AGN
divide, and they will occur plentiful in wide-field SKA/VLBI
surveys. Observations of such objects at any one wavelength will be
misleading, since they are heavily obscured, and only high-resolution
radio objects can reliably identify the inner emission mechanisms.
Another application to the study of star-formation and AGN is the
radio-infrared relation, which is a linear relation between the radio
and infrared flux densities of galaxies over many orders of
magnitude. It can be used to identify AGN when there is a large radio
excess, since the relation is thought to arise from star-forming
activity. However, a small fraction of AGN-dominated galaxies show a
radio excess, but many lower-power AGN appear to obey this
relation. Using the SKA, observations of VLBI cores in millions of
objects will allow one to investigate the effect of AGN on this
relation beyond the local Universe, and to calibrate it more
accurately.

\smallskip

{\bfseries\boldmath\small The evolution of AGN:~~}Existing observational evidence suggests that
low-luminosity radio sources correspond to a distinct type of AGN,
accreting through a radiatively inefficient mode (the so-called
``radio mode''), rather than the radiatively efficient accretion mode
typical of optically or X-ray selected AGN (the so-called ``quasar
mode''). The physical reasons behind these two different accretion
modes are unclear, but it is possible that ``quasar mode'' AGN accrete
cold gas, and that they contribute significantly to the buildup of
super-massive black holes predominantly in young (bluer)
galaxies. Traditional VLBI observations mostly target well-selected
(and bright) samples of objects, for example FRI and FRII galaxies,
quasars and blazars, but also weaker objects such as Seyfert
galaxies. In all these cases, a pre-selection has been made based on
other known properties of the objects, and so such surveys are limited
to particular classes of object. Sensitive, wide-field SKA/VLBI
surveys will contribute substantially to this issue, providing an
unbiased view on the pc-scale radio Universe as it evolves from z\,$\sim$\,3 to
the present day.

\smallskip

{\bfseries\boldmath\small Studies of particular classes of object; serendipity:~~} Surveying the
sky with high resolution and high sensitivity will enable studies of
particular classes of object, which alone would not warrant the
required observing time. Many traditional VLBI surveys target samples
of tens of objects, which must be sufficiently bright so that
detections can be made in reasonable times. Such studies will benefit
substantially from wide-field SKA/VLBI surveys because of the large
number of objects with generally much lower flux densities (which
therefore are more distant and younger). Furthermore, surveys
exploring new volumes of the parameter space have always returned new,
unexpected, and unpredictable results. We do not know what we are
going to find, but we are confident some of it is going to be
unexpected, interesting, and puzzling.\\

\bigskip

\subsubsection{Magnetic field amplification with the small-scale
  dynamo model: Implications for the SKA  {\scriptsize [D.R.G. Schleicher, R. Banerjee]}}

\noi {\bfseries\boldmath\small Probing the origin of magnetic fields:~~}Due to the ubiquity
of magnetic fields in the local Universe (e.g. Beck \etal\ 1996), the
origin of magnetic fields is a subject of considerable interest. With
the SKA, this topic can be explored by probing magnetic fields at
higher redshifts and at higher resolution. If high-redshift magnetic
fields have a primordial origin (e.g. Grasso \& Rubinstein 2001), one
might expect them to be coherent on rather large scales. However, even
in the absence of strong primordial fields, the small-scale dynamo may
produce strong small-scale fields already at high redshift (Beck 1996,
Arshakian \etal\ 2009, Schleicher 2010). These possibilities and the
implications for the SKA are discussed in the sections below.\\

\noi {\bfseries\boldmath\small Astrophysical magnetic fields: The small-scale dynamo:~~}In
the absence of strong primordial fields, astrophysical mechanisms for
the production of seed magnetic fields need to be considered. Such
possibilities include in particular the Biermann battery (Biermann
1950), which may create seed fields of the order $10^{-15}$~G in the
first star-forming halos (Xu \etal\ 2008). In the presence of shocks,
which are expected to occur during structure formation and the
virialization of halos, additional and potentially stronger fields may
be generated by the Weibel instability (Schlickeiser \& Shukla 2003,
Medvedev \etal\ 2004, Lazar \etal\ 2009).

\noi Regardless of the origin of the seed, rapid amplification is expected
in the presence of turbulence. The latter is released during the
virialization of the halo from the gravitational energy, giving rise
to a scenario termed ''gravity-driven turbulence'' (Elmegreen \&
Burkert 2010, Klessen \& Hennebelle 2010). The presence of sub-sonic
turbulence is visible in numerical simulations of the first
star-forming halos (e.g. Abel \etal\ 2002, Turk \etal\ 2009), while
supersonic turbulence is found in simulations of larger primordial
galaxies (Wise \& Abel 2007, Greif \etal\ 2008).\\

\bigskip

\parbox{\textwidth}{

\hspace{2.5cm}
\parbox{5cm}{
\begin{tabular}{ccc}
Model & $B_0$ [nG] & $f_*$  \\
\hline
$1$ & $0$ & $0.1\,\%$  \\
$2$ & $0.02$ & $0.1\,\%$ \\
$3$ & $0.05$ & $0.1\,\%$  \\
$4$ & $0.2$ & $0.1\,\%$  \\
$5$ & $0.5$ &$0.1\,\%$ \\
$6$ & $0.8$ & $0.1\,\%$  \\
$7$ & $0.8$ & $1\,\%$ \\
\hline
\end{tabular}\vspace{-4cm}}

\hspace{8.3cm}
\parbox{6cm}{Table 1$_{\rm Sc}$: The models used in the figure to calculate
  the redshift evolution of the mean 21\,cm brightness
  temperature. Given are both the co-moving field strength $B_0$ and
  the star formation efficiency $f_*$. We refer to (Schleicher \etal\ 2009) for further details.}
\vspace{1cm}}

\bigskip

\noi The theory of turbulent field amplification was originally proposed by
(Kazantsev 1968), and further refined by (Subramanian 1998). The
theory suggests magnetic field amplification on the timescale of
turbulent eddies at the resistive scale. This timescale is
considerably smaller than the dynamical timescales of the system,
providing a rapid amplification mechanism. It has been explored and
tested using numerical simulations at high resolution
(Haugen \etal\ 2004a, b).  While the initial studies typically
explore subsonic turbulence, it was recently shown that the
small-scale dynamo operates under a large range of conditions, even at
highly supersonic turbulence (Federrath \etal\ 2011), as well as during
gravitational collapse (Sur \etal\ 2010, Federrath \etal\ 2011).\\

\noi The SKA provides a unique opportunity to compare this prediction with
observations in a variety of systems. In the local Universe, dwarf
galaxies may resemble the conditions at high redshift and show a
magnetic field structure that is tangled on small scales. While
magnetic fields have indeed been detected in these systems
(Chy{\.z}y 2011), the SKA provides the unique opportunity to increase
the resolution by an order of magnitude. In particular, one may then
attempt to infer the slope of the power spectrum. While for a
saturated field, a scaling of $\sim$\,$k^{-1/3}-k^{-1/2}$ is expected for
Kolmogorov or Burgers type turbulence, the characteristic Kazantsev
slope of $k^{3/2}$ would occur during the kinematic phase of the
small-scale dynamo.

\bigskip
\bigskip

\parbox{\textwidth}{

\parbox{7cm}{ 
\hspace{-0.2cm}
\includegraphics[scale=0.3]{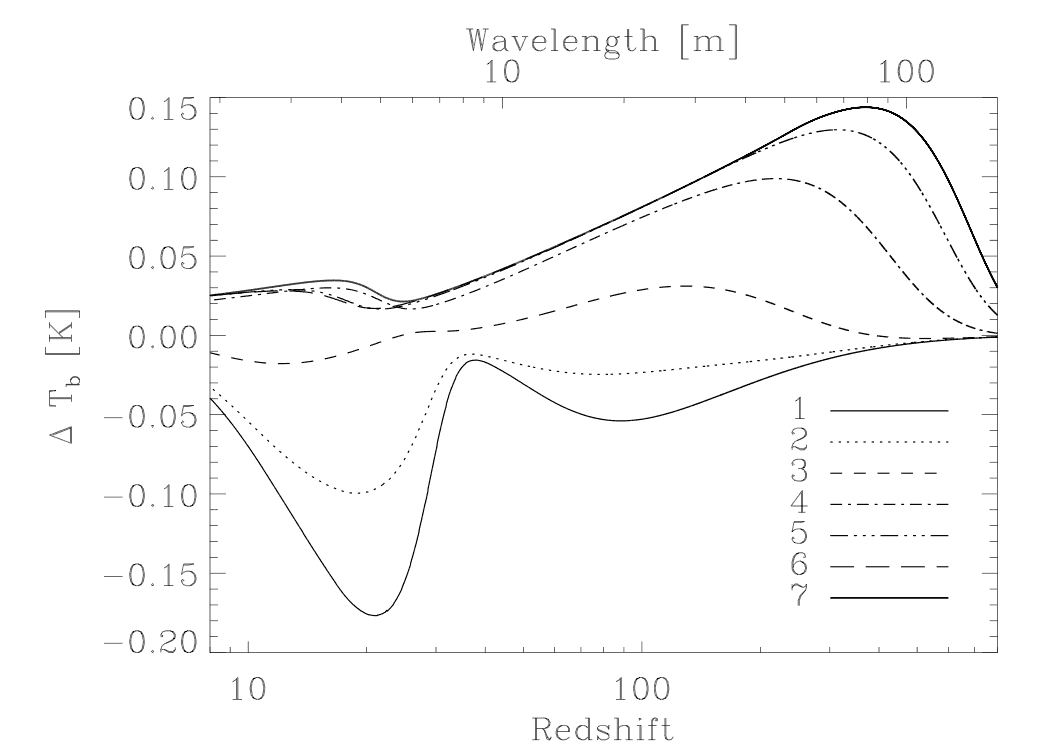}\vspace{-6.75cm}}

\hspace{10.8cm}
\parbox{6cm}{Figure\ 1$_{\rm Sc}$: The evolution of the mean brightness temperature fluctuation
   as a function of redshift, in the presence of primordial magnetic
   fields (Schleicher \etal\ 2009). The models are listed in the
   table.}
\vspace{3cm}}\\

\newpage

Another opportunity to explore the implications of turbulence is
provided by galaxy interactions, which were recently shown to
considerably amplify the tangled component of the magnetic field
(Drzazga \etal\ 2011). Also in this case, higher-resolution observations
with the SKA would provide a unique opportunity to probe these
magnetic structures on much smaller scales, in order to infer
characteristic properties of the small-scale dynamo and MHD
turbulence.

\noi Finally, we would like to point out that the small-scale dynamo was
also suggested to operate on large cosmological scales
(e.g. Ryu \etal\ 2008), and drive the magnetic field amplification in
galaxy clusters (Dolag \etal\ 1999, Subramanian 2006). The study of galaxy
clusters is another topic of intense study with the SKA, which will
shed additional light on the questions discussed here.\\

\noi {\bfseries\boldmath\small Primordial magnetic fields: Constraints from the epoch of
  reionisation:~~}An alternative scenario for the origin of magnetic fields is the
generation of primordial fields in the early Universe (Grasso \&
Rubinstein 2001). In the presence of an inverse cascade, their
integral scale can be shifted to kpc-scales due to the conservation of
magnetic helicity (Christensson \etal\ 2001, Banerjee \& Jedamzik
2003, Banerjee \& Jedamzik 2004). Strong primordial fields were even
suggested to influence the process of structure formation in the
Universe (Wasserman 1978, Kim \etal\ 1996) and could have a
considerable impact on the epoch of reionisation (Sethi \& Subramanian
2005, Tashiro \& Sugiyama 2006). As suggested by (Schleicher \etal\
2008), such effects can be used to derive constraints on the
primordial field strength. A recent analysis based on the observed UV
luminosity functions suggests a $2\sigma$ constraint of 2\,--\,3\,nG due to
the reionization optical depth (Schleicher \& Miniati 2011).

A more sensitive probe is provided with the SKA, which will measure
the 21\,cm signal during reionization up to z\,$\sim$18. The presence
of strong primordial fields could heat the high-redshift intergalactic
medium to temperatures up to $10^4$~K via ambipolar diffusion (Sethi
\& Subramanian 2005, Tashiro \& Sugiyama 2006, Schleicher \etal\
2008), which is reflected in the spin temperature and thus the \hi\
brightness temperature from reionization (Tashiro \etal\ 2006,
Schleicher \etal\ 2009, Sethi \& Subramanian 2009). The evolution of
the mean 21\,cm signal as a function of redshift, and its dependence
on the primordial field strength, is shown in the following 
figure. Additional characteristic signatures may
be inferred from the 21\,cm power spectrum from reionization. The SKA
activities on the epoch of reionisation will thus shed additional
light on the presence of primordial fields, and may help to infer
stronger upper limits or potential signatures. We are thus entering an
exciting epoch both with respect to the formation of the first
structures and the origin of magnetic fields.\\

\parbox{0.9\textwidth}{
\noi{References:}\\
\noi{\scriptsize
Abel T., Bryan G.L., Norman M.L.,  2002, Science, 295, 93;
Arshakian T.G., Beck R., Krause M., Sokoloff D.,  2009, A\&A, 494, 21;
Banerjee R., Jedamzik K.,  2003, Physical Review Letters, 91, 251301;
Banerjee R., Jedamzik K.,  2004, PRD, 70, 123003;
Beck R., \etal , 1996, ARA\&A, 34, 155;
Biermann L.,  1950, Zeitschrift Naturforschung Teil A, 5, 65;
Christensson M.,  Hindmarsh M., Brandenburg A.,  2001, PRE, 64, 056405;
Chy{\.z}y K.T., We{\.z}gowiec M., Beck R., Bomans D.J., 2011, A\&A,
529, A94;
Dolag K., Bartelmann M., Lesch H., 1999, A\&A, 348, 351;
Drzazga R.T., Chy{\.z}y K.T., Jurusik W., Wi{\'o}rkiewicz K., 2011, A\&A, 533, A22;
Elmegreen B.G., Burkert A.,  2010, ApJ, 712, 294;
Federrath C., \etal ,  2011, PRL, accepted;
Federrath C., \etal , 2011, ApJ, 731, 62;
Grasso D., Rubinstein H.R.,  2001, Phys. Rep., 348, 163;
Greif T.H., Johnson J.L., Klessen R.S., Bromm V., 2008, MNRAS, 387, 1021;
Haugen N.E.L., Brandenburg A., Dobler W.,  2004a, PRE, 70, 016308;
Haugen N.E.L., Brandenburg A.,  Dobler W.,  2004b, ApSS, 292, 53;
Kazantsev A.P., 1968, Sov. Phys. JETP, 26, 1031;
Kim E.-J., Olinto A.V., Rosner R., 1996, ApJ, 468, 28;
Klessen R.S.,  Hennebelle P.,  2010, A\&A, 520, A17;
Lazar M., Schlickeiser R., Wielebinski R., Poedts S., 2009, ApJ, 693, 1133;
Medvedev M.V.,  \etal ,  2004, Journal of Korean Astronomical Society, 37, 533;
Ryu D.,  Kang H., Cho J., Das S.,  2008, Science, 320, 909;
Schleicher D.R.G., Banerjee R., Klessen R.S.,  2008, PRD, 78, 083005;
Schleicher D.R.G., Banerjee R., Klessen R.S.,  2009, ApJ, 692, 236;
Schleicher D.R.G., \etal ,  2010, A\&A, 522, A115;
Schleicher D.R.G., Miniati F.,  2011, ArXiv e-prints 1108.1874;
Schlickeiser R., Shukla P.K.,  2003, ApJL, 599, L57;
Sethi S.K., Subramanian K.,  2005, MNRAS, 356, 778;
Sethi S.K., Subramanian K.,  2009, JCAP, 11, 21;
Subramanian K.,  1998, MNRAS, 294, 718;
Subramanian K.,  2006, Astron.~Nachr., 327, 403;
Sur S., \etal ,  2010, ApJL, 721, L134;
Tashiro H., Sugiyama N.,  2006, MNRAS, 368, 965;
Tashiro H., Sugiyama N., Banerjee R.,  2006, PRD, 73, 023002;
Turk M.J., Abel T., O'Shea B.,  2009, Science, 325, 601;
Wasserman I.~M., 1978, PhD thesis, Harvard University;
Wise J.H., Abel T.,  2007, ApJ, 665, 899;
Xu H.,  \etal ,  2008, ApJL, 688, L57
}}\\

\subsubsection{The origin and evolution of cosmic magnetism
  {\scriptsize [R. Beck]}}

\noi Magnetism is one of the four fundamental
forces. However, the origin of magnetic fields in stars, galaxies and
clusters is an open problem in astrophysics and fundamental
physics. When and how were the first fields generated? Are present-day
magnetic fields a result of dynamo action, or do they represent
persistent primordial magnetism? What role do magnetic fields play in
turbulence, star formation, cosmic ray acceleration and galaxy
formation? The SKA can deliver superb data which will directly address
these currently unanswered issues. The key data base is an {\em
  all-sky survey of Faraday rotation measures}, in which Faraday
rotation towards about $10^7$ background sources will provide a dense
grid for probing magnetism in the Milky Way, in nearby galaxies, and
in distant galaxies, galaxy clusters and protogalaxies. Using these
data, we can map out the evolution of magnetized structures from
redshifts z\,$>$\,3 to the present epoch, can distinguish between
different origins for seed magnetic fields in galaxies, and can
develop a detailed model of the magnetic field geometry of the
intergalactic medium and of the overall Universe. With the
unprecedented capabilities of the SKA, the window to the Magnetic
Universe can finally be opened.

\medskip

\noi Understanding the Universe is impossible without
understanding magnetic fields. They fill intracluster and interstellar
space, affect the evolution of galaxies and galaxy clusters,
contribute significantly to the total pressure of interstellar gas,
are essential for the onset of star formation, control the density and
distribution of cosmic rays in the interstellar medium (ISM), and
affect the propagation of the highest-energy cosmic rays which enter
the Milky Way. But in spite of their importance, the {\em evolution},
{\em structure}\ and {\em origin}\ of magnetic fields are all still
open problems in fundamental physics and astrophysics. Specifically,
we still do not know how magnetic fields are generated and maintained,
how magnetic fields evolve as galaxies evolve, what the strength and
structure of the magnetic field of the intergalactic medium (IGM)
might be, or whether fields in galaxies and clusters are primordial or
generated at later epochs. Ultimately, we would like to establish
whether there is a connection between magnetic field formation and
structure formation in the early Universe, and to obtain constraints
on when and how the first magnetic fields in the Universe were
generated.

\medskip

\noi Most of what we know about astrophysical magnetic
fields comes through the detection of radio waves. {\em Synchrotron
  emission}\ measures the total field strength, while its {\em
  polarisation}\ yields the orientation of the regular field in the
sky plane and also gives the field's degree of ordering (see
figure). {\em Faraday rotation}\ of polarisation vectors
provides a measurement of the mean direction and strength of the field
along the line of sight. Together this yields a full three-dimensional
view. Nevertheless, measuring astrophysical magnetic fields is a
difficult topic still in its infancy, still restricted to nearby or
bright objects. The Square Kilometre Array (SKA) has the power to
revolutionise the study of ``Cosmic Magnetism''.

\bigskip
\bigskip
\bigskip

\vspace{0cm}
\parbox{\textwidth}{

\hspace{2.5cm}
\parbox{5cm}{ \includegraphics[scale=0.2]{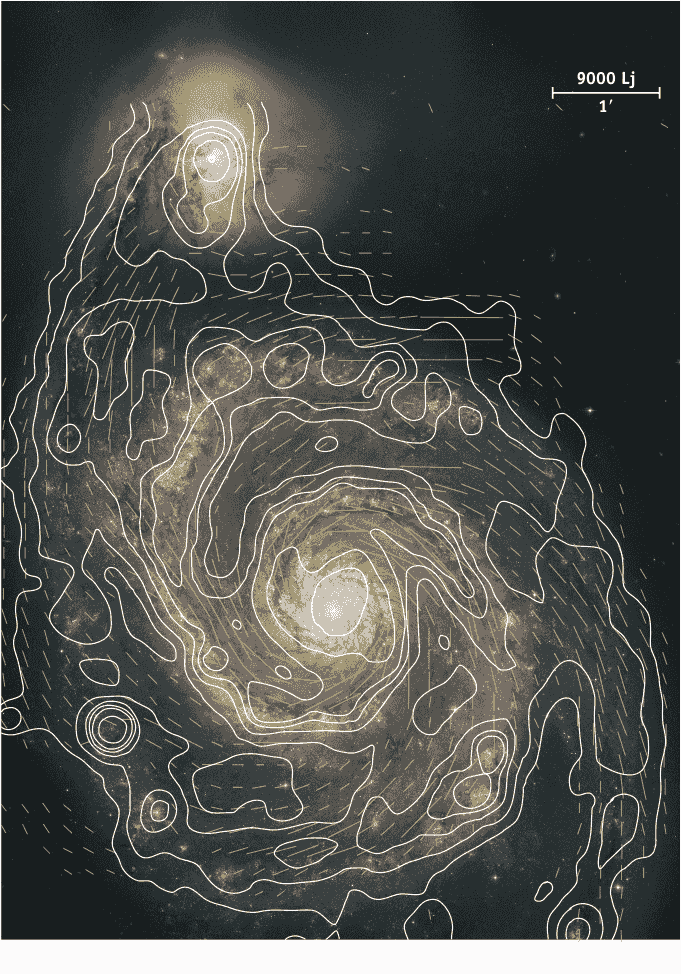}\vspace{-6.8cm}}

\hspace{8.2cm}
\parbox{6cm}{Figure 1$_{\rm Be}$: The
    magnetic field of the grand design spiral galaxy M\,51. The image
    shows an optical {\em HST}\ image, overlaid with contours showing
    the radio total intensity emission at 5\,GHz. The vectors show the
    orientation of the magnetic field, as determined from 5\,GHz linear
    polarisation measurements. Faraday rotation is small at this
    frequency (from Fletcher et al. 2011). }
}

\newpage

\noi Much of what the SKA can contribute to our understanding of
magnetic fields will come from its {\em polarimetric
  capabilities}. The main SKA specifications which enable this
capability will be {\bf \small high polarisation purity}\ and {\bf
  \small spectro-polarimetric capability}. Another important
requirement is a field of view at 1.4\,GHz of at least 1\,deg$^2$
which can be fully imaged at $1$\,arcsec resolution. At an observing
frequency of 1.4\,GHz with a fractional bandwidth of 25\,\%, a 1-min SKA
observation of a source with a linearly polarised surface brightness
of $\approx 10$\,$\mu$Jy~beam$^{-1}$ will yield a RM determined to an
accuracy $\Delta {\rm RM} \approx \pm 5$~rad~m$^{-2}$, and its
intrinsic position angle measured to within $\Delta \Theta \approx
\pm10^\circ$. Measurements of this precision will be routinely
available in virtually any SKA observation of a polarised source.
Currently $\sim$\,1200 extragalactic sources and $\sim$\,300 pulsars have
measured RMs. These data have proved useful probes of magnetic fields
in the Milky Way, in nearby galaxies, in clusters, and in distant
Lyman-alpha absorbers. However, the sampling of such measurements over
the sky is very sparse, and most measurements are at high Galactic
latitudes.

\medskip

\noi A key platform on which to base the SKA's studies of cosmic
magnetism will be to carry out an {\bf \small All-Sky RM Survey}, in
which spectro-polarimetric continuum imaging of 10\,000\,deg$^2$ of
the sky can yield RMs for approximately $2\times10^4$ pulsars and
$\times10^7$ compact polarised extragalactic sources in about a year
of observing time (Gaensler et al. 2004). This data set will provide a grid of RMs at a mean
spacing of $\sim$\,30\,arcmin between pulsars and just $\sim$\,90\,arcsec
between polarised extragalactic sources.  In our {\bf \small Milky
  Way}, the large sample of pulsar RMs obtained with the SKA, combined
with distance estimates to these sources from parallax or from their
dispersion measures, can be inverted to yield a complete delineation
of the magnetic field in the spiral arms and disk on scales
$\ge100$~pc (e.g. Stepanov et al. 2002). Small-scale structure and
turbulence can be probed using {\em Faraday tomography}, in which
foreground ionised gas produces complicated frequency dependent
Faraday features when viewed against diffuse Galactic polarised radio
emission. Magnetic field geometries in the Galactic
halo and outer parts of the disk can be studied using the
extragalactic RM grid (Sun et al. 2008).  In {\bf \small external
  galaxies}, magnetic fields can be directly traced by diffuse
synchrotron emission and its polarisation (Beck et al. 2004, see also
contribution by Marita Krause). 

\medskip

\noi In {\bf \small clusters of galaxies}, magnetic fields play a
critical role in regulating heat conduction (Chandran \etal 1998,
Narayan \& Medvedev 2001), and may also both govern and trace cluster
formation and evolution. Estimates of the overall magnetic field
strength come from inverse Compton detections in X-rays, from
detection of diffuse synchrotron emission, from cold fronts and from
simulations, but our only direct measurements of field strengths and
geometries come from RMs of background sources. Currently just
$\sim$\,1\,--\,5 such RM measurements can be made per cluster; only by
considering an ensemble of RMs averaged over many systems can a crude
picture of cluster magnetic field structures be established.  With the
SKA, the RM grid can provide $\sim$\,1000 background RMs behind a
typical nearby galaxy cluster; a comparable number of RMs can be
obtained for a more distant cluster through a deep targeted
observation (see contribution by Martin Krause). These data will allow
us to derive a detailed map of the field in {\em each}\ cluster. With
such information, careful comparisons of the field properties in
various types of cluster (e.g. those containing cooling flows, those
showing recent merger activity, etc.) at various distances can be
easily made. Furthermore, detailed comparisons between RM
distributions and X-ray images of clusters will become possible,
allowing us to relate the efficiency of thermal conduction to the
magnetic properties of different regions, and to directly study the
interplay between magnetic fields and hot gas.

\medskip

 \noi {\bf \small Galaxies at
  intermediate redshifts}\ ($0.1\le$\,z\,$\le 2$) are representative of
the local population but at earlier epochs. Measurements of the
magnetic field in such systems thus provides direct information on how
magnetised structures evolve and amplify as galaxies mature. As the
linearly polarised emission from galaxies at these distances will
often be too faint to detect directly, Faraday rotation thus holds the
key to studying magnetism in these distant sources.  A large number of
the sources for which we measure RMs will be quasars showing
foreground Lyman-alpha absorption; these absorption systems likely
represent the progenitors of present-day galaxies. If a large enough
sample of RMs for quasars at known redshift can be accumulated, a
trend of RM vs z can potentially be identified. The form of this
trend can then be used to distinguish between RMs resulting from
magnetic fields in the quasars themselves and those produced by fields
in foreground absorbing clouds; detection of the latter effect would
then directly trace the evolution of magnetic field in galaxies and
their progenitors.  At yet {\bf \small higher redshifts}, we can take
advantage of the sensitivity of the deepest SKA fields, in which we
expect to detect the synchrotron emission from the youngest galaxies
and proto-galaxies. The tight radio-infrared correlation, which holds
also for distant infrared-bright galaxies, tells us that magnetic
fields with strengths similar or even larger than in nearby galaxies
existed in some young objects, but the origin of these fields is
unknown. The total intensity of synchrotron emission can yield
approximate estimates for the magnetic field strength in these
galaxies.  

\medskip

\noi Fundamental to all the issues discussed above is the search
for {\bf \small magnetic fields in the intergalactic medium}. All of
``empty'' space may be magnetized, either by outflows from galaxies,
by relic lobes of radio galaxies, or as part of the “cosmic web”
structure. Such a field has not yet been detected, but its role as the
likely seed field for galaxies and clusters, plus the prospect that
the IGM field might trace and regulate structure formation in the
early Universe, places considerable importance on its discovery. To
date there has been no detection of magnetic fields in the IGM;
current upper limits on the strength of any such field suggest
$|B_{\rm IGM}| \le 10^{-9}$\,G. Indirect evidence for weak IGM fields
of $|B_{\rm IGM}| \ge 10^{-16}$\,G comes from bright galactic nuclei
(blazars) which have been detected in the TeV $\gamma$-ray regime, but
not in the GeV regime, possibly due to scattering of secondary
particles in the IGM field (Neronov \& Vovk 2010).  Using the SKA,
this all-pervading cosmic magnetic field may finally be identified
through the RM grid. If an overall IGM field with a coherence length
of a few Mpc existed in the early Universe and its strength varied
proportional to (1\,+\,z)$^2$, its signature may become evident at
redshifts of z\,$>$\,3. The RM distribution can provide the magnetic power
spectrum of the IGM as a function of cosmic epoch and over a wide
range of spatial scales. Averaging over a large number of RMs is
required to unravel the IGM signal. The goal is to detect an IGM
magnetic field of 0.1\,nG strength, which needs an RM density of
$\approx$\,1000 sources deg$^{-2}$ (Kolatt 1998). Such measurements will
allow us to develop a detailed model of the magnetic field geometry of
the IGM and of the
overall Universe.\\

\noi The SKA can provide exciting new insights into the
origin, evolution and structure of cosmic magnetic fields. The sheer
weight of RM statistics which the SKA can accumulate will allow us to
characterize the geometry and evolution of magnetic fields in
galaxies, in galaxy clusters and in the IGM from high redshifts
through to the present. We may also be able to provide the first
constraints on when and how the first magnetic fields in the Universe
were generated. Apart from these experiments which we can conceive
today, we also expect that the SKA will discover new magnetic
phenomena beyond what we can currently predict or even imagine.\\

\parbox{0.9\textwidth}{
\noi{References:}\\
\noi{\scriptsize
Beck R., Gaensler B.M., 2004, in \emph{Science with the Square
Kilometre Array}, eds. C.~Carilli and S.~Rawlings, New Astr. Rev.,
48, 1289;
Chandran B.D.G., Cowley S.C., 1998, Phys. Rev. Lett., 80, 3077;
Fletcher A., et al., 2011, MNRAS, 412, 2396;
Gaensler B.M., Beck R., Feretti L., 2004, in \emph{Science with the
 Square Kilometre Array}, eds. C.~Carilli and S.~Rawlings, New Astr. Rev., 48, 1003;
Kolatt T., 1998, ApJ, 495, 564;
Narayan R., Medvedev M.V., 2001, ApJ, 562, L129;
Neronov A., Vovk I., 2010, Science, 328, 73;
Stepanov R., \etal , 2008, A\&A, 480, 45;
Sun X.H., et al., 2008, A\&A, 477, 573}}\\

\subsubsection{Evolution of
  magnetic fields in galaxies and testing the galactic dynamo theory
  {\scriptsize [Marita Krause]}}

Future deep field observations of the synchrotron emission with the
SKA will make it possible for the first time to probe the radio
continuum emission of normal star forming galaxies out to the edge of
the Universe. Further, deep field polarisation observation will give
us for the first time the possibility of tracing the magnetic field
evolution and structure across cosmic time. From observations in the
nearby Universe we know that radio continuum emission and far infrared
emission are very tightly related to each other (known as ``radio-FIR
correlation'' ) which is still not fully understood. While the radio
emission depends on the magnetic field, the FIR emission is an
indicator of star formation. It is also known that magnetic fields are
needed for star formation and that star formation and its evolution
has important influences on the dynamo action which is regarded to be
the most important mechanism for magnetic field amplification and
structure formation. Hence, deep field radio continuum and
polarisation observations of star forming galaxies as planned with the
SKA will give the unique chance to follow both, star formation and
magnetic field evolution far backwards in time and by this will open a
new door in understanding their interplay.

\noi Polarised synchrotron emission from galaxies is a sign of
large-scale, coherent magnetic fields, even when the galaxies remain
unresolved. Stil et al. (2009) showed that the degree of polarisation
of unresolved spiral galaxies depends on inclination, uniformity of
the magnetic field, and galaxy luminosity (see Figure\ $_{\rm Kra}$). This result
implies that distant normal disk galaxies with flux densities
$\leq$\,100\,microJy will be an important population of polarised radio
sources for the SKA at frequencies $\geq$\,1\,GHz. The polarisation
angle of an unresolved spiral galaxy with an axisymmetric magnetic
field will be oriented along the apparent minor axis of the disk, thus
creating a polarised source with a (nearly) constant polarisation
angle despite significant internal Faraday rotation. RM synthesis
(Brentjens \& de Bruyn 2005), however, of unresolved spiral galaxies
will allow a reconstruction of the distribution of RMs in the disk
even though the polarisation angle of the integrated emission is
constant. The polarisation properties of a large sample of galaxies as
a function of redshift at the same rest frame frequency reveal the
evolution of magnetic fields and Faraday depolarisation in these
galaxies. The polarisation quantities of these galaxies can be related
to other observations to connect the evolution of the magnetic field
to the global star formation rate and other tracers of galaxy
evolution. Disk galaxies are common up to z\,$\sim$\,1, and exist up to
z\,$>$\,3 (e.g. Elmegreen et al. 2007). Observational evidence for the
existence of magnetic fields in galaxies up to z\,=\,2 has been found
from increased RM values for quasars with strong Mg\,II absorption
lines at a smaller redshift than the quasar itself (Bernet et
al. 2008, Kronberg et al. 2008).

\bigskip

\vspace{0.0cm}
\parbox{\textwidth}{

\hspace{1.5cm}
\parbox{5cm}{ \includegraphics[scale=0.3]{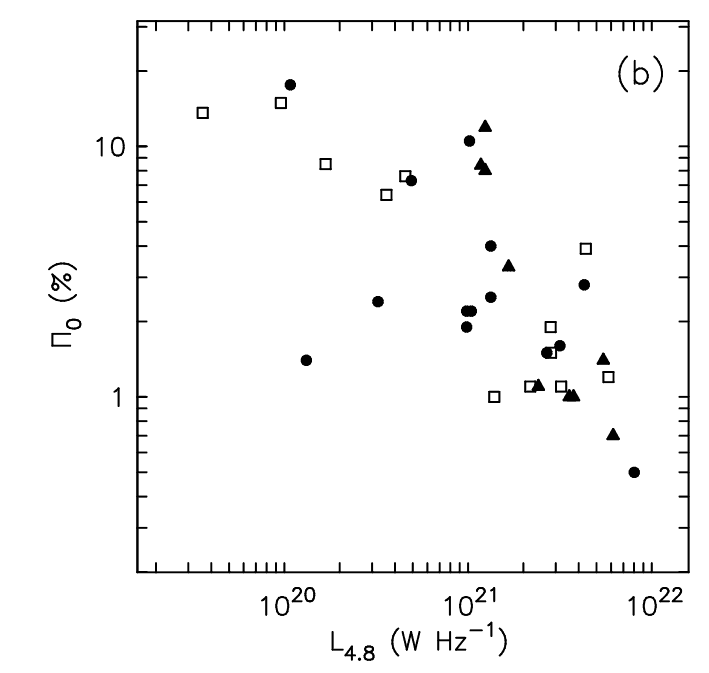}\vspace{-6.9cm}}

\hspace{9.3cm}
\parbox{6cm}{Figure 1$_{\rm Kra}$: Radio luminosity at 4.8\,GHz and degree of linear polarisation
of unresolved spiral galaxies in the local Universe (Stil et
al. 2009).} \vspace{5.5cm}}\\

\noi The existence of large-scale, coherent magnetic fields in galaxies are
only predicted by dynamo theory, according to which the axisymmetric
magnetic field structure in the galactic disk is expected to dominate
(e.g. Beck et al. 1996). This field structure has indeed mainly been
observed in the disk of nearby spiral galaxies. Large-scale dynamos
can order magnetic fields in Milky Way type galaxies on kpc scales for
redshift z\,$\leq$\,3. The regular field strength depends on galaxy
parameters and is expected to remain almost unchanged --with little
evolution-- until present (Arshakian et al. 2009). RMs from intervening
galaxies up to z\,=\,2 towards background sources (Kronberg et al. 2008,
Bernet et al. 2008) indeed imply magnetic field strengths of a few $\mu$G,
similar in strength to the large-scale fields found in spiral galaxies
today. At redshift z\,$\leq$\,3 significant evolutionary effects are only
expected for the magnetic field structure, as the coherence scale
varies as (1\,+\,z)$^{-1.5}$ (Arshakian et al. 2009). Fully ordered magnetic
fields in Milky Way type galaxies should be observed only at much more
recent times (z\,$<$\,0.5). Major mergers with other galaxies can distort
the field order and slow down the establishment of large-scale
magnetic fields.  The polarisation deep field as planned with the SKA
will test the dynamo theory by tracing the emergence of fully ordered
magnetic fields in galaxies as a function of redshift and galaxy size
for more than a million galaxies. Observations of polarised emission
from galaxy disks and RMs of background polarisation sources will
allow direct observational evidence for the emergence of coherent
magnetic field structures in galaxies over cosmic time.\\

\parbox{0.9\textwidth}{
\noi{References:}\\
\noi{\scriptsize
Arshakian T.G., Beck R., Krause M., Sokoloff D., 2009, A\&A, 494, 21;
Beck R., \etal , 1996, ARA\&A, 34, 155;
Bernet  M.L., \etal , 2008, Nature, 454, L302;
Brentjens M.A., de Bruyn A.G., 2005, A\&A, 441, 1217;
Kronberg P.P., \etal , 2008, ApJ, 676, 70;
Stil J.M., Krause M., Beck R., Taylor A.R., 2009, ApJ 693, 1392
}}\\

\subsubsection{Magnetic fields in spiral galaxies {\scriptsize
    [R. Beck, Marita Krause]}}

\noi The role of magnetic fields in the dynamical evolution
of galaxies and of the interstellar medium (ISM) is not well
understood. Radio astronomy provides the best tools to measure
galactic magnetic fields: synchrotron radiation traces fields
illuminated by cosmic-ray electrons, while Faraday rotation allow us
to detect fields in all kinds of astronomical plasmas, from lowest to
highest densities. Fundamental new advances in studying magnetic
fields in nearby galaxies can be made through observations with the
SKA. Mapping of diffuse polarised emission in many narrow bands over a
wide frequency range will allow us to carry out {\em Faraday
  tomography}, yielding a high-resolution three-dimensional picture of
the magnetic field, and allowing us to understand its coupling to the
other components of the ISM. The combination with Faraday rotation
data will allow us to determine the magnetic field structure in these
galaxies, and to {\em test both the dynamo and primordial field
  theories} for field origin and amplification.

\vspace{-0.5cm}
\parbox{\textwidth}{\hspace{-1cm}
\parbox{5cm}{ \includegraphics[scale=0.2,angle =-90]{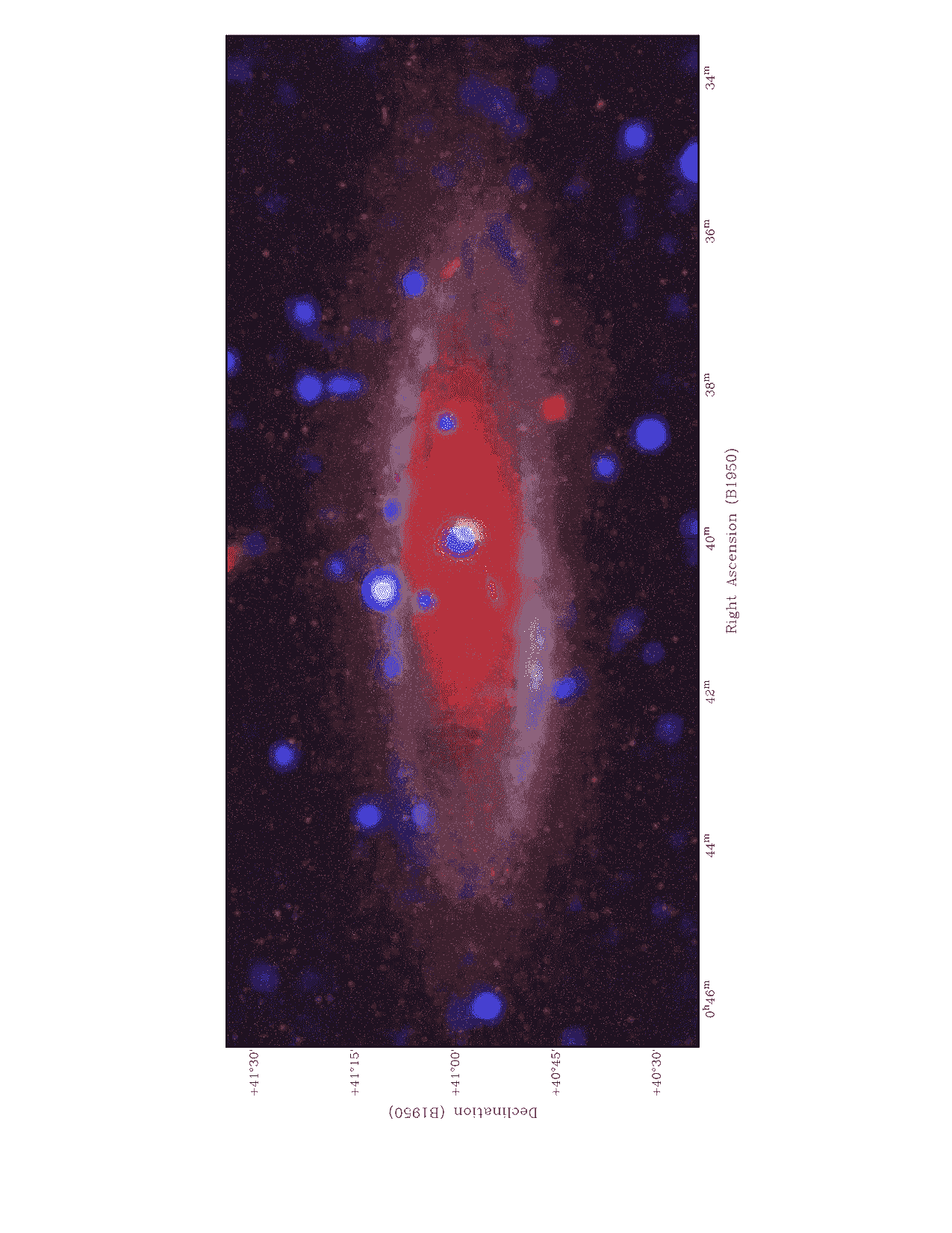}\vspace{-0.5cm}}
\hspace{6.5cm}
\parbox{6cm}{Figure 1$_{\rm Bec}$: Simulation of RMs towards background sources (white points) in the region
      of M\,31 observable with the SKA within 1\,hour. Optical emission
      from M\,31 is shown in red, diffuse radio continuum intensity in
      blue and diffuse polarised intensity in green (from Bryan
      Gaensler, priv. comm.). }
\vspace{0cm}
}\\

\noi A full understanding of the structure and
evolution of galaxies is impossible without understanding magnetic
fields. Magnetic fields fill interstellar space, contribute
significantly to the total pressure of interstellar gas, are essential
for the onset of star formation, and control the density and
distribution of cosmic rays in the interstellar medium (ISM). New
insights into galactic magnetic fields can be provided by the unique
sensitivity, resolution and polarimetric capabilities of the SKA.

\medskip

\noi Magnetic fields can be detected via synchrotron emission only if there
are cosmic-ray electrons to illuminate them. Cosmic rays are probably
accelerated in objects related to star formation. However, the radial
scale length of synchrotron emission in nearby galaxies is much larger
than that of the star formation indicators like infrared or CO line
emission (Beck 2007). Magnetic fields must extend to very large radii,
much beyond the star-forming disk. Field strengths in the outer parts
of galaxies can only be measured by Faraday rotation measures of
polarised background sources. Han et al. (1998) found evidence for
regular fields in M\,31 out to 25\,kpc radius. However, with only a few
detectable polarised sources per square degree at current
sensitivities, no galaxies beyond M\,31 could be mapped in this way.
The observation of large-scale RM patterns in many galaxies (Beck
2005) proves that the regular field in galaxies has a {\em coherent
  direction}\ and hence is not generated by compression or stretching
of irregular fields in gas flows. In principle, the dynamo mechanism
is able to generate and preserve coherent magnetic fields, and they
are of appropriate spiral shape (Beck et al. 1996) with radially
decreasing pitch angles. However, the physics of dynamo action is far
from being understood. Primordial fields, on the other hand, are hard
to preserve over a galaxy's lifetime due to diffusion and reconnection
because differential rotation winds them up. Even if they survive,
they can create only specific field patterns that differ from those
observed.  The widely studied {\em mean-field $\alpha$–$\Omega$ dynamo
  model}\ needs differential rotation and the $\alpha$ effect (see
below). Any coherent magnetic field can be represented as a
superposition of modes of different azimuthal and vertical
symmetries. The existing dynamo models predict that several azimuthal
modes can be excited (Beck et al. 1996), the strongest being $m=0$ (an
axisymmetric spiral field), followed by the weaker $m=1$ (a
bisymmetric spiral field), etc. These generate a Fourier spectrum of
azimuthal RM patterns. The axisymmetric mode with even vertical
symmetry (quadrupole) is excited most easily. Primordial field models
predict bisymmetric fields or axisymmetric fields with odd (dipole)
symmetry.  The SKA will dramatically improve the situation. Within the
fields of M\,31, the LMC or the SMC (a few square degrees each), a deep
observation could provide $>10^5$ polarised background sources
(Figure 1$_{\rm Bec}$), and thus allow fantastically detailed maps of the
magnetic structure. The SKA will be able to confidently determine the
Fourier spectrum of dynamo modes from high-resolution RM maps of the
diffuse polarised emission. An RM grid of background sources is even
more powerful: Already 10\,RM values are sufficient to identify a
single dominating dynamo mode (Stepanov et al. 2008). With the SKA,
galaxies out to about 100\,Mpc distance become observable. RM grids
towards nearby galaxies will allow a 3-d reconstruction of the field
pattern.\\

\vspace{0cm}
\parbox{\textwidth}{

\hspace{-0.3cm}
\parbox{5cm}{ \includegraphics[scale=0.25]{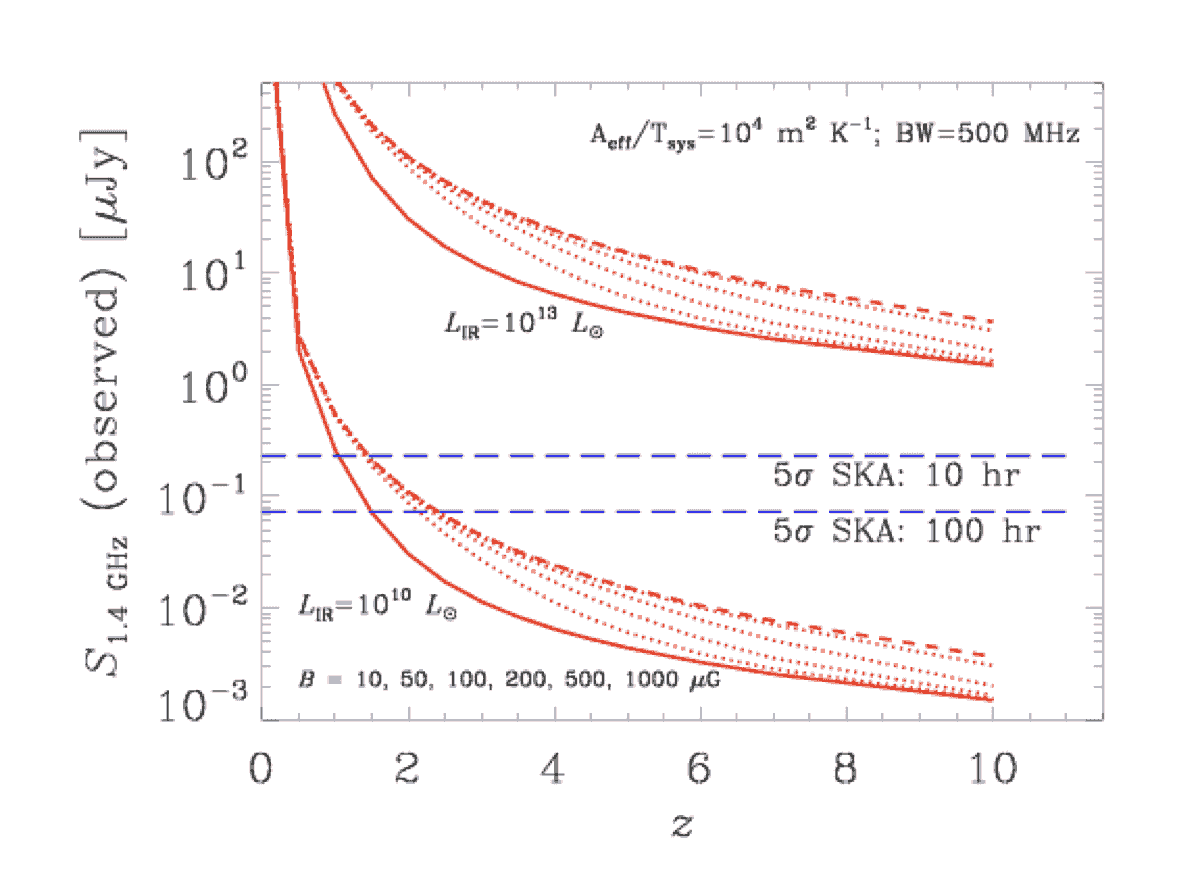}\vspace{-6.85cm}}

\hspace{10cm}
\parbox{6cm}{Figure 2$_{\rm Bec}$: Synchrotron emission at 1.4\,GHz as a function of redshift z
  and magnetic field strength $B$, and the $5\sigma$ detection limits
  for 10 and 100\,hrs integration time with the SKA (from Murphy
  2009).\vspace{4cm}}
}\\

\noi The SKA has the potential to increase the galaxy sample with
well-known field patterns by up to three orders of magnitude. The
conditions for the action of galactic dynamos can be clarified. For
example, strong density waves are claimed to support the $m=2$ mode
while companions and interactions may enhance the bisymmetric $m=1$
mode. A dominance of bisymmetric fields over axisymmetric ones would
be in conflict with existing dynamo models and would perhaps support
the primordial field origin. The lack of a coherent magnetic field in
a resolved galaxy would indicate that the timescale for dynamo action
is longer than the galaxy's age (Arshakian et al. 2009), or that the
mean-field dynamo does not work at all.  The detailed structure of the
magnetic fields in the ISM of nearby galaxies and in galaxy halos can
be observed. The turbulence power spectra of the magnetic fields can
be measured. Direct insight into the interaction between gas and
magnetic fields in these objects will become possible.  Faraday
rotation in the direction of bright quasars allows us to determine the
strength and pattern of a regular field in an intervening galaxy. This
method can be applied to distances of young quasars
(z\,$\simeq$\,5). Mean-field dynamo theory predicts RMs from regular
galactic fields at z\,$\le$\,3 (Arshakian et al. 2009).  Unpolarised
synchrotron emission, signature of turbulent magnetic fields, can be
detected with the SKA out to very large redshifts for starburst
galaxies, depending on luminosity and magnetic field strength
(Figure 2$_{\rm Bec}$). However, for fields weaker than $3.25~\mu$G
$(1+z)^2$, energy loss of cosmic-ray electrons is dominated by the
inverse Compton effect with CMB photons, so that their energy appears
mostly in X-rays and not in the radio range. On the other hand, for
strong fields the energy range of the electrons emitting at a 1.4~GHz
drops to low energies, where ionization and bremsstrahlung losses may
become dominant. In summary, the mere detection of synchrotron
emission at high redshifts will constrain the range of allowed
magnetic field strengths.\\

\parbox{0.9\textwidth}{
\noi{References:}\\
\noi{\scriptsize  
Arshakian T.G., Beck R., Krause M., Sokoloff D., 2009, A\&A, 494, 21;
Beck R., \etal , 1996, ARA\&A 34, 155;
Beck R., 2005, Astronomische Nachrichten, 326, 608;
Beck R., 2007, Advances in Radio Science, 5, 399;
Han J.L., Beck R., Berkhuijsen E.M., 1998, A\&A, 335, 1117;
Stepanov R., \etal , 2008, A\&A, 480, 45;
Murphy E., 2009, ApJ, 706, 482;
}}\\

\subsubsection{Structure and evolution of magnetic fields in star-forming galaxies
  {\scriptsize [T.G.~Arshakian]}}

\noi Presently, the magnetic fields of only few tens of nearby galaxies
are studied due to limited sensitivity of present-day radio
telescopes. With capabilities of the SKA it will become possible to
investigate the structure of magnetic fields in local star-forming
(SF) disk galaxies and evolution of magnetic fields in distant SF
galaxies.

\medskip

\noi Faraday rotation measures (RM) of polarised background sources
towards SF galaxies is shown to be a powerful tool to
\emph{recognise} and \emph{reconstruct} the regular magnetic field
patterns in galaxies, characterised by the spiral pitch angle
(Stepanov et al. 2008). Single and mixed modes, the amplitude of the
regular field, and the azimuthal phase can be recognised from a
limited sample of about 15\,RM measurements of polarised background
sources. Galaxies with strong turbulence and small inclination
angles need more background sources for a reliable recognition.
Future all-sky RM survey with the SKA at about 1\,GHz would allow
field recognition of about tens of thousands SF galaxies up to a
distance of 100\,Mpc.

\medskip

\noi A reliable reconstruction of the field structure would require at
least 20\,RM values on a cut along the projected minor axes, which
translates into thousands RM measurements towards a galaxy. The
reconstruction method is superior for galaxies inclined at $\gtrsim
70$\,degrees and is possible for the nearest galaxies up to 10\,Mpc
(M\,31, M\,33, IC\,342, NGC\,6946, etc.) but it will require tens to
a hundred hours of integration time to achieve a sufficient
detection limit of polarised intensity at 1.4\,GHz (Stepanov et al.
2008).

\medskip

\noi It follows to note that the integration time of the SKA to detect
polarised background sources strongly depends on the slope of the
number count function (uncertain by now) and on Faraday
depolarisation, which becomes stronger at lower radio frequencies
(Arshakian \& Beck 2011). Hence, high radio frequencies ($\gtrsim
1$\,GHz) are preferable for the reconstruction and recognition of
field structures. 

\medskip

\noi Very little is known about regular magnetic fields and their
evolution in distant galaxies. Dynamo theory is able to describe the
amplification and ordering of regular magnetic fields to a level seen
in present day starforming (SF) galaxies (Arshakian et al. 2009a).
Simulations of the total intensity, polarisation, and Faraday depth of
an evolving SF disk galaxy are shown in Figure 1$_{\rm Ar}$ (Arshakian
et al. 2011).  These simulations predict patchy field structures and
field reversals of regular fields, asymmetric and inhomogeneous RM
structures in younger galaxies (less than a few Gyrs); a weak regular
field and small ordering scale in young galaxies; polarisation
patterns and asymmetric structures in older galaxies, that are
frequency dependent due to depolarisation effects at low radio
frequencies; complicated field pattern in interacting and merging
galaxies. Predictions of present dynamo models (Arshakian et al. 2011,
Moss et al. 2012) and future sophisticated models can be tested with
upcoming/future sensitive radio telescopes such as ASKAP and the
SKA. Radio continuum and polarisation observations of distant disk and
dwarf galaxies will allow the merging rate, history of field
reversals, change of the magnetic field strength and coherence to be
determined up to a redshift z\,$\approx$\,3. Observations of RMs against
distant background polarised quasars (Kronberg et al. 2008) can be
applied to study the evolution of regular fields in SF disk galaxies
(identified by optical spectroscopic observations of absorption line
systems, see Bernet et al. 2008) even to high redshifts of z\,$\approx$\,5. 

\medskip

\noi Another promising tool to study the evolution of regular fields in
distant galaxies is the observation of RM of background sources
against gravitational lens systems (Narasimha \& Chitre 2004). 
The power of this method is that the
observed difference of RMs of a lens system originates in the
magneto-ionic medium of a lens galaxy and it does not depend on
Faraday rotation of our Galaxy and intergalactic medium. A large
number of lens systems is needed for a meaningful statistics of RM
differences. Thousands of lens systems per square degree will be
detected with the SKA down to a limiting rms sensitivity of
1\,$\mu$Jy (Koopmans et al. 2004). Many SF galaxies and radio-quiet
AGN are expected to be lens galaxies. This will allow the
cosmological evolution of magnetic fields in these galaxies to be
probed beyond z\,$\approx$\,1 (Arshakian et al. 2009b).

\bigskip
\bigskip

\hspace{0cm}
\parbox{\textwidth}{
\parbox{5cm}{ \includegraphics[scale=0.5]{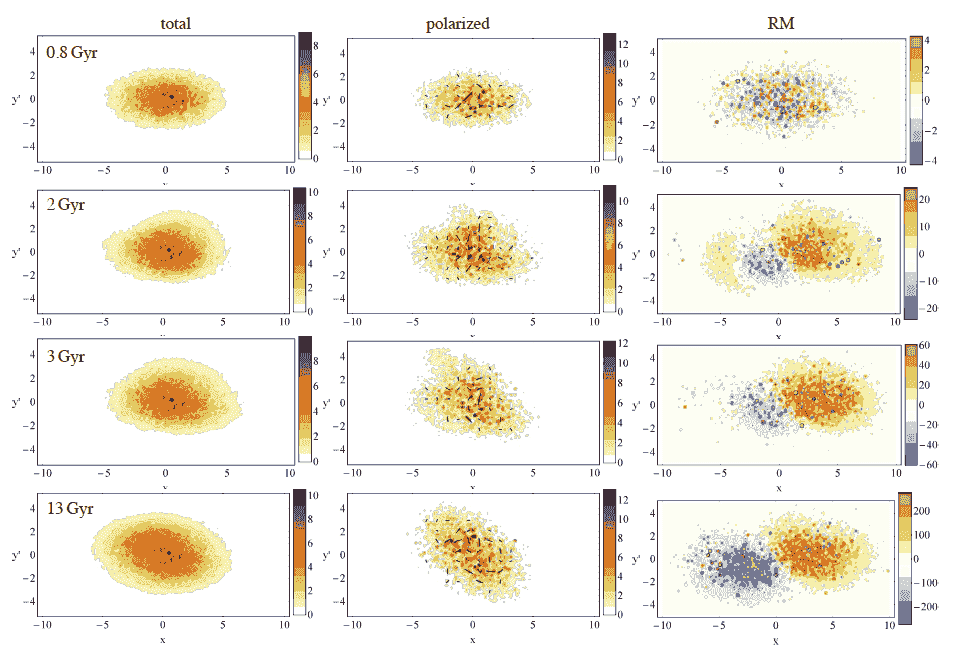}}

\hspace{1cm}
\parbox{0.8\textwidth}{Figure 1$_{\rm Ar}$: Simulations of a SF disk
  galaxy are shown at 0.8\,Gyr, 2\,Gyr, 3\,Gyr, and 13\,Gyr after disk
formation. The simulations are based on an evolving dynamo model. The observed total intensity
(\emph{left panel}), polarisation (\emph{middle panel}), and Faraday
rotation (\emph{right panel}) at 150\,MHz are simulated for a galaxy with an
inclination angle of 60\ndeg\ and star-formation rate of 10\,M$_{\sun}$ yr$^{-1}$. The frame units are given in kpc. The colour bars in the first and 
second columns (total and polarised intensity) are given given in arbitrary units. The colour bar of the third column (Faraday rotation measure) is given in units of rad\,m$^{-2}$.}
}

\bigskip
\bigskip

\noi This research will be done in close collaboration with the
magnetic field research group of the ``Max-Planck-Institut f\"ur
Radioastronomie'' in Bonn.\\

\parbox{0.9\textwidth}{
\noi{References:}\\
\noi{\scriptsize
Arshakian T.G., Beck R., 2011, MNRAS, 418, 2336;
Arshakian T.G., \etal , 2011, Astronomische Nachrichten, 332, 524;
Arshakian T.G.,\etal , 2009a, A\&A, 494, 21;
Arshakian T.G.,\etal , 2009b, in: Wide Field Astronomy and Technology for the Square Kilometre Array, eds. S.\,A. Torchinsky et al., p.~103;
Bernet, M.L., \etal ,2008, Nature, 454, 302;
Koopmans L.V.E., Browne I.W.A., Jackson N.J., 2004, New Astronomy Review, 48, 1085;
Kronberg P.P., \etal , 2008, ApJ 676, 70;
Moss D., \etal ,2012, A\&A, 537, A68;
Narasimha D., Chitre S.M., 2004, Journal of Korean Astronomical Society, 37, 355;
Stepanov R., \etal ,2008, A\&A, 480, 45
}}\\

\subsubsection{The SKA and the plasma Universe -- new insights into
  radio galaxies and galactic outflows {\scriptsize [Martin Krause]}}

Since the Cosmic re-ionisation at high redshift, the Universe is
predominantly ionised and therefore a plasma, carrying currents and
magnetic fields. In order to understand the physics of this plasma,
including the transport of heat and Cosmic Rays, magnetohydrodynamics,
turbulence and mixing with other gas phases, measurements of both,
particles and the magnetic fields are required. Ionised gas is now
being seen in X-rays and emission lines out to redshifts beyond
unity. The magnetic field is mainly accessible via synchrotron
emission and Faraday rotation in polarised radio emission. Currently,
we have this information essentially only for a few very nearby
clusters and groups at redshift z\,$<$\,0.1 (Krause \etal\ 2009),
with a few notable exceptions of rotation measures towards luminous
high redshift radio sources (O'Sullivan 2011, Broderick 2007).\\

\bigskip
\bigskip

\vspace{0cm}
\parbox{\textwidth}{

\parbox{5cm}{ \includegraphics[scale=0.28]{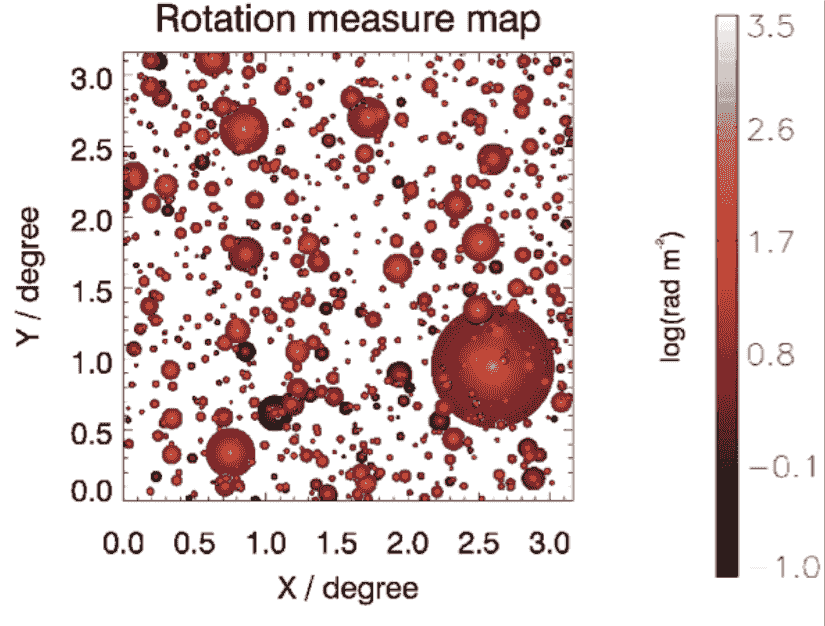}\includegraphics[scale=0.28]{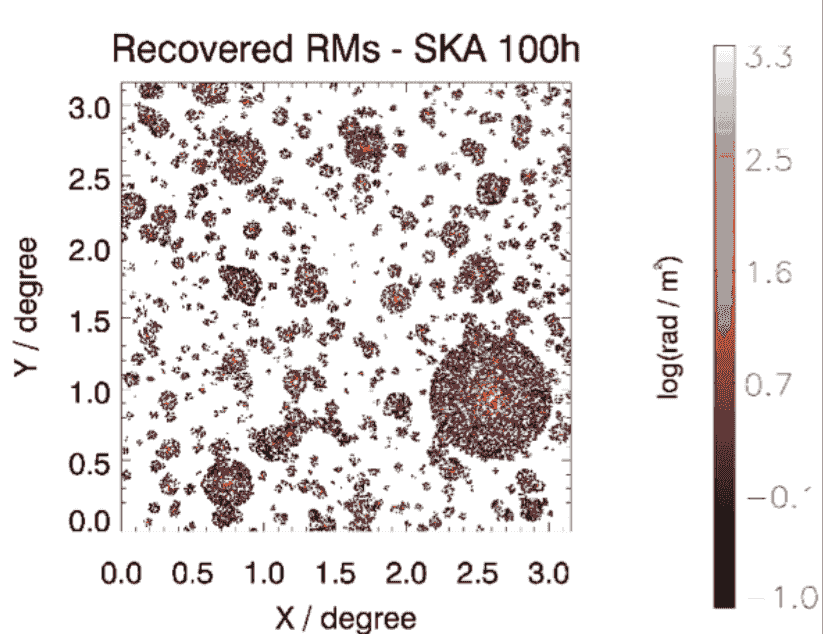}\vspace{0.5cm}}

\hspace{0.3cm}
\parbox{0.9\textwidth}{Figure 1$_{\rm Kra}$: left: Distribution of
  rotation measures from a Cosmological model.
 
right: Recovered
  rotation measure distribution for a simulated SKA observation using
  specs appropriate for the proposed mid-frequency array (AA-high,
  Schilizzi et al. 2007, Faulkner et al. 2010). Figures are from Krause et al. 2009.\\

}
\vspace{0.5cm}}\\

\noi The Square Kilometre Array (SKA)
has the potential to revolutionise the field, providing much more
detailed information on the magnetic field in nearby groups and
clusters (Krause \etal\ 2009), and extending the magnetic horizon
out to beyond a redshift of unity, comparable with current and planned
capabilities of X-ray missions. The impressive recovered rotation
measure distribution from a Cosmological calculation is shown in
Figure\ 1$_{\rm Kra}$ (Krause \etal\ 2009). A 100\,hrs SKA observation should yield
about 10\,000 polarised sources per square degree, resulting in an RM
grid with a spacing of about half a minute of arc. This will allow us
to follow the evolution of magnetic fields in the dense structures of
the Universe.  One expectation is that magnetic fields originate in
galaxies and are brought to the intra-cluster/group medium (and
possibly beyond) by galactic outflows. They are then further amplified
due to the gas kinematics (e.g. Ryu \etal\ 2008), which may be partly due to the
same outflows, but probably largely due to the Cosmic flow. The
outflows themselves will be a target of the SKA. \\

\noi Both, star-forming
galaxies and outflows related to the jets of Active Galactic Nuclei
are radio emitters. I focus here on the latter.  Massive galaxies at
high redshift frequently have powerful radio jets (Miley \& de Breuck 2008). They are
associated with gaseous outflows, the so-called emission line or Lyman-alpha halos, which remove a significant amount of ionised gas from the
host galaxies. Neutral hydrogen is frequently found in absorption. A
successful model that explains all the currently available data is the
following (Krause \etal\ 2005): First, intense star formation with 100s of solar masses
per year produces a galactic wind. The wind shell cools and hydrogen
recombines. Then the jet is started, hits the shell and is impeded for
a while. During that time it fills the wind shell (Figure\ 2$_{\rm Kra}$). 

\bigskip
\bigskip

\parbox{\textwidth}{

\hspace{1cm}
\parbox{5cm}{ \includegraphics[scale=0.43]{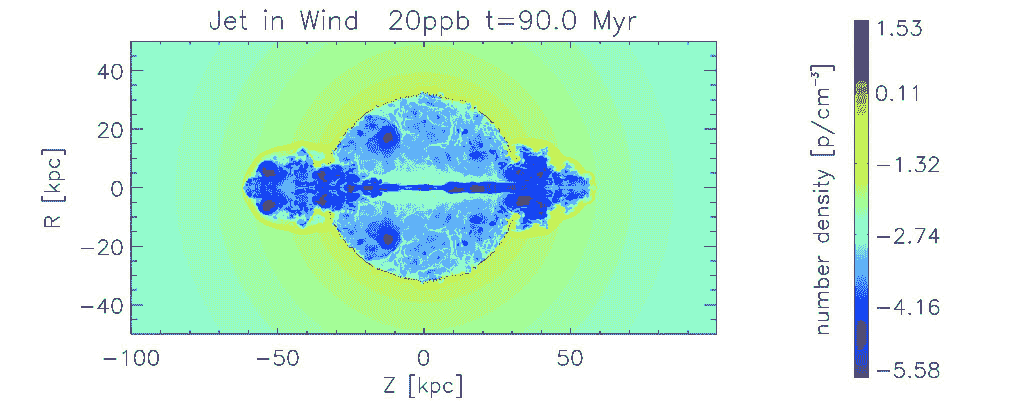}\vspace{0.5cm}}

\hspace{0.2cm}
\parbox{0.9\textwidth}{Figure 2$_{\rm Kra}$: This figure displays the
    simulation of an extragalactic jet breaking through the shell of a
    galactic wind caused by supernova activity (Krause 2005). The blue
    regions are radio emitting plasma, the thin red almost circular
    region is the galactic wind shell, which provides a screen for
    Faraday rotation. When the jet first hits the wind shell, it is
    considerably impeded, and the radio lobes quickly fill the wind
    shell. Radio sources of this kind are known at redshifts of 1\,--\,5,
    have polarised fluxes of order 10\,mJy, and angular diameters of a
    few to tens of seconds of arc. They will thus be well resolved and
    easily detected with the SKA.}\vspace{0.5cm}}

The acceleration by the excess pressure destroys the wind shell,
explaining why the corresponding absorbers are only found in small
radio sources. Faraday rotation of about 10$^4$\,rad\,/\,m$^2$ has been
measured for the source PKS B0529-549 at redshift z\,=\,2.6. The source
also has a prominent local neutral hydrogen absorber, which could be
due to such a wind shell. The rotation measure is consistent with a
line of sight magnetic field of 10\,$\mu$G and a neutral fraction of a
few per cent within the wind shell. The Faraday rotation could also be
due to the general interstellar medium in the host galaxy. The
polarised radio flux of PKS B0529-549 is about 10\,mJy, no issue for
SKA with a 1-hour sensitivity better than 1\,\muJy . Such a source will
therefore be resolved by the SKA in polarised flux. The SKA will therefore be
able to tell us

\begin{itemize}
\item[--]{if the shape of the radio emission is reminiscent of
a wind shell} 
\item[--]{if the absorbing column of neutral hydrogen agrees with
the measurements from Lyman-alpha absorption}
\item[--]{if the column density
increases, and the bulk velocity of the absorbers decrease towards the
edges, as expected for wind shells}
\item[--]{if the rotation measure increases
towards the edges as expected from the mainly tangential field
geometry expected in a compressed layer (compare with Huarte-Espinosa
\etal\ 2011)}
\end{itemize}

\noi Thus, we will be able to see and quantify the ejection of
magnetised plasma directly. \\

\noi In nearby clusters, cold fronts could be
mapped out (Alexander \etal\ 2007), and thus the SKA would be able to confirm, if the
necessary reduction in conductivity is really due to a tangential
magnetic field, as we would expect. Traditional radio sources still
present serious problems to plasma physics: While having a mean free
path which exceeds the source size by many orders of magnitude, they
are able to displace the surrounding cluster gas. Their lobe
kinematics seems to be describable by stretching and
magnetohydrodynamic turbulence (Huarte-Espinosa 2011b). The magnetic field at their
boundaries must be exactly tangential, otherwise the heat flux to the
cluster gas would not allow the radio lobes to grow. New plasma
physics simulations beyond ideal magnetohydrodynamics are required to
understand these issues. The SKA will provide the required high
resolution polarisation observations to compare with a regular
magnetic field along the line of sight. Extragalactic
radio sources confine their heat and particle content almost
perfectly, a quality that is intensely sought-after in laboratory
fusion plasma research (e.g. Wolf 2003). While one would hardly expect that
the solution for the problems in the lab will be found in space, one
should keep in mind that progress is being made in related areas of
physics, with possibilities of cross-fertilisation.\\

\parbox{0.9\textwidth}{
\noi{References:}\\
\noi{\scriptsize
Alexander P. \etal\ , 2007, Proceedings of Science (MRU) 052, 068, 525;
Broderick J. W., De Breuck C., Hunstead R. W., Seymour, N., 2007,
MNRAS, 375, 1059;
Faulkner A. \etal , 2010, SKA Memo, 122; 
Huarte-Espinosa M., Krause M., Alexander M., 2011, MNRAS, 417, 382;
Huarte-Espinosa M., Krause M., Alexander M., 2011b, MNRAS, 418, 1621;
Krause M., \etal , 2009, MNRAS, 400, 646;
Krause M., 2005, A\&A 436, 845;
Miley G., de Breuck C., 2008, A\&ARv, 15, 67; 
O'Sullivan S. P., Gabuzda D. C., Gurvits L. I. 2011, MNRAS, 415, 3049;
Ryu D., Kang H., Cho J., Das S., 2008, Science, 320, 909;
Schilizzi R.T. \etal\ , 2007, SKA Memo, 100;
Wolf R.C., 2003, Plasma Phys. Control. Fusion, 45, R1- R91
}}\\

\subsection{Galactic astronomy}

\subsubsection{Neutral hydrogen in the Milky Way and the local
  Universe  {\scriptsize [J. Kerp]}}

\noi{\bfseries\boldmath\small Abstract:~~}Our aim is to create a major data base for galactic astronomy for the next
decade. We propose to combine all-sky surveys from the worlds largest single
dish telescopes with SKA data for velocities -1000\,\kms\,$< |v|
<$\,6000\,\kms .  This range
covers the whole Milky Way neutral hydrogen (\hi ) radiation as well as all members of the local 
Galaxies, where zero spacing correction is urgently needed to provide a
comprehensive view of the complex gas physics.\\

\noi We propose to use SKA data acquired during wide-field area surveys to generate an all-purpose
database on the \hi\ distribution in the Milky Way and the local Universe at an
angular resolution of half an arcmin. This database will comprise all declinations
below $+$30\,degree, unbiased and homogeneous in its coverage. Because of the
proximity of the objects of interest, most of the warm gaseous phase of the \hi\
distribution will not be included within the SKA data. Combining the
SKA survey with the Parkes (HIPASS and GASS) and Effelsberg (EBHIS) data
will provide an unbiased physical view of the Milky Way, the accretion history
of the local group and the local large-scale structure.

\noi A Galactic \hi\ database is urgently needed for the exploration of the warm hot
intergalactic medium (WHIM).  The WHIM is thought to host about 80\,\% of the
total baryonic budget of the Universe. Due to the structure formation shocks
heat the gas of the Lyman-alpha forest to coronal temperatures in the range of
$10^6$\,K. It is difficult to observe this gas phase because its
radiation (E\,$\sim$\,0.2\,keV) is strongly attenuated by
photoelectric absorption traced by the \hi\ distribution. According to this, the nature of the WHIM in the local group and
around the Milky Way, in particular the spatial distribution on arcmin scales,
is currently nearly unexplored. The team aims to correlate the eROSITA
soft X-ray survey with high resolution short spacing corrected \hi\ data. The
strength of the photoelectric absorption is determined by the amount of warm
neutral gas, while the cold neutral gas causes strong but spatially well
defined (arcmin scales and below) X-ray shadows. \hi\ data covering all spatial frequencies
up to the arcmin scale are needed to open the window to the distant Universe.

\noi The exploration of the early Universe at high redshifts an SKA \hi\ survey is
even more important. The proposed SKA \hi\ survey database will open for the first
time large areas of the sky for X-ray astronomers. Future X-ray observatories
like IXO will have a significant fraction of their detection power in the soft
X-ray energy range below 1\,keV. The emission of active Galactic nuclei at
high redshifts (z\,$\sim$\,10) or the faint emission of clusters of galaxies at
moderate redshifts (3\,$\leq$\,z\,$\leq$\,5) is shifted
(E\,=\,$\frac{E}{\rm 1+z}$) to the
soft X-ray band where photoelectric absorption strongly modulates the X-ray
intensity of the objects. It is not possible to analyze the X-ray data
quantitatively without knowing in detail the distribution of the Galactic
ISM. The SKA will overcome the present situation that X-ray astronomy are
focused predominantly towards two low \hi column density windows of the Milky
Way (HDF, CDFS). To overcome the ``cosmic conspiracy'' it is necessary to open
the whole high Galactic latitude sky to X-ray astronomy. ASKAP, with short
spacing data based on LAB, GASS and EBHIS will provide an ideal data base for
this purpose.

\noi Investigations of the accretion history of the Milky Way halo gas will need
SKA \hi\ data. This includes a search for stream-like \hi\ structures, HVCs, and
IVCs. Within a research project our team could demonstrate that optical and UV
absorption line measurements correlated with single dish and interferometric
\hi\ data disclose the physical similarity of HVCs and IVCs with absorption line
systems at cosmological distances. Here, excitation conditions and chemical
composition appear to be pretty comparable. This offers the opportunity to
study in great detail, because of the proximity, dynamics and structure of the
metal-absorption line systems and connecting the accretion history of the
Milky Way to the cosmological evolution of the Universe as whole.\\

\subsubsection{Hydroxyl masers in the Milky Way and local group galaxies {\scriptsize [D. Engels]}}

In the Milky Way maser emission is often observed in the circumstellar
shells of red giant stars and in the surroundings of young stellar
objects. These are environments, which are cool enough to form
molecules and provide sufficient velocity coherence and density, so
that the masers naturally are excited. The strongest stellar masers
are those of oxygen-bearing molecules like hydroxyl (OH), water (H$_2$O)
and silicium-oxide (SiO). Methanol (CH$_3$OH) and to lesser extent
ammonia were detected in addition close to star-formation regions. The
major maser lines accessible with the SKA will be from OH, at
frequencies of 1612, 1665 and 1667\,MHz after completion of \skai\
construction and the CH$_3$OH maser line at 6.7\,GHz later on. Other
prominent maser lines at higher frequencies such as from H$_2$O at 22\,GHz
and from CH$_3$OH at 12.2\,GHz will not be accessible during the first two
construction phases.

\noi Currently $>$\,2000 stellar (Engels et al. 2010) and several hundred
interstellar OH masers are known in the Milky Way. Most were
discovered by surveys with single-dish radio-telescopes with typical
survey limits of several hundred mJy, but the ATCA-VLA survey
(Sevenster 2002) already showed that the number of detections still
increases strongly with decreasing sensitivity limits. The known OH
masers are typically located in the Galactic Plane at distances 2\,--\,8\,kpc from the Sun. Typical peak luminosities of stellar OH masers are
$3\times10^{13}$\,Watt/Hz. Their characteristic double-peaked profiles allow
the determination of the radial velocity of the star and the expansion
velocity of the circumstellar shell with high accuracy. OH masers have
been used to infer the presence and strength of magnetic fields in the
outer regions of the circumstellar shells of AGB stars (Vlemmings
2007).

\noi With the SKA several thousand OH maser sources can be discovered already
in \skai\ with shallow surveys ($\sim$\,100\,mJy). Such surveys will uncover
the population of OH masers beyond the Galactic Centre (GC), allowing the
study of the structure and the kinematics of stellar populations in
the parts of the Milky Way opposite to the Sun. As the OH maser
lifetimes might be limited (Engels \& Jimenez- Esteban 2007), repeated
sensitive surveys along the Galactic Plane will unearth
lower-luminosity and only temporarily present masers. With SKA's
sensitivity, polarisation studies of the OH maser lines of large
samples of AGB stars are feasible. The physical mechanism responsible
for the launch of the strong winds on the AGB and of non-axisymmetric
winds in the post-AGB phase is still not understood, and such
polarisation studies will help to understand the role of magnetic
fields for launching and shaping the winds. SKA studies of the OH
maser lines and ALMA studies of higher frequency masers (f.e. SiO,
H$_2$O) in the mm and submm wavelength will be complementary, as sources
showing maser lines from one molecule often have maser emission from
the other molecules as well.

\noi Only few maser sources with luminosities similar to their
Galactic analogs are known outside the Milky Way. In the Large
Magellanic Cloud about 10 OH masers (1612\,MHz) with peak-flux
densities 17\,--\,600\,mJy are known in late-type stars (Marshall et
al. 2004), and the number of star-formation sites with interstellar
masers is of the same number (Ellingsen et al. 2010). In the Local
Group M\,31, M\,33 and IC\,10 a few intrinsically bright H$_2$O masers
belonging to star-formation regions are known (Darling 2011). Taking
advantage of the compactness of maser sites, Brunthaler et al. (2005)
have shown that with VLBI and phase-referencing techniques, the proper
motions of local group galaxies can be measured on scales of tens of
microarcseconds per year. This technique will be applicable to the OH
and CH$_3$OH maser sources alike.  Galactic analog OH masers at the
distance of M\,31 are expected to have flux densities of
$\sim$\,400\,microJy, extending up to $\sim$\,15\,mJy for the
brightest. The bright end of the maser population can be detected at
the 5 sigma level already in \skai\ in $\sim$\,400\,s (bandwidth 5\,kHz),
while the average Galactic analog OH maser will be detectable in
\skaii\ in $\sim$\,40\,min. SKA surveys for OH masers in Local Group
galaxies will provide sufficient targets for proper motion studies of
the galaxies and provide key insight into the Local Group's historic
and future dynamical evolution. The detection of the maser population
is also helpful for modelling of their stellar populations in general,
because the maser properties can be used to clarify the nature of
those stars, which currently evolve in an optically obscured, infrared
bright phase, either as young stars or as late-type giants.\\

\parbox{0.9\textwidth}{
\noi{References:}\\
\noi{\scriptsize  
Brunthaler A., Reid M.J., Falcke H., et al., 2005, Science 307, 1440;
Darling J., 2011, ApJ 732, L2;
Ellingsen S.P., Breen S.L., Caswell J., et al., 2010, MNRAS 404, 779;
Engels D., Jimenez-Esteban F., 2007, A\&A 475, 941;
Engels D., Bunzel F., Heidmann B., 2010, Database of Circumstellar Masers v2.0,
http://www.hs.uni-hamburg.de/maserdb ;
Marshall J.R., van Loon J.T., Matsuura M., et al. 2004, MNRAS 355, 1348;
Sevenster M.N., 2002, AJ, 123, 2772;
Vlemmings W.H.T., 2007, in IAU Symp. 242, Eds. J.M. Chapman \& W.A. Baan. p. 37
}}\\

\subsubsection{The Galactic magnetic field  {\scriptsize [W. Reich]}}

\noi Magnetic fields are an importent constituent of the interstellar medium. 
Magnetic fields could not be directly observed, but are traced by changing 
the propagation of cosmic-rays or the polarisation plane of electro-magnetic 
waves in the presence of thermal gas. Radio polarisation provides the key 
information to reveal the properties of the Galactic magnetic field. Intensive
efforts were made in the past employing various observing techniques to trace 
its small-scale turbulent components on sub-parsec scales and its large-scale 
regular components on kpc-scales. Synchrotron emission dominates the radio sky 
at low frequencies, where its intensity depends on the strength of the magnetic 
field component perpendicular to the line-of-sight. The direction in the plane 
of sky is observed by linear polarisation when corrected for Faraday rotation 
effects. The measured linear polarisation percentage is significantly lower 
than its intrinsic value due to irregular magnetic fields, Faraday depolarisation 
and also beam averaging effects. Faraday rotation traces the magnetic field along 
the line-of-sight. Combining all data should in principle provide a
3-d reconstruction 
of the magnetic field and its properties. The following figure shows a currently widely 
accepted model of a regular axisymmetric-field with one proven reversal revealed by 
pulsar RM observations, which seems to agree qualitatively with almost all 
presently available observations.

\bigskip
\bigskip
\bigskip
\bigskip

\parbox{\textwidth}{

\hspace{1cm}
\parbox{5cm}{ \includegraphics[bb=0 0 270 270, scale=0.35]{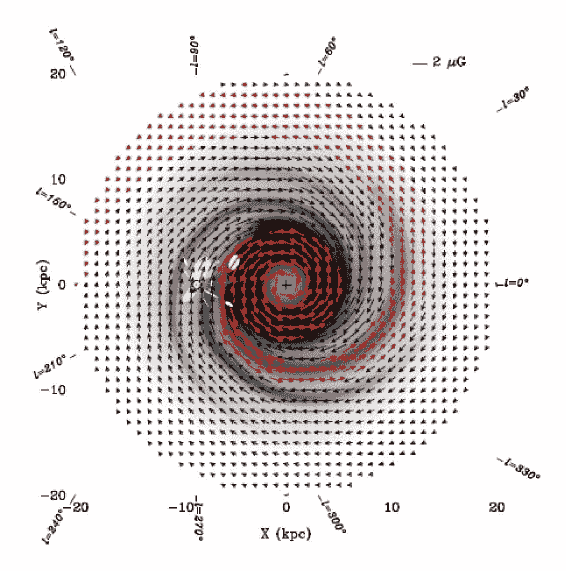}\vspace{-9.5cm}}

\hspace{9.4cm}
\parbox{6cm}{Figure 1$_{\rm Re}$: The regular Galactic magnetic field when seen face-on as
modeled by Sun et al. (2008). The magnetic field follow the spiral arms
based on the NE2001 model of the Galactic thermal electron density 
distribution by Cordes \& Lazio (2002). The position of the Sun is indicated 
and also some thermal features accounted for in the NE2001 model. }\vspace{0cm}
}\\

\newpage

\noi However, we know from magnetic field observations of nearby galaxies, that 
clear deviation exists from a simple spiral pattern assumed for the Galaxy. Discrete 
sources like supernova remnants and HII-regions highly influence the magnetic field 
structure locally on scales up to a hundred parsec or more. Only recently passive
Faraday screens were revealed as objects with strongly enhanced regular magnetic fields 
on scales of tens of parsecs and often have a reverse direction to the Galactic large 
scale field. They must be known and taken into account when talking about the large
scale field, but so far little is known about their number and origin. The turbulent 
magnetic field component at least as strong as the regular field, but its spectrum 
is not very well constrained. More high quality data are needed and are expected to 
be provides by the SKA. Most important is the SKA L-band Rotation Measure (RM) survey, 
which will result in several million RMs of extragalactic sources at about 1\,arcsec angular 
resolution (Beck \& Gaensler, 2004). This RM survey will improve available datasets 
by two orders of magnitude. Combining extragalactic RM data with arcsec SKA polarisation 
observations of diffuse Galactic emission will allow to separate any faint intergalactic 
magnetic field component from the Galactic foreground component. The Galactic foreground 
needs also to be taken into account when studying magnetic fields of resolved galaxies 
or cluster of galaxies by polarised emission. Extensive high resolution polarisation 
simulations were made by Sun \& Reich (2009) within SKADS at 1.4 GHz. Galactic emission 
patches at various Galactic latitudes were simulated, which were based
on the 3-d Galactic emission model by Sun et al. (2008). These global models are based on total intensity and 
polarisation all-sky maps together with extragalactic RM data. The turbulent magnetic field 
clearly dominates on arcsec scales and was assumed to be of Kolmogorov-type, where 
the outer scale and the length of the line-of-sight determine the slope of the polarisation
maps structure function. By changing the parameters the statistical properties can be adapted 
to those observed. The SKA is expected to provide a major step towards the understanding of 
the properties of the Galactic magnetic field.\\

\parbox{0.9\textwidth}{
\noi{References:}\\
\noi{\scriptsize  Cordes J.M., Lazio, T.J., 2002, astro-ph/0207156;
Sun X.H, Reich W., Waelkens A., Ensslin T.A. 2008, A\&A, 477, 573;
Sun X.H., Reich W. 2009, A\&A, 507, 1087 
}}\\

\subsubsection{Pulsars and the Galactic magnetic field {\scriptsize [A. Noutsos]}}

\noi One of the important but missing components of Galactic
astrophysical research is a reliable description of the free-electron density of
the Milky Way. Next-generation, low-frequency arrays (e.g. LOFAR) and ultimately
the SKA will play a major role in the reconstruction of the warm ionised medium
(WIM) using an array of observable effects like Faraday rotation, interstellar
dispersion and scattering of pulsar radio emission, scintillation, angular
broadening and extreme scattering events, neutral and ionised hydrogen emission,
and latest astrometric results and information on the Galactic spiral arm
structure. Using a combination of proven and new algorithms to analyse the
largest possible set of input data will enable researchers to construct the most
complete picture of the Galactic ionised ISM, yet. The derived model will shed
light on poorly understood aspects of the turbulent ISM, but will also find
applications in improving our knowledge of the Galactic magnetic field, distance
estimates for pulsars and transients and guide the exploration of current and
future telescopes. One of the advantages of this multi-disciplinary project, is
the demand for a close and frequent interaction between different research
groups, which would therefore allow for important cross-fertilisation between
groups interested in the interstellar medium and its physics.\\

\noi {\bfseries\boldmath\small Introduction:~~} Understanding the most basic properties of galaxies requires
understanding the interstellar medium (ISM) that fills the seemingly empty space
between the stars. It means to understand the composition, the physical
processes and the interplay between the different constituents, the radiation
fields and the galactic magnetic field. In particular, the exchange of energy
via turbulent processes between different length scales plays an important role
in determining the properties of the ISM that are linked through the cycle of
matter in the galaxy, from the formation to the death of stars. Many of these
properties and processes are still not fully understood or known, even though
the highly dynamical ISM is crucial for the interpretation of many astrophysical
phenomena such as star formation in general, the energy-mass feedback from
evolved stars or the threading of the magnetic field with interstellar matter.

\medskip

\noi The understanding of the ISM in the Milky Way is of primary importance for
studying the properties, formation and evolution of external galaxies. However,
the present knowledge of the Galactic ISM is still limited and constrains, for
example, our ability to obtain precise distance estimates based on interstellar
dispersion, or to accurately correct effects of interstellar weather in pulsar
timing data, which in turn hampers efforts to detect gravitational waves. A
particularly interesting constituent of the Galactic ISM that manifests itself
in many astrophysical observations is its free electron content. We need to
understand it for reasons both related to comprehending the physical processes
ongoing in the ISM, as well as for its use in astrophysical applications,
including:

\begin{itemize}

  \item[--]to decipher the fundamental turbulence processes and
  the distribution of energy from the largest to the smallest scales, (see, e.g. You et al. 2007).
  
  \item[--]to characterise the ``interstellar weather'' to take it into account
  during many experiments, (e.g. You et al. 2007)
  
  \item[--]to reveal the structure of the Milky Way including its small- and
  large-scale magnetic field (as done by Noutsos et al. 2008, amongst others),
  
  \item[--]to use it for precise distance measurements via an interaction of the ISM
  with electromagnetic radiation (as used, amongst others, in Verbiest et al.
  2010) and,
  
  \item[--]to help compare and constrain ``Dark Matter'' models of the Milky Way disc (as
  was done, amongst others, by Kalberla 2003, Kalberla et al. 2007).

\end{itemize}

\noi Free electrons are a component of the diffuse warm ISM. Separated
from their atoms, the electrons can interact with electromagnetic
radiation via a large number of observable effects, including Faraday
rotation, dispersion and scattering of pulsed radio emission,
scintillation, angular broadening and extreme scattering events
(excellent reviews can be found in Rickett 1977, Rickett 1990, Backer
et al. 1998). The average number density lies at only 0.03 cm$^3$, but
the distribution is far from homogeneous (see e.g. Walker et
al. 2008), with clumps and both local maxima (e.g. HII regions) and
minima. Given their many astrophysical manifestations, it is of utmost
importance to understand the distribution of free electrons on small
and large scales. Still, the available models for the free electron
density distribution (in particular Cordes \& Lazio 2003) are
incomplete and insufficient.

\noi The desired outcome is the derivation of an improved electron density model of
our Galaxy, taking into account a wide range of (old and new) observations and
detailed investigations of the turbulent processes that lead to the
inhomogeneities in the distribution, the measurable phenomena related to those
and the Galactic distribution of ISM constituents other than free electrons.

\noi A full Galactic ISM free electron model will not only shed light on little
understood aspects of the ISM, but will also aid in solving a large number of
astrophysical questions. It will find applications in improving our knowledge of
the Galactic magnetic field, it will help to understand the local ISM, it will
allow precise distance estimates for pulsars and transients and it will reveal
the location of astrophysically interesting objects like the location of unknown
HII regions and star forming regions, which are potential hiding places of
pulsar-black hole binaries. A reliable ISM model can be used to derive the most
promising observing strategies for delivering the Key Science proposed for the
future Square Kilometre Array (SKA).

\noi The modelling of the free-electron ISM requires input from a diverse range of
observables. On the front of Galactic magnetism, one can use pulsar-polarisation
data to perform fits to multi-parametric models and derive the optimal set of
parameters for various models of the Galactic magnetic field (Noutsos et al.
2008). Another aspect is pulsar timing, which can be used to measure parallax
distances. Measuring pulsar distances independently of an ISM model will be
crucial as to actually measure the column density along the line of sight
(Verbiest et al. 2008, Verbiest et al. 2010).

In addition, measurements of the neutral and ionised hydrogen as well
as hydroxyl (OH) maser measurements will be important
contributors. With a combination of \hi\ and OH ($\lambda$\,=\,21\,cm and
$\lambda$\,=\,18\,cm) absorption measurements, one can obtain the neutral
(\hi\ and H2) column densities. Together with dispersion measures of
pulsars, one can derive the fractional ionisation of the intervening
ISM, which is not only important for the free-electron model but also
an astrochemically interesting quantity.

Moreover, data from other disciplines like stellar dynamics, based on maser
velocities and an assumed Galactic gravitational potential, have revealed
bar-like features and voids in the ISM of the inner Galaxy. Such features are
important and must be included in the next ISM model of the Galaxy.

\noi Invaluable contributions to independent distance measurements can also come from
trigonometric parallaxes of methanol masers: parallax distances up to 10\,kpc
with an accuracy of $\sim$\,10\,\% have been measured (Rygl et al. 2010). Such
measurements can be expanded to pulsar distances, where VLBI can yield much more
accurate distances that will be an input to the model.

Many interstellar effects are stronger at low frequencies. The
next-generation low-frequency telescopes such as LOFAR are expected to play an
important role in ISM modelling efforts.

The revised distances and velocities from a new free-electron density model will
have direct implications on the kick-velocity distribution of core collapse
supernovae. Moreover, using the model predictions will identify the most
probable galactic regions to find previously hidden sources of particular
astrophysical interest, such as pulsar-black hole binaries.\\

\parbox{0.9\textwidth}{
\noi{References:}\\
\noi{\scriptsize 

Backer D.C., \etal , 1998, AAS, 30, 1151; 
Cordes J.M., Lazio T.J.W., 2002, arXiv:astro-ph/7156;
Kalberla P.M.W., 2003, ApJ, 588, 805;
Kalberla P.M.W., 2007, A\&A, 469, 511;
Noutsos A., Johnston S., Kramer M., Karastergiou A., 2008, MNRAS, 386, 1881;
Rickett B.J., ARA\&A, 1977, 15, 479;
Rickett B.J., ARA\&A, 1990, 28, 561;
Rygl K.L.J., \etal , 2010, A\&A, 511, 2; 
Verbiest J.P.W., \etal , 2008, ApJ, 679, 675;
Verbiest J.P.W. , Lorimer D.R., McLaughlin M.A., 2010, MNRAS, 405, 564;
Walker M.A., et al., 2008, MNRAS, 388, 1214;
You  X.P., et al., 2007,  MNRAS, 378, 493}}\\

\subsubsection{Linking theoretical aspects of the interstellar and
  intergalactic medium to radio \mbox{observations} {\scriptsize [D.~Breitschwerdt, M.A. de Avillez]}}

{\bfseries\boldmath\small High resolution simulations:~~} The Interstellar Medium (ISM) of
star forming galaxies is a highly turbulent and compressible medium
(v. Weizs\"acker 1951). This leads morphologically to a coupling of
structures on all scales and to a filamentary (see left figure below)
gas distribution (e.g. Avillez and Breitschwerdt 2004). 

\medskip

\noi It can be shown that the main driver of turbulence is supernova
(SN) explosions (MacLow and Klessen 2004), and to a lesser extent
stellar winds. Energy is fed in at an integral scale of about 75\,pc
(Avillez and Breitschwerdt 2007) and, in purely hydrodynamical
turbulence, cascading down through an inertial range until dissipation
takes place on the viscous scale. Heating and cooling give rise to a
multiphase medium, as thermal instability controls the evolution of
the gas. However, contrary to classical textbook models (e.g. McKee
and Ostriker 1977), a significant amount of mass resides in thermally
unstable regions, again a consequence of turbulence. The complexity of
the thermal structure of the ISM is increased due to the difference in
atomic time scales ionization and recombination. As a consequence the
gas is in general not in ionization equilibrium (NEI), leading e.g. to
an overionized appearance of gas because of delayed
recombination. This also modifies the cooling function, which now
varies in space and time. This important aspect has been neglected in
large-scale ISM simulations until now (Avillez et al. 2011). In
addition, the electron density distribution (see right figure above)
in the ISM is different for an NEI plasma. Radio observations at high
spatial resolution together with pulsar dispersion measures can give
important constraints. Preliminary calculations point to a bimodal
overall probability density function (pdf), with individual
temperature range pdfs, such as the warm neutral medium (WNM) being
lognormally distributed, in agreement with observations (Berkhuijsen
and Fletcher 2008).

\bigskip
\bigskip

\parbox{\textwidth}{

\hspace{-0.2cm}
\parbox{5cm}{ \includegraphics[scale=0.24]{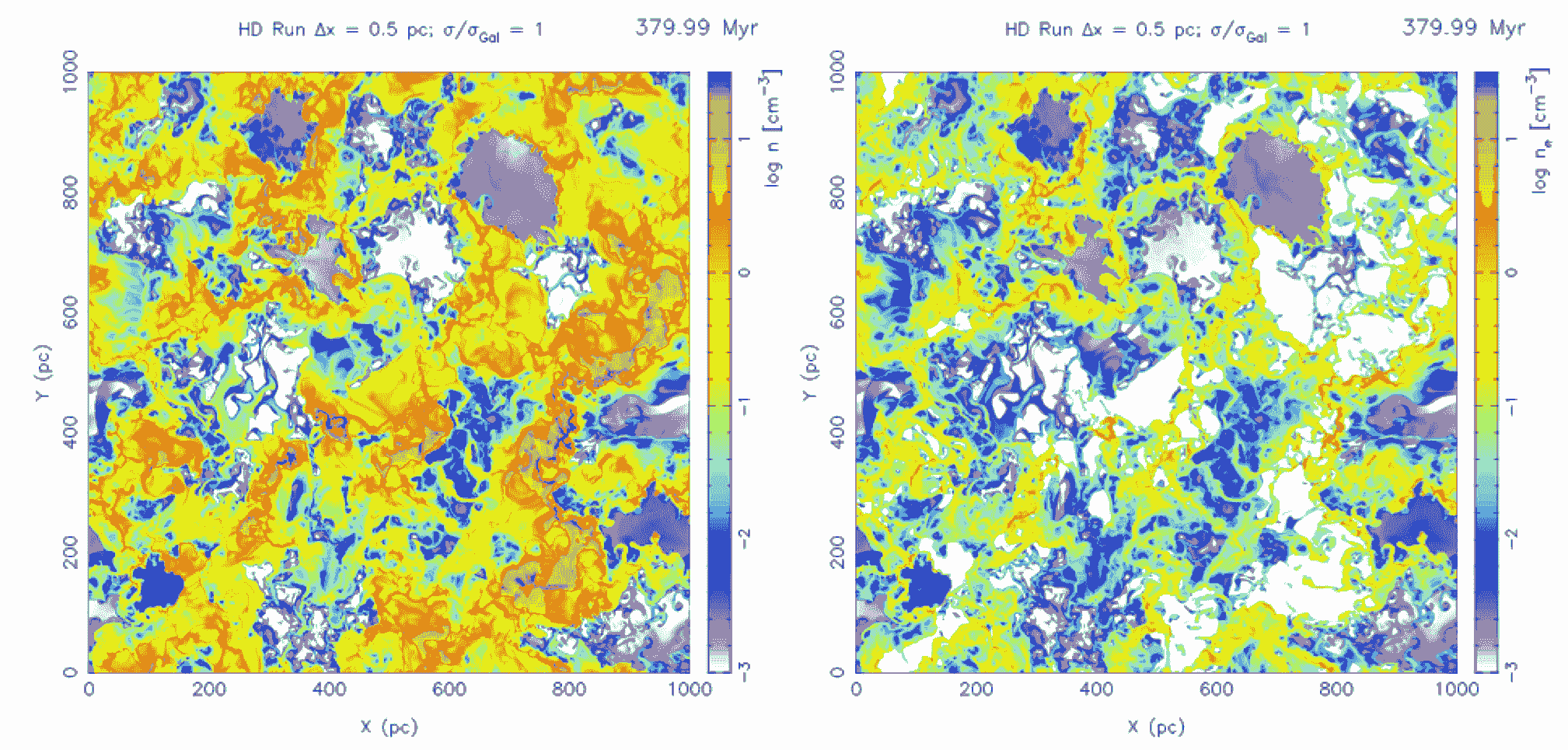}\vspace{0.5cm}}

\hspace{0.2cm}
\parbox{0.9\textwidth}{Figure 1$_{\rm Br}$: High resolution simulations of an ISM plasma in
  non-equilibrium ionization (NEI), showing slices through the
  Galactic mid-plane with a highest resolution of 0.5 pc (left panel: total number density, right panel: electron density) after 380 Myr of evolution (Avillez and Breitschwerdt 2011).}
\vspace{0.2cm}
}

\bigskip
\bigskip

\noi {\bfseries\boldmath\small Radiocontinuum emission from galactic halos:~~} Combined SN activity in star clusters can generate an outflow
perpendicular to the galactic disk, which is driven by hot plasma and
cosmic rays (CRs; see Breitschwerdt et al. 1991). High resolution
simulations show that the flow is ram pressure dominated, and the
magnetic field lines are drawn out into the halo (see Avillez and
Breitschwerdt 2005). High resolution and high sensitivity radio
observation will be able to reveal the field topology in and off the
disk. CR nucleons escape the galaxy eventually together with
electrons, which suffer strong synchrotron and inverse Compton losses,
thus making them an ideal tracer of the halo profile. CR transport
calculations show that a wind advects electrons further out into the
halo, resulting in a shallower radio spectral index. Thus SKA
observations will be able to unveil the dynamical structure of
galactic halos. If the wind interacts with the intergalactic gas,
resulting shocks accompanied by particle acceleration may be observed
as well.\\

\parbox{0.9\textwidth}{
\noi{References:}\\
\noi{\scriptsize  
Avillez M.A., Breitschwerdt D., Manoel N.C. 2011, A\&A (submitted);
Avillez M.A., Breitschwerdt D., 2007, ApJ 665, L35;
Avillez M.A., Breitschwerdt D., 2005, A\&A 436, 585;
Avillez M.A., Breitschwerdt D., 2004, A\&A 425, 899;
Berkhuijsen E.M., Fletcher A., 2008, MNRAS 390, 19;
Breitschwerdt D., McKenzie J.F., V\"olk H.J., 1991, A\&A 245, 79;
MacLow M.M., Klessen R.S, 2004, RvMP 76, 125  ;
McKee C. F., Ostriker J. P., 1977, ApJ 218, 148;
von Weizs\"acker C. F., 1951, ApJ 114, 165
}}\\

\subsubsection{Magnetic field structures and statistics: Unraveling the inner
  workings of magnetic \mbox{dynamos} {\scriptsize [T.~En{\ss}lin]}}

{\bfseries\boldmath\small Abstract:} The three-dimensional structures of cosmic magnetic
fields in our Galaxy, external galaxies, galaxy clusters and the
inter-galactic medium were shaped by dynamo processes, turbulent gas
motions and shock waves. The fields are remnants of violent
hydrodynamical processes, and need to be studied in great detail in
order to unravel the field's origin. The SKA, with its unprecedented
combination of sensitivity, resolution, and frequency coverage, will
play a leading role in charting magnetic structures via polarimetric
measurements. Detailed two- and three-dimensional information on field
configurations in various galactic and intergalactic environments
resulting from synchrotron continuum measurements, Faraday rotation
grids, and Faraday rotation synthesis will become available. A number
of statistical analysis methods tailored to be applied to such data
are already developed, largely by the German community. These methods
include the measurement of magnetic power spectra as a signature of
turbulent cascades, the statistics of magnetic tension forces as a
window into the inner workings of magnetic dynamos, the testing for
magnetic helicity as a key ingredient in galactic large-scale field
generation, and the reconstruction of the three-dimensional magnetic
field of the Milky Way as a key input for astronomy with ultra-high
energy cosmic rays.

\medskip

\noi {\bfseries\boldmath\small Introduction:} The observed cosmic magnetic fields in
galaxies, clusters of galaxies and the yet to be detected fields in
the wider intergalactic medium are believed to be produced by magnetic
dynamos. Two broader classes of dynamos have to be distinguished, the
small-scale and the large-scale dynamo. Direct proof of their
influence, and observations of their detailed inner working, are
unfortunately still sparse. The SKA is expected to change this
situation dramatically.

\medskip

\noi {\bfseries\boldmath\small The small-scale dynamo} amplifies magnetic fields by random field
line stretching, and thereby imprints properties of the driving
turbulence into the magnetic field statistics.  In particular, there
should be similarities between the magnetic power spectrum and the
hydrodynamical power spectrum driving the dynamo.  This seems already
to been observed in the Kolmogorov-like magnetic spectrum as inferred
for the cool core region of the Hydra-A galaxy cluster (En{\ss}lin \&
Vogt 2003, Vogt \&En{\ss}lin 2005, Kuchar \& En{\ss}lin 2011). There, the magnetic power spectrum was
measured with novel inference methods exploiting the correlation
structure in the extended Faraday rotation maps of the Hydra A radio
galaxy. The probed fields reside in front of the radio emitting plasma
and therefore are typical for the cool core region of this cluster,
but not representative for its wider intra-cluster medium. Other
Faraday rotation based magnetic spectrum estimates in less central
regions of galaxy clusters are also consistent with Kolmgorov-like
magnetic spectra (Bonafede \etal\ 2010). However, due to a less
uniform probing of the cluster volume in these cases, the picture is
far from being conclusive (Govoni \etal\ 2006).

\noi The SKA will provide dense grids of Faraday rotation measurements
through many clusters of galaxies. This will allow for the inference
of the magnetic field power spectra, not only in a large number of
clusters, but also as a function of cluster position (core region,
outer region, cool cores) and dynamical state (relaxed cluster,
merging cluster).  Studies of the magnetic part of cluster weather
phenomena will provide unique insight into the interior of the largest
virialized objects in the Universe.

\medskip

\noi {\bfseries\boldmath\small The large-scale dynamo:~~}The more ordered large-scale
galactic fields are believed to result from some sort of mean field
dynamo, which relies on the differential rotation in spiral galaxies.
Although various scenarios for these dynamos have been proposed, a
common key ingredient of all theories seems to be magnetic helicity
(Brandenburg 2009, Shukurov \etal\ 2006, Sokoloff 2007).  The build-up
of large-scale helicity must be accompanied by the dissipation or
expulsion of small-scale helicity, due to helicity conservation in
magnetohydrodynamics.  Any way to study magnetic helicity on large or
small galactic scales would therefore be extremely important to
unravel the inner workings of the Galactic dynamo.

\noi Since magnetic helicity manifests itself in all three components of
the magnetic field instantaneously, those have to be probed
simultaneously. Suitably constructed combinations of Faraday rotation
and synchrotron polarisation data probe parallel and perpendicular
field components in such a way as to be sensitive to magnetic helicity
(Kahmiashvili \& Ratra 2005, Junklewitz \& En{\ss}lin 2011).  A straight-forward
application of such a helicity sensitive statistic to galactic Faraday
rotation data (Taylor \etal\ 2009) and synchrotron polarimetric
data (Page \etal\ 2007) has not yet revealed the presence of
magnetic helicity (Oppermann \etal\ 2012).  However, this was
expected since the highly structured galactic thermal electron
distribution severely affects these measurements and requires
correction.\\

\bigskip
\bigskip
\bigskip

\parbox{\textwidth}{
\parbox{\textwidth}{

\hspace{1cm}
\parbox{5cm}{\includegraphics[bb=0 0 1204 888, scale=0.1]{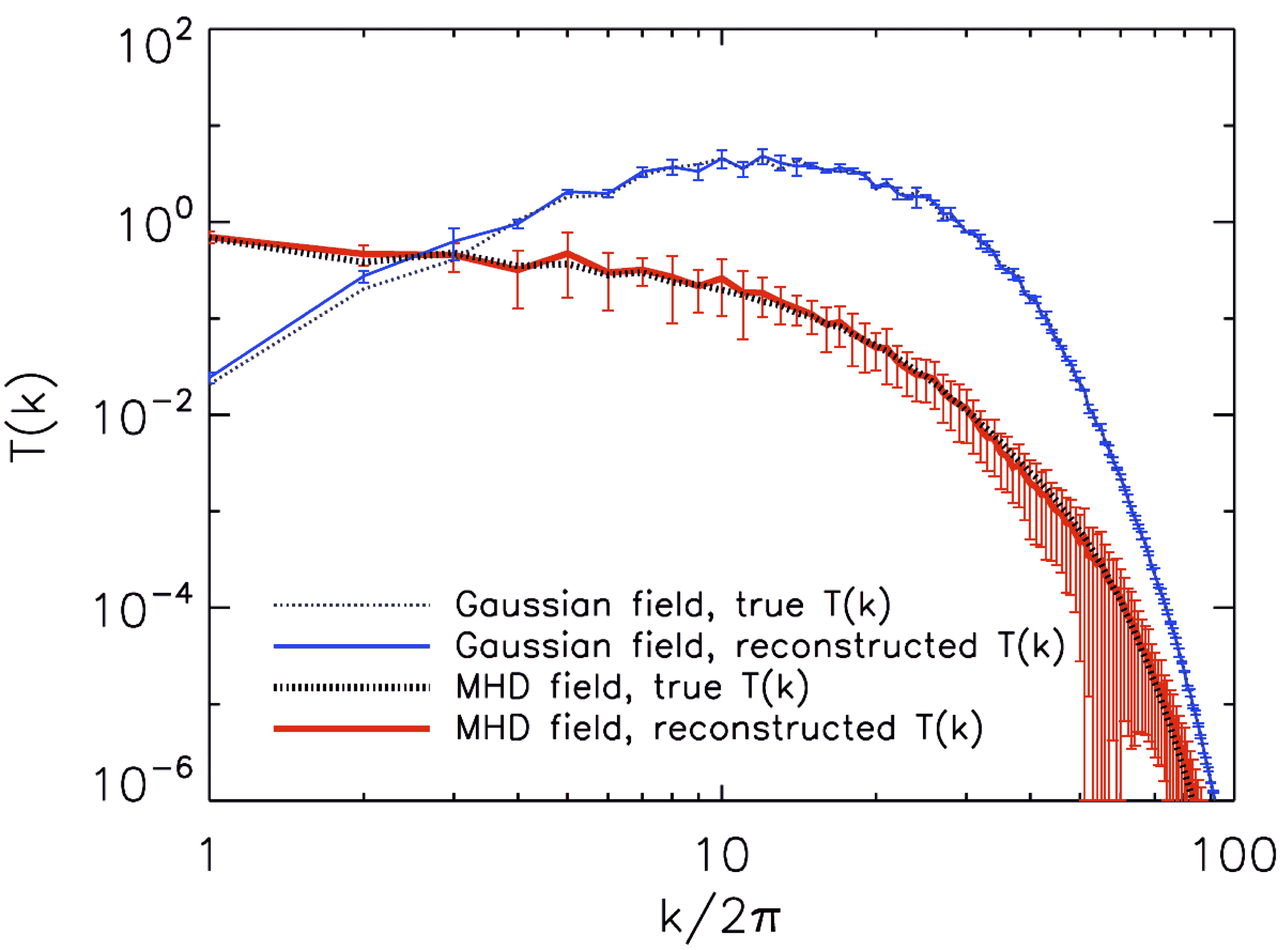}\vspace{11.5cm}}
\hspace{-4.5cm}
\parbox{5cm}{\includegraphics[bb=0 0 862 452, scale=0.125]{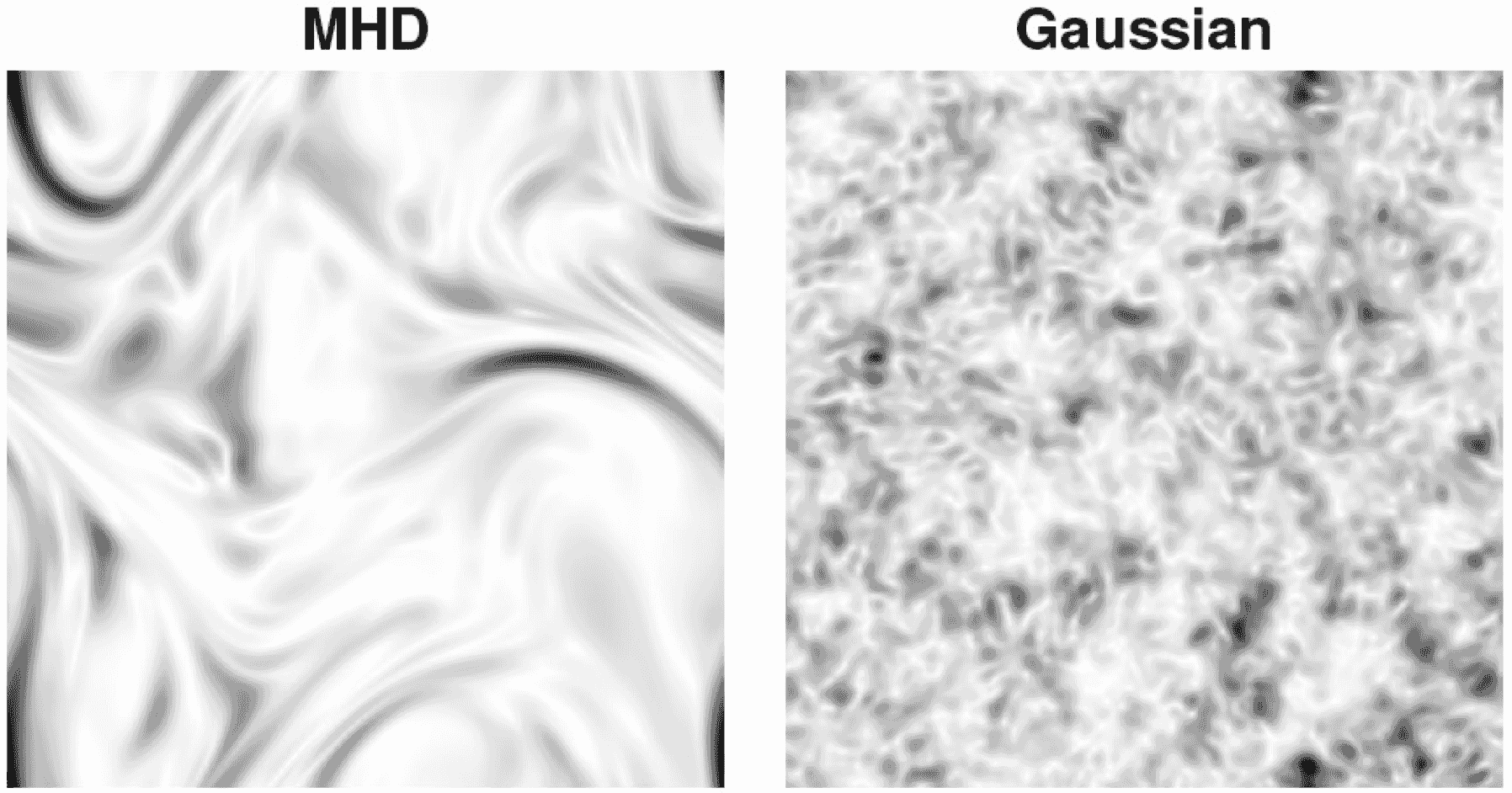}}
\vspace{-21.8cm}
}
\parbox{\textwidth}{

\hspace{10cm}
\parbox{5cm}{Figure 1$_{\rm En}$: Top: tension force spectrum reconstructed from mock
  polarimetry data using the method of Stokes correlators (from
  Waelkens \etal\ 2009b). The Gaussian random field was constructed to
  exhibit the same magnetic energy spectrum as the
  magneto-hydrodynamical simulation, but it has a different
  fourth-order statistic as measured by the Stokes
  correlators. \\

Bottom: slices through the tension force fields in the
  two cases, showing that the Gaussian field has more small-scale
  tension forces than the more realistic MHD simulation, although both
  have the same magnetic energy spectra.}
}
\vspace{1cm}}

\bigskip
\bigskip

\noi The SKA will significantly increase the sensitivity of such tests for
helicity.  It will provide a much denser grid of Faraday rotation
measurements, high resolution polarimetric observations, and, most
importantly, exquisite data on the galactic thermal electron structure
from pulsar observations. The latter will allow for the construction
of an accurate, three-dimensional model of the thermal electron
distribution that can be used to correct for the impact of the
electron structure in tests for magnetic helicity.\\

\noi {\bfseries\boldmath\small The tension force spectrum} of turbulent magnetic fields highlights the back-reaction of the magnetic fields on the hydrodynamics. Tension forces should be especially strong in the saturated states of magnetic dynamos, since their action on the gas flows should compensate for the otherwise exponential increase of magnetic energy. 
Tension force spectra should therefore be very sensitive to the specific dynamo scenario in operation. 
Tension force spectra are encoded in high resolution radio polarimetry
data (Waelkens \etal\ 2009b) as well as
ordinary power spectra (Spangler 1983, Spangler 1982, Junklewitz \&
En{\ss}lin 2011, Eilek 1989, Eilek 1989b).  The SKA will be a powerful
tool to  discriminate between different magneto-hydrodynamical
scenarios (see Figure\ 1$_{\rm En}$). \\

\medskip

\noi {\bfseries\boldmath\small 3-d Galaxy:~~}The three-dimensional magnetic field structure of our galaxy can be
reconstructed tomographically using the combination of several data
sets. The following data sets will substantially improve with SKA
measurements:
 \begin{itemize}
\item[--]Faraday rotation probes by extragalactic sources provide the
  line-of-sight (LOS) integrated LOS magnetic field component
  (weighted with the thermal electron density), see Figure\ 2$_{\rm En}$.
\item[--]Faraday rotation probes by galactic pulsar emission provide similar information, but for shorter path-lengths through our Galaxy.  This allows for the discrimination between magnetic structures at different physical depths.
\item[--]Pulsar measurements also provide dispersion measures, i.e. LOS integrated thermal electron densities.  This is important auxiliary information if one hopes to identify the specific contribution of magnetic fields to Faraday rotation measurements. 
\item[--]Galactic radio synchrotron emission in total intensity and
  polarisation (at high, Faraday-rotation-free frequencies) shows the
  LOS projected perpendicular magnetic field components (see Figure\ 3$_{\rm En}$). 
\item[--]Polarised galactic synchrotron emission at low frequencies,
  which is affected by intrinsic galactic Faraday rotation (see
  Figure\ 3$_{\rm En}$),
 carries depth information about the
  different emission regions. Using the novel Faraday synthesis method
  to reconstruct the emission as a function of Faraday depth results
  in 3-d images of polarised intensity (Brentjens \& de Gruyn 2006, Brentjens \& de Gruyn 2005).
\item[--]The most prominent features expected in such 3-d Faraday
  synthesis images are Faraday caustics (Bell \etal\ 2011). These sheets of singularities in the polarised emissivity are caused by physical regions in which the LOS component of the magnetic field is close to  zero, i.e. due to a field reversal. Since all polarised radiation emitted in this region has basically the same Faraday depth, the Faraday caustics appear as boundaries of galactic regions with opposite polarity (with respect to the LOS).
\end{itemize}

\bigskip

\vspace{4cm}
\parbox{\textwidth}{

\hspace{1cm}
\parbox{\textwidth}{
\parbox{5cm}{\includegraphics[bb=0 0 1004 620, scale=0.11]{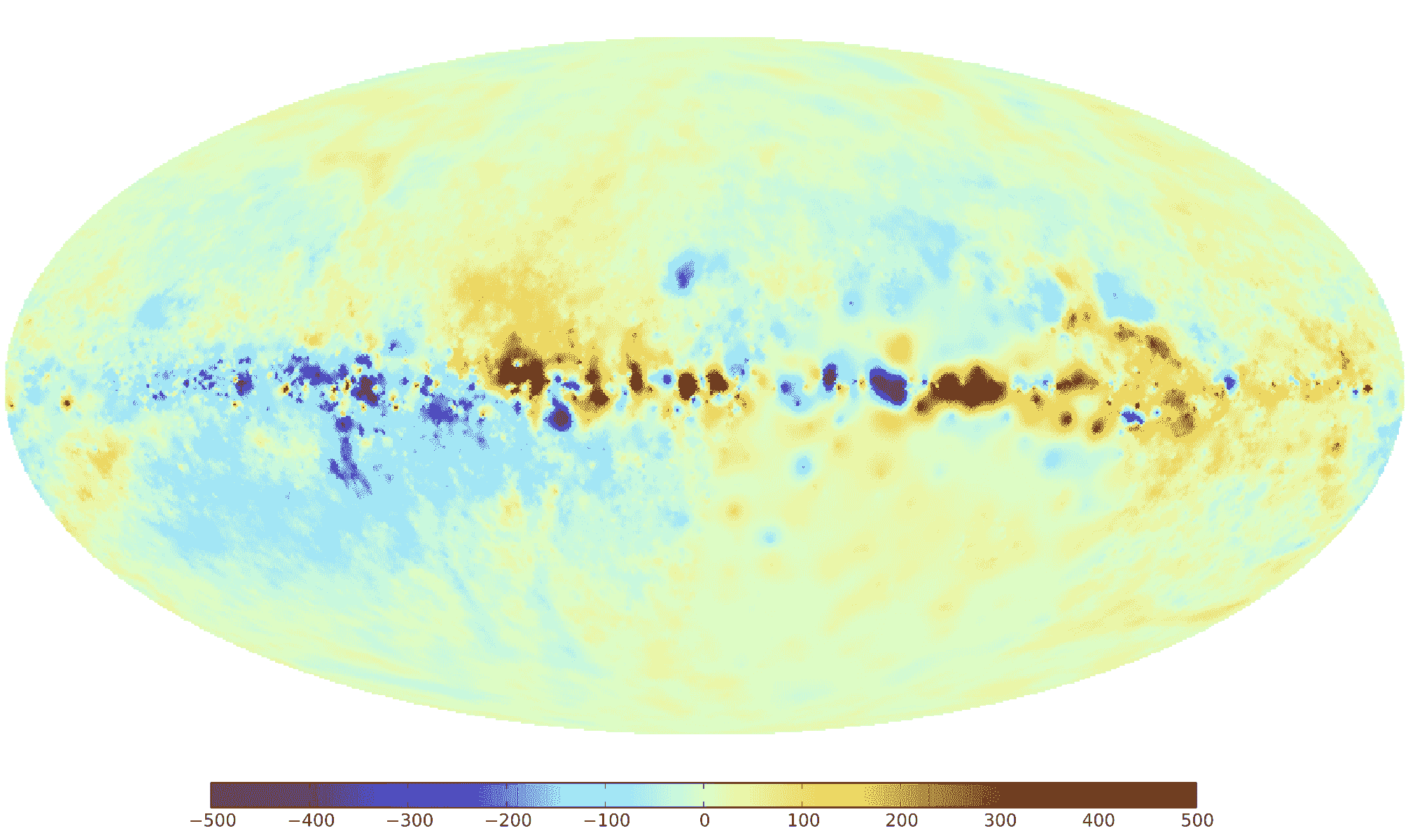}\vspace{12.5cm}}
\hspace{-5.25cm}
\parbox{5cm}{\includegraphics[bb=0 0 1004 620, scale=0.11]{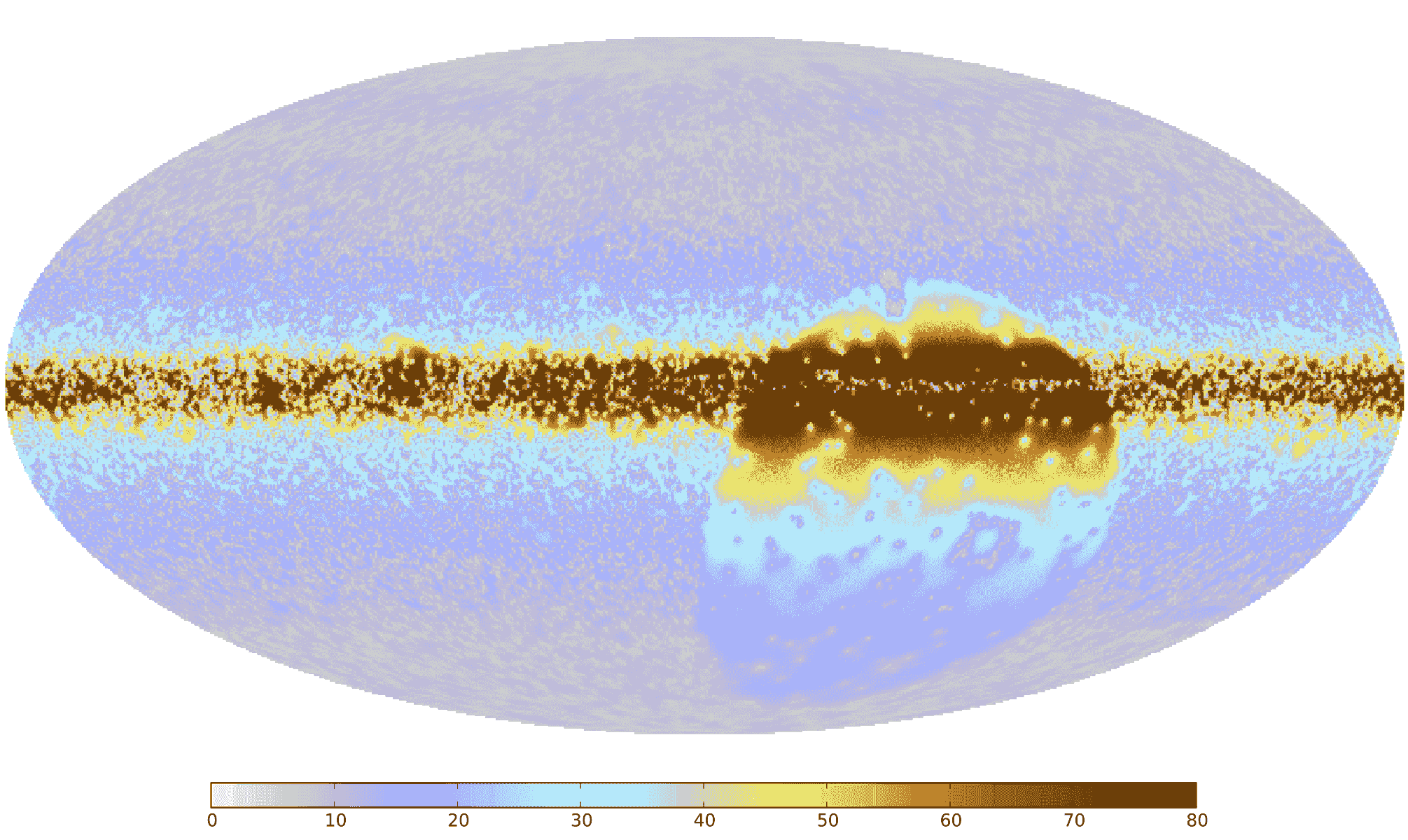}}
\vspace{-13cm}
}
\parbox{\textwidth}{

\hspace{10.5cm}
\parbox{6cm}{Figure 2$_{\rm En}$: Faraday sky (top) and its uncertainty (bottom) from extragalactic Faraday rotation measurements (Oppermann \etal\ 2012,
Taylor \etal\ 2009).
}
}
\vspace{3cm}
}

\parbox{\textwidth}{
\vspace{3cm}

\hspace{1cm}
\parbox{\textwidth}{
\parbox{5cm}{\includegraphics[bb=0 0 1004 620, scale=0.12]{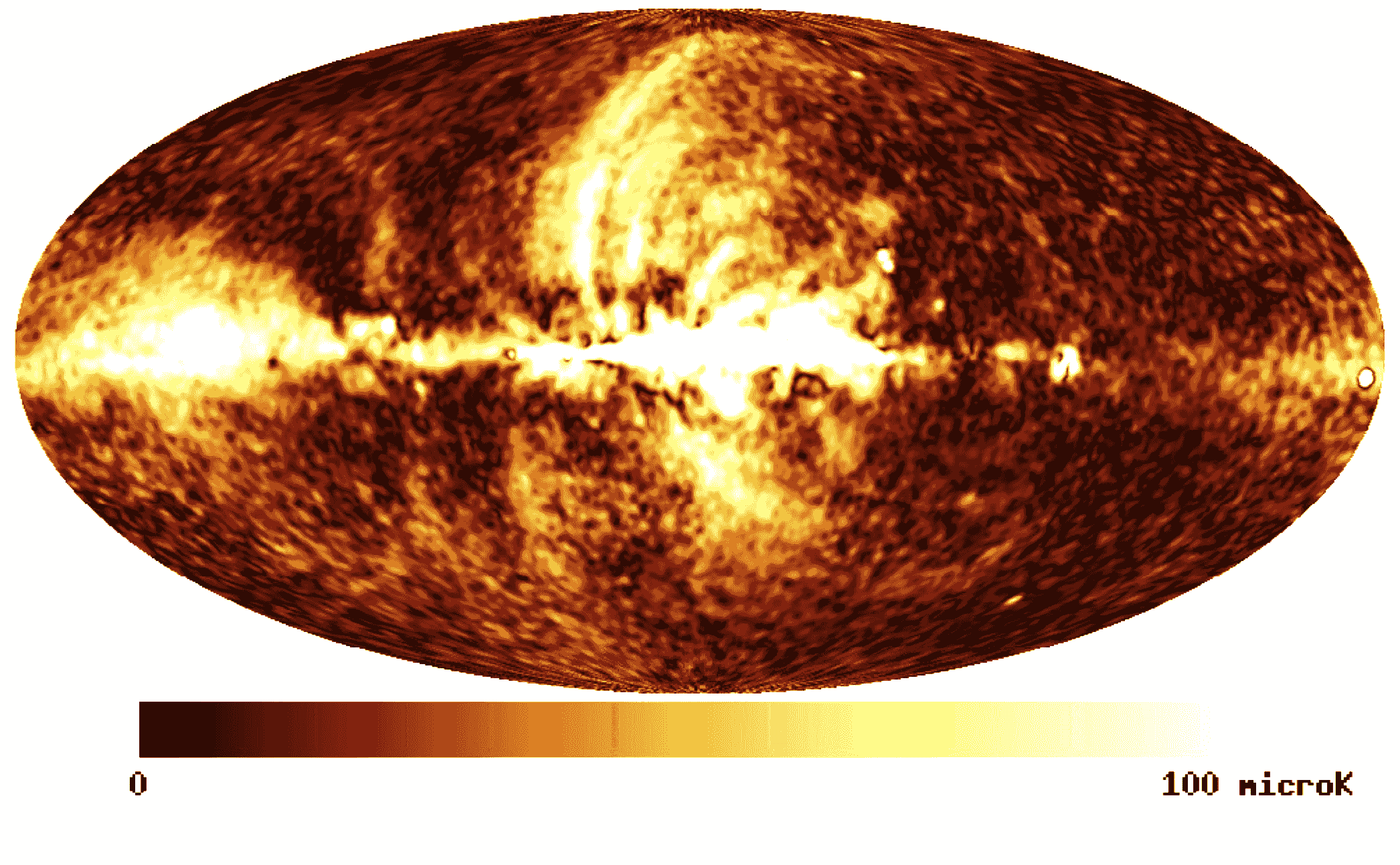}\vspace{9cm}}
\hspace{-5.15cm}
\parbox{5cm}{\includegraphics[bb=0 0 589 294, scale=0.20]{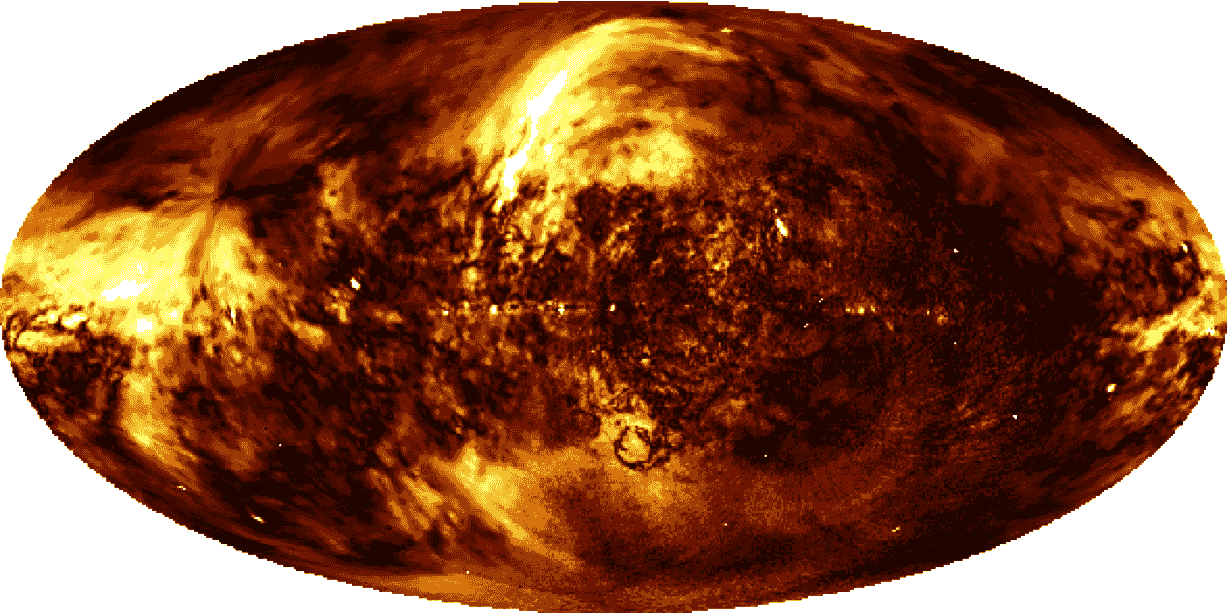}}
\vspace{-13cm}
}
\parbox{\textwidth}{

\hspace{10.5cm}
\parbox{6cm}{Figure 3$_{\rm En}$: The WMAP 22.8\,GHz all-sky polarised intensity map (upper
panel) and the 1.4\,GHz all-sky polarised intensity map (lower panel).
The polarised intensities are shown greyscale coded from 0 to 100\,$\mu$K
for 22.8\,GHz and from 0 to 570\,mK for 1.4\,GHz. Galactic Faraday-depolarisation structures 
are visible in the lower frequency map. Data from (Wolleben \etal\
2006, Page \etal\ 2007, Testori \etal\ 2008) and figures from (Sun
\etal\ 2008).
}
}
\vspace{3cm}
}

\noi There has been substantial progress in combining such data sets into
coherent pictures of the 3-d galactic magnetic field structure,
although a unique model has not yet been reached (Sun \etal\ 2008,
Waelkens \etal\ 2009, Jaffe \etal\ 2010, Jiang \etal\ 2010, Pshirkov
\etal\ 2011, Sun \& Reich 2010 ).
However, with the superb data expected from the SKA, a detailed
mapping of our magnetic galaxy will become feasible.

\noi This will not only be important for understanding the large-scale
dynamo operating in our galaxy, or to reveal the role magnetic fields
play in the galactic metabolism of hot gas, cold clouds and star
formation, but it will also be essential for the correction of the
galactic magnetic screen that inhibits extragalactic astronomy with
high energy charged particles (Sigl \& Lemoine 1998, Waelkens \etal\
2009, Giacinti \& Semikoz 2011b, Giacinti \& Semikoz 2011).\\

\parbox{0.9\textwidth}{
\noi{References:}\\
\noi{\scriptsize  
Bell M.R., Junklewitz H.,En{\ss}lin T.A.., 2011, arXiv:1105.2693;
Bonafede A., \etal , 2010, A\&A, 513, 30;
Brandenburg A., 2009, Plasma Physics and Controlled Fusion, 51(12):124043;
Brentjens M.A.,  de Bruyn A.G., 2005, A\&A, 441, 1217;
Brentjens M.A.,  de Bruyn A.G., 2006, Astronomische Nachrichten, 327, 545;
Eilek J.A., 1989, AJ, 98, 244;
Eilek J.A., 1989b, AJ, 98, 256;
En{\ss}lin T.A.,  Vogt C., 2003, A\&A, 401, 835;
Giacinti G., Semikoz D.V., 2011, Phys. Rev. D, 83(8), 083002;
Giacinti G., Semikoz D.V., 2011b, arXiv:1107.2356;
Govoni F., \etal , 2006, A\&A, 460, 425;
Jaffe T.R., \etal , 2010, MNRAS, 401, 1013;
Jiang Y.-Y., \etal , 2010, ApJ, 719, 459;
Junklewitz H., En{\ss}lin T.A., 2011, A\&A, 530, 88;
Kahniashvili T., Ratra B., 2005, Phys. Rev. D, 71(10), 103006;
Kuchar P., En{\ss}lin T.A., 2011, A\&A, 529, 13;
Oppermann N., \etal , 2012, A\&A, 542, A93;
Page L., \etal , 2007, ApJS, 170, 335;
Pshirkov M.S., Tinyakov  P.G., Kronberg P.P., Newton-McGee K.J.,
2011, arXiv:1103.0814;
Shukurov A., Sokoloff D., Subramanian K., Brandenburg A., 2006, A\&A,
448, 33;
Sigl G, Lemoine M., 1998, Astroparticle Physics, 9, 65;
Sokoloff D., 2007, Plasma Physics and Controlled Fusion, 49, 447;
Spangler S.R., 1982, ApJ, 261, 310; Spangler S.R., 1983, ApJL, 271, 49;
Sun X.-H., Reich W., 2010, Research in Astronomy and Astrophysics, 10,
1287;
Sun X.-H., Reich W., Waelkens A.H., En{\ss}lin T.A., 2008, A\&A, 477, 573;
Taylor A.R., Stil J.M., Sunstrum C., 2009, ApJ, 702, 1230;
Testori J.C., Reich P., Reich W., 2008, A\&A, 484, 733;
Vogt C.,  En{\ss}lin T.A., 2005, A\&A, 434, 67;
Waelkens A.H.,  \etal , 2009, A\&A, 495, 697;
Waelkens A.H.,, Schekochihin A.A., En{\ss}lin T.A.,2009b, MNRAS, 398, 1970;
Wolleben M., Landecker T.L., Reich W., Wielebinski R., 2006, A\&A,
448, 411
}}\\

\subsubsection{Accurate distance measurements: Astrometry
  {\scriptsize [A. Brunthaler]}}

The SKA will have even better astrometric capabilities than current Very Long 
Baseline Interferometry (VLBI) arrays (see Fomalont \& Reid 2004). 
Since systematic errors from the atmosphere or the array geometry (e.g. antenna 
positions, earth orientation parameters) scale directly with the angular separation 
of the reference and target sources on the sky, it is essential to use as close as 
possible reference sources. The superior sensitivity will allow the use of much 
weaker (and therefore much closer) background reference sources.\\

\noi {\bfseries\boldmath\small Trigonometric Parallaxes:~~} Many astrophysical properties of objects in star forming regions, like
the size, luminosity, and mass, depend strongly on the distance. Therefore,
a true understanding of star formation and a detailed comparision of
theoretical models and observations is only possible if accurate distances
are known to star forming regions. Current galactic distance estimates are
often affected by dust obscuration (photometric distances) or highly model
dependend (kinematic distances). A completely unbiased method to estimates
distances is the trigonometric parallax, which requires extremely precise
astrometric measurements in the range of a few micro-arcseconds ($\mu$as).
While GAIA will measure parallaxes of a billion stars in the Galaxy,
it will not probe parts of the Galaxy that are obscured by dust, i.e. large
parts of the plane of the Milky Way beyond 1 or 2\,kpc inward from the Sun
and in particular not the deeply obscured regions where new stars are forming.
However, radio observations at cm-wavelengths are not hindered by dust and can
provide a view of the Milky Way that is complementary to GAIA.

Currently, astrometric radio observations of methanol and water masers in star
forming regions with VLBI surveys like the ``Bar and 
Spiral Structure Legacy'' (BeSSeL) survey (e.g. Brunthaler et al. 2011)
can reach parallax accuracies of up to 6\,$\mu$as (Reid et al. 2009a,
Hachisuka et al. 2009). Therefore, these observations have
the potential to measure accurate distances of most Galactic star forming
regions, to map the spiral structure of a large part of the Milky Way,
and to determine important parameters such as the rotation velocity and the
distance to the Galactic centre with high accuracy (Reid et al. 2009b).

\noi The SKA in Phase\,1 can observe the important 6.7\,GHz methanol maser
line. This line is a tracer of high mass star formation, widespread in
the Galaxy, and has been already used for astrometric observations
(Rygl et al. 2010). In Phase\,3 (for which up to this point is no
defined schedule), the SKA will also cover the 22\,GHz
water maser line, the strongest maser line in star forming
regions. Furthermore, the large continuum sensitivity of the SKA
allows the observation of radio stars at much larger distances than
the few hundred parsec currently possible (Menten et al. 2007, Loinard
et al. 2007). With parallax accuracies
approaching $\sim$\,1\,$\mu$as, the SKA can measure distances to
sources in 10\,kpc with 1\,\% accuracy.

Since the Large Magellanic Cloud (LMC) is known to host many methanol and water 
masers, the SKA can even measure a 5\,\% accurate trigonometric parallax to this 
important first step on the extragalactic distance ladder. Since the LMC is 
receeding with $\sim$\,280\,km\,s$^{-1}$ from the Sun, the separation between masers 
on different sides of the galaxy will shorten at a rate of a few 10\,$\mu$as~yr$^{-1}$. 
Since this apparent motion will be easily detectable with the SKA, one can see
the LMC shinking as it recedes.\\

\noi {\bfseries\boldmath\small Galaxy Motions: ~~} Most galaxies in the Universe are not isolated objects, but are found in
groups or clusters. Thus, large-scale structure formation and galaxy formation
are closely connected topics. In the concordance $\Lambda$CDM cosmological
model, galaxies grow hierarchical in mass by accreting smaller galaxies. This
cosmic cannibalism can be witnessed even today: Small galaxies falling into
the gravitational potential of the dark matter halo of a more massive galaxy
experience strong tidal interactions that can lead to strong disturbances or
even tearing apart the smaller galaxy. Having a detailed description of this
process is therefore a key ingredient in understanding galaxy formation and
evolution.

\noi Nearby examples of galaxy interactions are found in the Local Group (LG)
and in nearby galaxy groups and clusters. However, a description of galaxy
interactions cannot be complete without knowing the full 3 dimensional space
motions of the interacting galaxies. Additionally, the flow of galaxy groups
and clusters is strongly connected to the distribution of matter in large-scale
structures. For example, it is widely believed that the motion of the Milky Way 
relative to the cosmic microwave background (CMB), which is of the order of 500\,\kms, 
is induced by mass concentrations within 150\,Mpc of the LG, but there is a discrepancy 
between the direction of the motion and the distribution of visible mass in the 
local Universe (see e.g. Loeb \& Narayan 2008, and references therein).

While current VLBI arrays are already able to measure proper motions of galaxies 
througout the Local Group (Brunthaler et al. 2005, Brunthaler 
et at. 2007), the SKA can go far beyond this and measure motions 
of galaxies with smaser emission or weak AGN in nearby galaxies groups and even 
nearby galaxy clusters (e.g. Virgo, Fornax). Therefore, the  SKA will be able to 
measure the large-scale 3-d velocity field of galaxies out to nearby clusters.\\

\parbox{0.9\textwidth}{
\noi{References:}\\
\noi{\scriptsize  
Fomalont E., Reid M., 2004, NewAR 48, 1473;
Brunthaler A., et al., 2011, AN, 332, 461;
Reid M., et al., 2009a, ApJ, 693, 397;
Hachisuka K., et al., 2009, ApJ, 696, 1981;
Reid M., et al. , 2009b, ApJ, 700, 137;
Rygl K.L.J., et al., 2010, A\&A, 511, A2;
Menten K., et al., 2007, A\&A, 474, 515;
Loinard L., et al., 2007, ApJ, 671, 546;
Loeb A., Narayan R., 2008, MNRAS, 386, 2221;
Brunthaler A., \etal , 2005, Science, 307, 1440;
Brunthaler A., \etal , 2007, A\&A, 462, 101
}}\\

\subsubsection{Prospects for accurate distance measurements of pulsars
  {\scriptsize [N.~Wex, M.~Kramer]}}
  \label{mknw}

\noi Parts of the following contribution are taken from Smits et
al. 2011. 

\smallskip

\noi The Square Kilometre Array (SKA) will be an ideal instrument for measuring the 
distances of pulsars, having a very high sensitivity as well as baselines 
extending up to several thousands of kilometres. In particular, 
the SKA will be able to perform parallax measurements on a great number of 
pulsars either by direct parallax measurements or, in case of millisecond 
pulsars, timing observations, allowing accurate determination of their distances 
from the Earth. 

\noi The determination of accurate pulsar distances is vital for various physical 
and astrophysical reasons. Just to give a few examples: 

i) In this way, the interstellar electron model can be calibrated and
will not only provide reliable estimates of the DM distances of
pulsars without a parallax measurement in return, but will
simultaneously provide a map of the free electron content in the Milky
Way that can be combined with \hi\ and HII measurements to unravel the
Galactic structure and the distribution of ionised material. Combined
with Faraday rotation measurements of the same pulsars, the Galactic
magnetic field can also be studied in much greater detail as today,
since the precise distance measurements potentially allow us to
pinpoint field-reversals occurring with some accuracy.

ii) The determination of accurate pulsar distances is also important for those 
pulsars that are part of binary systems, in particular for those with another 
compact object, such as the ``Double Pulsar'' (Burgay et al.\ 2003; Lyne et al.\ 
2004). Relativistic effects can be used to determine the distance to some of 
these systems when the validity of general relativity is assumed. In reverse, to 
perform precision tests of general relativity, kinematic effects have to be 
removed for which it is often required to know the distance precisely (Lorimer \& 
Kramer 2005). 

iii) The SKA has the potential of detecting the gravitational waves from 
individual super-massive binary black hole systems, and determining the location 
of the source in the sky with high precision. For this it is important to 
properly account for the ``pulsar term'' in the gravitational wave signal, i.e.\ 
the impact of the gravitational wave on the emission of the pulsar signals at the 
pulsar. This is only possible if the distance between Earth and pulsar is known 
with high precision (Lee et al.\ 2011).

\smallskip

\noi {\bfseries\boldmath\small Imaging parallax:~~}
The straightforward method for measuring the parallax of pulsars is by measuring 
the position of the pulsar on the sky over time by means of imaging. Fomalont \& 
Reid (2004) estimate that the astrometric accuracy that the SKA can potentially 
obtain is 15\,$\mu$as at the frequency of 1.4\,GHz and a 3000\,km SKA baseline.
However, for many pulsars the limiting factor will be given by the limited SNR of 
the pulsar detection. Smits et al.\ (2011) find that the SKA can potentially 
measure the parallaxes for $\sim$\,9000 pulsars with an error of 20\,\% or smaller. 
This includes pulsars out to a distance of 13\,kpc. The imaging parallax depends 
only on the observed strength of the radio emission, not the rotation 
characteristics of the pulsar. 

\smallskip

\noi {\bfseries\boldmath\small Timing parallax:~~}
A second set of methods of performing astrometry of a pulsar involves accurate 
timing of the pulse time-of-arrival (TOAs) at the telescope. These methods are a) 
parallax measurements using the Earth orbit, b) parallax measurements for binary 
pulsars using the Earth orbit and that of the pulsar and c) distance estimates of 
binary pulsars based on the comparison of observed orbital parameters with those 
predicted by general relativity.

The ``classical'' timing parallax measurement (a) utilises the fact
that the wave front curvature of a pulsar signal is directly related
to the distance of the source. The curvature of the wavefront
introduces an annual periodic change in the apparent direction, hence
a six-monthly periodicity in the TOAs (Lorimer \& Kramer 2005). The
apparent change in direction is more easily measurable for low
ecliptic latitudes --- in contrast to an imaging parallax. For
millisecond pulsars, Smits et al. (2011) show that timing
observations of pulsars with the SKA can yield a parallax precision
that exceeds that of an imaging parallax by at least an order of
magnitude in terms of both relative and absolute precision. Such
timing precision suggest that distances can be measured with a
precision of at least 10\,\% to a distance of about 10\,kpc.

If a pulsar happens to be in a binary system, the pulsar orbit will be viewed 
under slightly different angles from different positions of the Earth's annual 
orbit. The result is a periodic change in the observed longitude of periastron 
and the projected semi-major axis. This effect, known as the annual orbital 
parallax (b), depends on the distance and therefore can also lead to an accurate 
pulsar distance measurement for a binary pulsar (Kopeikin 1995). It will require 
the timing precision of the SKA, to convert this effect into a precise
distance measurement of the binary system (Smits et al. 2011).

If the intrinsic decay rate of the orbital period $P_b$ of a binary pulsar is 
determined purely by gravitational wave damping, one can predict a change in 
$P_b$ if the pulsar and companion masses are obtained via pulsar timing thanks to 
the measurement of post-Keplerian parameters. For a pulsar at a finite distance 
$d$ with an observed proper motion in the sky, the predicted change in $P_b$ will
differ from the observed one due to the the kinetic contribution of the 
Shklovskii term (Lorimer \& Kramer 2005), which is proportional to $d$. The 
superb timing precision of the SKA will allow for high precision measurements of 
the binary parameters and the pulsar's proper motion in the sky. Using general 
relativity to calculate the intrinsic change of $P_b$, the Shklovskii term can be 
determined with high precision, which converts into a very accurate distance 
estimation (c), which can easily supersede the ``classical'' timing parallax
(Smits et al.\ 2011).\\

\noi {\bfseries\boldmath\small Summary:~~}
From the above, it is clear that the SKA will become a superb astrometry 
instrument for pulsars, providing high precision pulsar distances. These 
measurements feed directly back into astrophysical questions, for instance, 
related to the distribution of the ionised gas in the Galaxy and the structure of 
the Galactic magnetic field. Imaging and timing parallax measurements can help us 
significantly with precision tests of general relativity to correct for kinematic 
effects. This is of vital importance for tests of the fundamental properties of 
gravity, like the emission of gravitational waves and a time dependent 
gravitational constant. Furthermore, accurate pulsar distances are key parameters 
in a pulsar timing array to study the emission of nano-Hertz gravitational waves 
by super-massive black hole binaries.\\

\parbox{0.9\textwidth}{
\noi{References:}\\
\noi{\scriptsize  
Burgay M., et al., 2003, Nature, 426, 531;
Fomalont E., Reid M., 2004, NewAR 48, 1473;
Kopeikin S., 1995, ApJ, 439, L5;
Lee K., et al., 2011, MNRAS, 414, 3251;
Lorimer D., Kramer M., 2005, Handbook of Pulsar Astronomy (Cambridge Univ.\ 
Press, Cambridge, UK);
Lyne A.~G., et al., 2004, Science, 303, 1153;
Smits R., et al., 2011, A\&A, 528, A108;
}} \\

\subsubsection{The dynamic radio sky  {\scriptsize [M.~Kramer, E. Keane, D. Champion]}}

\noi The radio sky reveals variable and transient phenomena which are
relevant for a huge range of (astro-)~physical questions. Repeated
monitoring can reveal new, previously unseen objects which are
important for understanding cosmic explosions, population studies or
the connections between radio and high-energy processes. Often radio
observations can trigger searches and studies in other observational
windows. Perhaps one of the most exciting such possibilities is the
triggering of gravitational wave searches for neutron star
oscillations after the detection of a pulsar glitch.

\noi Excellent recent examples for the scientific potential still hidden in
the dynamic Universe are the discovery of Rotating Radio Transients
(RRATs, McLaughlin et al. 2006) or the discovery of
radio bursts of potential extragalactic origin (Lorimer et
al.~2007). Whether in these examples these objects are
indeed representing a completely different class of neutrons stars or
whether they are really at cosmological distances, respectively, is
not only interesting but similar studies promise a huge potential in
changing our mindset about processes in the Universe.

\noi Radio monitoring with good cadence is clearly essential to discovery
new phenomena and to eventually study the implications. Indeed, when
attempting to extract the extreme physics from the observed signals,
it is important to not only obtain a snapshot of the sky but to follow
the transient sky systematically. With its multi-beam capability to
observe vastly different areas of the sky simultaneously, the SKA will
be a unique instrument in studying transient phenomena and their
physics over a wide range of radio frequencies. In oder to demonstrate
this, we consider as an example the discovery that radio pulsars
appear to be able to change the structure of current flow in the
magnetosphere in an seemingly instantaneous way. This recognised class
of ``intermittent pulsars'' (Kramer et al. 2006), seem to be active for a
period of time before switching off completely for a further period of
time, where the change in observed radio output is correlated with a
change in pulsar spin-down rate. In the case of PSR B1931+24, if the
pulsar is emitting radio waves, the spin-down rate is about 50\,\%
larger than when the pulsar is off. This faster spin-down rate is
caused by an extra torque that is given by the electric current of the
plasma that also creates the radio emission. If some or all of the
plasma is absent, the radio emission is missing together with the
additional torque component, so that the spin-down is
slower. 

\noi Recently, Lyne et al.\ (2010) realised that intermittent
pulsars are actually the extreme form of a more common phenomenon, in
which the restructuring of the plasma currents does not necessarily
lead to a complete shutdown in the radio emission, but can be observed
as changes between distinct pulse shapes. Lyne et al.\ showed that
particular profiles are indeed correlated with specific values for the
spin-down rates, confirming the previous picture. The times when the
switch in the magnetospheric structure occurs are for some pulsars
quasi-periodic but in general difficult to predict. The resulting
change between typically two spin-down rates leads to seemingly random
timing residuals that have been in the past classified as ``timing
noise''. The observations by Lyne et al.~therefore simultaneously
connect the phenomenon of timing noise to that of intermittent pulsars
and that of ``moding'' and ``nulling'' (see e.g.~Lorimer 
\& Kramer 2005).  An
interesting aspect is the possibility of determining the exact times
of the switch between magnetospheric changes by precisely measuring
the pulse shapes with high-sensitivity, high-cadence monitoring with
the SKA. In this case, the changes in spin-down rate can be taken into
account and the pulsar clock can be ``corrected''. This would offer
the opportunity to use not only the MSPs for timing experiments, but
to utilize also the 20\,000 to 30\,000 normal pulsars that will be
discovered in a Galactic census described below.  Even though the
precision will not be as high as for MSPs, the large number of pulsars
may help to detect, for instance, gravitational waves.

\noi Monitoring specifically radio pulsars can also lead to the discovery
of pulsar ``glitches''. A pulsar glitch is a sudden increase in spin
frequency of the neutron star caused by an internal reconfiguration of
the star's interior structure (e.g.\ Lyne \& Smith 2004). The relaxation of
the glitch gives information about the super-fluid interior of the
pulsar and enables us to do neutron star seismology. It is expected
that such an event may also cause the neutron star to oscillate, which
for certain vibration modes, should result in the emission of
gravitational waves (e.g.\ van Eysden \& Melatos 2008). Knowing the exact moment of
the glitch through dense radio monitoring, gravitational wave data can
be searched accordingly. The statistical information obtained from
such radio monitoring with the SKA will be used to study the glitch
mechanism and to answer questions as to whether a possible bimodal
distribution of the glitch sizes means that two different types of
glitch processes are acting (e.g.\ Espinoza 2010).\\

\noi Overall, the SKA will revolutionize our understanding of cosmic time
variable processes by providing us with a snapshot of this dymamic and
non-static Universe.\\

\parbox{0.9\textwidth}{
\noi{References:}\\
\noi{\scriptsize  McLaughlin M.A., \etal , 2006, Nature, 439, 817;
Lorimer D.R., \etal , 2007, Science, 318, 777;
Lyne  A.G., Smith F. G. , 2004, {\em Pulsar Astronomy, 3rd ed.} (Cambridge University Press, Cambridge, 2004);
van Eysden C.A. , Melatos A., 2008, Classical and Quantum Gravity, 25, 225020;
Espinoza C., 2010, PhD thesis, University of Manchester, UK;
Kramer M., \etal , 2006, Science, 312, 549;
Lyne A., \etal , 2010, Science, 329, 408;
D.~R. Lorimer D.R., Kramer M., 2005, {\em Handbook of Pulsar
  Astronomy} (Cambridge University Press, Cambridge, 2005)
}}\\

\subsubsection{Multi-wavelength astronomy in the age of the SKA and
  CTA  {\scriptsize [D.I. Jones, F.A. Aharonian, R.M.~Crocker]}}

\noi {\bfseries\boldmath\small Introduction:~~}The past decade has seen an explosion in the
number of sources detected at very high energies (VHE; E $\ge$ 0.1\,GeV) as what was a promising new window on the electromagnetic
spectrum has metamorphosised into a fully-fledged observational
science. This transformation has primarily been brought about by
construction of a new generation of Imaging Atmospheric Cherenkov
Telescopes (IACTs), such as the High Energy Stereoscopic System
(HESS). These telescopes collect Cherenkov photons produced by the
interaction of astrophysical gamma-rays with the Earth's
atmosphere. With the planned construction of the Cherenkov Telescope
Array (CTA), VHE gamma-ray astronomy is moving into its second
``golden age'' at a time when -- serendipitously -- radio astronomy is
also planning a renaissance. In this section, we point out the
synergies that such a fortuitous happenstance can bring from (i) the
standpoint of the detection of supernova remnants (SNRs) and (ii) the
study of phenomena at the Galactic centre (GC). These are, of course,
just two examples of many such problems that a coordinated approach to
astrophysics in truly broadband terms can
explore.\\

\bigskip
\bigskip

\hspace{3cm}
\parbox{\textwidth}{
\parbox{5cm}{ 
\begin{tabular}{|l|c|c|c|c|}\hline \hline 
Telescope & Resolution & Sensitivity & Field of view \\
& [arcsec FWHM] & [erg cm$^{-2}$ s$^{-1}$] & [deg at FWHM] \\ \hline
JVLA     & $<$\,1 & 10$^{-18}$ & 0.5 \\
ATCA    & $\sim$\,1 & 10$^{-17}$ & 0.5 \\ \hline
ASKAP & 10 & 10$^{-19}$ & 30 \\
SKA & $<<$\,1 & 10$^{-20}$ & 200 \\ \hline
HESS & 678 & 10$^{-13}$ & 5 \\
VERITAS & 678 & 10$^{-13}$ & 5 \\
CTA &120 & 10$^{-14}$ & 3\,--\,4 \\ \hline
\end{tabular}
\vspace{0.8cm}
}

\hspace{-2.9cm}
\parbox{0.9\textwidth}{Table 1$_{\rm J}$: The characteristics of present and future IACT and radio telescopes in common units. The sensitivity for the radio is the r.m.s. sensitivity for 1\,hr integration, whereas the gamma-ray sensitivity is for 50\,hrs (and assuming a 5$\sigma$ detection). For the resolution, the figures are the best available for the frequency (energy) range covered, as well as for the sensitivity – both of which are frequency (energy) dependent for the radio (gamma-ray) telescopes. The sensitivity is computed at a frequency of 1\,GHz and include the best available bandwidth for the radio telescope. Note that the figures for ATCA are for the recent Compact Array Broadband Backend (CABB) upgrade and the SKA and ASKAP FoV values are actually in units of square degrees. }
\vspace{0.5cm}
}\\

\bigskip

\noi {\bfseries\boldmath\small Connecting Galactic supernova detection in VHE gamma-rays
and radio domains:~~}Since low-frequency radio emission (i.e. $\lesssim$2\,GHz) traces the MeV\,--\,GeV
part of the Galactic CR spectrum through synchrotron emission, it has
long been the method-of-choice for the detection of SNRs. The past few
decades has begun to change this due the advent of sensitive,
(relatively) high resolution gamma-ray astronomy. The new generation
of radio and gamma-ray telescopes promises to change this view. Given
the imminent construction of telescopes such as the CTA and the SKA
pathfinders, such as ASKAP, it is instructive to connect the search
for SNRs at radio frequencies, with the search for SNRs with VHE
gamma-ray telescopes.  It is well-known that the number of SNRs
detected in the radio domain is generally much less than that expected
statistically (we expect $\grtsim$1000 Galactic SNRs -- based on pulsar birth
rates, OB star counts, Fe abundance, etc; (Brogan et al. 2006) and
know of 274; (Green 2009)). This suggests that there is a population of
faint SNRs that radio surveys are blind to. Indeed the traditional way
of searching for SNRs -- based on their low frequency radio continuum
emission -- skews the statistics so that we only observe the brightest
remnants (Green 2009). Therefore, the limited availability of time on
high-resolution, large collecting area telescopes, such as the VLA,
has limited our ability to identify SNRs to large angular sizes, and
down to a surface-brightness limit, $\Sigma_{1} \sim 10^{-20}$\,W m$^{-2}$
\,Hz$^{-1}$\,sr$^{-1}$, measured at a fiducial frequency of 1\,GHz (Green
2009 and references therein). The SKA (and to a lesser extent its
pathfinder instruments such as LOFAR and ASKAP), with its mammoth
collecting area and huge instantaneous fields of view (FoV; see table), will change this.\\

\noi {\bfseries\boldmath\small A new era of SNR detection: simultaneous SNR surveys in the
  radio and gamma-rays bands:~~}The table shows the present and future capabilities of radio and
gamma-ray telescopes. It illustrates that although gamma-ray telescope
sensitivities are improving (e.g. the order-of-magnitude improvement
in sensitivity between HESS/VERITAS and CTA), there is still a gulf
between gamma-ray and radio telescopes in this regard.  

\noi However, there are important lessons to be learnt from the gamma-ray
community’s treatment of SNRs – which is to think of them in terms of
the SNR evolution, that is, a particle physics view of a particular
SNR. Specifically, consideration of the evolutionary stage of a
particular remnant in terms of the particle acceleration and diffusion
shows that gamma-ray telescopes, such as HESS, are very good at
observing very young SNRs. This follows because higher energy
particles radiate their energy more quickly than those at lower
energies.

\noi The above statement is, in fact, a logically equivalent to the
argument in the preceding section that radio continuum surveys miss a
population of SNRs because of observational biases. Therefore, in the
coming era of large, sensitive gamma-ray and (low-frequency
especially) radio telescopes, there is a very compelling argument to
undertake simultaneous searches at GHz and TeV frequencies and
energies using telescopes such as the SKA and CTA. This will uncover
many in this hitherto unknown population and thus close the gap
between the expected and observed number of SNRs within our Galaxy.\\

\noi {\bfseries\boldmath\small Synergy between SKA and gamma-ray telescopes and high-energy
  particle astrophysics; applications to understanding the Galactic
  nucleus:~~}The SKA will offer an unprecedented sensitivity to
extended, low-surface-brightness features in the low frequency radio
sky. Such large but relatively dim structures are features of:

\begin{itemize}
\item[--] The local ISM e.g. nearby supernova remnants whose detection/characterization is important
for a full understanding of the local cosmic ray spectrum, particularly that of cosmic ray electrons and positrons
\item[--] The Galaxy-at-large e.g.the recently-discovered ``Fermi Bubbles''
(Su\ \etal\ 2010; and described in detail below)
\item[--] Extra-galactic structures like giant radio galaxy lobes, galaxy
cluster radio halos, etc, which are, again, important potential
sources of cosmic rays at the highest energies.
\end{itemize}

\noi The SKA will provide crucial new data on the non-thermal particle
populations of these disparate regions (through low-frequency radio
continuum observations of synchrotron emission from relatively
low-energy electrons) in addition to their ISM conditions (magnetic
field strength and structure, gas density, etc). In any case, aside
from the intrinsic interest they hold, understanding all these
features of the ``local'' Universe is, of course, important to properly
understanding the ``foregrounds'' to cosmological measurements.

\noi There is a great and continuing interest in the Galactic centre and
the apparently-related Fermi Bubbles. The latter were discovered in
gamma-ray data collected at $\sim$\,GeV energies by the orbiting Fermi
gamma-ray telescope and extend fully 10\,kpc north and south from the
Galactic plane above the Galactic centre. These structures are
coincident with a non-thermal microwave ``haze'' found in WMAP data
and an extended region of X-ray emission detected by ROSAT. In the
broadest terms, the Bubbles, while enigmatic, are likely related to
the nuclear feedback processes that have acted to limit star-formation
in the inner Galaxy and ultimately control the size of the Galactic
bulge.

\noi The SKA will contribute vitally to understanding the Bubbles,
their relationship to processes occurring in the Galactic centre (the
sustained star-formation activity that has occurred there, potential
Seyfert-like outbursts of the resident super-massive black hole), and
to the wider Galaxy. Crucially-important characteristics of the
instrument in this connection are its southern location, sensitivity,
and wide field-of-view (up to tens of degrees).

\noi The SKA will significantly contribute in at least three ways, namely,
by offering the capability to perform:

\smallskip

\begin{itemize}

\item[1.] Sensitive rotational measure
synthesis studies of background polarised radio sources that could be
used to interrogate the magnetic field structure of the Bubbles;

\item[2.] Very high resolution low frequency radio continuum morphology
studies that show the detailed connection between the Galactic centre
and the Bubbles (i.e. that can probe whether the radio continuum
spurs visible to current instruments extend down to the super-massive
black hole in particular or, rather, originate in star-forming regions
in the Galactic centre); 

\item[3.] Low frequency radio continuum spectral studies of synchrotron
emission that could be used to constrain the low-energy component of
the cosmic ray electron population inhabiting the Bubbles. This latter
is of particular relevance for the question of whether the Bubbles
non-thermal emission is generated by electrons or protons. In the
proton case, the observed, hard-spectrum, non-thermal microwave
emission (at tens of GHz) from the Bubbles is due to secondary
electrons created in {\it pp} collisions. For kinematic reasons the spectrum
of such secondaries begins to diverge from a pure power below
$\sim$\,1\,GeV in electron energy finally cutting off below
$\sim$\,100\,MeV in electron energy. The primary electron distribution,
on the other hand, would display no such features. The low-energy
secondary electron spectral features are mapped by synchrotron
emission on to the radio continuum spectrum produced by these
particles which starts to diverge from a pure power law at frequencies
below 100\,MHz (B/10\,$\mu$G).  
\end{itemize}

Beyond these particular applications,
we would also like to draw attention here to the possibilities that
may open up with the simultaneous analysis of data taken by the SKA
and that taken at other wavelengths (and with quite different
instrumental modalities, such as IACT observations discussed
above). For instance, again in the context of the Galactic centre, we
were recently able to demonstrate (Crocker et al. 2010) a novel
technique for measuring magnetic field amplitude through the
simultaneous and self-consistent analysis of GHz radio continuum data
with GeV gamma-ray data.\\

\noi {\bfseries\boldmath\small The future and conclusions:~~}We have shown here, that both
the fields of gamma-ray and radio astronomy are entering new, ``golden
ages'' simultaneously. We have shown that, using a coordinated
approach, fundamental astrophysical questions -- such as the number and
distribution of SNRs within our Galaxy or the nature and origin of the
Fermi Bubbles -- may be answered in the near future. This is, of
course, over and above the exciting
discoveries that one, as yet, cannot anticipate.\\

\parbox{0.9\textwidth}{
\noi{References:}\\
\noi{\scriptsize Brogan C. L., \etal , 2006, ApJL, 639, 25; 
Green D. A., 2009, Bull. Astr. Soc. India, 37, 45;
Su M., Slatyer T. R., Finkbeiner D. P., 2010, ApJ, 724, 1044;
Crocker R. M., \etal , 2010, Nature, 463, 65
}}\\

\bigskip


\subsection{Solar system science}

\subsubsection{The Sun {\scriptsize [G. Mann]}}

The Sun is an intense radio source in sky. During solar
eruptions as flares and coronal mass ejections (CME),
the intensity of the solar radio emission is strongly enhanced
over a broad spectrum from the GHz down to the MHz range.
Just this frequency range is covered by the SKA (70\,MHz\,--\,10\,GHz).
Therefore, the SKA will be of great interest for solar physics.

\noi The observation of the solar radio radiation is of great importance
for studying the flare process. Observationally, a flare is defined as
a sudden enhancement of the Sun's emission of electromagnetic radiation
covering a broad spectrum from the radio up to the $\gamma$-ray range.
That can be demonstrated by the example of the solar event on October
28, 2003.\\

\parbox{\textwidth}{

\hspace{1.3cm}
\parbox{5cm}{ \includegraphics[scale=0.1]{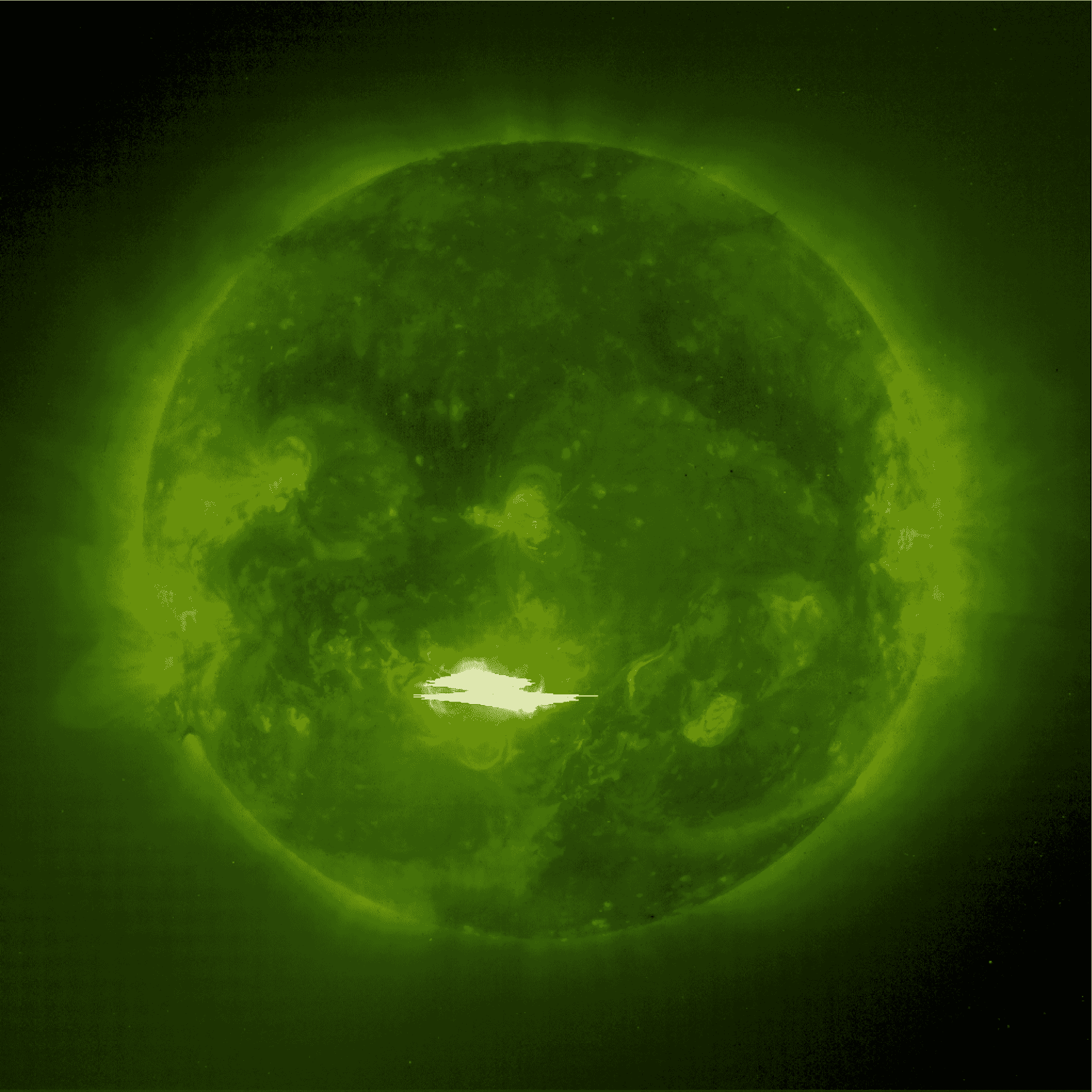}\vspace{-7.5cm}}

\hspace{9.3cm}
\parbox{6cm}{Figure 1$_{\rm Ma}$: The Figure shows an EUV image of the Sun at the solar event on October
28, 2003. The location of the flare is seen as the flash in the EUV light.}
\vspace{6cm}
}\\

\noi It was one of the hugest solar eruptions ever observed. The
Figure\ 1$_{\rm Ma}$
shows the image in the EUV range. The flare is obviously seen on the
disc of the Sun by a local flash in the EUV range. The temporal
behaviour of the hard X- and $\gamma$-ray photon flux is presented at the
top of the Figure\ 2$_{\rm Ma}$ for the same event. The simultaneous radio spectrum
in the range 200\,--\,400\,MHz is shown at the bottom of Figure\ 2$_{\rm Ma}$. During the
impulsive phase of the flare, i.e. at 11:02 UT, the emission of the
electromagnetic radiation from the radio (200\,--\,400\,MHz) up to the
gamma-ray (7.5\,--\,10\,MeV) range is strongly enhanced. Thus, there is
a strong correlation between the radio and the hard X-ray radiation.
That indicates the generation of energetic electrons during solar
flares. It is one of the basic questions in solar physics,
how a huge number of electrons (i.e. 10$^{36}$) are accelerated up
to high energies (i.e. $>$\,30\,keV) per second during solar flares.
Solar radio observations are an important tool for answering
this question.\\

\noi During solar flares, a huge amount of stored magnetic field energy
is suddenly released and transferred into a local heating of the
coronal plasma, mass motions (e.g. jets) and particle acceleration
(solar energetic particle; SEP). Furthermore, a large amount of coronal
material can be ejected in to the interplanetary space. That is usually
called coronal mass ejections. All these phenomena of solar activity
have their special signatures in the Sun's radio radiation. Thus, the
study of the solar radio radiation provides very important
information on
\begin{itemize}
\item[--] magnetic energy release
\item[--] electron acceleration
\item[--] plasma jets
\item[--] coronal shock waves
\item[--] coronal mass ejections
\item[--] transport of energetic particles from the Sun into the interplanetary
 space
\item[--] solar energetic particle events
\end{itemize}

\noi It should be emphasised that all these processes also happens at other
places in space, e.g. in other stellar coronae, supernova remnants etc.,
but they can studied in the best way anywhere than on the Sun as our
nearest star.

\parbox{\textwidth}{

\hspace{-0.6cm}
\parbox{5cm}{ \includegraphics[scale=0.1]{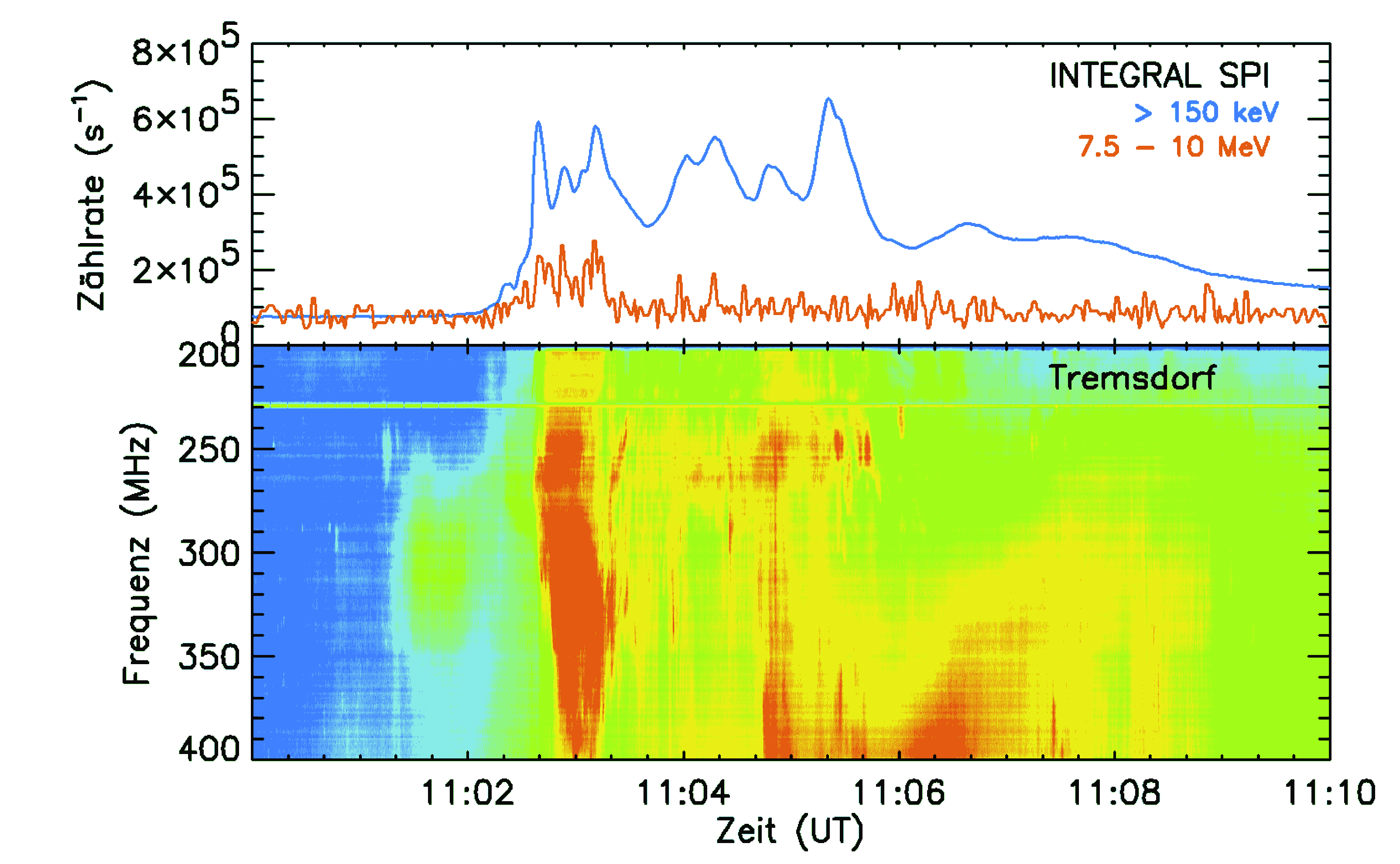}\vspace{-6.4cm}}

\hspace{10.5cm}
\parbox{6cm}{Figure 2$_{\rm Ma}$: At the top, the temporal behaviour of the photon count rates are
presented for the energies $>$\,150\,keV (blue line, hard X-rays)
and 7.5\,--\,10\,MeV (red line, $\gamma$-rays). The simultaneous
dynamic radio spectrum in the range 200\,--\,400\,MHz is shown at the
bottom. The radio intensity is colour coded.}
\vspace{2.7cm}
}

\noi The activity of our Sun is influencing our Earth's environment
and technical civilization. For instance, solar flares are affecting the
global communication. That is usually called {\it Space Weather}.

The solar radio emission is emanating from the corona. Especially in the
dm- and m-wave range, the radio radiation of the Sun is generated by
plasma emission, i.e. it is emitted at the local electron plasma
frequency and/or its harmonics.
Due to the gravitational density stratification the radio waves
at different frequencies come from different heights in the corona.
Since the SKA will be able to provide radio maps of the Sun at different
frequencies (70\,MHz\,--\,10\,GHz), the SKA will allow a 3-d tomography of the
solar activity in the corona. Because of the large SKA dishes and
the high precision of the polarisation measurements, the SKA will
enable the measurement of the longitudinal  component of the magnetic
field in the corona. That will also allow magnetic measurements to be
extended beyond sunspots.\\

\noi In summary, the SKA is a very important radio facility of solar physics.
It can deliver complimentary information of the Sun in close
collaboration to other solar facilities as, for instance, GREGOR on
ground as well as Hinode, SDO, and Proba 3 in space.

\bigskip

\subsubsection{Ionospheric science and space weather {\scriptsize [D. Innes, L. Gizon, N. Krupp]}}
\vspace{0cm}

\noi {\bfseries\boldmath\small Space weather:~~} The radio sources on the Sun change from
day to day and the most interesting, flares and coronal mass ejections
occur with at most three days warning. This makes solar observations
of flares difficult to schedule. It is nevertheless urgent to
understand the origin and signatures of solar energetic particles
(SEPs) and CMEs, not only to get at the heart of these fascinating
features and events, but also because they are the principle causes of
space weather responsible for changing the properties of the Earth’s
ionosphere and thus affecting satellite orbits and global
communications.  Flares and shocks are thought to be the primary
sources of SEPs but neither source can explain recent STEREO
detections of SEPs with an angular spread of 180 degrees from single
flares (Innes \& Buick 2011). The SKA will be
able to image the source and propagation of high energy electrons as
they travel trough the corona with unparalleled high resolution maps
of 1\,--\,10\,GHz bursts. In addition to the high resolution, the
sensitivity of the SKA will enable the direct imaging of CME plasma,
visible through its gyroresonance and free-free emission. Analysis of
the radio spectra gives plasma density, temperature, magnetic
morphology and propagation speed in all three dimensions. The aim
would be to see where and when particle acceleration occurs in
relation to shocks and the bulk plasma’s density and magnetic
configuration.  Less spectacular but more frequent and involving the
same basic physics are active region jets and their associated radio
bursts. It has recently been suggested that the electron beams
responsible for interplanetary radio bursts originate in reconnection
regions above sunspot umbra and may be triggered by running penumbral
waves (Innes et al. 2011). The SKA can
observe the sunspot waves at high frequency ($>$\,5\,GHz), the onset of
the jets (500\,--\,100\,MHz) and track their propagation using images at
lower frequencies.\\

\noi {\bfseries\boldmath\small Magnetic field measurements:~~}The main hindrance to
progress in coronal physics is the relative lack of measurement of the
coronal magnetic field. Observations in the radio wavelength range
provide great promise to change this frustrating situation. The large
SKA dishes and high polarisation precision will enable the measurement
of the longitudinal component of the coronal magnetic field in weak
field (10\,G) regions using observations of free‐free emission above 10\,GHz. This will allow magnetic field measurements to be extended beyond
sunspots and the most active regions (to which current measurements
are restricted) to the quieter parts of the solar atmosphere.\\

\noi {\bfseries\boldmath\small Future complementary solar facilities in space:~~}The SKA
will be operational at a time when we hope that the Proba‐3
coronagraph will be giving uncontaminated white light images of the
solar corona from as close as 1.05 solar radii. This innovative
mission will have two spacecraft flying in formation with the sunward
spacecraft carrying the occulter and the other the
detectors. Combining the coronagraph images with SKA observations will
give valuable insight into how and where electron acceleration occurs
with respect to the bulk plasma.\\

\parbox{0.9\textwidth}{
\noi{References:}\\
\noi{\scriptsize
Innes D. , Cameron , Solanki, 2011, A\&A 531, L13;
Innes and Buick, 2011, EGU General Assembly
}}\\

\subsubsection{Determining the masses of the planets {\scriptsize [D. Champion]}}
\vspace{0cm}

\noi For the most stable pulsars, pulsar timing can predict the time of
arrival (TOA) of a pulse at the Solar System Barycentre (SSB) to
within a few 100s of nanoseconds. To convert these to arrival times at
the observatory a Solar System ephemeris is used to determine the
relative positions of the SSB and Earth. If the Solar System
ephemeris was incorrect this would lead to an unmodelled signal in the
pulsar timing, this can be effect can be reversed to measure the
masses of the planets. This analysis was done using the four longest
and most precise data sets taken for pulsar timing (Champion et
al. 2010), in all cases, these measurements are consistent with the
best-known measurements. For the Jovian system, the measurement
improves on the \emph{Pioneer} and \emph{Voyager} spacecraft
measurement and is consistent with the mass derived from observations
of the \emph{Galileo} spacecraft. Pulsar timing has the potential to
make the most accurate measurements of planetary system masses and to
detect currently unknown solar system objects such as trans-Neptunian
objects.

\noi Pulsars are extremely stable rotators and any effect that
causes a delay or advance of their TOA can be measured precisely. The
pulses observed during a single observation are added together to
produce a single, high signal-to-noise ratio pulse profile. The
number of complete rotations between these pulses is then determined
using a model. The difference between the model and the observations
is minimised by fitting for the model parameters. This technique
produces a phase-connected timing solution which accounts for every
rotation of the pulsar over the span of the data set, of ten many
years.

\noi Before the pulsar model can be fit, the effect of the
Earth's orbit must be removed. Depending upon the position of the
pulsar there may be as much as 16\,mins light travel time delay over
the orbit of the Earth. This is equivalent to 500\,000 pulses for a
fast pulsar. To remove this delay the time of the pulse arrival at the
observatory is converted to the arrival time at the SSB. To do this a
Solar System ephemeris is used to determine the position of the Earth
relative to the SSB, the most common used being from NASA's Jet
Propulsion Laboratory.

\noi These ephemerides are constructed by numerical integration
of the equations of motion and adjustment of the model parameters to
fit data from optical astrometry, astrolabe measurements, observations
of transits and occultations of the planets and their rings, radar
ranging of the planets, radio astrometry of the planets using very
long baseline interferometry, radio ranging and Doppler tracking of
spacecraft, and laser ranging of the Moon. These observations
constrain the motion of Solar System bodies with respect to the Earth,
however they do not tightly constrain the planetary masses. This is
reflected in the fact that the planetary/solar mass ratios are
normally held fixed in the fit.

\noindent If the relative positions of the SSB and observatory are not
correctly calculated then an unmodelled signal will be present in the
residuals of the fit. For example, an underestimation of the mass of
the Jovian system will result in sinusoidal timing residuals with a
period of Jupiter's orbit. The identification of such residuals
therefore provides a method to limit or detect planetary mass errors
in the ephemeris.

\noindent This technique was used by Champion et al. 2010 to measure
the masses of the planets from Mercury to Saturn using data taken as
part of the International Pulsar Timing Array project (Hobbs et
al. 2010). The four pulsars were selected based upon the precision of
their measured TOAs, the magnitude of timing irregularities and on the
length of the data set. The resulting mass measurements are listed in
the table.

\bigskip
\bigskip

\begin{centering}
\begin{tabular}{|l|l|c|l|l|}\hline \hline
System & Best-Known Mass (M$_\odot$)& Ref. & This Work
(M$_\odot$) &$\delta_j/\sigma_j$ \\ \hline
Mercury & 1.66013(7)$\times 10^{-7}$     & 1 & 1.6584(17)$\times 10^{-7}$  & 1.02\\
Venus   & 2.44783824(4)$\times 10^{-6}$  & 2 & 2.44783(17)$\times 10^{-6}$ & 0.05\\
Mars    & 3.2271560(2)$\times 10^{-7}$   & 3 & 3.226(2)$\times 10^{-7}$    & 0.58\\
Jupiter & 9.54791898(16)$\times 10^{-4}$ & 4 & 9.547921(2)$\times 10^{-4}$ & 1.01\\
Saturn  & 2.85885670(8)$\times 10^{-4}$  & 5 & 2.858872(8)$\times
10^{-4}$ & 1.91\\ \hline
\end{tabular}

\vspace{0.5cm}

\hspace{0.9cm}{\scriptsize (1) Anderson et al. 1987; (2) Konopliv et al. 1999; (3) Konopliv et
al. 2006; (4) Jacobson et al. 2000; (5) Jacobson et al. 2006.}
\label{tab:results} 
\end{centering}
\vspace{0.5cm}

\bigskip

\noi All these results were consistent with the best current
measurements. Our current data sets are sensitive to mass differences
of approximately $10^{-10}$\,M$_\odot$, independent of the
planet. Consequently, our most precise fractional mass determination
is for the Jovian system. While the result for the Jovian system is
more precise than the best measurement derived from the \emph{Pioneer}
and \emph{Voyager} spacecraft by a factor of $\sim$\,4, the result from
the \emph{Galileo} spacecraft is still better by a factor of $\sim$\,20.

To improve upon these results requires higher precision of the TOAs
and longer data sets for more pulsars. For example, after
$\sim$\,7\,years of observations a pulsar timing array of 20 pulsars,
regularly sampled every two weeks, with an rms timing residual of
100\,ns will surpass the \emph{Galileo} measurement for Jupiter; see
the figure below. With $\sim$\,13\,years of data, the uncertainty of
the current \emph{Cassini} measurement for Saturn is reached. Although
it is unlikely that this timing precision will be reached for 20
pulsars using current telescopes the SKA will be able to time large
numbers of pulsars to better than 100\,ns and will likely find even
more stable pulsars that can be used as part of the
array.\\

\vspace{0cm}
\parbox{\textwidth}{

\hspace{0cm}
\parbox{5cm}{ \includegraphics[scale=0.175]{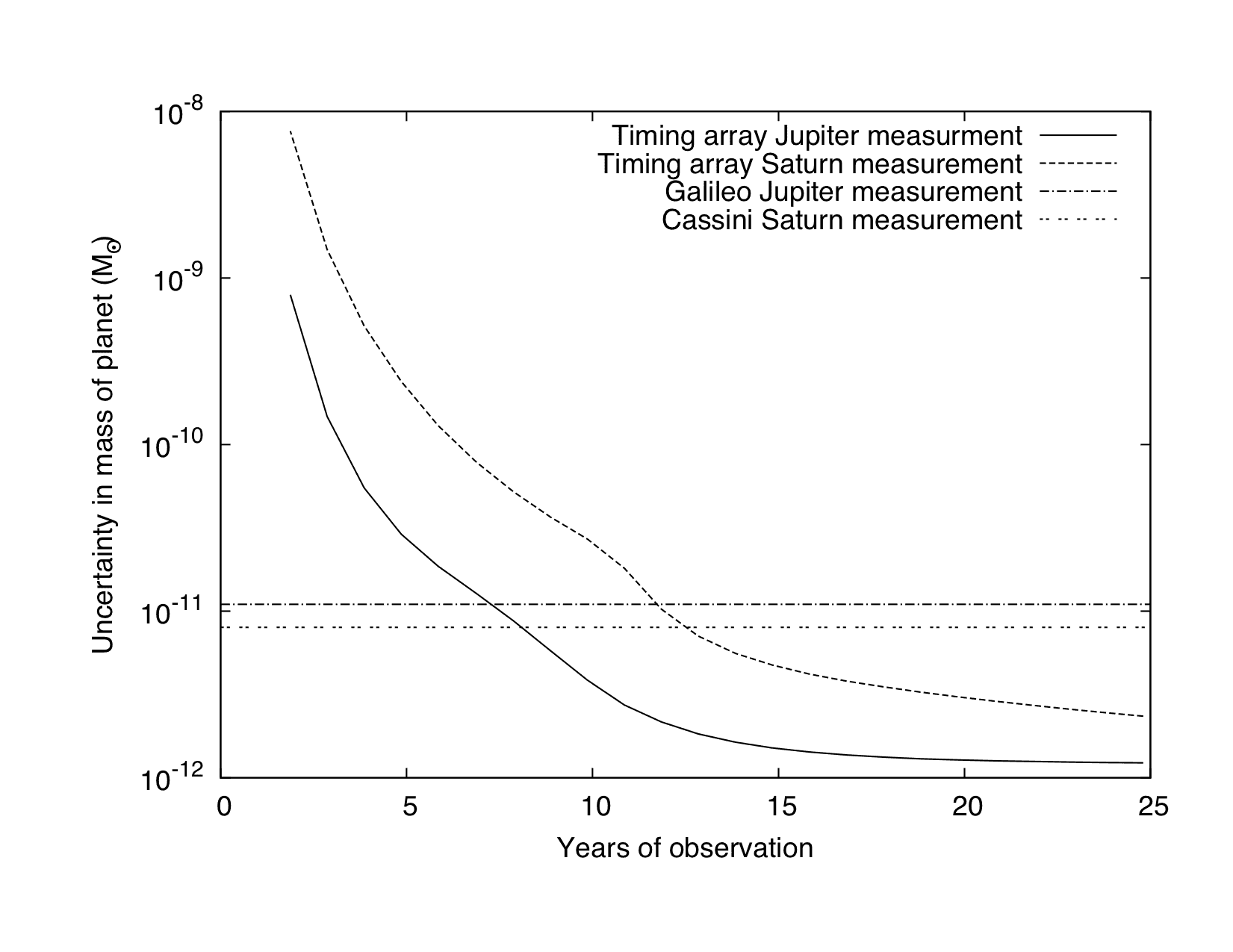}\vspace{-6.75cm}}

\hspace{10cm}
\parbox{6cm}{Figure 1$_{\rm Ca}$: Mass uncertainties for the Jovian and Saturnian system using
  simulated data from an array of 20 pulsars sampled every 14 days,
  timed with an rms timing residual of 100\,ns for different data
  spans. Also plotted are the current best mass measurements for
  Jupiter (Jacobson et al. 2000) and Saturn (Jacobson et al. 2006).}
\vspace{2cm}}

\noi Although spacecraft measurements provide precise mass measurements for
most of the planets, it should be noted that the pulsar measurements
are independent with different assumptions and sources of
uncertainty. Independent methods are particularly important for
high-precision measurements where sources of systematic error may not
be well understood. In addition, the spacecraft measure the mass of
the body being orbited and do not measure the mass of the whole
planetary system (planet and any satellite masses). When combined with
the spacecraft measurements this can provide a measure of the mass
undetermined by spacecraft observations.

\noi The pulsar timing technique is also sensitive to currently unknown
masses in the Solar System, e.g. trans-Neptunian objects
(TNOs). Pulsar timing arrays comprising a large number of stable
pulsars with a wide distributions on the sky will be sensitive to any
error in the Solar System ephemeris, including those induced by
currently unknown TNOs.\\

\parbox{0.9\textwidth}{
\noi{References:}\\
\noi{\scriptsize
Champion D.J., \etal , 2010, ApJ, 720, 201;
Hobbs G., \etal , 2010, Classical and Quantum Gravity, 27, 084013;
Anderson J.D., \etal , 1987, Icarus, 71, 337;
Konopliv A.S., Banerdt W.B., Sjogren W.L., 1999, Icarus, 139, 3;
Konopliv A.S., \etal , 2006, Icarus, 182, 23;
Jacobson R.A., Haw R.J., McElrath T.P., Antreasian P.G., 2000, J. Astronaut. Sci., 48, 495;
Jacobson R.A., \etal , 2006, AJ, 132, 2520
}}\\

\subsection{Fundamental physics}
\label{fuphy}
\subsubsection{Fundamental physics with weakly interacting particles {\scriptsize [A. Lobanov]}}

\noi Modern physics finds itself on the brink of a whole new understanding
of the most fundamental laws governing particles and forces in the
Universe. Investigations of dark matter, dark energy, and new
particles span across many fields of research, from accelerator
experiments, to astroparticle physics studies, and astrophysical
observations.

\smallskip

\noi Most extensions of the standard model of particle physics predict the
existence of a ``hidden'' sector of particles which interact only
weakly with the visible sector particles (standard model particles):
with the weakly interacting massive particles (WIMP, with masses $m
\gtrsim 100$\,GeV; e.g. neutralinos with masses in the GeV to
TeV range) and ultralight weakly interacting sub-eV particles (WISP;
e.g. axions, axion-like particles and hidden photons with
masses in the sub-eV range) positioned as the most promising
candidates for the dark matter (DM) and dark energy (DE) particles.

\smallskip

\noi Providing observational evidence for WIMP and WISP or setting limits on
their physical properties are of paramount importance for modern
cosmology, particle physics, and fundamental physics --- and the SKA will
bring an enormous impact on this field.  Dedicated SKA studies will
provide an excellent potential for detecting directly the electromagnetic
signatures of neutralinos, axions, and hidden photons, and for
exploring truly unique areas of the parameter space for each of these
particles.\\

\noi {\bfseries\boldmath\small WIMP neutralinos.}~Recent observations of excess cosmic ray
$e^{\pm}$ flux (cf., Aharonian et al. 2008, Abdo et al. 2009) provide
compelling arguments for decaying or self-annihilating dark matter and
put forth WIMP neutralinos as leading dark matter candidates.  For
both hadronic and leptonic final states of neutralino annihilation,
electromagnetic (EM) signatures are expected to be detectable in the
high-energy regime through inverse-Compton emission and in the radio
band through synchrotron emission from $e^{\pm}$ annihilation products
(Colafrancesco et al. 2007).

\smallskip

\noi The annihilation signal is expected to be very compact in the
high-energy bands, while in the radio band it should manifest itself
as a smooth and extended halo.  This brings a unique advantage for
radio observations with the SKA which would be capable of resolving
spatially the annihilation signal, hence obtaining much
more strignent constraints on mass distribution in DM halos and clumps.

\smallskip

\noi SKA searches for the EM signature of DM annihilation can be made in a
number of astrophysical objects, including the Galactic Centre,
globular clusters, diffuse galactic emission, nearby dwarf spheroidal
and spiral galaxies, and in galaxy clusters. Dwarf spheroidal (dSph)
galaxies feature the largest mass-to-light ratios of all these objects
(indicative of being dominated by dark matter). The dSph galaxies also
have the smallest contamination by astrophysical sources (star
formation, SNR, interstellar gas), making them attractive targets for
DM searches.  Due to an almost two order of magnitude sensitivty
improvement provided by the SKA, even non-detection of the
annihilation signal will enable a substantial improvement of the
limits on the annihilation cross-section and ruling out particular
classes of the DM halo models.

\smallskip

\noi An estimate made for arguably the most difficult case (a globular
cluster with $M_\mathrm{cl} = 10^{5}\,M_\odot$,
$M_\mathrm{cl}/L_\mathrm{cl} = 5$, $r_\mathrm{cl} = 10$\,pc located at
$d_\mathrm{cl} = 10$\,kpc) yields a total 1\,GHz flux density of $\sim
50$\,mJy distributed over a solid angle of $\sim$\,0.1\,degrees
(assuming a magnetic field $B = 1\,\mu$G, a neutralino mass
$m_{\chi}=100$\,GeV and an annihilation cross-section $\langle\sigma_a
v\rangle = 3 \times 10^{-23}\, \mathrm{cm}^3\, \mathrm{s}^{-1}$.  For
the compact core of the SKA, this yields a peak brightness of
$\sim$\,2\,mJy/beam.  With the imaging sensitivity of the SKA reaching a $\sim
$\,1\,$\mu$Jy/beam, the DM signal could therefore be detectable even for
much smaller (and more physically plausible) values of $\langle\sigma_a
v\rangle$.

\bigskip
\bigskip

\parbox{\textwidth}{

\hspace{-0.7cm}
\parbox{5cm}{\includegraphics[width=0.45\textwidth]{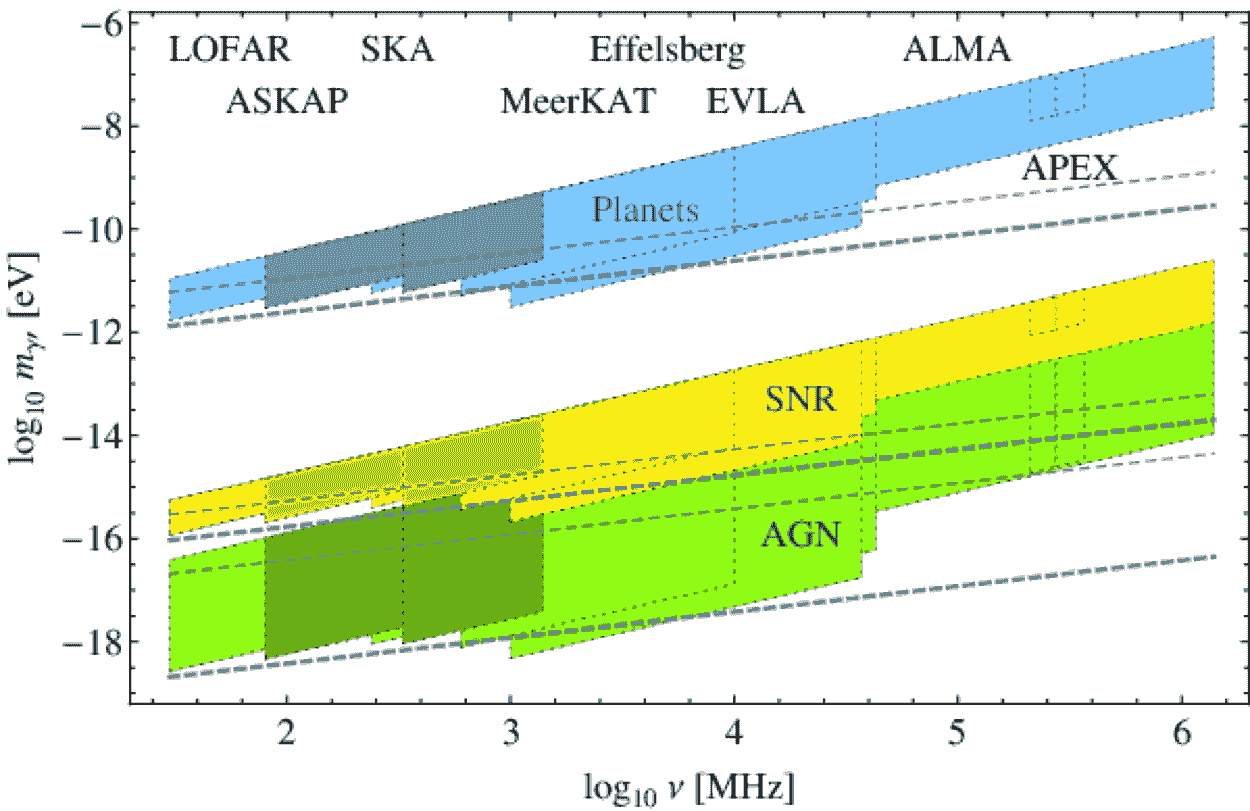}\includegraphics[width=0.54\textwidth]{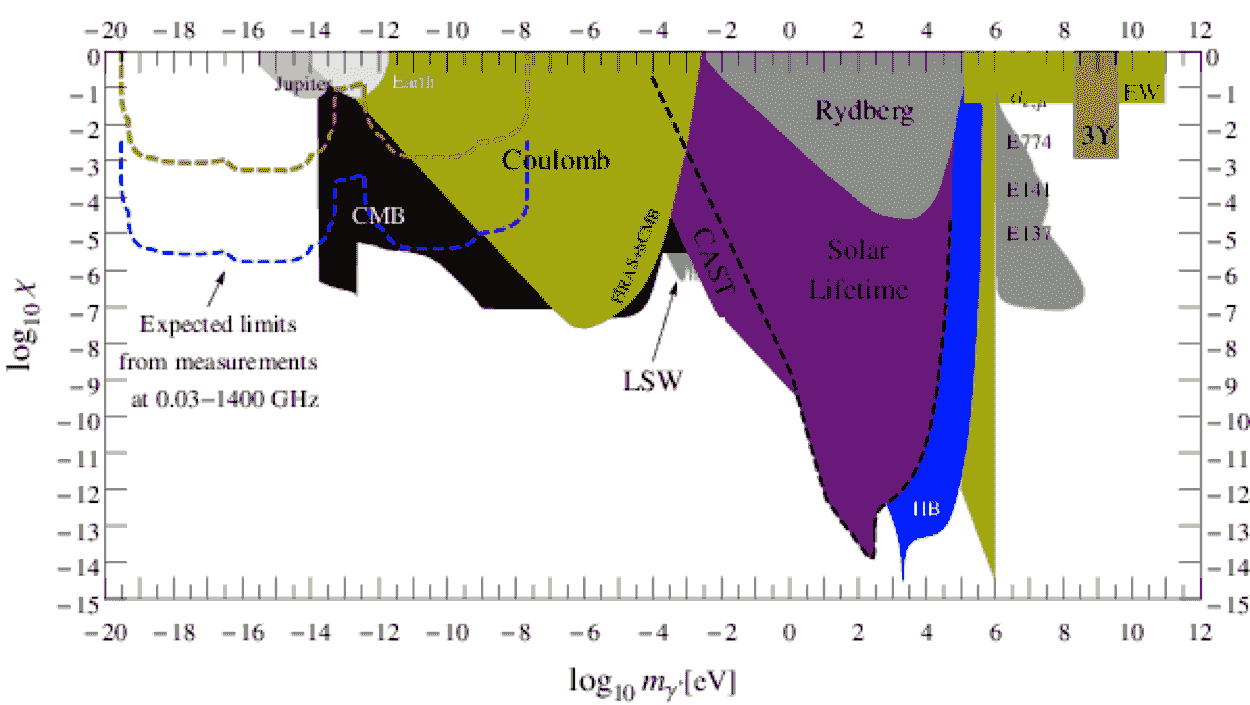}\vspace{0.5cm}}

\hspace{0.2cm}
\parbox{0.9\textwidth}{Figure 1$_{\rm Lo}$:~\small {\it left:}~Ranges of hidden photon mass,
$m_{\gamma^{\prime}}$, that can be probed by different radio
astronomical instruments at different observing frequencies and with
different targets. For each colour, darker shades mark the mass ranges
accessible to the SKA. The calculations assume typical instrumental setups
and generic ranges of distances to planets (0.5\,--\,10\,au), supernova
remnants (1\,--\,10\,kpc) and active galaxies (0.02\,--\,3\,Gpc). For each
group of targets, dotted lines show minimum detectable
$m_{\gamma^{\prime}}$ as a function of highest observing frequency and
distance to a given target, ranging from the minimum (thin lines) to
the maximum (thick lines) distances used in the calculation. Actual
sensitivity of measuerements and resulting bounds on $\chi$ may vary
substantially between different instruments and different observing
bands. ~{\it right:}~Ranges of hidden photon mass $m_{\gamma^{\prime}}$ and
kinetic mixing parameter $\chi$ excluded by various experiments. The
best cumulative bounds expected to be obtained from observations in
the radio regime are shown with dashed-line curves for single object
(red) and stacked (blue) observations of a number of objects. The SKA
measurements provide the best bounds for hidden photon with masses of
$10^{-10}$\,--\,$10^{-12}$\,eV and below $10^{-14}$\,eV (frame adapted
from Jaeckel \& Ringwald 2010).
}
\vspace{0.7cm}
}\\

\noi {\bfseries\boldmath\small WISP axions/ALPs and hidden photons.}~Compelling evidence for
existence of ultralight (sub-eV) particles arises from a number of
recent experiments, with axions and axions-like particles (ALPs;
Raffelt \& Stodolsky 1988, Jaeckel et al. 2007), and massless or
light, hidden {\em U}(1) gauge bosons (hidden photons; Okun 1982,
Jaeckel \& Ringwald 2010) positioned as plausible candidates.

\smallskip

\noi A fundamental feature predicted for WISP is the possibility of their
mixing with normal photons.  Following this prediction, photons should
oscillate between their normal state ($\gamma$) and a ``hidden'' state
($\gamma^{\prime}$) in which they propagate along time-like geodesics
but do not interact with any normal matter (hence properties of a
hidden photon can be completely described by its mass
$m_{\gamma^{\prime}}$ and the kinetic mixing parameter, or mixing
angle, $\chi$; with $\chi \ll 1$ expected). The axions can convert to
photons in the presence of an ambient magnetic field, with the
conversion strength determined by the axion-photon coupling constang
$g_\mathrm{a}$.  Experiments and measurements performed so far have
yielded bounds on the electro-magnetic coupling for a range of axion
masses extending down to $m_{\mathrm{a}} = 1\times 10^{-15}$\,eV
(Arias et al. 2010) and hidden photon masses down to
$m_{\gamma^{\prime}} = 2\times 10^{-14}$\,eV (Redondo 2010).

\smallskip

\noi Radio astronomy measurements made with the SKA will extend axion searches
to masses below $\sim$\,10$^{-9}$\,eV and probe the axion coupling
strength down to $g_{\mathrm{a}} = 10^{-14}$\,GeV$^{-1}$ (cf., Chelouche
et al. 2008). This will expand the experimentally probed range of
parameters by about five orders of magnitude.  

\smallskip

\noi For hidden photons in particular, SKA
observations at frequencies below 3\,GHz will offer an excellent (if
not unique) tool for placing bounds on $\chi$ for $m_{\gamma^{\prime}}
< 10^{-14}$\,eV.  Oscillations of flux density induced by the hidden
photons and occurring at fractional frequency intervals $\Delta
\nu/\nu = 2\,\nu\, m_{\gamma^{\prime}}^{-2} L^{-1}$ will enable detecting
kinetic mixing with $\chi \ge \sigma_\mathrm{rms}^{1/2}$ up to a
distance $L_\mathrm{m} = 2 (\nu\, \Delta \nu)^{1/2}
m_{\gamma^{\prime}}^{-2} \sigma_\mathrm{rms}^{-1}$.  For radio
observations made with a spectral resolution $\Delta\nu/\nu$ and
reaching $\sigma_\mathrm{rms}$, this implies $\chi(\nu) =
(\nu\,\Delta\nu)^{1/4} m_{\gamma^{\prime}}^{-1} L^{-1/2}$.
Improvements of $\sigma_\mathrm{rms}$ (factor of $\sim$\,100) and
$\Delta\nu/\nu$ (factor of $\sim$\,100) that will be provided by the
SKA, as well as the extension (factor of $\sim$\,10) to lower
frequencies, will enable exploring a new range of
$m_{\gamma^{\prime}}$ below $5\times 10^{-16}$\,eV and obtaining
substantially better bounds on the kinetic mixing of hidden photons
with masses below $10^{-14}$\,eV and (as illustrated in
the figure; Lobanov et al. in prep.).\\

\parbox{0.9\textwidth}{
\noi{References:}\\
\noi{\scriptsize
Abdo  A.A. et al., 2009, Phys. Rev. Lett., 102, 181101; 
Aharonian F., et al., 2008, Phys. Rev. Lett., 101, 261104; 
Arias P. et al., 2010, Phys. Rev. D, 82, 5018; 
Chelouche D., et al., 2009, ApJS, 180, 1;
Colafrancesco S. et al., 2007, Phys. Rev. D, 75, 023513; 
Jaeckel  J. et al., 2007, Phys. Rev. D, 75, 013004;
Jaeckel  J., Ringwald A., 2010, ARNPS, 60, 405; 
Okun  L.B., 1982, Sov. Phys. JETP, 56, 502; 
Raffelt  G., Stodolsky L., 1988, Phys. Rev. D, 37, 1237;
Redondo J., 2010, arXiv:1002.0447
}}\\

\subsubsection{Measuring neutrino masses and primordial non-Gaussianity
  {\scriptsize [J. Niemeyer]}}

The statistical distribution of cosmological structures contains a
wealth of information about the history, geometry, and matter
inventory of the Universe. It also serves as a valuable probe of the
laws of gravity and the physics of inflation during which, according
to currently favored theories, the primordial fluctuations were
produced. Much of our knowledge to date has been gained from
observations of the cosmic  microwave background (CMB; e.g Komatsu
et al. 2008), combined with galaxy redshift surveys (e.g. Tegmark et
al. 2004) and investigations of the Lyman-alpha forest (e.g.
McDonald et al. 2006). There are plausible reasons to believe that
future radio surveys of neutral hydrogen (\hi ) emission, especially in
combination with planned optical or near-IR redshift surveys (such as
BOSS, BigBOSS, HETDEX, Euclid, etc.), will be an ideal probe of
fundamental physics during the next two decades. Apart from sharpening
constraints on the time dependence of dark energy using BAO features
in the cosmological power spectrum (see Section\ 8.1.1 and 8.1.2), measuring the neutrino mass and
determining the extent of primordial non-Gaussianities are among the
most prominent goals for future cosmological precision measurements.

\noi Cosmological redshift surveys are ultimately limited by sampling
variance, i.e. the number of independent modes accessible for
statistical analysis. A non-thresholded \hi -survey, or \hi\ intensity
mapping, by the SKA can probe a 3-d volume instead of the 2-d surface
seen in the CMB, and it can extend to redshifts z\,$\sim$\,6 far
exceeding those of optical or IR galaxy surveys (z\,$\lesssim$\,3\,--\,4). Hence, the uncertainties in determining the power spectrum can
theoretically become as low as $O(10^{-4})$ (Loeb \& Wyithe 2008,
Rawlings 2011), i.e. more than an order of magnitude improvement over
current limits. Furthermore, at redshifts z\,$\lesssim 1$, \hi -surveys
are expected to detect a sufficiently large number density of emitters
to measure the power spectrum with negligible shot noise errors
(Abdalla et al. 2010). These forecasts rely on estimates of the
evolution of the \hi\ mass fraction in galaxies obtained from
semi-analytic models (Marin et al. 2010, Power et al. 2010, Kim et
al. 2011) or hydrodynamical cosmological simulations (Duffy et
al. 2011a). Observations of the SKA  pathfinder programmes, combined
with advances in theoretical modeling and available computing power,
are likely to produce substantial progress in our understanding of the
systematic uncertainties of the cosmological evolution of neutral
hydrogen before the onset of SKA operations (Duffy et al. 2011b).

\noi The effect of neutrinos on the power spectrum is well understood:
neutrino free streaming produces a mild suppression of the power
spectrum on small scales, depending on the time when the neutrinos
became non-relativistic and on their total energy density. The effect
is difficult to predict precisely at low redshifts, since the scales
whose growth is suppressed are in the weakly to fully nonlinear regime
where the power spectrum cannot be reliably computed by perturbation
theory. Full N-body simulations that explicitly include neutrinos as
an additional particle component are still extremely challenging
(Brandbyge et al. 2008). Current constraints from cosmological data
yield a sum over neutrino masses of about 0.3\,eV (Thomas et
al. 2010), while experimental bounds from neutrino oscillations show
that at least one neutrino has a mass of at least $\sim$\,0.05\,eV. An
SKA \hi -galaxy survey out to z\,$\sim\,2$ combined with CMB data
would be sensitive to the entire allowed mass range at the 3\,$\sigma$
level, and it would allow measuring the number of massive neutrinos
(and hence distinguish between a normal and inverted hierarchy) down
to 0.25\,eV (Abdalla \& Rawlings 2007). At higher redshifts,
z\,$\sim$\,2\,--\,3, most of the scales affected by neutrinos are
still in the linear or weakly nonlinear regime accessible to
perturbative calculations; here, cross-correlation of SKA observations
with optical surveys such as HETDEX could yield a signal-to-noise
ration of $\sim$\,300 at z\,=\,2 (Rawlings 2011). Determining the
power spectrum at z\,$>$\,2 is also interesting for a cleaner
separation of curvature and dark energy, which becomes dominant at
lower redshifts.

\noi The standard assumption for the statistics of primordial perturbations
is a Gaussian distribution, i.e. the power spectrum contains the
entire available information. It is well justified for the simplest
(single field, slow roll) models of inflation which predict an
unmeasurable level of primordial non-Gaussianity (not to be confused
with non-Gaussianities produced by gravitational collapse in the
nonlinear regime). A clear detection of primordial non-Gaussianity
could therefore  falsify an entire class of inflationary
models. Furthermore, different classes of extensions beyond the
simplest model produce unique signatures in the bispectrum which in
principle allow to distinguish between e.g. multi-field, DBI, or
excited initial state scenarios. It has recently been shown that
primordial non-Gaussianity gives rise to a scale-dependent bias in the
galaxy power  spectrum that is strongest on large scales, i.e. outside
of the nonlinear regime (Dalal et al. 2008). Using this effect, galaxy
surveys have a potential to constrain $f_{\rm nl}$ (the commonly used
dimensionless amplitude of primordial non-Gaussianity of the local
type) that is already competitive with bounds  from the CMB (Slosar et
al. 2008). Current 1\,$\sigma$ bounds on $f_{\rm nl}$ are of the order
of 25. Large galaxy redshift surveys may become sensitive to $f_{\rm
nl}$ of order unity by optimally using the information in the galaxy
power spectrum and bispectrum (Jeong \& Komatsu 2009). Again, for \hi\
surveys with the SKA  an improved observational and theoretical
understanding of the clustering properties of \hi\ emission at different
redshifts will be  crucial. Roughly speaking, the effects of
primordial non-Gaussianity are strongest for the  most strongly biased
tracers of the density field. The current estimates for the bias of
\hi -emitting galaxies of $b \sim 1.5$ at z\,$\sim$\,2 (Kim et al. 2011)
make it seem plausible that the SKA can significantly contribute to
the search for primordial non-Gaussianity. Moreover, combining
multiple sets of differently biased tracers, such as BigBOSS or Euclid
and SKA galaxies, further enhances the sensitivity by canceling the
sampling variance (Seljak 2009). 

\noi Neutrino properties and primordial non-Gaussianities are only two
prominent examples from fundamental physics that can be explored with
the SKA. Aside from dark energy, further
fundamental questions that are potentially accessible include
modifications of General Relativity measured by the cosmological
growth rate (e.g. Daniel \& Linder 2010) or other forms of warm dark
matter (Viel et al. 2011). The degree to which the SKA will contribute
to the determination of these parameters will depend on its final
realisation. In the optimistic case that an all sky \hi\ ``billion
galaxy'' survey will become feasible, its constraining power will be
competitive with, or even superior to, other planned 
cosmology probes, e.g. the Euclid mission (Myers et al. 2009). \\

\parbox{0.9\textwidth}{
\noi{References:}\\
\noi{\scriptsize
Abdalla F., Rawlings S., 2007, \mnras , 381, 1313;
Abdalla F., Rawlings S. , 2010, 2010, \mnras , 401, 743;
Brandbyge J., Hannestad S., {Haugb{\o}lle} T., Thomsen B., 2008,
J. Cosmology Astropart. Phys. 8:20;
Dalal N., Dor{\'e}  O., Huterer  D., Shirokov A., 2008,
Phys. Rev. D., 77(12);
Daniel S.F., Linder E.V., 2010, Phys. Rev. D., 82(10);
Duffy  A.R., \etal , 2012, \mnras , 420, 2799;
Duffy  A.R., Moss A., Staveley-Smith, 2011, arXiv:1103.3944;
Jeong D., Komatsu E,, 2009, \apj , 703, 1230;
Kim  H.-S., \etal , 2011, \mnras , 414, 2367;
Komatsu E., \etal , 2008, \apjs , 192, 18;
Loeb A., Wyithe J.S.B., 2008, Phys. Rev. Letters, 100(16);
Marin  F.A., Gnedin N.Y., Seo H.-J., Vallinotto, A., 2010, \apj , 718,
972;
McDonald P., \etal , 2006, \apjs , 163, 80;
Myers S.T., \etal , 2009, arXiv:0903.0615;
Power C., Baugh C.M., Lacey C.G., 2010,  \mnras  , 406, 43;
Rawlings S., 2011, arXiv:1105.6333;
Seljak U., 2009, Phys. Rev. Letters, 102(2);
Slosar A., \etal , 2008, J. Cosmology Astropart. Phys. 8:31;
Tegmark \etal , 2004, Phys. Rev. D. 69(10);
Thomas S.A., Abdalla F.B., Lahav O., 2010, Phys. Rev. Letters, 105(3);
Viel M., Markovi{\v c} K., Baldi M.,Weller J., 2012, \mnras ,  421, 50
}}\\

\subsubsection{Ultra-High-Energy particles: Neutrinos and cosmic rays  {\scriptsize [C. James]}}

\noi {\bfseries\boldmath\small Introduction:~~}The mystery of what is producing the
highest-energy particles in nature, the
ultra-high-energy\footnote{Commonly, UHE is used to refer to events
  above 10$^{18}$\,eV. Here, the main region of interest is at and
  above approximately 5~10$^{19}$\,eV, where the GZK mechanism is
  expected to kick-in for protons, and the Pierre Auger observatory
  sees the greatest statistical anisotropy (The Pierre Auger
  Collaboration 2007).} cosmic rays, remains unsolved. These extreme
particles impact the Earth only once per km$^2$ per century, and
detectors even larger than the 3000\,km$^2$ Pierre Auger experiment are
required to gather enough statistics to determine their origin (The
Pierre Auger Collaboration 2007). UHE neutrinos -- the long-predicted
by-products of UHE cosmic-ray acceleration and propagation -- are also
thought to hold the key. As-yet unobserved, these elusive particles
also require the advent of a larger detector to allow their discovery,
and to shed light on the UHE cosmic-ray mystery.

\noi By using the ``Lunar technique'' (Askar'yan 1962, Askar'yan 1965,
Dagkesamanskii \& Zheleznykh 1989) the SKA promises to do both. The
Lunar technique utilises ground-based radio-telescopes observing the
Moon to search for UHE particle interactions in the Moon's outer
layers. Via the Askaryan effect (Askar'yan 1962, Askar'yan 1965), the
interactions of these particles will produce coherent pulses of
radio-wave radiation, which can escape the Moon if the interaction is
near the surface. Thus the entire visible surface of the Moon can be
used as a 10\,000\,000\,km$^2$
particle detector. \\

\noi {\bfseries\boldmath\small The Benefit of the SKA:~~}Several previous attempts
have refined Lunar the technique, but so far no observations have
successfully identified a UHE particle signal. This is because only a
telescope with the raw sensitivity of the SKA -- here, the figure of
merit is the product of collecting area and instantaneous bandwidth -
will be able to detect the known flux of UHE cosmic rays and beyond,
and test models of the flux of ``cosmogenic'' neutrinos (James \&
Protheroe 2008).

\noi The size of Lunar cascades (of order 10\,cm in width and 3\,m in length,
Alvarez-Muniz \& Zas 1998) means that signal coherency can extend up to a few GHz, but for
typical geometries, the coherency will extend to a maximum frequency
$\nu_{\rm max}$ of only a few hundred MHz or less due to absorption and
decoherence effects. Given that peak signal power increases as $\nu_{\rm max}$,
high frequencies are best suited to detecting a larger flux of lower
energy particles, and low frequencies to a rarer flux at higher
energies. Since the UHE neutrino flux, and the UHE cosmic-ray particle
flux above few 10$^{20}$\,eV, is currently unknown, the broad frequency
coverage of the SKA is critical to maximising discovery
potential.

\noi As well as increasing the sensitivity, the large
instantaneous bandwidth of the SKA will enable the detection of
spectral downturns in detected pulses, which can then be used to gauge
the observer's angle to the shower axis. The long SKA baselines will
also be able to pinpoint the position of signal origin on the Lunar
limb. Combined with the observed signal polarisation, this information
can be used to reconstruct the nature and arrival direction of the
primary particle, and the cascade energy  (James \& Protheroe
2009). These properties are
also critical for the confidant rejection of radio-frequency
interference (RFI). Taken together, this will enable the SKA not 
only to be a UHE particle detector, but also to perform true UHE particle
astronomy.\\

\noi {\bfseries\boldmath\small Detection Technique:~~}Observing nanosecond-scale
pulses with a giant radio array however requires novel
techniques. Only tied-array beams (coherent addition of signals from
each element) at full time resolution will achieve the necessary
sensitivity to detect signals coming from the Moon. Each of these
beams must be ``de-dispersed'' to account for the effects of the
Earth's ionosphere. Due to computational constraints, only the inner
part of the array will be used to form real-time detection beams:
small segments of data from the outer stations will be returned upon
the detection of a candidate event. This detection paradigm is
currently being developed by the NuMoon collaboration for LOFAR's
``UHEP'' mode (Singh \etal\ 2012), and the LUNASKA collaboration using
Parkes and ATCA (James \etal\ 2010).  However, enabling the SKA to
work in such a unique observation mode poses one of the great
challenges of this technique.\\

\noi {\bfseries\boldmath\small Summary:~~}Using the Lunar technique, the SKA will be
a superb instrument of UHE particle astronomy. If the full SKA can be
utilised, the known flux of UHE CR will be observable (James \etal\
2011, ter Veen \etal\ 2010) with an event rate at least ten times
greater than for Pierre Auger South (Dagkesamanskii \& Zheleznykh
1989). Lunar observations would be a significant undertaking for the
SKA. If the more pessimistic predictions of a cosmogenic neutrino flux
prove true, then of order 2000\,hours of observations may be required
to detect perhaps only a handful of neutrino events. The likely
payoff?  Solving the UHE cosmic-ray mystery, and the opening of a new
high-energy window on the
Universe.\\

\parbox{0.9\textwidth}{
\noi{References:}\\
\noi{\scriptsize Alvarez-Muniz J., Zas E., 1998, Phys.~Lett. B, 434, 396;
Askar'yan G.A., 1962, Sov.~Phys. ~JETP, 14, 441; 
Askar'yan G.A., 1965, Sov.~Phys. ~JETP, 48, 1988;
Dagkesamanskii R.D., Zheleznykh  I. M., 1989, Sov.~Phys.~ JETP Let.~50, 233;
James C.W. , Protheroe R.J., 2008, Astropart.~Phys., 30, 318;
James C.W., Protheroe R.J., 2009, Astropart.~Phys., 31, 392;
James C.W.,  et al., 2010, Phys. Rev. D, 81, 042003;
James C.W., Falcke H., Huege T., Ludwig M., 2011, Phys. Rev. E., 84, 056602;
Singh K., et al., 2012, Nuclear Instruments and Methods in Physics
Research Section A, 1, 171;
ter Veen S., et al., 2010, Phys. Rev. D, 82, 103014 M1;
The Pierre Auger Collaboration, 2007, Science, 318, 938
}}\\

\subsubsection{Neutron star science with combining radio and X-ray data  {\scriptsize [W. Becker]}}

Neutron stars represent unique astrophysical laboratories which allow
us to explore the properties of matter under the most extreme
conditions observable in nature (although black holes are even more
compact than neutron stars, they can only be observed through the
interaction with their surroundings). Studying neutron stars is
therefore an interdisciplinary field, where astronomers and
astrophysicists work together with a broad community of
physicists. Particle, nuclear and solid-state physicists are strongly
interested in the internal structure of neutron stars which is
determined by the behavior of matter at densities above the nuclear
density $\rho_{\rm nuc} = 2.8\times 10^{14} \mbox{g cm}^{-3}$.

Neutron stars are observable as pulsars, i.e. rapidly spinning,
strongly magnetized neutron stars which are radiating at the expense
of their rotational energy. With some more basic assumptions
(cf. Becker 2009) this allows one to estimate a neutron star's age by
measuring its period and period derivative. Knowing the age of this
objects then supports to study all kinds of evolutionary effects like
the thermal evolution of neutron stars.

Neutron stars are formed at very high temperatures of $\sim$\,10$^{11}$\,K,
in the imploding cores of supernova explosions. Much of the initial
thermal energy is radiated away from the interior of the star by
various processes of neutrino emission (mainly, Urca processes and
neutrino bremsstrahlung), leaving a one-day-old neutron star with an
internal temperature of about $10^9$\,--\,10$^{10}$ K.  After $\sim$\,100\,yrs (typical time of thermal relaxation), the star's interior
(densities $\rho>10^{10}$\,g\,cm$^{-3}$) becomes nearly isothermal,
and the energy balance of the cooling neutron star is determined by
the following equation:

\[
   C(T_i)\,\frac{{\rm d}\,T_i}{{\rm d}\,t} = - L_\nu(T_i) - L_\gamma(T_s)
   + \sum_k H_k ,
\]

\noi where $T_i$ and $T_s$ are the internal and surface temperatures,
$C(T_i)$ is the heat capacity of the neutron star (cf.\ Becker
2009). Neutron star cooling thus means a decrease of thermal energy,
which is mainly stored in the stellar core, due to energy loss by
neutrinos from the interior ($ L_\nu=\int Q_\nu\,{\rm d}V$, $Q_\nu$ is
the neutrino emissivity) plus energy loss by thermal photons from the
surface ($L_\gamma=4\pi R^2 \sigma T_s^4$).  The relationship between
$T_s$ and $T_i$ is determined by the thermal insulation of the outer
envelope ($\rho<10^{10}$\,g\,cm$^{-3}$), where the temperature
gradient is formed.  The cooling rate might be reduced by heating
mechanisms $H_k$, like frictional heating of superfluid neutrons in
the inner neutron star crust or some exothermal nuclear reactions.

\medskip

\noi The fact that the thermal evolution of neutron stars is very sensitive
to the composition and structure of their interiors, in particular, to
the equation of state at super-nuclear densities means that measuring
surface temperatures of neutron stars is an important tool to study
super-dense nuclear matter. Since typical temperatures of such neutron
stars correspond to the extreme UV -- soft X-ray range, the thermal
radiation from cooling neutron stars can be observed best with X-ray
detectors sufficiently sensitive at $E \lessapprox 1$\,keV.

\parbox{\textwidth}{

\hspace{-0.5cm}
\parbox{5cm}{ \includegraphics[scale=0.35]{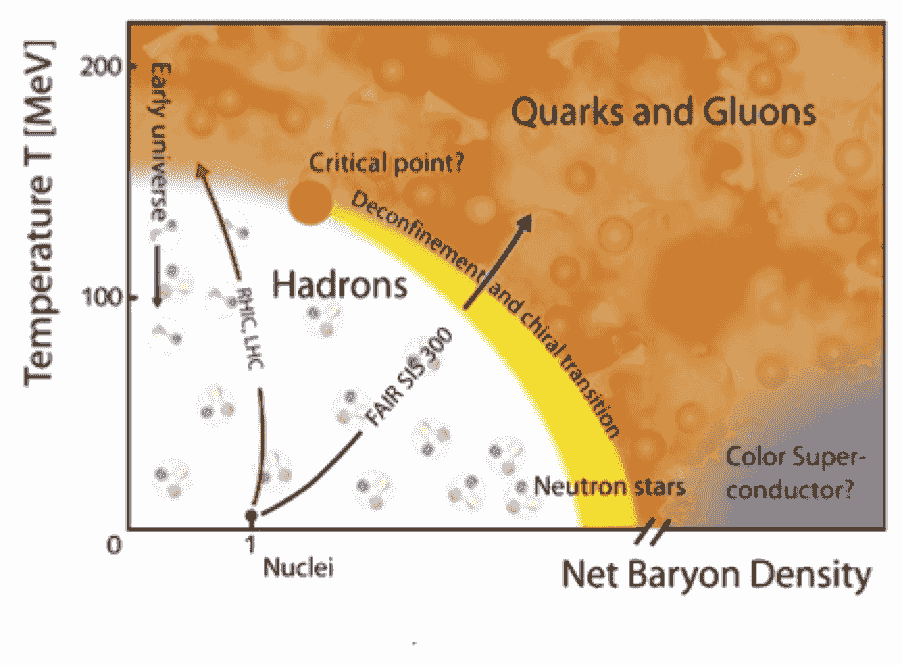}\vspace{-8.0cm}}

\hspace{10.8cm}
\parbox{6cm}{Figure 1$_{\rm Beck}$: Density vs. Temperature of cold condensed matter. Neutron stars probe the low temperature high density
                   region of the QCD phase diagram. Fundamental questions 
                   in physics like: what is the equation of state of matter at 
                   sub-nuclear density?, what is the nuclear interaction potential 
                   of nuclei at this density? And whether Quarks are able to maintain 
                   colour superconductivity can be answered by exploring this density 
                   region in e.g. studying the thermal evolution of neutron stars.
                   Todays laboratory experiments (LHC, RHIC, FAIR) will not reach this density
                   level at this low temperature.

}
}\\

\bigskip
\bigskip

\noi Of the 2000 radio pulsars known today only about 1\,--\,2\,\% turn out
to be suitable to study cooling effects using X-ray telescopes. This
is mostly because pulsars which are best suited to study cooling
effects should be nearby and in the age bracket between (0.5\,--\,1)\ 10$^6$\,yrs.  With a telescope like the SKA the number of radio pulsars
expected to be found is about 15 times the number of pulsars detected
today. Among them will be a large fraction of middle-age pulsars well
suited to study their thermal evolution.  The SKA thus will help to
enlarge the sample of cooling neutron stars by a significant factor
and thus improving todays results which are limited by small number
statistics of the detected objects.

\noi Pulsar research also has its application in autonomous spacecraft
navigation, by making use of the pulsar's characteristic timing
information (cf. Bernhard et al. 2011 and references
therein). Rotation-powered pulsars, especially millisecond pulsars,
are particularly interesting for two reasons: First, they provide
characteristic temporal signatures that can be used as natural
navigation beacons; second, the stability of their rotation
frequencies is comparable to, or even better than, the timing
stability of atomic clocks, which is most important because the
temporal resolution is the limiting factor for the performance of this
navigation technique.

\noi A pulsar-based navigation system in principle can be designed for any
energy band of the electromagnetic spectrum, but X-rays are preferable
for several reasons: Whereas radio pulsars are usually very faint and,
therefore, are observed by large radio antennas with diameters of
typically 50 to 100 metres, X-ray telescopes can be built relatively
compactly.  Furthermore, radio waves are subject to interstellar
dispersion, an effect that degrades the temporal resolution of pulsar
data due to pulse smearing. In contrast, X-ray propagation through the
interstellar medium does not affect pulsar timing.

\noi A sensitive spacecraft navigation system based on the X-rays
observation of millisecond pulsars relies on having suitable pulsars
in the field of view, regardless of the direction the instrument is
pointed to in the sky. Currently, however, only about 10\,\% of the
radio pulsars are millisecond pulsars, giving tight constrains to this
navigation approach. The SKA is supposed to increase the number of
radio millisecond pulsars by a factor of more than 10 compared with
the number detected today. This in turn will allow more millisecond
pulsars to be detected in the X-ray domain which will be of great
advantage for using these sources as natural navigation beacons in
space craft navigation.\\

\noi In view of the observational capabilities previous instruments
provided so far, and the intense neutron star research made over a
period of more than 42\,years, there are still fundamental questions
which have not be answered. Questions like ``How are the different
manifestations of neutron stars related to each other?'', ``What are
the physical parameters which differentiate AXPs/SGRs/CCOs/XDINs and
rotation-powered pulsars?'', ``What is the maximal upper bound for the
neutron star mass and what is the range of possible neutron star
radii?'', ``Is there any exotic matter in neutron stars?'', ``Do
strange stars exist?'', and ``What are the physical processes
responsible for the pulsars' broad band emission observed from the
infrared to the gamma-ray band?'' are just the most striking ones.
They all can be addresses with multi-wavelengths observations using
telescopes like the SKA in the radio and future X-ray observatories
in the high-energy band. New observatories are supposed to bring a
major improvement in sensitivity, making them even more suitable for
pulsar and neutron star astronomy than the instruments we have
available today.\\

\parbox{0.9\textwidth}{
\noi{References:}\\
\noi{\scriptsize Becker, W., 2009, Astrophysics and Space 
          Science Library, Vol.357, Springer; Bernhardt, M.G., Becker, W., Prinz, T., Breithuth, F.M., Walter, U. 2011, 
          ``Autonomous Spacecraft Navigation Based on Pulsar Timing Information''
          in IEEE proceedings of 2nd International Conference on Space
          Technology", Athens, Greece, 2011, Editors M. Petrou,
          M. Gargalakos, N. Uzunoglu, ISBN 9 7811457 718724}}\\

\subsubsection{Testing theories of gravity with binary pulsars and
  black holes  {\scriptsize [M. Kramer, N. Wex, K. Liu, G.~Sch\"afer]}}

\noi Previous experience has proven that finding a large number of new
pulsars will inevitably lead to the discovery of rare objects which
push our understanding of their formation or which can be used as
unique laboratories for fundamental physics. Currently, we know of
about 2000 radio pulsars of which about half were discovered in a
single survey, i.e.~the Parkes Multibeam Survey for Pulsars
(Manchester \etal\ 2001).  In comparison, the SKA's combination of
sensitivity and field-of-view will provide a Galactic census of
pulsars, which will essentially include every pulsar (visible from the
SKA location) that is beaming towards Earth. The 20\,000 to 30\,000 pulsars 
to be discovered should join the about 2000 millisecond pulsars 
(most of which will be in binaries), about 100 relativistic binaries and 
eventually the rare objects like pulsar-black hole systems (Kramer \etal\ 2004, 
Smits \etal\ 2009).

\noi In binary pulsars hosting a pulsar and a compact companion, we
essentially encounter two gravitational test masses of which (at
least) one is fitted with a precise cosmic clock. By tracing the
motion of this clock in the curved spacetime of its companion, we can
test the predictions of general relativity (GR) and alternative
theories of gravity. The advantage of using binary pulsars lies in the
combination of having access to a system of compact masses that only
interact gravitationally, so that their motions should be fully
explained by the theory of gravity to be tested, and the ranging
capabilities of a pulsar timing experiment.  As a result, binary
pulsars already provide the best tests of theories of gravity for
strongly self-gravitating bodies (Kramer \etal\ 2006).

\noi With the full SKA, we can expect an improvement in timing precision by
a factor of $\sim$\,100. Consequently, we expect that tests of theories
of gravity with binary pulsars will greatly surpass the precision of 
current gravity tests. Concerning the known systems, in particular the 
continued observations of the Double Pulsar will derive important constraints 
for testing alternative theories of gravity and the validity of specific concepts
in strong gravitational fields (Kramer \& Wex 2009). Moreover, this improvement 
in precision will for many of the relativistic effects allow to probe higher 
order and spin contributions predicted by GR (e.g.\ Damour \& Sch\"afer 1988; 
Blanchet \& Sch\"afer 1989).

Most importantly, however, with the SKA we will be able to probe the properties 
of black holes and compare those to the prediction of GR for Kerr black holes. 
With pulsars orbiting the super-massive black hole in the Galactic centre and the
discovery of binary pulsars with stellar-mass black hole companions, we will be 
able to measure the mass, spin and quadrupole moment of the black holes. These 
measurement will allow us to test the {\em cosmic censorship conjecture} as well 
as the {\em no-hair theorem} (Wex \& Kopeikin 1999, Kramer \etal\ 2004).  

\noi The cosmic censorship conjecture states that every
astrophysical black hole, which is expected to rotate, has within GR an event
horizon that prevents us from looking into the central
singularity. However, the event horizon disappears for a given value
of the black hole spin, so that we expect the measured spin to be
below the maximum allowed value. The no-hair theorem makes the
powerful statement that the black hole has lost all features of its
progenitor object, and that all black hole properties are determined
by only the mass and the spin (and possible charge). Therefore, if the
no-hair theorem is valid, the expected quadrupole moment of the black
hole can be uniquely determined from the mass and the spin. With a
measurement of all three quantities, this theorem can be tested
(Kramer \etal\ 2004, Liu 2012).

\noi With the chance to perform these experiments with the super-massive
black hole in the Galactic centre, stellar black holes and, possibly,
intermediate mass black holes in globular clusters, a whole black hole
mass range can be studied. As already pointed out by Damour \&
Esposito-Far\`ese (1998), a pulsar-black hole system will be  a
superb probe of gravity, even for theories which make the same
predictions for black holes as GR.

\noi In particular, the discovery of radio pulsars in compact orbits around
Sgr\,A$^*$ would allow an unprecedented and detailed investigation of the
spacetime of the supermassive black hole. Timing of even a single
pulsar could provide novel tests of GR. Recently, Liu
et al. 2012 presented a method that uses a phase-connected
timing solution to enable the determination of the mass of Sgr\,A$^*$, the
frame dragging caused by its rotation, and its quadrupole moment. They
show that a simultaneous measurement of the mass, the spin (magnitude
and orientation), and the quadrupole moment of Sgr\,A$^*$ is possible
by observing a single pulsar in a six months orbit over a period of a
few years.  Due to the strong pulse broadening caused by the
interstellar medium near Sgr\,A$^*$, only rather slow pulsars are expected
to be discovered in surveys, even at frequencies above 10
GHz. Considering a pulsar with 0.5\,s spin period we show that the
optimal timing frequency is above 15 GHz, and that uncertainties of
100\,$\mu$s in the arrival times are realistic with the SKA. If Sgr\,A$^*$ is spinning rapidly, weekly timing observations
over five years would lead to a measurement precision of 10$^{-3}$ for the
spin and 10$^{-2}$ for the quadrupole moment, if the
orbital period of the pulsar is a few months. These numbers convert
directly into a 1\,\% test of the no-hair theorem for Kerr black
holes. This method is capable of identifying perturbations caused by
distributed mass around Sgr\,A$^*$, thus providing high confidence in
these gravity tests.

\noi In summary, with its unique sensitivity the SKA will push the current
binary pulsar strong-field tests to unprecedented levels. More 
importantly, however, with the abilities of the SKA to also find more 
extreme systems and pulsars orbiting black holes, the observatory will also 
provide new tests that are qualitatively very different from what is possible 
today. In particular, the SKA will be able to test GR's description of black
holes with precise measurements, providing one of the best tests of gravitational 
theories imaginable.\\

\parbox{0.9\textwidth}{
\noi{References:}\\
\noi{\scriptsize 
Blanchet L., Sch\"afer G., 1989, MNRAS, 239, 845;
Cordes J.~M., \etal , 2004, New Astr., 48, 1413
Damour T., Esposito-Far{\`e}se G., 1998, Phys.\ Rev.\ D, 58, 1;
Damour T.,  Sch\"afer G., 1988, Nuovo.\ Cimento, 101, 127;
Kramer M., \etal , 2004, New Astr., 48, 993;
Kramer M., \etal , Science, 314, 97;
Kramer M., \& Wex N., Class.\ Quantum Grav.\ 26, 073001;
Liu K., 2012, PhD thesis, Manchester University;
Liu K., \etal , 2012, Astrophys.~J., 747, 1 ;
Manchester R.~N., \etal , 2001,  MNRAS, 328, 17;
Smits R., \etal , 2009, A\&A, 493, 1161;
Wex N., Kopeikin S., 1999, ApJ, 514, 388;
}}\\

\subsubsection{Magnetars  {\scriptsize [K. Kokkotas]}}

High-density equation of state is a holy grail for astrophysics,
relativity and nuclear physics. There is a continuous effort to use
all available data from observations of neutron stars in radio, X- or
gamma-rays in order to constrain the neutron star parameters and to
unveil their internal structure (Lattimer \& Prakash 2007). Apart from
observations in the electromagnetic spectrum, the elusive
gravitational waves offer an alternative opportunity window
(Andersson \& Kokkotas 1998, Gaertig \& Kokkotas 2011).

\noi An alternative/complementary way to reach this is through the study of
oscillations from magnetars (Colaiuda \& Kokkotas 2011). These
oscillations can be observed in the gamma and X-ray part of the
spectrum but radio observations can not be excluded while it is
possible to observe them via gravitational waves. Two such events have
been observed up to now, that is the giant flare of the 2004 December
27 from SGR\,1806-20, observed with the X-Ray Timing Explorer (RXTE) and
the 1998 giant flare of another magnetar the SGR\,1900+14.

\noi Magnetars are thought to be neutron stars with very strong
($>10^{14}$G) surface magnetic fields. 21 of them have so far been
observed in X-rays
(http://www.physics.mcgill.ca/$\sim$pulsar/magnetar/main.html). A
groundbreaking discovery was made recently by Gavriil et al. (2008)
who found that the outburst of a rotation-powered pulsar in Kes 75
(PSR J1846-0258) was typical of magnetars. This was evident through
the sudden and significant change in braking index and glitch in its
timing behavior (Kuiper \& Hermsen 2009). Also five magnetar-like
X-ray bursts were observed during this outburst, making this a
promising candidate for observing oscillations when sensitivity is
high enough (Gavriil \etal\ 2008). After the outburst the source
went back to normal rotation-powered activity, with radio pulses being
however still absent from this system (Archibald \etal\ 2008).

\bigskip

\parbox{\textwidth}{

\hspace{2cm}
\parbox{5cm}{ \includegraphics[scale=0.15]{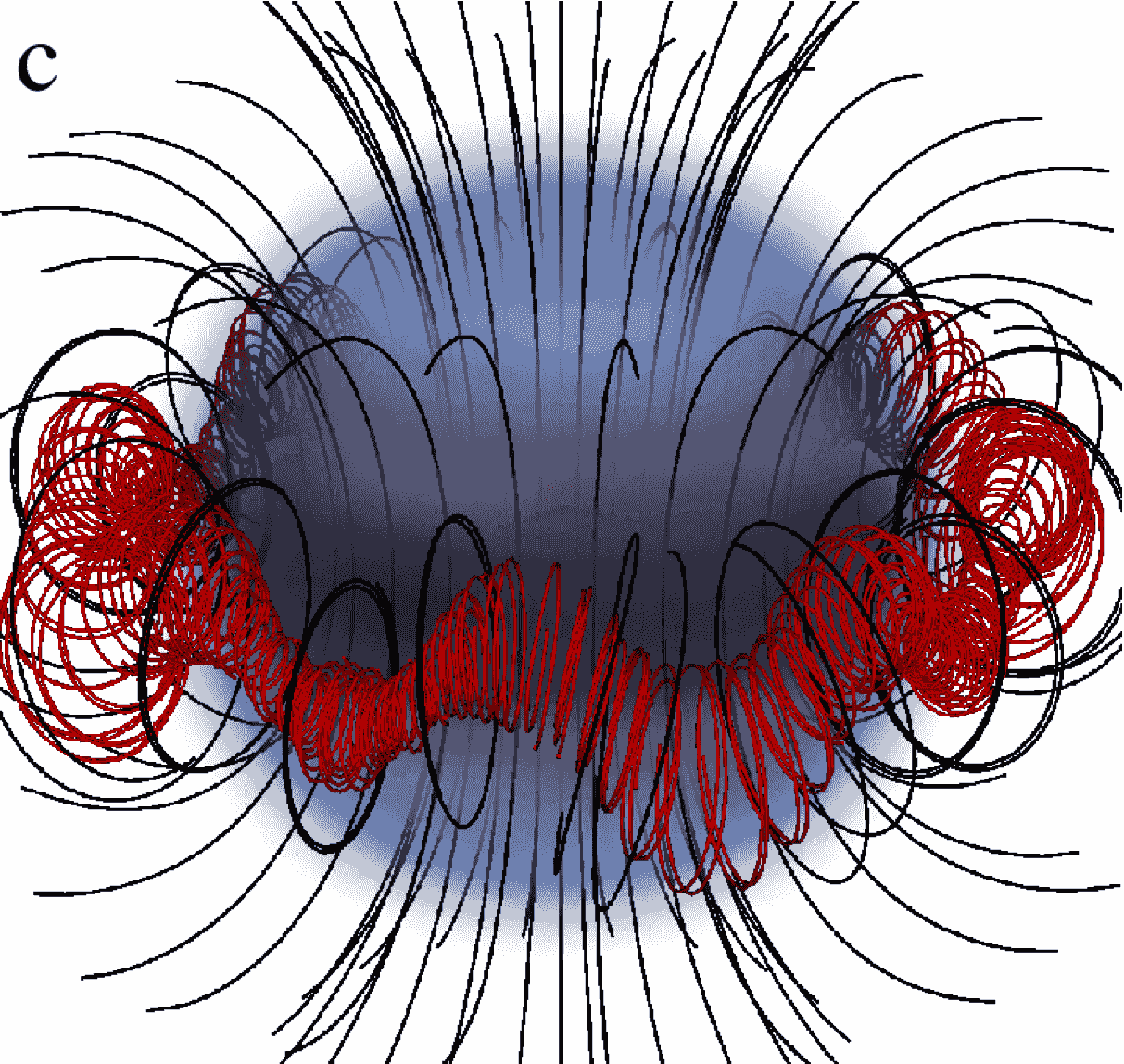}\vspace{-6cm}}

\hspace{9cm}
\parbox{6cm}{Figure 1$_{\rm Ko}$: The growth of an MHD instability in the magnetar interior
     as depicted by the non-linear simulation in (Lasky et al.
     2011). This type of instabilities maybe responsible for the
     catastrophic rearrangement of the magnetic field and even the
     production of weak gravitational waves.
 }
\vspace{3cm}}\\

\noi Observing this type of systems with a next generation radio telescope,
like the SKA, will provide important insight in the behavior of this type
of enigmatic sources. The fact that such events are not very long
lasting provides food for speculation that this source (PSR
J1846-0258) is not an exception and more rotation-powered pulsars show
magnetar like activity (Ng \& Kaspi 2010). This may be
supported by the radio pulsations seen from three magnetars
(Rea \& Esposito 2011). These pulses have some different
characteristics than the typical ones for rotation-powered pulses but
also some similarities (Kramer \etal\ 2007). Searches and
observations with the SKA could determine the origin of the emission and
clarify this open question. Monitoring known pulsars with the SKA and
discovering new ones will enlarge this population and expectedly
discover other sources with switching behavior. This will extend the
number of known magnetars and allow for a deeper study of their
activity. Near-simultaneous observations of these sources in X-rays
and radio waves will be critical for these studies.\\

\parbox{0.9\textwidth}{
\noi{References:}\\
\noi{\scriptsize 
Andersson N., Kokkotas K.D., 1998, MNRAS, 299, 1059;
Archibald A.M., \etal , 1998, ApJ, 688, 550; 
Colaiuda A., Kokkotas K.D., 2011, MNRAS, 414, 3014;
Gaertig E., Kokkotas K.D., 2011, Phys. Rev. D, 83, 064031
Gavriil F. P., \etal , 2008, Science, 319, 1802;
Kramer M., \etal , 2007, MNRAS, 377, 107;
Kuiper L., Hermsen W., 2009, A\&A, 501, 1031;
Lasky P.D., Zink B., Kokkotas  K.D., Glampedakis K., 2011, ApJL, 735, L20;
Lattimer J.M., Prakash M., 2007, Phys. Reports, 442, 109;
Ng C.Y., Kaspi V.M. 2010, arxiv:astro-ph.HE, 1010.4592;
Rea N., Esposito P., 2011,  in \emph{High-Energy Emission from Pulsars
  and their Systems}, eds. D.F..~Torres and N.~Rea, 247
}}\\

\subsubsection{The SKA and compact objects  {\scriptsize [J. Schaffner-Bielich, A. Sedrakian]}}

The structure and composition of compact objects is uniquely
determined by the equation of state (hereafter EoS) of matter in their
interior. This is strictly true if General Relativity is the
correct theory of gravity. We need to keep in mind, however, that
alternative theories of gravity may become an option if we find a
drastic disagreement between theory and observations.  Then,
for any given equation of state, one can compute the so-called global
(or integral) parameters of an compact star, typically the mass, the radius,
and the moment of inertia.

\noi Among the global parameters of neutron stars their masses are most
sensitive to the equation of state at high densities. Therefore,
pulsar mass measurements provide one of the key experimental
constraints on the theory of ultra-dense matter.  The masses measured
in the pulsar binaries, after the discovery of the first millisecond
pulsar (MSP) in 1982, are clustered around the value 1.4\,$M_{\sun}$
and have been considered as ``canonical'' for a long time.  However, in
recent years mounting evidence emerged in favor of substantially
heavier neutron stars with $M\ge 2M_{\sun}$.  In particular, the
recent discovery of a compact star with a mass of $1.97~M_{\sun}$ through
the measurement of Shapiro delay provides an observationally ``clean''
lower bound on the maximum mass of a compact star. This single result
revolutionises our view on the dense matter and provides a clear
evidence that the EoS of dense matter is stiff.

\noi Indeed, theoretically, it is now well established that emergence of
new degrees of freedom at high densities softens the EoS of
matter. For example, allowing for hyperons to appear in dense matter
can reduce the maximum mass of a sequence of compact stars below the
canonical mass 1.4\,$M_{\sun}$. A similar reduction in the maximum mass
occurs if kaon condensation takes place. The situation for quark
matter is less clear, as it constitutes a new form of matter and not
an additional degree of freedom. An extended mixed phase of normal
matter and quark matter will reduce the maximum mass, but quark matter
can also help to stabilize massive compact stars by providing an
additional counterpressure in the core.  The observation of a
2$M_{\sun}$ mass neutron star is certainly an evidence that the
ultra-dense matter in neutron stars can not be soft, \ie, agents that
will substantially soften the equation of state are potentially
excluded. In particular, this observation already rules out models of
dense matter which advocate kaon condensation with typical maximum
masses of around $1.5\,M_{\odot}$. Also, model parameters for quark
matter being potentially present in the core of compact stars start to
be substantially constrained.

\noi The fact that the SKA will be able to measure a large number ($\sim$\,3000)
of MSPs, a considerable fraction of which will certainly be in a
binary orbit, will largely increase the chance to observe objects
drawn from the high-mass tail of the pulsar mass distribution. It is
important to note that a single ``clean'' observation of a high-mass MSP
could strongly constrain the EoS.  Such constraints will have profound
impact on our understanding of dense matter and on the problem of
deconfinement transition from nuclear matter to quark matter - one of
the ``holy grails'' of elementary particle physics. For example, it has
been demonstrated that the observation of ``twins'' - stars having same
mass but different moments of inertia - could be a clear signal in
favor of hybrid configuration, i.e., stars featuring quark matter in
their interiors. The high-density EoS serves as a crucial input for
simulations of core-collapse supernovae and neutron star mergers which
are considered to be the prime sites for the nucleosynthesis of heavy
elements in our Universe. The investigation of the high-density and
high temperature EoS is a key for our understanding of the early
Universe shortly after the big bang and relativistic heavy-ion
collisions, as probed by the Relativistic Heavy Ion Collider (RHIC), the
Large Hadron Collider (LHC) and the Facility for Antiproton and Ion
Research FAIR at GSI Darmstadt.

\noi We can go on and ask what if accurate measurements of a pair of global
parameters were possible?  The discovery of the double-pulsar system
PSR J0737-3039 already focused our attention on the possibility of
simultaneous measurements of the pulsar masses and moment of
inertia. A single such measurement could place a point on the mass
versus moment of inertial graph through which any candidate EoS should
pass. The moment of inertia is roughly proportional to the product of
the star's mass and its radius squared, therefore such measurement
will give us also an estimate of the radius of the star.  Observations
of several pairs of global parameters for different pulsar masses will
allow us to map out portions of the mass versus moment-of-inertia diagram
and thus even further constrain the EoS. In fact, some have argued
that the nuclear equation of state can be reconstructed (inverted)
from the mass versus radius (or equivalently moment of inertia)
observations (Lindblom 1992). Although such
reconstruction will not give us the EoS chosen by nature per se,
it will provide an extremely useful guidance in the studies of dense
matter; we can anticipate that the number of models of dense matter,
which is inflated today up to a few tens will narrow down to a
few. Furthermore, there is a potential of discovering a new branch of
compact objects - self-bound strange stars - which have a mass-radius
(or mass-moment of inertia) dependence that is different from the one
of the objects bound by gravity. 

\noi As compared to other channels of observation of compact objects the
SKA programme is extremely valuable because of the precision at which
the global parameters of compact objects can be measured. These could
be compared with e.g. the constraints on neutron star radii from
x-ray bursters and thermal radiation from cooling neutron stars that
are relatively weak by itself.

\noi Our current knowledge suggests that the pulsar masses in binaries have
a certain distribution which has its origin either in the long-term
evolutionary processes (mass accretion from a companion) or short-term
dynamics (e.g. creation in a collapse.) The promise held by the SKA to
observe a large number ($\sim$\,3000) MSPs and about 100 relativistic
binaries will lead to a potential break-through in our understanding
of the evolutionary path undergone by MSPs. Large statistics will allow
us to build and test statistical models of the evolution of MSPs, to find
reliable statistical correlations between different parameters, such
as the mass, moment of inertia, $B$-field, spin and its
derivatives. These observations could improve our understanding of
pulsar genesis and evolution in the $P-\dot P$ diagram, i.e., either
confirm the present paradigm of spin-up of millisecond pulsars by
accretion or suggest new evolutionary avenues.

\noi Increasing the statistics of millisecond pulsars through the SKA
programme means increased probability of finding fast rotating
pulsars. Each EoS has its unique mass-shedding limit frequency for
each fixed mass star. The limit refers to the frequency beyond which
the centrifugal forces cannot be balanced by the gravity and a mass
loss starts from the equator. Thus, some models can be excluded on the
basis of an observation of a stable, rapidly rotating millisecond
pulsar. An increase in the currently observed maximal frequency by a
factor of two will already place strong constraint on the EoS. Measuring
a pulsar close to its Keplerian frequency together with its mass can
provide constraints of the same quality as those provided by the mass
versus radius or mass versus moment of inertia measurements.\\

\parbox{0.9\textwidth}{
\noi{References:}\\
\noi{\scriptsize
Lindblom L., 1992, ApJ, 398, 569
}}\\

\subsubsection{Motion of particles and light rays in strong
  gravitational fields  {\scriptsize [C. L\"ammerzahl, E. Hackmann,
    M.~List, V. Perlick, D. Puetzfeld, J. Kunz, V. Kagramanova, B. Hartmann]}}

{\bfseries\boldmath\small Expertise:~~}The expertise of the Bremen and Oldenburg gravity groups is (a) to
find analytical and also numerical black hole solutions of the
Einstein equations in standard Einstein General Relativity as well as
in generalized gravitational models, and (b) also to find analytical
and numerical solutions of the equations of motion for a large class
of axially symmetric space- times. For the analytical solutions of the
geodesic equation this includes the equation of motion of charged
particles and for light rays in the whole class of Pleba\'nski-Demia\'nski
space-times (Hackmann \& L\"ammerzahl 2008a, Hackmann \& L\"ammerzahl
2008b, Hackmann \etal\ 2009). Also the motion of particles in space-time geometries
with cosmic strings has been treated by us. We are also solving
equations of motion for particles with spin (Obukhov \& Puetzfeld 2011) and are also approaching
solutions for test particles with mass quadrupole moment (Steinhoff \& Puetzfeld 2010), all are
described by the Mathisson-Papapetrou-Dixon equations. Our method can
be applied to the effective one-body formalism describing binary
systems in GR in terms of a power series, too. We also started to
consider the motion of test particles including the influence of the
gravitational self force.

\medskip

\noi {\bfseries\boldmath\small Research programme:~~}We are very
interested in applying our theoretical expertise to astrophysical
observations of the SKA. By that we would like to (i) confront
analytical results for observables for orbits of test objects in
gravitational fields with high precision observations, (ii) to
calculate orbits taking into account the emission of gravitational
waves whose form we also would like to calculate, and (iii) use high
precision observations for tests of alternative theories of gravity.

\smallskip

{\bfseries\boldmath\small I. Orbits of stellar objects in strong gravitational fields:~~}It is only by the observation of test particles (realised by
light rays and small astrophysical objects like stars or not too big
black holes) that one can determine the space-time geometry. This
space-time geometry influences the timing, frequency, and direction of
the emitted radio and gravitational waves. This has to be taken into
account in the analysis of radio data. Since objects are not only
pointlike but in general have an internal structure (in astrophysics
all stellar objects are rotating and, thus, possess spin and also
axisymmetric higher order mass multipole moments), the influence of
these internal structures on the motion also has to be taken into
account. Once the gravitational field is known one can calculate and
interpret further effects with light rays and radio waves which then
are used for consistency considerations. What we would like to treat
in detail is: a. Analytic calculation of orbits and orbital effects
like perihelion shift, Lense-Thirring effect, light bending,
etc. (e.g. Hackmann \& L\"ammerzahl 2012)
b. Influence of the spin of test particles on the orbits and on the
mentioned orbital effects. For particular configurations like parallel
spin which might be realised in astrophysics, one may find analytic
solutions and analytic expressions for the orbital effects. Otherwise
we would like to study chaos which in general occurs for the motion of
particles with internal structure.  c. The same holds for the
influence of (axially symmetric) mass quadrupole effects on orbits and
the orbital effects. Again, analytic solutions in particular cases and
also study of chaos is aimed for. Topic b and c also represent a test
of the Mathisson- Papapetrou-Dixon equation for spinning and extended
objects.  d. As one further application of the bending of light we
consider the analytical calculation of the shadow of black holes, or
higher order images (also called relativistic images). From this one
can deduce the properties (mass, rotation parameter) of black holes.
e. We would like to derive analytic expressions for timing formulas,
based on the analytic representation of orbits and light rays. This
should in particular be applicable to binary pulsar systems and
compared with high precision observations. With these analytic
representations we can test approximation results and also can compare
it with results from numeric calculations.  f. Also the structure and
the behaviour of an accretion disc of a dense object results from the
dynamics of particles in strong gravitational fields. In this context
the Jacobi equation (dynamics of the relative distance of neighbouring
points) plays an important role. This is also a starting point for an
analytical treatment of the disruption of objects in strong
gravitational fields.

\smallskip

{\bfseries\boldmath\small II. Orbits and the emission of gravitational waves:~~}The motion of masses in general will create gravitational
waves. These emitted gravitational waves directly depend on the orbits
of the masses (also obtained in the first part of this study) and also
influence the motion of the body through the loss of energy and
angular momentum of the system. Therefore a consistent picture has to
be obtained for the motion of the bodies and the form of the
gravitational waves. This can be treated in the extreme mass ratio
inspiral (EMRI) case where one of the objects is nearly a test
particle, or within the effective one-body formalism. Here we would
like to calculate: a. The inspiraling orbits and the correspondingly
emitted gravitational waves through orbiting EMRI objects. From that
the orbital parameters and effects and the timing formulae and other
features can be calculated and compared with observations.  b. The
inspiraling orbits and the created gravitational waves for near equal
mass binary systems in the framework of the effective one-body
formalism.  c. The inspiraling orbits as well as created gravitational
waves for a particle which spirals into a boson star. From the
measured data of gravitational waves one can identify the object's
mass, spin and multipole moments. Due to these physical properties
especially their ratio towards each other one should be able to
distinguish a boson star from a black hole as the central gravitating
object.  The resulting orbits then can be observationally confirmed
through the observation of orbital parameters or of the timing of
pulsar signals. The effect of the radiated gravitational waves also
should be compared with results from self-force calculations for the
motion of near test particles in gravitational fields.

\smallskip

{\bfseries\boldmath\small III. Testing alternative gravitational
  theories:~~}Any motion of particles and light rays can be used for
testing alternative theories of gravitation. Here we restrict to
particular tests including black holes and similar high density
objects and objects with spin polarisation. We aim at considering
three aspects through the observation of the motion of test particles
and light rays: (i) To proof the validity of certain properties of
black holes encoded in the famous uniqueness and scrutinize boson
stars (which might be a consequence of the standard model of particle
physics). (ii) To compare the motion of particles and light rays in
black hole space-times versus the motion in space-times obtained for
boson stars in order to decide how good high density objects which
today are considered to be black holes may also be represented by
boson stars (which may be a consequence of the standard model of
particle physics).  (iii) To search for unusual properties of black
holes related to generalized theories of gravity where we have, for
example, black holes pierced by cosmic strings (Hackmann \etal\ 2010),
black hole with hairs (including scalar hairs), counterrotating
horizons etc. These properties of black holes also can be tested using
the motion of test particles, light rays, and radio waves in the
vicinity of black holes, see e.g. Enolski et al. (2012).  a. One way to investigate properties of
black holes is to analyze higher order images (intensity, angular
sequence, ...) at mostly axisymmetric black holes as well as the
shadow of black holes (Kerr black hole, black holes pierced with
cosmic string, ...), and boson stars (Perlick 2012). Here we expect major information
concerning the very interesting issue of clearly characterising black
holes and boson stars from observations of a central object due to
their physical properties e.g. mass, spin and multipole moments.
b. We aim to test the no-hair theorem through adding not only mass
multipoles but arbitrary axisymmetric perturbations to standard
General Relativity and also of generalized gravity models. Then the
motion of test bodies and light rays within these black hole
geometries could be determined. Additionally tests of black holes with
hair, including scalar hair, for generalized theories of gravity are
considered as they are predicted e.g. from low energy effective
Lagrangians from resulting from string theory.  c. In the context of
the issue of looking for a possibility of boson stars mimicking black
holes numerical methods have to be used. Boson star solutions of the
Einstein field equations can be obtained only through numerical
methods -- no analytical solution is known. Correspondingly, the motion
of test particles and light rays in these space- times also has to be
calculated numerically (Kleihaus \& Kunz 2005, Kleihaus \& Kunz 2008). The results for the orbits, the orbital
parameters, the light rays in terms of shadows, higher order images,
etc. have to be compared with measurements and could lead to a better
understanding and specification of gravitating sources.\\

\parbox{0.9\textwidth}{
\noi{References:}\\
\noi{\scriptsize
Enolski V., \etal  ,2012 , J. Math. Phys., 53, 012504;
Hackmann E., L\"ammerzahl C., 2008a, Phys. Rev. Lett., 100, 171101;
Hackmann E., L\"ammerzahl C., 2008b, Phys. Rev. D, 78, 024035;
Hackmann E., Kagramanova V., Kunz J., L\"ammerzahl C., 2009, Phys. Rev. D, 81, 044020;
Hackmann E., L\"ammerzahl C., Hartmann B., Sirimachan P., 2010, Phys. Rev. D, 82, 044024;
Hackmann E., L\"ammerzahl C., 2012, Phys. Rev. D, 85, 044049;
Kleihaus B., Kunz J., List M., 2005, Phys. Rev. D, 72, 064002;
Kleihaus B., Kunz J., List M., Schaffer I., 2008, Phys. Rev. D, 77, 064025;
Obukhov, Y., Puetzfeld D., 2011, Phys. Rev. D, 83, 044024;
Perlick V., 2004, Living Rev. Relativity, 7, 9;
Steinhoff J., Puetzfeld D., 2010, Phys. Rev. D, 81, 044019
}}\\

\bigskip

\subsubsection{Gravitational wave astronomy: Precision pulsar timing
  {\scriptsize [K.J. Lee, N. Wex, M. Kramer]}}

\noi The observed orbital decay in binary pulsars detected via
precision timing experiments so far offers the only evidence for the
existence of gravitational wave (GW) emission. Intensive efforts are
therefore on-going world-wide to make a direct detection of GWs that
pass over the Earth. Ground-based detectors like GEO600, VIRGO, and
LIGO use massive mirrors, the relative distance of which are measured
by a laser interferometer set-up, while the future space-based LISA
detector uses formation flying of three test-masses that are housed in
satellites.  The change of the space-time metric around the Earth also
influences the arrival times of pulsar signals measured at the
telescope, so that high-precision timing of millisecond pulsars (MSPs)
can also potentially directly detect GWs. Because pulsar timing
requires the observations of a pulsar for a full Earth orbit before
the relative position between pulsar, Solar System Barycentre and
Earth can be precisely determined, only GWs with periods of more than
a year can usually be detected. In order to determine possible
uncertainties in the used atomic clocks, planetary ephemerides used,
and also since GWs are expected to produce a characteristic quadrupole
signature on the sky, several pulsars are needed to make a
detection. The sensitivity of such a ``Pulsar Timing Array'' (PTA)
increases with the number of pulsars and should be able to detect
pulsars in the nano-Hz regime, hence below the frequencies of LIGO
($\sim$\,kilo-Hz and higher) and LISA ($\sim$\,milli-Hz).  A number of PTA
experiments are ongoing, namely in Australia, Europe and North America
(see Hobbs \etal\ 2010 for a summary). The currently derived upper
limits on a stochastic GW background (e.g.\ Jenet \etal\ 2006, Ferdman
\etal\ 2010) are very close to the theoretical expectation for a
signal that originates from binary super-massive black holes expected
from the hierarchical galaxy evolution model (Sesana \etal\ 2008,
Sesana \& Vecchio 2010).

\noi Demonstrating the power of PTA experiments, Champion et
al. 2010 recently used data of PTA observations to determine
the mass of the Jovian system independently of the space-craft data
obtained by fly-bys. Here, the idea is that an incorrectly known
planet mass will result in an incorrect model of the location of the
Solar System Barycentre (SSB) relative to the Earth. However, the SSB
is the reference point for pulsar arrival time measurements, so that a
mismatch between assumed and actual position would lead to a periodic
signal in the pulsar data with the period being that of the planet
with the ill-measured mass. This measurement technique is sensitive to
a mass difference of two hundred thousand million million tonnes --
just 0.003\,\% of the mass of the Earth, and one ten-millionth of
Jupiter's mass.
 
\noi While progress is currently made, GW astronomy using pulsars really
requires the SKA (Kramer \etal\ 2004). A first detection is virtually guaranteed with 
Phase\,1 of the SKA.  But the science that can eventually be done with
the full SKA goes far beyond simple GW detection -- a whole realm of
astronomy and fundamental physics studies will become possible. For
instance, it will be possible to study the properties of the graviton,
namely its spin (i.e.~polarisation properties of GWs) and its mass
(note that in general relativity the graviton is massless,  Lee \etal\
2008, Lee \etal\ 2010a).  This is
achieved by measuring the degree of correlation in the arrival time
variation of pairs of pulsars separated by a certain angle on the sky.
A positive correlation is expected for pulsars in the same direction
or 180\ndeg\ apart on the sky, while pulsars separated by 90\ndeg\ should
be anti-correlated. The exact shape of this correlation curve
obviously depends on the GW polarisation properties (Lee \etal\ 2008) but
also on the mass of the graviton (Lee \etal\ 2010a). The latter becomes clear
when we consider that a non-zero mass leads to a dispersion relation
and a cut-off frequency $\omega_{\rm cut}=m_{\rm g}c^2/\hbar$, below
which a propagation is not possible anymore, affecting the degree of
correlation possible between two pulsars. With a 90\,\% probability,
massless gravitons can be distinguished from gravitons heavier than
$3\times 10^{-22}$\,eV (Compton wavelength $\lambda_{\rm g}=4.1
\times 10^{12}$ km), if bi-weekly observation of 60 pulsars are
performed for 5 years with pulsar RMS timing accuracy of 100\,ns. If
60 pulsars are observed for 10 years with the same accuracy, the
detectable graviton mass is reduced to $5\times 10^{-23}$\,eV
($\lambda_{\rm g}=2.5 \times 10^{13}$ km; Lee \etal\ 2010a).

\noi In addition to detecting a {\em background} of GW emission, the
probability of detecting a {\em single} GW source increases from a few
percent now to well above 95\,\% with the full SKA. We can, for instance,
expect to find the signal of a single super-massive black hole
binary. Considering the case when the orbit is effectively not
evolving over the observing span, we can show that, by using
information provided by the ``pulsar term'' (i.e.~the retarded effect
of the GW acting on the pulsar's surrounding space time), we can
achieve a rather astounding source localization. For a GW with an
amplitude exceeding $10^{-16}$ and PTA observations of 40 pulsars with
weekly timing to 30\,ns, precision one can measure the GW source position to an
accuracy of better than $\sim$\,1\,arcmin  (Lee \etal\ 2010b). With such an error
circle, an identification of the GW source in the electromagnetic
spectrum should be easily feasible. We note that in order to achieve
such a result, a precise distance measurement to the pulsars is
needed, which in turn can then be improved further during the fitting
process that determines the orbital parameters of the GW
source. Fortunately, the SKA will be a superb telescope to do
astrometry with pulsars (discussed in Section~\ref{mknw}).

\noi Astrometric parameters can be determined in two ways for
pulsars. Firstly, using the telescope array as an imaging
interferometer, the pulsar can be treated as point source while
boosting the signal-to-noise ratio by gating the correlator to use
only signals during the few percent duty cycle when the pulsar is
actually visible. With images spread over a period of time, it will be
possible to measure parallaxes for nearly 10\,000 pulsars with an
accuracy of 20\,\% or less (Smits \etal\ 2010). Secondly, pulsar timing can
also be used for a precise determination of the position, proper
motion and parallax as all these parameters affect the arrival time of
the pulsars at our telescope on Earth. Here, MSPs with their higher
timing precision can be used more readily. Distances are retrieved via
a ``timing parallax'' which essentially measures the variation in
arrival time at different positions of the Earth is orbit due to the
curvature of the incoming wavefront. In contrast to an imaging
parallax, the sensitivity is highest for low ecliptic latitudes and
lowest for the ecliptic pole.
 Interestingly, it is still possible to measure a parallax
with finite precision at the ecliptic pole, in contrast to first
expectations. The origin of this is the small eccentricity of the
Earth orbit which allows us to still detect a variation of the arrival
time at different times of the year (Smits \etal\ 2010). We expect that we
can measure distances of 20\,kpc with a precision better than 20\,\% for
about 300\,MSPs, while for some sources distances of 20 to 40\,kpc can
be measured to 10\,\% or better, enabling the single GW source
studies described above.

\bigskip

\noi In summary, the SKA will be a superb observatory for GW astronomy
with numerous exciting applications in fundamental physics.\\

\parbox{0.9\textwidth}{
\noi{References:}\\
\noi{\scriptsize
Kramer M., \etal , 2004, New Astr., 48, 993;
Smits R., \etal , 2009, A\&A, 493, 1161;
Hobbs G., \etal , 2010, Classical and Quantum Gravity, 27, 084013;
Jenet F.A., \etal , 2006, ApJ, 653, 1571;
Ferdman R.D., \etal , 2010,  Classical and Quantum Gravity, 27, 084014;
Sesana A., Vecchio A., Colacino C.N., 2008, MNRAS, 390, 192;
Sesana A.,  Vecchio A., 2010, Classical and Quantum Gravity, 27, 084016;
Champion D., \etal , 2010,  ApJL, 720, L201;
Lee K.J., Jenet F.A., Price R.H., 2008, ApJ, 685, 1304;
Lee K.J., \etal , 2010a, ApJ, 722, 1589;
Lee K.J., \etal , 2010b, MNRAS, 424, 325;
Smits R., \etal , 2010, A\&A, 528, 108
}}\\

\bigskip

\subsubsection{Gravitational waves with the SKA {\scriptsize [A. Sesana]}}

In the next decade gravitational wave (GW) astronomy will open a new
window on the Universe. While signals coming from compact stars and
binaries fall in the observational domain of operating and planned
ground based interferometers (such as LIGO, VIRGO, GEO and the
proposed Einstein Telescope (ET)), massive black hole (MBH) binaries
are expected to be among the primary actors on the upcoming low
frequency stage, where the $10^{-4}-10^{-1}$ Hz window could be
probed by the European Laser Interferometer Space Antenna
(eLISA). Precision timing ($\lesssim 100$ns) of an ensemble of
millisecond pulsar (forming a so called pulsar timing array, PTA) will
offer the unique opportunity to detect GWs in the $10^{-9}-10^{-7}$Hz
frequency window (e.g. Detweiler 1979), complementing eLISA
observations (see the figure below). Infact, while eLISA will be
sensitive to individual mergers of relatively light binaries, PTA
observations will probe the genuine MBH ($M\grtsim10^8\msun$) at low
redshift (z\,$<$\,1) in their GW driven inspiral phase. Although current
PTA projects, such as the EPTA (Janssen \etal\ 2008), the PPTA
(Manchester 2007), NANOGrav (Jenet \etal\ 2009), and the IPTA (Hobbs
\etal\ 2010), may succeed in the quest of detection, but it is 
the SKA that makes GW astronomy at nHz frequencies 
at all possible (e.g. using MBH binaries).

\bigskip

\hspace{-0.7cm}
\parbox{\textwidth}{
\parbox{5cm}{ \includegraphics[scale=0.225]{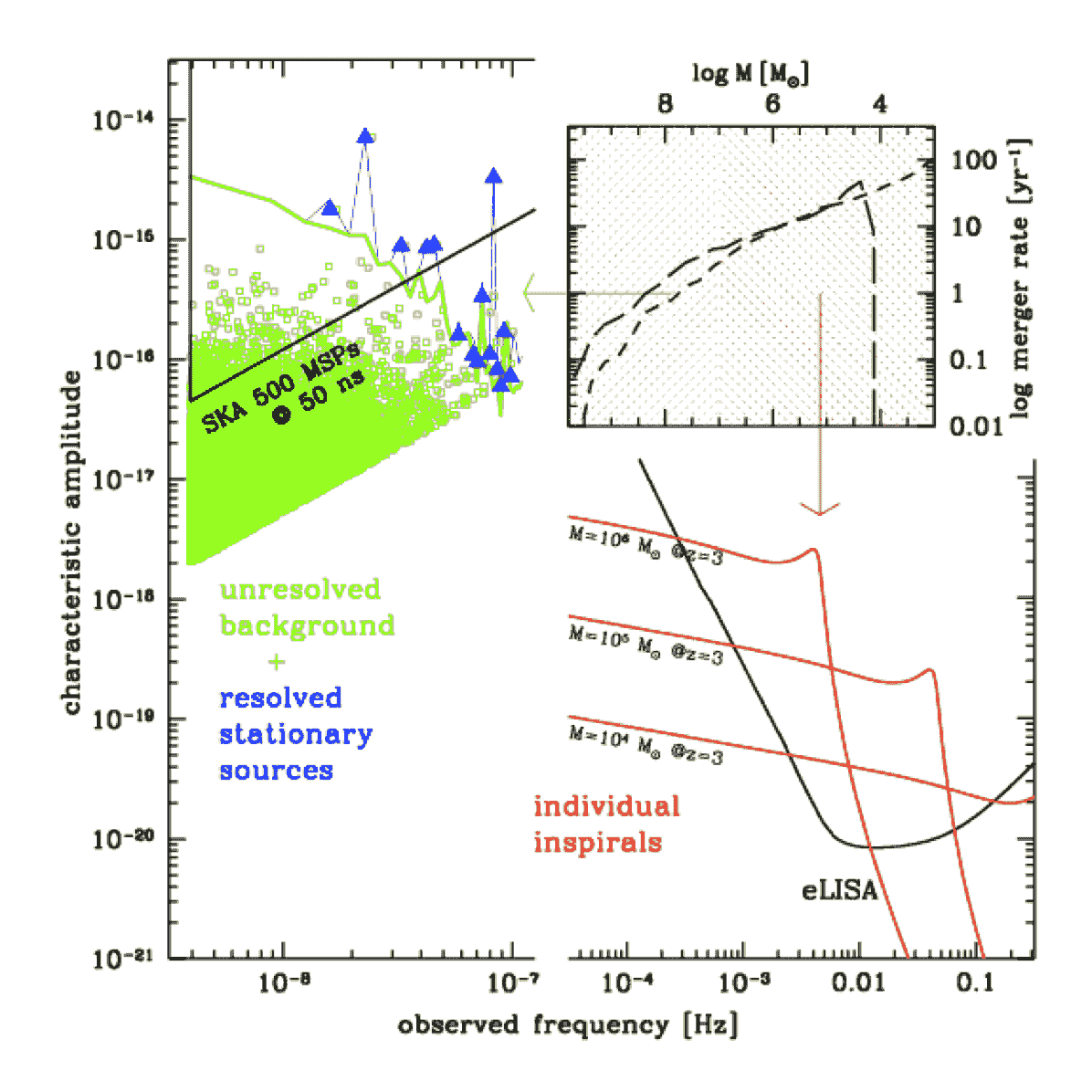}}
\hspace{4.3cm}
\parbox{8cm}{Figure 1$_{\rm Se}$: The MBH binary landscape in the GW window. Individual inspirals (red
solid lines) could be traced by eLISA (black solid line) in the mHz frequency
band. At much lower frequencies, the unresolved background (green solid line)
generated by the superposition of the signal coming from the cosmic population 
of MBH binaries (green opens squares) will be detected by the SKA (solid black 
line). Particularly bright sources (blue triangles) will be individually 
resolvable. The upper right inset shows the mass function of merging
MBH binaries, demonstrationg the complementarity of eLISA and 
SKA observations: the former will probe the infancy of MBH binaries by 
detecting light binaries mostly at z\,$>$\,3 (red shaded area); the latter
will detect the genuine MBH binaries at low redshift (green shaded area). }
}\\

\bigskip

\noi By combining observations of hundreds of millisecond pulsars with a timing
accuracy of $\sim$\,50\,ns the SKA will achieve a spectacular nominal sensitivity of 
few ns. At such level, the confusion noise generated by the superposition
of the signal coming from the cosmic population of MBH binaries will be easily
detected and characterized (Sesana 2008), and several sources will be individually 
resolved (Sesana \etal\ 2009, Boyle \& Pen 2010). The latter will offer the unique opportunity of 
combining GW and conventional electromagnetic observations, opening a new multimessenger 
astronomy era, based on the synergy between radio observations, GW detections
and Optical/X-ray followup monitoring. Counterparts to GW sources can be 
identified through periodicity or peculiar spectral features (Sesana
\etal\ 2011, Tanaka \etal\ 2011), 
both in Optical/UV and in X-ray by upcoming all sky surveys such as
LSST (LSST Science Collaborations \etal\ 2009) and eRosita (Predehl
\etal\ 2010).
The identification of the host galaxy of a MBH binary merger will (i)
improve our understanding of the nature of the galaxy hosting coalescing MBH binaries
(e.g. galaxy type, colours, morphology, etc.); (ii) help to reconstruct the dynamics
of the merging galaxies and of their MBHs; and (iii) offer the possibility of studying
accretion phenomena onto systems of known mass and spin (which, also in PTA
observations, can be measured from the GW signal e.g. Sesana \&
Vecchio 2010, Corbin \& Cornish 2010).\\

\parbox{0.9\textwidth}{
\noi{References:}\\
\noi{\scriptsize Boyle L.,  Pen U.,  2010, arXiv:1010.4337;
Corbin V.,  Cornish N.J.,  2010, arXiv:1008.1782;
Detweiler S.L. 1979, ApJ, 234, 1100;
Hobbs G., \etal , 2010, Classical and Quantum Gravity, 27, 084013;
Jenet F. \etal , 2009, arXiv:0909.1058;
Janssen G.H. et al., 2008, in  \emph{40 YEARS OF PULSARS: Millisecond
Pulsars, Magnetars and More. AIP}, Conference Proceedings, Volume, 983,  633;
LSST Science Collaborations, \etal , 2009, arXiv:0912.0201;
Manchester R.N., 2008, in  \emph{40 YEARS OF PULSARS: Millisecond
Pulsars, Magnetars and More. AIP}, Conference Proceedings, Volume, 983,  584
Predehl P. \etal ,  2010, in \emph{Society of Photo-Optical
  Instrumentation Engineers (SPIE) Conference Series of Presented at
  the Society of Photo-Optical Instrumentation Engineers (SPIE) Conference}, 7732;
Sesana A.,  Vecchio A., Colacino C.N., 2008, MNRAS, 390, 192;
Sesana A.,  Vecchio A., Volonteri M., 2009, MNRAS, 394, 2255;
Sesana A.,  Vecchio A.,  2010, Phys. Rev. D, 81, 104008;
Sesana A., Roedig C., Reynolds M.T., Dotti M., 2011, arXiv:1107.2927; 
Tanaka T., Menou K., Haiman Z., 2011, arXiv:1107.2937 
}}

\newpage

\section{The SKA high performance computing challenge: A German perspective}

\noi The SKA will be the World's premier imaging and surveying
telescope with a combination of unprecedented versatility and
sensitivity.  The technical challenges and the upcoming data products will
dramatically change the way astronomical research is done
and to make the SKA a reality demands a revolutionary break from
the traditional framework designs. The SKA will
drive technology development particularly in information and
communication technology.

\noi In that respect the SKA is truly a next generation software
telescope, in which high performance computing (HPC) is an essential
part, allowing the handling and the processing of measurements in real
time.  The data rates produced by the SKA can not be handled by
today's HPC facilities. In the following these aspects are considered
by: a first view of the challenge, the future technical development,
and the basic consideration to address the aspects of the upcoming
demands.

\subsection{The challenge {\scriptsize [B. Allen]}}

\noi The data analysis for the SKA is a significant computational and
organisational challenge.  This comes about for two reasons.  First,
the total volume of SKA data is very large. Second, the total number
of computations which must be performed to extract the science from
the data is also very large.  So it presents two challenges, one of
which is data distribution and storage, and the second of which is
data reduction and analysis.

\noi The total volume of data from SKA can only be estimated, because the
instrument configuration is not yet frozen.  However one can estimate
that the order-of-magnitude is about 4 Terabytes (Tb) per second in the initial
configuration, and a factor of a few more in the final
configuration. Moreover, this data rate is not ``peak'' but ``average'',
since the instrument is expected to be operating on a 24 hours times 7
days basis.  To
put this into perspective, at the time that this contribution was 
written (fall, 2011) the market price for a 2 Tb consumer storage disk
is about 100 Euros. The raw SKA data would fill one of these disks
every 4 seconds; the raw disk costs alone would be about 90\,000 Euro/hour.
Over the past 8 years the unit cost of disk storage has dropped by a
factor of 25; if similar improvements take place over the coming
8 years, then by 2019 this storage would cost about 40\,000 Euro/hour or of
order 30 million Euro/year.  Note that this does not include the cost of
electrical power, the systems hosting the disks, etc.  This means that
it is impractical to store all the data.  The data can only be stored
for some period of time, and must be analysed and processed within
that window.

\noi The computational challenge arises in two ways.  First, simply
distributing and accessing such a large data set is a computational
challenge.  To set the scale, the data rate is several times larger
than the data rate passing through the Amsterdam Internet Exchange,
the largest data hub in Europe.  Second, the computational work that
needs to subsequently be done on the data is very significant.
Synthesising the beams (summing together the data streams from the
different antenna elements) is a Petaflop-scale problem that would
challenge the largest supercomputers available today.  And searching
all of these beams (for example to locate all of the radio pulsars in
the Galaxy) is an Exaflop-scale problem.  If computing power continues
its exponential growth for the coming 8 years, we can expect that the
largest purpose-built supercomputers might be capable of handling this
load.

\noi At the moment, the growth in computer power (the so-called ``Moore's
Law'') is being maintained by the development of multi-core computing
architectures, such as Graphics Processor Unit systems containing
~1000 cores and capable of Teraflops/s of computation, and the newly
announced Intel Many Integrated Core (MIC) architecture, which has
about an order-of-magnitude fewer compute cores but broader vector
floating-point units.  It is expected that over the coming 8 years,
these systems will continue to envolve and improve, though it is also
clear that exponential growth in compute capacity at fixed power and
cost can not continue indefinitely.  However to use these systems
effectively for SKA data analysis, a new generation of programming
methodologies and software must be developed.

\noi The challenges of SKA data distribution and storage, and data
reduction and analysis, contain many interesting research problems and
possible creative solutions. To cite one example, estimates indicate
that it might be possible to solve both of these problems by adopting
a ``Public Volunteer Computing'' such as the one that has been used by
Einstein@Home during the past 6 years.  Here, the data would be stored
(redundantly) on computer hosts which were ``signed up'' by the general
public; the computing could also be done on those machines.  If the
number of volunteers were about one order of magnitude larger than the
number who are active in Einstein@Home, this would provide a very
cost-effective solution to both these challenges!

\noi More conventional solutions require advances in networking, data
storage, and power infrastructure in remote locations; these present a
variety of other interesting research challenges.\\

\subsection{Technological developments toward the SKA {\scriptsize [T. Fieseler]}}

Over the past years, several precursor and pathfinder telescopes have
been designed and setup to prepare SKA technology and to serve as
design studies. In these experiments, the current network, storage,
and computing technologies to transport, process and store data are
pushed their limits and new technologies are applied. Even these
smaller sized experiments like the LOFAR telescope will exhaust large
amounts of resources in the final stage of expansion. Current
requirements for LOFAR are about tens of Gigabytes (Gb) per second
network bandwidth, tens of Petabytes (Pb) of disk and tape storage and
a few Petaflops/s computation power. These requirements are not static
but will grow with the development of the experiment, as the surveys
and analysis which will be performed during the life cycle of the
telescope will constantly seek the boundaries of the resources which
are available at the time.  The requirements of the SKA will be an
order of 100 or 1000 higher than the requirements of experiments like
LOFAR. In this respect, the SKA will rely on the most powerful High
Performance Computing (HPC) technologies which are available. The
order of magnitude of the requirements will be in the TB/s range for
network bandwidth, in the Exabyte range for disk and tape storage and
in the Exaflop/s range for computing.  The provision of resources
fulfilling these requirements will be a challenge for the following
years. The J\"ulich Supercomputing Centre (JSC) hosts leadership-class
supercomputing systems with different architectures. The current Blue
Gene installation JUGENE was Europe's first petaflop system and will
be further extended to a few Petaflop/s in the near future. The
general purpose supercomputer JUROPA is a co-development of the JSC
and industrial partners, based on INTEL processors. Furthermore,
special cluster solutions like the Cell based cluster QPACE or GPU
enhanced clusters are developed and operated. The investigation and
operation of a variety of different architectures is the key to
finding the best-suited architectures for the SKA challenges. Today,
it is unknown which architectures will be the best choice in 10 to 15
years, when the telescope is fully operational. Thus, it is mandatory
to gain experience and expertise with all promising candidates of
architectures for the SKA.  Apart from the mere provision of resources
in these huge dimensions, the energy efficiency of the resources will
be one of the limiting factors. At present the energy consumption of a
Petaflops/s system is about a few megawatts. For example, the power
consumption of the 1\,Petaflop/s Blue Gene system JUGENE is about
2\,megawatt (MW). The leading system of the June 2011 Top 500 list has
a power consumption of 10\,MW for a peak performance of 8
Petaflops/s. With the current technologies, an Exapflop/s system would
have a power consumption in the gigawatt scale, an order of magnitude
which would require a dedicated power station for the operation of the
supercomputer. Therefore, the development of energy efficient
architectures for exascale supercomputing systems is crucial for SKA
computer system candidates. Developments like the QPACE cluster (rank
1 in the Green 500 November 2009 and June 2010) and the upcoming
BlueGene/Q systems (rank 1 in the Green 500 lists since November 2010)
are examples of architectures with an outstanding low power
consumption.  The development of energy efficient supercomputing
systems requires an holistic approach optimising the energy efficiency
at all levels. This includes the use of energy efficient processor
architectures, whose performance can be optimally exploited by the
given application, the reduction of losses in power conversion and
distribution as well as the implementation of energy efficient cooling
concepts at system level. The research in this direction will dominate
the development of prototype systems which could evolve into future
SKA technologies. An example is the European exascale research project
DEEP coordinated by J\"ulich. This project aims to exploit the
performance of a new type of many-core processors which provide a
significantly higher computing performance within a power envelope
similar to current multiprocessor nodes.\\

\subsection{SKA data processing and management: Basic considerations {\scriptsize [H. Enke]}}

\noi {\bfseries\boldmath\small LOFAR as a starting point:~~}From
LOFAR\footnote{See page~\pageref{subslofar} were LOFAR is referred to as a
  precursor for SKA.} a basic design schema for a software radio
telescope emerges:

\begin{itemize}
\item [--]distributed antennae stations with on-site compute power to reduce the station signals,
\item [--]a fiber optics network for signal transportation to a central processing facility (CPF),
\item [--]the CPF for correlating the station signals and various  basic observation modes,
\item [--]additional compute  and short term storage facilities for more sophisticated post processing,
\item [--]a long term storage facility for the science ready data,
\item [--]distributed facilities for scientific work on the data.
\end{itemize}
The above list only names the key hardware elements of the
system (Begeman \etal\ 2011). Control and management components have to be
added. For the LOFAR project, these components are implemented in a
hierarchical design, which consists of three ``tiers''. Since the
basic schemas for the SKA are similar to LOFAR, a similar schema can be
applied. Dedicated data processing centres (DPC) can be considered the
building blocks in this context. For the SKA, the geographical location of
the DPC needs careful consideration.\\

\noi{\bfseries\boldmath\small Tier 0 Data Processing Centre (DPC$_0$):~}The SKA-CPF has to be located onsite, near the antennae stations, and
connected to them by custom fiber optics network.  With an estimated
data flow of a few Terabytes/s from the stations to the CPF there is no
feasible alternative to this setup. Also, the control and management
components of the SKA need to be co-located with the CPF, and thus located
onsite as well.

LOFAR-CPF requires processing power of many Teraflop/s (for full
operation), for the SKA this scales to many Petaflop/s or even more. The
DPC$_0$ will consist of a PetaScale computer for correlation and a
file server with sufficient storage capacity to take the short term
data products and for buffering the raw data for special processing
modes (pipelines).\\

\noi{\bfseries\boldmath\small Tier 1 Data Processing Centre (DPC$_1$):~}The data produced by DPC$_0$ has to be transferred from the CPF into a
dedicated storage facility. The data proces sing within DPC$_0$ might
reduce the amount of data to store by a factor of 100 or 1000, but
still requires a bandwidth of the order of some 100 Gbit/s from
DPC$_0$ to DPC$_1$. The storage capacity of the DPC$_1$ archive should
be able to store the whole data product of DPC$_0$ for one year or
more. In addition, it might be used as a buffer for reprocessing data
at DPC$_0$. Therefore the DPC1 needs a dedicated low latency fiber
optics network connection to DPC$_0$.

For efficiency, the DPC$_1$ should feature compute facilities with
suitable environments for the scientists to remotely work with the
data. The technology to provide such environments on demand (virtual
machines with customized scientific tools, collaborative virtual
infrastructures) is currently maturing.\\

\noi{\bfseries\boldmath\small Tier 2 Data Processing Centre (DPC$_2$):~}The SKA data has to be distributed and replicated. Not all of the
scientific interesting processing can be done at DPC$_1$. For the
various key science areas, specialized archives will store only the
particular subset of data the project is working on and offer
additional processing capabilities. Parts of the data might need
reprocessing on a long term basis, combining data from various
observations. Additional data products will result from the key
science areas, which will be located at more than one DPC$_2$.
Another reason for replication is to insure against data loss.

Several DPC$_2$ will be located all over the world. The transport of
the data to these locations will have lower demands on network
bandwidth than from DPC$_0$ to DPC$_1$. The network connection,
however, remains a challenging task with the given intercontinental
network structure.  Furthermore, the DPC$_2$ can provide
access to SKA data to all interested parties and act as provider for
data publication.
  
A distributed structure makes the data output of the SKA manageable, but
requires in turn careful procedures for bookkeeping of the data sets,
keeping their integrity etc. Cataloguing and validating the archived
data, as well as tracking their replications in other archives could
become a shared responsibility of DPC$_1$ and DPC$_2$.

Since both DPC$_0$ and DPC$_1$ have to be co-located with the antennae
field, one of the challenges will be to build these facilities onsite,
where the supply chains for hardware and skilled labor is very
different from Europe.\\

\noi{\bfseries\boldmath\small Conclusion} This distributed schema for
the SKA data processing flow would allow to decouple the demands on
network bandwidth for ``real-time'' processing of the incoming antennae
data, getting first results on the one hand, and on the other hand
transporting the (reduced) data to DPC$_2$ facilities in Europe and
all over the world for scientific processing on longer time scales.\\

\parbox{0.9\textwidth}{
\noi{References:}\\
\noi{\scriptsize
Begeman K., \etal\ , 2011, \emph{LOFAR Information System in Future Generation Computer Systems}, 27, Elsevier B.V., 319
}}

\newpage

\section{The SKA and industry interaction}

\noi Radio astronomy provides a demanding, yet open, development and
test environment for state-of-the-art devices, systems and
algorithms. Construction of the SKA over a 6-year period is
the equivalent of building and commissioning a 100-m radio telescope every
20 days, a task far beyond the World's astronomical community. Large
industry contracts will therefore be necessary to build the
SKA.  Even before the construction phase, many of the \rd\
programmes needed for SKA demonstrators require industry know-how,
especially in crucial areas such as economic mass production and
system engineering.  The scale of the project makes it certain that
industry collaboration will be significant for the SKA in general and can be
divided into 4 phases in which interaction between science and
industry are crucial: 1. Test, Development \& Design, 2. Engineering \&
Manufacturing, 3. Construction, 4. Exploitation.

\noi The SKA has the opportunity to lead in the development of new
techniques for mega-project management and effective global research
collaboration, being an enabler for improved global-science-industry
links for German industry. The nature of the project will
inspire individuals, research groups, industrial partners and
governments to be part of a global endeavour that will endure beyond
their involvement. Furthermore, the profit and benefits to all those
involved will be realised over a long timescale and in the broadest
sense, building capacity and kudos for those who engage. Effective
collaborations have demonstrated pathways for talented individuals and
institutions to develop skills and capacity that can be applied
domestically and globally. Links and co-locations (technology
parks) between industry and science foster innovation and commercially
exploitable patents, bring wealth and create jobs.

\noi The SKA project, and its associated national and international
consortia programmes, welcomes interest from potential industry
partners. In general terms, any joint research and development is
viewed as a shared-risk endeavour, with SKA consortia and industry
each contributing to defined activities.The SKA has an agreed policy
on intellectual property (IP) developed under its aegis. Broadly,
industry partners exploit their own IP contributions in arenas outside
the SKA project, but innovations are available to the SKA project free
of any licensing charge.\\

\noi {\bf \small SKA's potential business opportunities:}

\begin{itemize}
\item[--]technological overview of the SKA telescope
\item[--]system integration
\item[--]telescope mass production
\item[--]telescope controlling, monitoring and maintenance
\item[--]infrastructure development
\item[--]complete electrical solutions for the SKA infrastructure
\item[--]chip design \& electronics 
\item[--]software development
\item[--]high performance \& GRID Computing
\item[--]data transport, management, storage, processing and analysis
\item[--]energy production and distribution
\item[--]operational expenses and sustainability
\end{itemize}

\noi In order to gain visibility of the SKA project for industry an
expression-of-interest (EoI) process was launched in May
2012. Based on the EoI process a formal request for business proposals
will be issued mid 2012.

\smallskip

\noi During first steps of the EoI process, various collaboration opportunities
have been engaged to build a German collaboration that may form a
consortia.\\

\subsection{A global model for renewable energy
  {\scriptsize[M. Vetter, E. Weber]}}

\noi {\small \bf Abstract:~~}The energy consumption of the SKA
telescope and its infrastructure will be of the order of
50\,--\,100\,megawatt (MW). Such energy demands has to be provided at
locations with little or no access to a power grid. While such demands
could only be produced by atomic power plants some 50 years ago, it is
now within the reach of photovoltaic systems. Regardless of the
numerous challenges to be met to power the SKA via a solar energy
solution the promised rewards are manifest. In particular, the future
location, the practical demands, and the technical requirements of the
SKA are ideal to pioneer strategies of green power production and
handling. Such studies would impact and accelerate technology
development in the areas of scalable energy generation and storage,
distribution, efficiency and demand reduction, and furthermore provide
a launch pad for commercialisation of innovative green energy
technologies. Especially in the developing nations the spin-off
benefits of such a project will impact on human quality of life
(e.g. 1.5 billion people are not connected to the power grid). The
economics of remote communities will thrive and prosper with access to
cheaper (renewable) power. In particular it will push the development
of infrastructure directly improving the public transport, the
education, and the medical health care.

\noi A study of a pathfinder project, to investigate the possibilities
to generate, supply, and store renewable energy of the order of 100\,MW, has
been
initiated by CSIRO Astronomy and Space Science involving the Fraunhofer Institute of Solar Energy and the ``Max-Planck-Institut f\"ur
Radioastronomie'' (MPIfR).\\


\noi{\small \bf Introduction:~~}A common problem for private, public
and large-scale industrial consumers all over the world is that they
are often situated far from the grid, or are not supplied with
sufficient or permanent energy capacity.  Concepts to supply and
distribute power that seems to work for the German or for large parts
of the European grid infrastructure, turns out to be not applicable to
remote locations and less developed countries in the world. 

To cope with an inefficient power supply it is common practise to use
diesel generator sets to power e.g. the ALMA array in Chile or the
mines in Australia. However the use of fossil fuel has a negative
impact on the CO$_{2}$ footprint of these countries. In addition, the
increasing and at the same time fluctuating fuel costs do not only
constitute an extremely high cost factor, but it turns out to be an
unpredictable element. The Indonesian fish-processing
industry may serve to illustrate this issue: most of the fish
canneries are run far from the grid (Indonesia consists of more than
17\,000 islands, which would be comparable to the number of individual
stations of the SKA telescope). In times of very high crude oil
prices the factories have stopped production directly affecting the life of
thousands of people  in the region.

\noi The energy needed to run the SKA is of the order of 100\,MW which
needs to be available 24 hours every day, since radio observatories
compared to optical observatories can observe day and night. Due to
the SKA infrastructure 50\,\% of the energy needs to be distributed
within the central region, whereas the other 50\,\% needs to be
distributed to remote stations up to 1000\,km distant. Building a
central power plant and a grid infrastructure to power the SKA is on
economic reasoning not the preferred solution (e.g. 10\,\% of the power
could be lost during transport). Instead, using solar photovoltaic systems
to power the SKA seems to be possible.  Figure\ IX shows the intensity
of the solar radiation on the earth. Both SKA sites in South
Africa and in Australia are placed in regions of  highest solar
radiation impact and are very well situated to cover a large fraction
of the energetic needs via solar energy.

\parbox{\textwidth}{ 

\hspace{0.1cm}
\parbox{0.4\textwidth}{ \includegraphics[scale=0.18,
  angle=0]{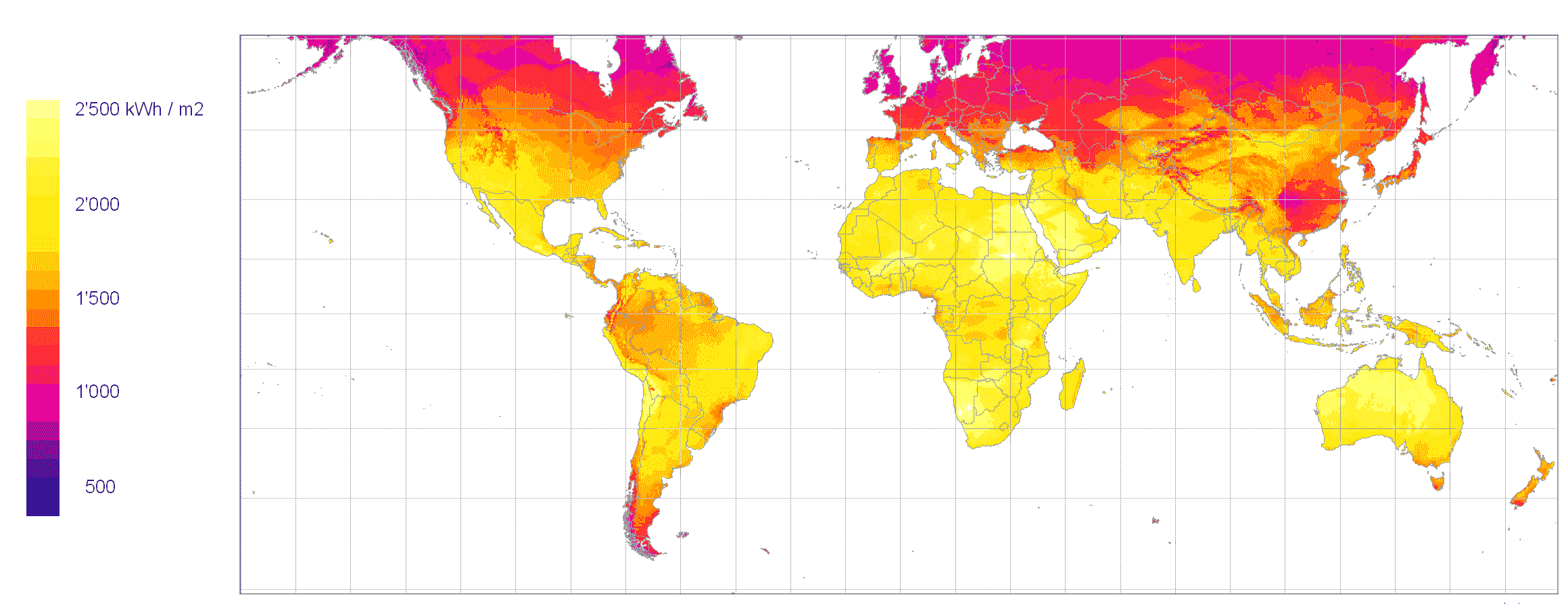}\vspace{0.5cm}}

\hspace{0.3cm}
\parbox{0.9\textwidth}{Figure IX: Earth with solar
      radiation impact shown in colour. The radiation impact is based on the yearly sum of the
      direct normal irradiance obtained between 1981 to 2000. (Image
      credit: \copyright\ METEOTEST; based on {\small [www.meteonorm.com]})}
\vspace{0.5cm}}\\

\noi The power requirements and the special setup of the SKA are an
ideal testbed to develop cost efficient power management
infrastructure for remote locations, that could also be applied to
remote societies. Such a renewable strategy would accelerate
technology development in the areas of scalable energy generation and
storage, distribution, efficiency and demand reduction and provide a
launch pad for commercialisation of innovative green energy
technologies.\\

\noi{\small \bf Pathfinding:~}The challenges to power the SKA via a
solar energy solution are numerous, but the low running costs
unaffected by fluctuations in global fuel price, and the technical
benefits and know-how transfer is worth investigation. A Memorandum of
Understanding (MoU) to foster collaboration between the Fraunhofer
Institute of Solar Energy, the ``Max-Planck-Institut f\"ur
Radioastronomie''  and Australian CSIRO Astronomy and Space Science was signed on 7 April 2011 in
Berlin during a workshop on ``Renewable Energy Concepts for
Mega-Science Projects demonstrated by the SKA and its Pathfinders''. A
key focus of this MoU is to promote scientific and research
cooperation in renewable energy capture, storage and management for
the SKA between Australia and experts from
Germany and the rest of the world. The MoU also looks to advance
collaboration between Fraunhofer ISE, the MPIfR and CSIRO in the development of renewable energy
systems for the Murchison Radio-astronomy Observatory (MRO) and the
Australian SKA Pathfinder (ASKAP) instrument as an SKA precursor
facility.

\smallskip

\noi The main aims and R\&D challenges can be summarised as
follows:

\begin{itemize}
\item[--]assessment of energy storage possibilities (thermal or electric)

\item[--]developing concepts of concentrated solar power (CSP)
 and concentrating photovoltaic (CPV) coupling

\item[--]developing concepts of electric storage systems with CPV 

\item[--]developing concepts to include thermal storage 

\item[--]development of a reliable energy management system; including
strategies for operational management of CPV-CSP hybrid systems to
cope with high electrical and thermal loads.

\item[--]development and testing of highly reliable and efficient power
electronics

\item[--]evaluation of the electronic needs of an astronomical
observatory with respect to the day and night phases in order to
minimise the needed electronic storage

\end{itemize}

\bigskip

\noi One of the major keystones of the MoU is to develop a test
concept that will be evaluated on the Effelsberg site to investigate
the practicality of the concepts within an astronomical RFI sensitive
environment.\\

\noi{\small \bf Summary:~~}The specific requirements of
the SKA will challenge current green-power-generation concepts and
will pioneer the continuous use of renewable energy and its remote
generation.  The main aim of the project is to develop
concepts to use solar energy to provide a reliable and stable power
solution for the SKA. These concepts will be tested during the
development phases of the SKA and in particular in their pathfinder
instruments like MeerKAT and ASKAP. \\

\subsection{Connecting Europe: High data transport rates via satellites  {\scriptsize [J. Kerp]}}

\noi Astrophysical research suffers on a ``digital divide''. The most
advanced present and future observatories will be located on the
Earth's southern hemisphere, like the SKA, while most of the
scientific community as well as the high-tech computing facilities are
located in the north e.g. in Northern America, Europe, Japan, or
China. The daily amount of data needed to be scientifically analyzed
increased by two magnitudes in just the last decade. Today most of
the observational data is roughly inspected by searching for known
objects. Only these ``post stamps'' are stored and transfered for
serious scientific analyses, but data mining for the
``exploration of the unknown'' is currently out of reach.\\

\noi This digital divide needs to be overcome by linking the
high-performance-computing centres to the observatories by using
state-of-the-art satellite communication systems. Today the satellite
market is underdeveloped and deep-sea fiber connections transfer more
than 90\,\% of the data. This technology is cost intensive and subject
for distortions due to terrestrial events like earth-quakes, wars
etc. The financial value of scientific data is unknown and considered
to be insignificant. Because of this, fiber connections between the
upcoming next generation observatories like the SKA and the super computer
facilities on the northern hemisphere are not considered.
However high
frequency satellite communication could be considered a ``key'' to
change this situation entirely, using data transfer at 60\,GHz
to 400\,GHz radio frequencies to transfer hundreds of gigabytes
within the next decade. The low photon energy of the radio radiation
allows us to transfer --at a high signal-to-noise-- large amounts
of data with a minimum of energy. This is feasible by combining
multi-feed ground stations and sophisticated data compression
technologies with high radio frequency inter-satellite
links. Such inter-satellite links use frequencies between 100\,GHz and
400\,GHz, which is standard technology for astrophysicists and
just needs to be adapted to the requirements of the SKA and
possible industrial useage.\\

\noi Using today's latest scientific equipment in radio astronomy it
is feasible to transfer this technology to the commercial
market. Here, the technology needs to simplified (i.e. from cooled to
uncooled systems) and make use of the ``of-the-shelf'' systems. The new
to open commercial market is huge, because ground stations are easy to
maintain and can be located in nearly all environments. In combination
with the long term evolution (LTE) on the mobile communication
standard it is feasible to make these data
accessible world wide.

\newpage

\section{Positive impact on human capital development and employment}

\noi Astronomy is the study of everything beyond Earth and it is a
science that enters our daily lives directly. As a science it is
driven by observations, with links to mathematics, physics, chemistry,
computer science, geophysics, material science and biology. Astronomy
is important for society and culture, and helps attract young people
into natural science and technology.

\parbox{\textwidth}{
\hspace{-1cm} 
\vspace{-4.1cm}
\parbox{5cm}{ \includegraphics[scale=0.52]{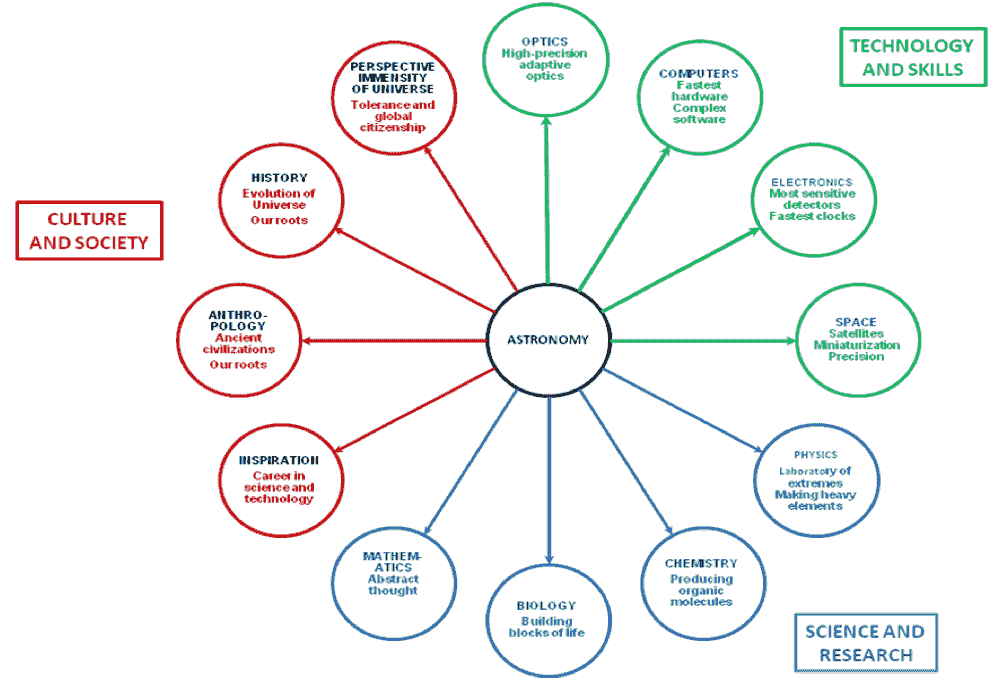}\vspace{14.5cm}}
\hspace{-3.5cm}
\parbox{0.8\textwidth}{Figure X: Overview on how astronomy impacts on science \& research,
  technology \& skills, and culture \& society (credit: SKA Project Execution
Plan).}
\vspace{-8.3cm}
}

In the last two decades astronomy has made particularly impressive
advances, technically, observationally and theoretically. In spite of
a downturn in intake for natural sciences at university level in
general, young students have continued to enter the field at steady
rate, driven by their innate curiosity, and motivated to
contribute directly to advances in knowledge. Excited by astronomy,
young and gifted minds are frequently attracted to related
scientific disciplines, so that astronomy acts as a springboard and
catalyst for wider scientific enquiry. A summary on how astronomy has
been integrated into our society is shown in the Figure\ X.

\smallskip

\noi In this spirt the SKA, with its scale and scope, has the potential
to inspire generations of young people with natural science.  It can do so not
only because astronomy appeals to our natural curiosity but also
because it is a stepping stone to many other fields of science and
technology development including computing, engineering, aerospace,
mathematics and natural sciences.

\noi The SKA can provide long-standing benefits to society, proportional to the
long-term investment it requires. Furthermore, the construction and operation of
the SKA facilities will impact on local and regional skill developments
in all fields of science.

\newpage
\pagestyle{empty}
\noi{\bf \,}

\pagestyle{plain}
\section{Abbreviations}

\parbox{\textwidth}{
\noi The list of abbreviations covers most of the acronyms used
in the SKA project and in this ``white paper''.

\smallskip

\hspace{1.2cm}
\parbox{\textwidth}{
{\tiny 
\begin{tabular}{l c l}

ACT                  & & Atacama Cosmology Telescope\\
AGN                   & & Active Galactic Nuclei \\
AIP                    & & Advanced Instrumentation Program \\
ALMA                 & & Atacama Large Millimeter/submillimeter Array\\
ALP                   & & Axions-like Particle\\
APERTIF            & & APERture Tile In Focus \\
ASKAP               & & Australian SKA Pathfinder \\
AXP                   & & Anomalous X-ray Pulsar\\

&& \\

BAO                   & & Baryon Acoustic Oscillation \\
BH                      & & Black Hole \\
BOSS                  & & Baryon Oscillation Spectroscopic Survey\\

&& \\ 

CABB                  & & Compact Array Broadband Backend\\
CCO                 & & Central Compact Object \\
CDFS               & &  Chandra Deep Field-South \\
CDM                & & Cold Dark Matter \\
CMB                & & Cosmic Microwave Background \\
CMBPOL           & & CMB POLarisation \\
CME                 & & Coronal Mass Ejection \\
CO                   & & Carbon Monoxide \\
COBE               & & Cosmic Background Explorer\\
CODEX            & & Cosmic Dynamics Experiment\\
COSMOS          & & Cosmic Evolution Survey\\
CPV                  & & Concentrating Photovoltaic\\
CR                    & & Cosmic Ray \\
CSP                   & & Concentrated Solar Power\\
CTA                 & & Cherenkov Telescope Array \\

&& \\

DE                   & & Dark Energy \\
DES                   & & Dark Energy Survey\\
DM                   & & Dark Matter \\
DPC                  & & Data Processing Centres\\

&& \\

EBHIS                & & Effelsberg Bonn \hi\ Survey \\
E-ELT               & & European Extremely Large Telescope \\
eEVN                 & & European VLBI Network\\
EM                     & & Electro Magnetic \\
e-MERLIN           & & Multi-Element Radio Linked Interferometer Network\\
EMRI                 & & Extreme Mass Ratio Inspiral \\
EoS                     & & Equation of State \\
EoR                    & & Epoch of Reionisation   \\
EoI                     & & Expression of Interrest  \\
eROSITA           & & X-ray \& gamma-ray satellite programme\\
ESA                     & & European Space Agency \\
ESFRI                 & & European Strategy Forum for Research
Infrastructures \\
ESO                     & & European Southern Observatory \\
ET                     & & Einstein Telescope \\
Euclid               & & ESA Cosmic Space Mision \\
EUV                   & & Extreme Ultraviolet\\
EVLA                  & & Expanded Very Large Array \\
&& - re-named in
2012  to Karl G. Jansky Very Large Array (JVLA)\\

&& \\

FoV                   & & Field of View \\
FR                     & & Fanaroff Riley Galaxy (type I and II)\\
&& \\

GAIA                  & & ESA Cosmic Space Mision \\
GC                      & & Galactic Centre \\
GEO600               & & Gravitational Wave Detector\\
GLOW                 & & German LOng Wavelength \\
GO-SKA               & & Global Organisation for the SKA\\
GR                     & & General Relativity \\
GW                    & & Gravitational Wave \\

&& \\ 

HDF                   & & Hubble Deep Field \\
HETDEX             & & Hobbly-Eberly Telescope Dark Energy eXperiment (optical)\\
HI                      & & Neutral Hydrogen \\
HII                     & & Ionised Region  (Str\"omgren spheres)\\
HIPASS              & & \hi\ Parkes All Sky Survey \\
HPC                   & & High Performance Computing\\
HST                    & & Hubble Space Telescope\\
HVC                   & & High Velocity Clouds \\
&& \\

ICM                    & & Inter Cluster Medium\\
IGM                    & & Inter Galactic Medium\\
INTEGRAL          & & INTErnational Gamma-Ray Astrophysics Laboratory\\
IP                       & & Intellectual Property \\
ISSC                    & & International SKA Steering Committee \\
ISM                     & & Interstellar Medium \\
ISPO                   & & International SKA Project Office  \\
ISRF                  & & Interstellar Radiation Field \\
IVC                   & & Intermediate Velocity Clouds \\

\end{tabular} \vspace{-20.5cm}}
}

\hspace{10.5cm}
\parbox{\textwidth}{
{\tiny 
\begin{tabular}{l c l}

JDEM                  & & Joint Dark Energy Mission (space mission)\\
JWST                   & & James Webb Space Telescope \\
JVLA                    & & Karl G. Jansky Very Large Array see EVLA\\
&& \\

KSP                     & & Key Science Project \\
KM3NeT              & & A Multi-km$^3$ sized Neutrino Telescope \\

&& \\
LF                       & & Luminosity Function \\
LG                      & & Local Group \\
LHC                   & & Large Hadron Collider \\
LIGO                  & & Laser Interferometer Gravitational Wave Observatory \\
(e)LISA               & & (European) Laser Interferometer Space Antenna \\
LMC                   & & Large Magellanic Cloud \\
LOFAR               &  & Low Frequency Array\\
LoS                    & & Line Of Sight \\
LSST                  & & Large Synoptic Survey Telescope\\
LTE                     & & long term evolution\\
&& \\

MeerKat             &  & Karoo Array Telescope \\
MHD                 & & Magnetohydrodynamics \\
MIC                  & & Many Integrated Core \\
MM                   & & MegaMaser \\
MoA                  & & Memorandum of Agreement \\
MoU                  & & Memorandum of Understanding\\
MSP                   & & Millisecond Pulsar \\

&& \\

NEI                   & & Non-Equilibrium Ionisation\\

&& \\
OH                    & & Hydroxyl \\
&& \\

PAF                    & & Phased Array Feed \\
PO                      & & Participating Organisations \\
PrepSKA             & & Preparatory phase proposal for the SKA\\
PSF                   & & Point Spread Function\\
PTA                   & & Pulsar Timing Array \\

&& \\

QCD                  & & Quantum Chromodynamics \\
QSO                  & & Quasi Stellar Object (quasar)\\

&& \\

R\&D                & & Research and Development \\
RFI                    & & Radio Frequency Interference \\
RHIC                 & & Relativistic Heavy Ion Collider \\
RM                    & & Rotation Measurement \\
RRAT                 & & Rotating Radio Transient \\
&& \\

SEP                    & & Solar Energetic Particles \\
SETI                   & & Search for Extraterrestrial Intelligence \\
SDSS                  & & Sloan Digital Sky Survey \\ 
SF                       & & Star Forming \\
SGR                    & & Soft Gamma Repeater \\
SKA                   & & Square Kilometre Array \\
SKADS               & & SKA design study \\
(S)MBH                 & & (super) massive black hole \\
SMC                   & & Small Magellanic Cloud \\
SMG                    & & Sub-Millimetre Galaxies \\
SN(R)                     & & Supernova(remnants) \\
SOWG                & & Site Options Working Group \\
SPDO                 & & SKS Program Development Office\\
SPICA                  & & ESA Cosmic Space Mision \\
SPT                        & & South Pole Telescope \\
SSB                      &  & Solar System Barycentre \\
SSEC                   & & SKA Science and Engineering Committee \\

&& \\

TNO                    & & Trans-Neptunian Objects\\
TOA                    & & Time Of Arrival \\

&& \\

UHE                    & & Ultra High Energy \\
(U)LIRG                & & (Ultra) Luminous Infrared Galaxies\\
URSI                    & & International Union of Radio Science \\
UV                       & & Ultraviolet \\

&& \\
VLA                     & & Very Large Array\\
VLBI                    & & Very Long Baseline Interferometry \\ 

&& \\

WHIM               & & Warm-Hot Intergalactic Medium\\
WiggleZ            & & dark energy survey\\
WIM                  & & Warm Ionised Medium \\
WIMP                & & Weakly Interacting Massive Particle\\
WISP                & & Weakly Interacting Sub-eV Particle\\
WMAP               & & Wilkinson Microwave Anisotropy Probe \\ 
WNM                & & Warm Neutral Medium \\
WP                    & & Work Package \\
WSRT                & & Westerbork Synthesis Radio Telescope \\

&& \\
XDIN               & & X-ray Dim Isolated Neutron Stars \\

\end{tabular}
}
\vspace{15cm}
}
}

\newpage
\pagestyle{empty}
\noi{\bf \, }

\newpage
\pagestyle{plain}
\section{Appendix}
\label{appen}
\hspace{1cm}
\parbox{\textwidth}{\parbox{15cm}{Figure\ XI: Overall SKA project
    timeline (release v11 01-06-2011, credit: SKA Project Execution
Plan)).\vspace{0.5cm}}
\hspace{2cm}
\parbox{0.8\textwidth}{ \includegraphics[angle=90,scale=0.105]{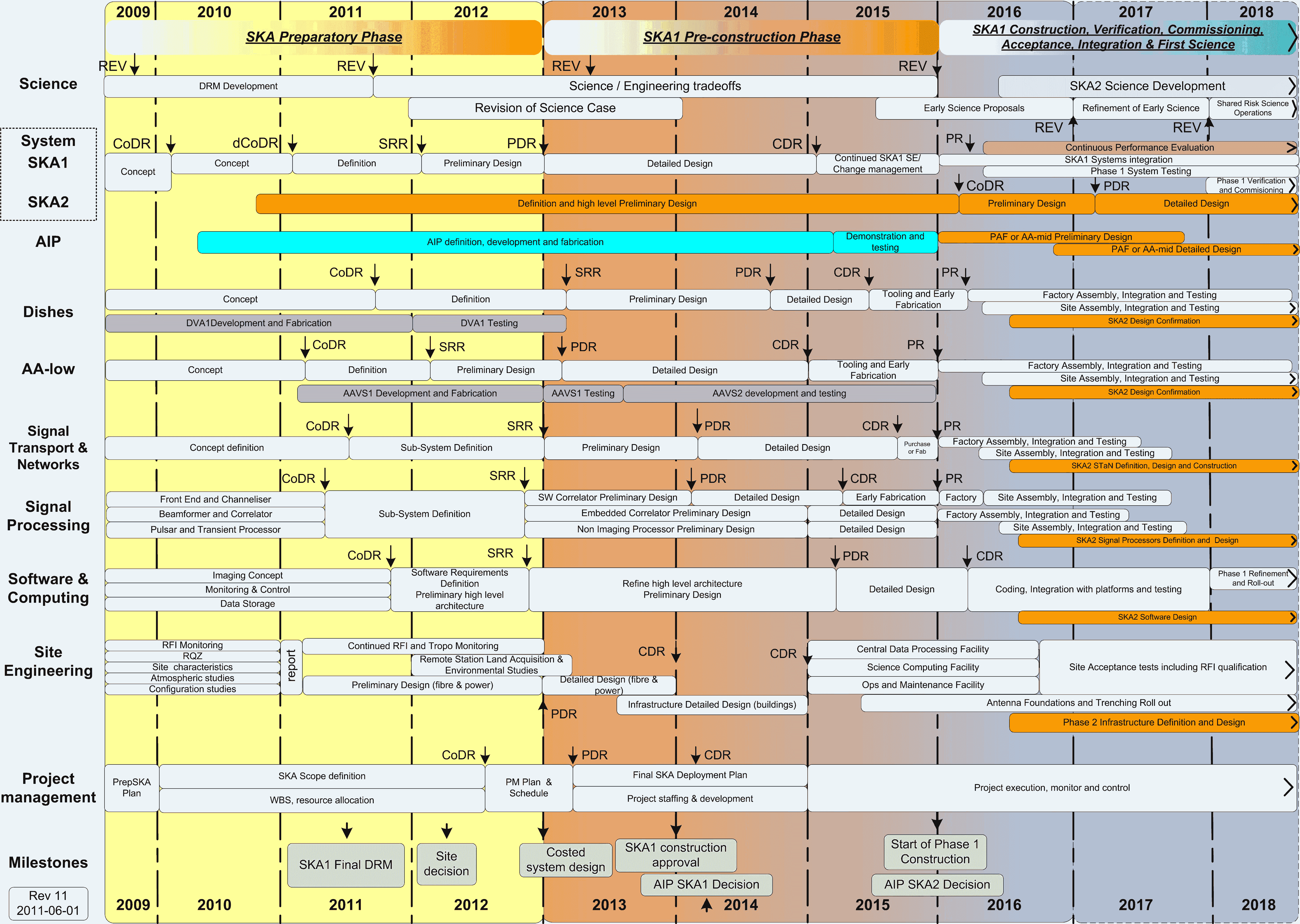}\vspace{0cm}}
}

\newpage
\pagestyle{empty}
\noi{\bf \, }

\newpage
\pagestyle{plain}
\section{Bibliography, References, \& Image credits}
\label{refs}

\bigskip

\noi {\bf Bibliography}

\smallskip

\noi Most of the SKA particulars shown and discussed in this ``white
paper'' are based on the information given in numerous SKA documents
and memos, which are available via the SKA homepage:

\smallskip

{\bfseries\boldmath\small www.skatelescope.org/publications}

\bigskip

\noi General information about the SKA can be found via:

{\bfseries \boldmath\small http://www.scholarpedia.org/article/Square\_kilometre\_array}

\bigskip

\noi The SKA project receives globale recognition in the media. Also 
several German newspapers have covered the SKA project and the
side discussion of the SKA. A record of
newspaper articles, online interviews, and further information can be found via:

\smallskip

{\bfseries\boldmath\small  http://www3.mpifr-bonn.mpg.de/public/pr/links-ska.html}\\

\noi Information on the German SKA collaboration can be obtained via:

\smallskip

{\bfseries\boldmath\small www.mpifr-bonn.mpg.de/seso/GESKA/GeSKA.html}

\bigskip

\bigskip

\noi {\bf References}

\smallskip

\noi The following documents are listed because they are references, 
are mentioned in the text, or are of particular
interest for a general understanding of the SKA project and its
impact.

\medskip
\parbox{0.7\textwidth}{\small 

Carilli C., Rawlings S., 2004, New Astronomy Reviews, Elsevier, 48

\smallskip

Dewdney P., et al., 2010, SKA Memo, 130

\smallskip

Faulkner A.J., et al., 2010, ``SKADS white paper''

\smallskip

\noi Garrett  M.A., et al., 2010, SKA Memo, 125 

\smallskip

\noi Kl\"ockner H.-R., Rawlings S., Jarvis M., Taylor A., Editors of
``Cosmology, Galaxy Formation and Astroparticle Physics on the Pathway
to the SKA'', 2006, ASTRON, ISBN 978-90-805434-4-7

\smallskip

Noordam J.E., Braun R., de Bruyn A.G., 1991, ASTRON Notes, https://www.astron.nl/documents/Notes/ASTRON-NOTE-585.pdf

\smallskip

Br\"uggen M., Falcke H., Beck R., En{\ss}lin T., Editors of ``German LOFAR -
white paper'', 2005, \mpifr , http://www.mpifr-bonn.mpg.de/public/pr/white.paper.oct6.pdf

\smallskip

Oort J.H., 1932, Bulletin of the Astronomical Institutes of the
Netherlands, 6, 249

\smallskip

Trimble V., Ceja J.A., 2008, Astron. Nachr., 329, 6, 632 

\smallskip

\noi Schilizzi  R.T., et al., 2011, SKA document, Project Execution Plan 

\smallskip

\noi Schilizzi  R.T., et al., 2007, SKA Memo, 100 

\smallskip

\noi Torchinsky S.A., van Ardenne A., van den Brink-Havinga, van Es
A.J.J., Faulkner A.J.  Editors of
``Wide Field Astronomy \& Technology for the Square Kilometre Array'', 2009, ASTRON, ISBN 978-90-805434-5-4

\smallskip

\noi Verschuur G.L., 2007, 2nd Edition, ``The Invisible Universe'', Springer, New
York, USA

\smallskip

\noi Wilkinson P.N., 1991, Astronomical Society of the Pacific, ``Radio interferometry: Theory,
techniques, and applications: Proceedings of the 131th IAU
Colloquium'', 428

\smallskip

\noi de Zeeuw P.T., Molster F.J., Editors of ``A Science Vision for
  European Astronomy'', 2007, ASTRONET, ISBN 978-3-923524-62-4, ``ASTRONET Science Vision''

\smallskip

\noi Yun M.S., Ho P.T.P, Lo K.Y., 1994, Nature, 372, 530

\smallskip

\noi Yun M.S., 1997, ``Galaxy
Interactions at Low and High Redshift: Proceedings of the 186th IAU Symposium, Editor D. Sanders''

\smallskip

\noi Zwicky F., 1933, Helvetica Physica Acta, 6, 110
}

\newpage

\noi {\bf Image credits}
\label{imcred}

\begin{itemize}
\item[] SKA related telescope and array images are obtained from the SKA homepage and are
produced by SPDO/Swinburne Astronomy (cover page and Figures\ VI, VII, VIII).

\item[] The image in Figure\ VIII of the SKA core and inner region within Berlin is
based on Goggle Maps.

\item[] The images in Figure\ I are based on the images provided
via ({\small http://chandra.harvard.edu/photo/2002/0157/})
Credit: X-ray (NASA/CXC/M. Karovska et al.); neutral hydrogen image
(NRAO/VLA/ van Gorkom, Schminovich et al.), radio continuum image
(NRAO/VLA/ Condon et al.); optical image (Digitized Sky Survey U.K. Schmidt Image/STScI).

\item[] The images in Figure\ II are based on the optical image by
  J. Gallego {\small [www.astrosurf.com/jordigallego]}, the \hi\ image
  is reproduced, based on data by Yun et al. 1994, and the simulation image is from Yun 1997).

\item[] The image in Figure\ X is a copy of Figure\ 11 of the SKA Project Execution
Plan (MGT-001.005.005-MP-001).

\item[] The image in Figure\ XI is a copy of the figure in the
  Appendix\ 3 of the SKA Project Execution
Plan (MGT-001.005.005-MP-001).
\end{itemize}

\noi The images of the individual science articles are provided by the
authors themselves and, if necessary, the image credits are given in
the individual figure captions.

\end{document}